\documentclass[12pt]{iopart}
\usepackage{iopams}
\usepackage{graphicx}

\newcommand{\intr}{\mathrm{intr}}
\newcommand{\ud}{\mathrm{ud}}
\newcommand{\FOW}{\mathrm{FOW}}
\newcommand{\lw}{\mathrm{lw}}
\newcommand{\tb}{\mathrm{tb}}
\newcommand{\nc}{\mathrm{nc}}
\newcommand{\grad}{\mathrm{grad}}
\newcommand{\acc}{\mathrm{acc}}
\newcommand{\dd}{\mathrm{d}}

\newcommand{\eq}[1]{(\ref{#1})}
\newcommand{\bun}{\hat{\mathbf{b}}}
\newcommand{\eun}{\hat{\mathbf{e}}}

\newcommand{\phiave}{\overline{\phi}}
\newcommand{\phiwig}{\widetilde{\phi}}

\newcommand{\boldr}{\mathbf{r}}

\newcommand{\bv}{\mathbf{v}}
\newcommand{\bw}{\mathbf{w}}
\newcommand{\bk}{\mathbf{k}}
\newcommand{\bR}{\mathbf{R}}

\newcommand{\bJ}{\mathbf{J}}
\newcommand{\bH}{\mathbf{H}}

\newcommand{\bB}{\mathbf{B}}
\newcommand{\bE}{\mathbf{E}}
\newcommand{\bV}{\mathbf{V}}

\newcommand{\matrixtop}[1]{\buildrel\leftrightarrow\over{#1}}
\newcommand{\matI}{\matrixtop{\mathbf{I}}}

\newcommand{\gsim}{ {\scriptstyle {{_{\displaystyle >}}\atop{\displaystyle \sim}}} }
\newcommand{\dotcross}{ \raise 0.65ex\hbox{${\scriptstyle {{_{\displaystyle \cdot}}\atop\times}}$} }
\newcommand{\crossdot}{ \raise 0.5ex\hbox{${\scriptstyle {{_\times}\atop{\displaystyle \cdot}}}$} }
\newcommand{\rhobf}{\mbox{\boldmath$\rho$}}

\newcommand{\kappabf}{\mbox{\boldmath$\kappa$}}
\newcommand{\zetabf}{\mbox{\boldmath$\zeta$}}

\newcommand{\zun}{\hat{\zetabf}}

\newcommand{\sumsig}{ \raise -1.3ex\hbox{${{\displaystyle \sum}\atop{\scriptstyle \sigma}}$} }
\newcounter{appnumb}

\begin{document}
\title{Intrinsic rotation in tokamaks: theory}
\author{Felix I Parra and Michael Barnes}
\address{Rudolf Peierls Centre for Theoretical Physics, University of Oxford, Oxford, UK}
\address{Culham Centre for Fusion Energy, Abingdon, UK}
\address{Plasma Science and Fusion Center, Massachusetts Institute of Technology, Cambridge, MA, USA}
\eads{\mailto{f.parradiaz1@physics.ox.ac.uk}}

\begin{abstract}
Self-consistent equations for intrinsic rotation in tokamaks with small poloidal magnetic field $B_p$ compared to the total magnetic field $B$ are derived. The model gives the momentum redistribution due to turbulence, collisional transport and energy injection. Intrinsic rotation is determined by the balance between the momentum redistribution and the turbulent diffusion and convection. Two different turbulence regimes are considered: turbulence with characteristic perpendicular lengths of the order of the ion gyroradius, $\rho_i$, and turbulence with characteristic lengths of the order of the poloidal gyroradius, $(B/B_p) \rho_i$. Intrinsic rotation driven by gyroradius scale turbulence is mainly due to the effect of neoclassical corrections and of finite orbit widths on turbulent momentum transport, whereas for the intrinsic rotation driven by poloidal gyroradius scale turbulence, the slow variation of turbulence characteristics in the radial and poloidal directions and the turbulent particle acceleration can be become as important as the neoclassical and finite orbit width effects. The magnetic drift is shown to be indispensable for the intrinsic rotation driven by the slow variation of turbulence characteristics and the turbulent particle acceleration. The equations are written in a form conducive to implementation in a flux tube code, and the effect of the radial variation of the turbulence is included in a novel way that does not require a global gyrokinetic formalism.
\end{abstract}

\submitto{\PPCF}
\maketitle

\section{Introduction} \label{sec:intro}

Rotation quenches large-scale MHD instabilities \cite{deVries96}, and a moderate rotation shear can reduce turbulence levels \cite{barnes11a, highcock10, parra11b}. Tokamak plasmas can rotate freely around their axis of symmetry \cite{hinton85, catto87}. They can reach velocities of the order of the sound speed when large amounts of momentum are injected into them, but momentum injection is not the only way to obtain rotation in tokamaks. It has been observed that tokamak plasmas rotate even in the absence of any obvious external sources of momentum. This natural rotation is known as intrinsic rotation. The origin of intrinsic rotation is a redistribution of momentum within the tokamak.

While intrinsic rotation is routinely observed in tokamaks, its characteristics are far from universal. The experimental observations of intrinsic rotation show complex dependences on magnetic geometry, boundary conditions, and heating sources \cite{rice99, rice05, bortolon06, scarabosio06, degrassie07, duval07, rice07, eriksson09, incecushman09, lin09, camenen10, solomon10, mcdermott11, rice11a, rice11b, parra12a}. The direction of the rotation depends on different parameters such as the density and the plasma current \cite{bortolon06, rice11b}, and it is not the same across the plasma. There is ample evidence that gradients in density and temperature play an important role in deciding the rotation magnitude \cite{scarabosio06, rice07, solomon10, rice11a, parra12a}, but they do not seem to be the only causes since different RF heating and current drive mechanisms cause different changes in the intrinsic rotation profile \cite{rice99, degrassie07, eriksson09, incecushman09, lin09, mcdermott11}. Magnetic geometry and boundary conditions affect both the direction and the magnitude of the rotation \cite{rice05, duval07, camenen10}. Any comprehensive modeling effort needs to consider the effect of the gradients, the heating and the magnetic geometry, and must allow them to compete to obtain the variety of intrinsic rotation profiles observed in experiments.

Theory and simulation efforts attempting to explain intrinsic rotation have proliferated in recent years. There have been several effects that have been proposed as causes of intrinsic rotation: RF heating and current drive \cite{perkins01, lee12}, up-down asymmetry of the magnetic flux surfaces \cite{camenen09b, camenen09c, ball14}, the radial variation of the gradients of density and temperature \cite{diamond08, gurcan10, waltz11, camenen11}, the neoclassical flows of particles and heat parallel to the flux surface \cite{parra10a, parra11d, barnes13, lee14a, lee14b, lee14c}, finite orbit width effects \cite{parra10a, parra11d}, and the poloidal variation of the turbulence characteristics \cite{sung13}. The main problem with modeling intrinsic rotation is the need to use reduced kinetic equations that are accurate to an order higher in the small parameter $\rho_\ast = \rho_i/a \ll 1$ than is usual in the literature \cite{parra10b, parra12b}. Here $\rho_i$ is the characteristic ion gyroradius, and $a$ is the minor radius of the tokamak. The lowest order reduced kinetic equations that describe the turbulent fluctuations, the gyrokinetic equations \cite{catto78, frieman82}, satisfy a symmetry in up-down symmetric tokamaks that precludes momentum redistribution. This symmetry was first discovered for the linear equations \cite{peeters05}, and later extended to the nonlinear equations \cite{parra11c, sugama11a}. We need the $O(\rho_\ast)$ corrections to the fluctuations and their relative phases to break the symmetry and find net momentum flux from one flux surface to the next. The gyrokinetic equations correct to next order in $\rho_\ast$ are cumbersome, having a large number of terms of different origin in them. As a result, previous work on intrinsic rotation has focused on individual effects. In some cases, the models were not derived from first principles and just gave a simple physical interpretation of one of the effects \cite{diamond08, gurcan10}; in other cases, the equations were derived from first principles, but the results were obtained using a quasilinear approach that cannot give the turbulence amplitude saturation \cite{camenen09b, camenen09c, camenen11, sung13}; and, finally, there are nonlinear simulations that do not have all the effects included \cite{waltz11, barnes13, lee14a, lee14b, lee14c}. 

This article and its companion \cite{barnes13b} present a complete, self-consistent treatment of intrinsic rotation in the core. We have simplified the equations using an expansion in $B_p/B \ll 1$ similar to the one proposed in \cite{parra10a, parra11d}. Here $B$ is the magnitude of the magnetic field and $B_p$ is the poloidal magnetic field; in most tokamaks $B_p/B \sim r/qR \sim 0.1$, where $q$ is the safety factor, $R$ is the major radius, and $r$ is the minor radius of the flux surface of interest. The expansion in $B_p/B \ll 1$ can be relaxed to include tokamaks with $B_p/B \sim 1$, i.e., spherical tokamaks and regular tokamaks with strong shaping that have large poloidal magnetic fields in regions of the plasma. The equations for tokamaks with $B_p/B \sim 1$ are in \cite{parra11a, calvo12, calvo14}. The differences between the equations presented in this article and the results in \cite{parra10a, parra11d}, which were the first attempt to find a self-consistent formulation for $B_p/B \ll 1$, are five: first, we consider different ion species; second, we include the effect of ion-electron collisions that lead to a rotation drive proportional to the difference between the electron and ion temperatures; third, we include the effect of the injection of energy and momentum; fourth, the new equations are derived for a turbulent eddy size that can range from the ion gyroradius $\rho_i$ to the ion poloidal gyroradius $(B/B_p) \rho_i$ \cite{barnes11b}, whereas the model in \cite{parra10a, parra11d} only considered eddies of the order of the ion gyroradius; and fifth, we present a new treatment of the corrections to the spatial derivatives due to the slow radial and poloidal variation of the turbulence characteristics. Our treatment of the slow derivatives is useful because it makes the model implementable in flux tube gyrokinetic codes \cite{dorland00, candy03, dannert05, peeters09c}, which are the least computationally costly tokamak turbulence codes. The implementation of these equations in a flux tube code is described in the companion paper \cite{barnes13b}. 

In addition to computational efficiency, the equations derived in this article give a natural classification for the different intrinsic rotation drives. Using the new equations, we calculate how the rotation driven by each mechanism scales with different parameters (turbulent eddy size, $B/B_p$, $\rho_\ast$...). Importantly, to determine these scalings, one needs a new symmetry of the lowest order gyrokinetic equations different from the one given in \cite{parra11c, sugama11a} and valid for turbulent eddies smaller than the ion poloidal gyroradius. Finally, we also give physical pictures to explain the scalings that we have obtained. These physical pictures can be used to guide the analysis of experimental data.

The remainder of the article is organized as follows. In section \ref{sec:momflux} we give the equation for plasma rotation, which depends on the momentum flux across flux surfaces. Most of the momentum flux will be due to turbulence. In section \ref{sec:ordering} we discuss the separation of time and length scales between turbulence and background density, temperature and electric field profiles. We present a formalism for momentum flux that includes in a natural way this separation of scales in sections \ref{sec:radialflux} and \ref{sec:gyrokinetics}, where we give a simple form of the momentum flux and the equations for the distribution function and the potential. The symmetry of the lowest order gyrokinetic equations given in \cite{parra11c} plays an important role in the derivation because it is the reason for needing next order corrections in the gyrokinetic equation. Thus, we review this symmetry in section \ref{sec:symmetry}. In the rest of  the article we expand in $B_p/B \ll 1$ to simplify the equations. To do so, we need the scaling of the turbulence amplitude and characteristic length scales with $B_p/B$. In section \ref{sec:expansion} we describe the two turbulence regimes that we consider: a regime in which the turbulent eddies are of the order of the ion poloidal gyroradius, studied in \cite{barnes11b}, and a regime in which the turbulence characteristic lengths are of the order of the ion gyroradius, considered in \cite{parra10a, parra11d} and justified in \cite{yoo15}. In section~\ref{sec:expansion} we also give the new symmetry of the lowest order gyrokinetic equations valid for turbulence with eddies much smaller than the ion poloidal gyroradius. In section \ref{sec:secondordertotal}, we find the equations for intrinsic rotation valid in the two turbulence regimes described in section~\ref{sec:expansion}. Finally, in section \ref{sec:pictures} we identify the possible intrinsic rotation drives in the limit $B_p/B \ll 1$, and we provide physical pictures for them. We also give the scaling of the intrinsic rotation driven by these mechanisms with different important parameters. We conclude with a summary of the results in section~\ref{sec:conclusion}.

The main results of this article are the equations given in section \ref{sec:secondordertotal}, and the discussion of the different effects in section \ref{sec:pictures}. A reader interested in the final equations and not the details of their derivation may simply read sections \ref{sec:secondordertotal} and \ref{sec:pictures}.

\section{Rotation in a tokamak} \label{sec:momflux}

In this section we discuss how to calculate the rotation in a tokamak. We first present the basic equations for the plasma in subsection \ref{sub:basicequations}. We only treat low $\beta$ plasmas in which magnetic field fluctuations are negligible. After presenting the basic equations, we give the toroidal angular momentum equation in subsection \ref{sub:momequation}, and we discuss the accuracy needed to calculate the radial flux of toroidal angular momentum.

\subsection{Basic equations} \label{sub:basicequations}

The distribution function for species $s$, $f_s (\boldr, \bv, t)$, is determined by the Fokker-Planck equation,
\begin{equation} \label{eq:FPequation}
\fl \frac{\partial f_s}{\partial t} + \bv \cdot \nabla f_s + \frac{Z_s e}{m_s} \left ( \bE + \frac{1}{c} \bv \times \bB  \right ) \cdot \nabla_v f_s = \sum_{s^\prime} C_{ss^\prime} [ f_s, f_{s^\prime} ] + Q_s.
\end{equation}
Here $\bE$ and $\bB$ are the electric and magnetic fields, $m_s$ and $Z_s e$ are the mass and charge of species $s$, $c$ is the speed of light, $e$ is the charge of the proton, $Q_s$ are sources and sinks of particles, momentum and energy representing the effect of the transformer electric field, neutral beams, pellet injection and RF heating and current drive, and $C_{ss^\prime} [f_s, f_{s^\prime}]$ is the Fokker-Planck collision operator for collisions between species $s$ and $s^\prime$,
\begin{equation} \label{eq:FPoperator}
\fl C_{ss^\prime} [ f_s, f_{s^\prime} ] = \frac{\gamma_{ss^\prime}}{m_s} \nabla_v \cdot \left [ \int \dd^3 v^\prime\, \nabla_w \nabla_w w \cdot \left ( \frac{f_{s^\prime} (\bv^\prime)}{m_s} \nabla_v f_s (\bv) - \frac{f_s (\bv)}{m_{s^\prime}} \nabla_{v^\prime} f_{s^\prime} (\bv^\prime) \right ) \right ].
\end{equation}
Here $\gamma_{ss^\prime} = 2\pi Z_s^2 Z_{s^\prime}^2 e^4 \ln \Lambda$, $\ln \Lambda$ is Couloumb logarithm, $\bw = \bv - \bv^\prime$, $w = |\bw|$, $\nabla_w \nabla_w w = (w^2 \matI - \bw \bw)/w^3$ and $\matI$ is the unit matrix.

We assume an axisymmetric magnetic field,
\begin{equation} \label{eq:Bdef}
\bB = I \nabla \zeta + \nabla \zeta \times \nabla \psi,
\end{equation}
where $\psi (\boldr)$ is the poloidal magnetic flux, $\zeta(\boldr)$ is the toroidal angle, $\nabla \zeta = \zun/R$, $\zun$ is the unit vector in the toroidal direction, $R$ is the major radius, $I (\psi) = RB_\zeta$ is a flux function to lowest order, and $B_\zeta$ is the toroidal component of the magnetic field. The poloidal magnetic flux $\psi$ is determined by the Grad-Shafranov equation, and the time evolution of the function $I$ is controlled by the current diffusion. The form for the magnetic field in \eq{eq:Bdef} is appropriate for very low $\beta$ plasmas in which the turbulence is basically electrostatic and the magnetic field is axisymmetric to a very high order. Based on this assumption, we find that the electric field is
\begin{equation} \label{eq:Edef}
\bE = - \nabla \phi,
\end{equation}
where $\phi$ is the electrostatic potential. The quasineutrality equation determines the electrostatic potential,
\begin{equation} \label{eq:QNequation}
\sum_s Z_s \int \dd^3 v\, f_s = 0.
\end{equation} 
To lowest order in the expansion parameter $\rho_\ast \ll 1$, the quasineutrality equation does not determine the long wavelength, flux surface averaged piece of the potential \cite{parra08, parra09b, calvo12}; this piece of the potential can only be calculated in up-down symmetric tokamaks using the quasineutrality equation if the expansion in $\rho_\ast$ is performed to fourth order \cite{parra09b, parra10b}. Thus, from here on it would be useful to think of the quasineutrality equation as only determining the potential up to a long wavelength, flux surface averaged piece that must be obtained from the conservation of toroidal angular momentum. Proceeding in this manner, the expansion in $\rho_\ast$ only needs to be performed to second order, as we show in section~\ref{sec:radialflux}.

Note that in equation \eq{eq:QNequation}, we have neglected the term $\nabla^2 \phi/4 \pi e$ because we assume that the Debye length $\lambda_D = \sqrt{T_e/4 \pi e^2 n_e}$ is smaller than the characteristic scale length of the problem. Here $T_e$ is the electron temperature and $n_e$ is the electron density. In our case, the Debye length has to be compared to the ion gyroradius $\rho_i$, i.e., 
\begin{equation}
\frac{\lambda_D}{\rho_i} \ll 1. 
\end{equation}
In the following sections we will keep in the equations terms that are small, but among these small terms we will never consider the term $\nabla^2 \phi/4 \pi e$. The reason is that this term does not break the symmetry described in section \ref{sec:symmetry} and as a result, it does not lead to momentum redistribution.

The Fokker-Planck equation \eq{eq:FPequation} for each species and the quasineutrality equation \eq{eq:QNequation} are in principle all we need to describe the plasma, but it will be convenient to take moments of the Fokker-Planck equation to find transport equations for particles, energy and momentum. These equations are of the form
\begin{equation} \label{eq:transportequation}
\frac{\partial h}{\partial t} + \nabla \cdot \bH = S_h,
\end{equation}
where $h(\boldr, t)$, $\bH (\boldr, t)$ and $S_h (\boldr, t)$ are functions of space and time. In these equations and in the Fokker-Planck equation, we will use flux coordinates that follow the surfaces parallel to the magnetic field. We choose the flux coordinates $(\psi, \theta, \zeta)$, where $\theta(\boldr)$ is a poloidal angle. The determinant of the Jacobian of the transformation $\boldr(\psi, \theta, \zeta)$ is
\begin{equation} \label{eq:jacobian}
\mathcal{J} (\psi, \theta) = \frac{\partial \boldr}{\partial \psi} \cdot \left ( \frac{\partial \boldr}{\partial \theta} \times \frac{\partial \boldr}{\partial \zeta} \right ) = \frac{1}{\nabla \psi \cdot (\nabla \theta \times \nabla \zeta)} = \frac{1}{\bB \cdot \nabla \theta}.
\end{equation}
Transport equations such as \eq{eq:transportequation} will become
\begin{eqnarray} \label{eq:transpfluxvariables}
\fl \left. \frac{\partial h}{\partial t} \right |_{\psi, \theta, \zeta}  + \frac{1}{\mathcal{J}} \frac{\partial}{\partial \psi} \left ( \mathcal{J} \bH \cdot \nabla \psi  \right ) + \frac{1}{\mathcal{J}} \frac{\partial}{\partial \theta} \left ( \mathcal{J} \bH \cdot \nabla \theta \right ) + \frac{1}{\mathcal{J}} \frac{\partial}{\partial \zeta} \left ( \mathcal{J} \bH \cdot \nabla \zeta \right )= S_h
\end{eqnarray}
in the new variables. On several occasions we will use the flux surface average, defined as
\begin{equation}
\langle \ldots \rangle_\psi = \frac{1}{V^\prime} \int \dd \theta\, \dd \zeta \, \mathcal{J}( \ldots ),
\end{equation}
where
\begin{equation}
V^\prime = \int \dd \theta\, \dd \zeta \, \mathcal{J}
\end{equation}
is the derivative with respect to $\psi$ of the volume $V(\psi)$ bounded by the flux surface $\psi$. The flux surface average applied to \eq{eq:transpfluxvariables} gives
\begin{equation} \label{eq:transpfluxaverage}
\frac{\partial \langle h \rangle_\psi}{\partial t} + \frac{1}{V^\prime} \frac{\partial}{\partial \psi} \left ( V^\prime \left \langle \bH \cdot \nabla \psi \right \rangle_\psi \right ) = \langle S_h \rangle_\psi.
\end{equation} 

In section \ref{sec:gyrokinetics} equations \eq{eq:FPequation}, \eq{eq:FPoperator} and \eq{eq:QNequation} will be expanded order by order in the small parameter $\rho_\ast \ll 1$. Before doing so, it is useful to write the equation that determines the rotation in a tokamak because the size of the different terms in this equation will guide the decisions made in the rest of the paper.

\subsection{Conservation of toroidal angular momentum} \label{sub:momequation}

We will see in subsection \ref{sub:lwfirstorder} that the poloidal rotation is damped and hence only the toroidal rotation is of interest. For this reason, we focus on the conservation of toroidal angular momentum. Multiplying equation \eq{eq:FPequation} by $R m_s \bv \cdot \zun$, integrating in velocity space, flux surface averaging (see \eq{eq:transpfluxaverage}), and summing over species, we find
\begin{eqnarray} \label{eq:torangmomv1}
\fl \frac{\partial}{\partial t} \left \langle \sum_s R n_s m_s \bV_s \cdot \zun \right \rangle_\psi + \frac{1}{V^\prime}\frac{\partial}{\partial \psi} \left ( V^\prime \left \langle \sum_s R m_s \int \dd^3v\, f_s (\bv \cdot \zun) (\bv \cdot \nabla \psi) \right \rangle_\psi \right ) \nonumber \\  = \frac{1}{c} \langle \bJ \cdot \nabla \psi \rangle_\psi + \left \langle \sum_s R m_s \int \dd^3v\, Q_s \bv \cdot \zun \right \rangle_\psi,
\end{eqnarray}
where $n_s \bV_s = \int \dd^3v\, f_s \bv$ is the flow of species $s$, and $\bJ = \sum_s Z_s e n_s \bV_s$ is the current density. To obtain \eq{eq:torangmomv1} we have used that $\nabla (R\zun) = \nabla R \zun - \zun \nabla R$ is an antisymmetric tensor. From quasineutrality we see that $\langle \bJ \cdot \nabla \psi \rangle_\psi = 0$. In addition, we average over the time and length scales of the turbulence using the coarse grain average
\begin{equation} \label{eq:coarsegrain}
\langle \ldots \rangle_\mathrm{T} = \frac{1}{\Delta t \Delta \psi \Delta \theta \Delta \zeta} \int_{\Delta t} \dd t \int_{\Delta \psi} \dd \psi \int_{\Delta \theta} \dd \theta \int_{\Delta \zeta} \dd \zeta ( \ldots ),
\end{equation}
where $\Delta t \ll \tau_E$, $\Delta \psi \ll a R B_p$, $\Delta \theta \ll 2\pi$ and $\Delta \zeta \ll 2\pi$ are the time and length scales of the turbulence that are much smaller than the energy transport time $\tau_E$ and the size of the device $a$ \cite{mckee01}. Then, the conservation of toroidal angular momentum becomes
\begin{equation} \label{eq:torangmom}
\frac{\partial}{\partial t} \left \langle \left \langle \sum_s R n_s m_s \bV_s \cdot \zun \right \rangle_\psi \right \rangle_\mathrm{T}+ \frac{1}{V^\prime} \frac{\partial}{\partial \psi} ( V^\prime \Pi )  = T_\zeta,
\end{equation}
where
\begin{equation} \label{eq:torquedefinition}
T_\zeta = \left \langle \left \langle \sum_s R m_s \int \dd^3 v\, Q_s \bv \cdot \zun \right \rangle_\psi \right \rangle_\mathrm{T}
\end{equation}
is the toroidal torque due to neutral beams, RF and other external sources, and
\begin{equation} \label{eq:Pidefinition}
\Pi = \left \langle \left \langle \sum_s R m_s \int \dd^3v\, f_s (\bv \cdot \zun) \left ( \bv \cdot \nabla \psi \right ) \right \rangle_\psi \right \rangle_\mathrm{T}
\end{equation}
is the radial flux of toroidal angular momentum.

Our objective is to calculate $\Pi$ to the accuracy necessary to solve equation \eq{eq:torangmom}. We will see in section \ref{sec:pictures} that in an up-down symmetric tokamak without torque ($T_\zeta = 0$) the ion velocity is of order 
\begin{equation} \label{eq:Vestimate}
V_i \sim \frac{B}{B_p} \rho_\ast v_{ti},
\end{equation}
where $v_{ti} = \sqrt{2T_i/m_i}$ is the ion thermal speed, $T_i$ is the ion temperature, and $m_i$ is the ion mass. The characteristic time scale of variation of the background rotation is the energy transport time $\tau_E$, which we take to be bounded by
\begin{equation} \label{eq:tauEestimate}
\rho_\ast^2 \frac{v_{ti}}{a} \lesssim \frac{1}{\tau_E} \lesssim \frac{B}{B_p} \rho_\ast^2 \frac{v_{ti}}{a}.
\end{equation}
The two bounds on $\tau_E$ are considered in detail in section \ref{sec:secondordertotal} (see equation \eq{eq:tauEtbtotal}). 

In the first sections of this article, we consider $\rho_\ast \ll B_p/B \sim 1$ so that we first expand in $\rho_\ast$ (section \ref{sec:gyrokinetics}) and only later perform a subsidiary expansion in $B_p/B$ (sections \ref{sec:expansion} and \ref{sec:secondordertotal}). If we impose that all the terms in \eq{eq:torangmom} are of the same order in $\rho_\ast$, we find for $B_p \sim B$
\begin{equation}
\frac{\partial}{\partial t}  \left  \langle \left \langle \sum_s R n_s m_s \bV_s \cdot \zun \right \rangle_\psi \right \rangle_\mathrm{T} \sim \frac{1}{\tau_E} R n_e m_i \rho_\ast v_{ti} \sim \rho_\ast^3 \frac{R p}{a},
\end{equation}
\begin{equation} \label{eq:torqueestimate}
T_\zeta \sim \rho_\ast^3 \frac{R p}{a}
\end{equation}
and
\begin{equation} \label{eq:Piestimate}
\Pi \sim \rho_\ast^3 p R |\nabla \psi|,
\end{equation}
where $p \sim n_e T_e$ is the plasma pressure, and we have assumed that all the plasma species have comparable temperatures, i.e., $T_s \sim T_{s^\prime}$ for any $s$, $s^\prime$. According to \eq{eq:Piestimate} we need to go to third order in $\rho_\ast \ll 1$ relative to $p R |\nabla \psi|$ to calculate the intrinsic rotation profile. The rest of this article gives the equations to calculate $\Pi$ to this order, with the help of the subsidiary expansion $B_p/B \ll 1$. We order $T_\zeta$ as in \eq{eq:torqueestimate} so that the effect of intrinsic rotation is not lost. If the torque is larger than the estimate given in \eq{eq:torqueestimate}, it overwhelms the mechanisms that drive intrinsic rotation. The equations in this article permit torques larger than the estimate in \eq{eq:torqueestimate} as long as $V_i \ll v_{ti}$ because we have assumed that the velocity is subsonic, and as a consequence we are missing the effect of the centrifugal force \cite{casson10}. If the torque dominates, the estimate for the ion velocity is
\begin{equation}
V_i \sim \frac{T_\zeta \tau_E}{R n m} \sim \frac{T_\zeta a}{\rho_\ast^2 R n m_i v_{ti}} \gsim \rho_\ast v_{ti}.
\end{equation}

We proceed to derive a useful expression that gives the radial flux of toroidal angular momentum $\Pi$ with the least possible work (see section \ref{sec:radialflux}). We also write the equations that are needed to calculate $f_s$ to the correct order (see section \ref{sec:gyrokinetics}). In fact, in sections~\ref{sec:radialflux} and \ref{sec:symmetry} we will show that in an up-down symmetric tokamak the moment approach used here only requires $f$ through second order in $\rho_\ast$, while a direct solution of the gyrokinetic and quasineutrality equation requires $f$ through fourth order in \cite{parra09b, parra10b}.

\section{Orderings} \label{sec:ordering}

We see from \eq{eq:Piestimate} that we need to calculate the radial flux of toroidal angular momentum to third order in $\rho_\ast$. To do so, we obtain an expression for the radial flux of toroidal angular momentum $\Pi$ in section \ref{sec:radialflux} that is more convenient than \eq{eq:Pidefinition}. This equation for $\Pi$ depends on certain pieces of the distribution functions $f_s$ and the potential $\phi$ that are determined by the equations given in section \ref{sec:gyrokinetics}. To simplify $\Pi$ and obtain the equations for the distribution function and the potential, we need to order the different time and length scales and the different pieces of the distribution function and the potential. We assume $B_p/B \sim 1$ initially. We impose $B_p/B \ll 1$ as a subsidiary expansion later, in sections \ref{sec:expansion} and \ref{sec:secondordertotal}.

We are interested in the tokamak core, where
\begin{equation} \label{eq:collisionmax}
\nu_{ii} \lesssim \frac{v_{ti}}{a}.
\end{equation}
We assume that there are many collisions in a transport time scale $\tau_E$; in particular,
\begin{equation} \label{eq:collisiontauE}
\nu_{ee} \sim \nu_{ei} \gg \nu_{ii} \sim \nu_{zi} \gg \frac{1}{\tau_E} \sim \rho_\ast^2 \frac{v_{ti}}{a}, 
\end{equation}
where $\nu_{ee} \sim n_e \gamma_{ee}/m_e^{1/2} T_e^{3/2}$ is the electron-electron collision frequency, $\nu_{ei} \sim n_e \gamma_{ei}/m_e^{1/2} T_e^{3/2}$ is the electron-ion collision frequency, $\nu_{ii} \sim n_e \gamma_{ii}/m_i^{1/2} T_i^{3/2}$ is the ion-ion collision frequency, and $\nu_{zi} \sim n_e \gamma_{zi}/m_z^{1/2} T_z^{3/2}$ is the impurity-ion collisional thermalization frequency. Here $m_e$ is the electron mass, and $m_z$ and $T_z$ are the impurity $z$ mass and temperature. The collision frequencies $\nu_{ee}$ and $\nu_{ei}$ are much larger than any other collision frequencies because $\sqrt{m_e/m_i} \ll 1$ and $\sqrt{m_e/m_z} \ll 1$. Assumption \eq{eq:collisiontauE} implies that the distribution functions of the ions, electrons and impurities are Maxwellians to lowest order. We will not consider $Z_z \gg 1$ or $\sqrt{m_i/m_z} \ll 1$, giving
\begin{equation} \label{eq:collisionI}
\nu_{zz} \sim \nu_{zz^\prime} \sim \nu_{iz} \lesssim \nu_{zi} \sim \nu_{ii}
\end{equation}
for any impurities $z$ and $z^\prime$. Here the frequency for collisions between impurity $z$ particles, $\nu_{zz} \sim n_z \gamma_{zz}/m_z^{1/2} T_z^{3/2}$, the frequency for collisions between impurities $z$ and $z^\prime$, $\nu_{zz^\prime} \sim n_{z^\prime} \gamma_{zz^\prime}/m_z^{1/2} T_z^{3/2}$, and the frequency for ion-impurity $z$ collisions, $\nu_{iz} \sim n_z\gamma_{iz}/m_i^{1/2} T_i^{3/2}$, are smaller than $\nu_{ii}$ when the impurity density $n_z$ is smaller than the ion density $n_i \sim n_e$. Since we assume $\sqrt{m_i/m_z} \sim 1$, the impurity temperatures and the ion temperature are the same, i.e., $T_z = T_i$ for every $z$ \footnote{In \cite{barnes12} the assumption $T_i = T_z$ is relaxed for partially ionized, heavy impurities.}. Combining \eq{eq:collisionmax} -\eq{eq:collisionI}, we obtain
\begin{equation} \label{eq:collisionorder1}
\rho_\ast^2 \ll \frac{\nu_{ii} a}{v_{ti}} \sim \frac{\nu_{zi} a}{v_{tz}} \sim \frac{\nu_{ee} a}{v_{te}} \sim \frac{\nu_{ei} a}{v_{te}} \lesssim 1
\end{equation}
and
\begin{equation} \label{eq:collisionorder2}
\frac{\nu_{zz} a}{v_{tz}} \sim \frac{\nu_{zz^\prime} a}{v_{tz}} \sim \frac{\nu_{iz} a}{v_{ti}} \lesssim \frac{\nu_{ii} a}{v_{ti}}
\end{equation}
for any impurities $z$ and $z^\prime$. Here $v_{te} = \sqrt{2T_e/m_e}$ and $v_{tz} = \sqrt{2T_z/m_z}$ are the electron and impurity $z$ thermal speed. Assumptions \eq{eq:collisionorder1} and \eq{eq:collisionorder2} are modified to include factors $B_p/B \ll 1$ in section \ref{sec:secondordertotal}.

Even though $T_i = T_z$, we want an ordering that permits different ion and electron temperatures, $T_i \neq T_e$. For this reason, we order $\sqrt{m_e/m_i} \sim \sqrt{m_e/m_z}$ as small. In particular, we require that the collisional temperature equilibration time be of the same order as the transport time scale,
\begin{equation} \label{eq:tequilibrationorder}
\frac{m_e}{m_i} \nu_{ei} \sim \frac{1}{\tau_E} \sim \rho_\ast^2 \frac{v_{ti}}{a}.
\end{equation}
Then,
\begin{equation} \label{eq:massorder}
\sqrt{\frac{m_e}{m_i}} \sim \sqrt{\frac{m_e}{m_z}} \sim \frac{v_{ti}}{a \nu_{ii}} \rho_\ast^2 \ll 1.
\end{equation}
The consequences of this ordering are a simplification of the electron-ion and electron-impurity collision operator, and the neglect of the ion-electron and impurity-electron collisions in several places. The equations that we give in this article are correct as long as  $\sqrt{m_e/m_i} \sim \sqrt{m_e/m_z} \ll 1$. The detailed balance in \eq{eq:massorder} is only a way to make $T_e \neq T_i$ possible. If for example, $\sqrt{m_e/m_i} \gg (v_{ti}/a \nu_{ii}) \rho_\ast^2$, our equations will just give $T_e \simeq T_i$.

We order the sources $Q_s$ in \eq{eq:FPequation} to be consistent with the transport time scale $\tau_E$,
\begin{equation} \label{eq:Qordering}
Q_s \sim \frac{f_{Ms}}{\tau_E} \sim \rho_\ast^2 f_{Ms} \frac{v_{ti}}{a},
\end{equation}
where we have used that to lowest order $f_s \simeq f_{Ms}$ is a Maxwellian $f_{Ms}$, given below in \eq{eq:fMsdef}. Note that we must order the momentum input term in $Q_s$ as smaller by $\rho_\ast$ than the energy and particle source terms since according to \eq{eq:torqueestimate}
\begin{equation}
\int \dd^3v\, Q_s \bv \sim \rho_\ast^3 \frac{v_{ti}}{a} v_{ts}^4 f_{Ms} \ll Q_s v_{ts}^4.
\end{equation}

Before ordering the distribution functions and the electrostatic potential, we use the coarse grain average \eq{eq:coarsegrain} to split the distribution functions and the electrostatic potential into long wavelength and turbulent components. For example, the long wavelength potential is
\begin{equation} \label{eq:philwdef}
\phi^\lw = \langle \phi \rangle_\mathrm{T},
\end{equation}
and the turbulent piece is
\begin{equation} \label{eq:phitbdef}
\phi^\tb = \phi - \langle \phi \rangle_\mathrm{T} = \phi - \phi^\lw.
\end{equation}
A similar separation applies to the distribution functions, $f_s = f_s^\lw + f_s^\tb$, although in the case of $f_s$, the scale separation will be performed in guiding center space (see subsection \ref{sub:fsordering}). The turbulent pieces describe the turbulent fluctuations in density, temperature, velocity and electric field, characterized by being small compared to the background density, temperature, velocity and electric field profiles \cite{mckee01} that are contained in the long wavelength pieces.

It is useful to discuss the assumptions for the size of the gradients and time derivatives of the potential and the distribution functions separately because the distribution function is most conveniently written in gyrokinetic variables, and the assumptions must be expressed in terms of those variables. For this reason, we give the assumptions for the electrostatic potential in subsection \ref{sub:phiordering}, and we use them to present the gyrokinetic coordinates in subsection \ref{sub:gkvariables}. Then, with these variables, we will give our assumptions for the size of the gradients and the time derivatives of the distribution function in subsection \ref{sub:fsordering}. Our assumptions for the distribution functions are expressed in terms of gyrokinetic coordinates, but in subsection \ref{sub:fsordering} we also discuss the implications of our ordering in real space coordinates.

\subsection{Electrostatic potential} \label{sub:phiordering}
We write the potential as
\begin{equation} \label{eq:philwexpansion}
\fl \phi^\lw (\boldr, t) = \phi_0 (\psi(\boldr), t) + \phi_1^\lw(\psi(\boldr), \theta(\boldr), t) + \phi_2^\lw(\psi(\boldr), \theta(\boldr), t) + \ldots
\end{equation}
and
\begin{equation} \label{eq:phitbexpansion}
\fl \phi^\tb (\boldr, t) = \phi_1^\tb (\boldr, t) + \phi_2^\tb (\boldr, t) + \ldots,
\end{equation}
where
\begin{equation} \label{eq:phi0order}
\frac{e\phi_0}{T_e} \sim 1
\end{equation}
and
\begin{equation} \label{eq:phinorder}
\frac{e\phi_n^\lw}{T_e} \sim \frac{e\phi_n^\tb}{T_e} \sim \rho_\ast^n
\end{equation}
for $n \geq 1$. Note that the turbulent potential fluctuations are small compared to the background potential. The lowest order long wavelength potential is a flux function, $\phi_0 (\psi, t)$, due to quasineutrality. The long wavelength higher order corrections $\phi_1^\lw (\psi, \theta, t)$ and $\phi_2^\lw (\psi, \theta, t)$ are axisymmetric, but unlike $\phi_0(\psi, t)$, they depend on $\theta$. The toroidal velocity and $\phi_0 (\psi, t)$ are related to each other (see \eq{eq:Omegadef}), and as a result, $\phi_0 (\psi, t)$ is determined by the conservation of toroidal angular momentum \eq{eq:torangmom}, where the radial flux of toroidal angular momentum \eq{eq:Pifinal} is comprised of the contributions \eq{eq:Piminusudtbtotal} to \eq{eq:Pi0QFOWtotal}. The higher order components $\phi_1^\lw$, $\phi_1^\tb$, $\phi_2^\lw$ and $\phi_2^\tb$ are calculated as functions of $\phi_0$ and the background gradients of density and temperature.

The long wavelength components have characteristic length scales of the size of the machine and time scales of the order of the transport time scale $\tau_E$. Then,
\begin{equation} \label{eq:ddtphilworder}
\frac{\partial}{\partial t} \ln \phi^\lw_n \sim \rho_\ast^2 \frac{v_{ti}}{a}
\end{equation}
and
\begin{equation} \label{eq:gradphilworder}
\nabla \ln \phi^\lw_n \sim \frac{1}{a}.
\end{equation}

For the turbulent piece of the electrostatic potential, we need to distinguish between the two directions perpendicular to the magnetic field line and the direction parallel to it. The reason is that the size of the turbulent structures is of order $\rho_i$ in the directions perpendicular to the magnetic field, and of order $a \sim R$ in the parallel direction. We use the spatial coordinates $\psi$ and $\alpha$, which are defined such that $\bB = \nabla \alpha \times \nabla \psi$, as the coordinates for the two directions perpendicular to the magnetic field line. The coordinate $\psi$ determines the flux surface, and the coordinate $\alpha$ determines the magnetic field line within the flux surface,
\begin{equation} \label{eq:alphadef}
\alpha = \zeta - \int^\theta \dd \theta^\prime\, \frac{I (\psi)  \mathcal{J} (\psi, \theta^\prime)}{[R(\psi, \theta^\prime)]^2}.
\end{equation}
The Jacobian $\mathcal{J}$ is defined in \eq{eq:jacobian}. The third spatial coordinate that determines the position along the magnetic field line is the poloidal angle $\theta$. In the perpendicular direction, we assume statistical periodicity, i.e., the plasma properties at points that are one correlation length apart should be statistically the same, and as a result, we can use Fourier analysis, giving
\begin{eqnarray} \label{eq:phitbscales}
\phi^\tb_n (\boldr, t) = \sum_{k_\psi, k_\alpha} \underline{\phi}^\tb_n (k_\psi, k_\alpha, \psi (\boldr), \theta (\boldr), t) \exp ( i k_\psi \psi (\boldr) + i k_\alpha \alpha (\boldr) ).
\end{eqnarray}
From here on we use the underline $\underline{\;\;}$ to denote Fourier coefficients of any function with turbulent pieces. We assume that 
\begin{equation} \label{eq:kpsiorder}
k_\psi \sim \frac{1}{\rho_i R B_p} 
\end{equation}
and 
\begin{equation} \label{eq:kalphaorder}
k_\alpha \sim \frac{R}{\rho_i},
\end{equation}
giving characteristic perpendicular lengths of order $\rho_i$ for the turbulence. The turbulent fluctuations also have long characteristic length scales that show the dependence of the turbulent fluctuations on the slow background profiles of density and temperature. The slow length scales are captured by the dependence of $\underline{\phi}_n^\tb$ on $\psi$ and $\theta$, 
\begin{equation} \label{eq:dphidpsiorder}
\frac{\partial}{\partial \psi} \ln \underline{\phi}^\tb_n \sim \frac{1}{a R B_p}
\end{equation}
and 
\begin{equation} \label{eq:dphidthetaorder}
\frac{\partial}{\partial \theta} \ln \underline{\phi}^\tb_n \sim 1.
\end{equation}
The dependence on $\theta$ also gives the length of the turbulent eddies along magnetic field lines. The turbulent time scale is assumed to be
\begin{equation} \label{eq:ddtphitborder}
\frac{\partial}{\partial t} \ln \underline{\phi}^\tb_n \sim \frac{v_{ti}}{a}.
\end{equation}
We will use the slow derivatives in $\psi$ and $\theta$, ordered as in \eq{eq:dphidpsiorder} and \eq{eq:dphidthetaorder}, to capture the effect of the slow variation of the turbulence characteristics on intrinsic rotation.

With the form in \eq{eq:phitbscales}, the perpendicular component of the gradient of $\phi^\tb$ is
\begin{equation} \label{eq:gradbotphitborder}
\fl \nabla_\bot \phi^\tb_n = \sum_{k_\psi, k_\alpha} \left ( i \bk_\bot  \underline{\phi}^\tb_n + \nabla \psi \frac{\partial  \underline{\phi}^\tb_n}{\partial \psi} + \nabla_\bot \theta \frac{\partial \underline{\phi}^\tb_n}{\partial \theta} \right ) \exp ( i k_\psi \psi + i k_\alpha \alpha ) \sim \frac{\phi_n^\tb}{\rho_i},
\end{equation}
where the term proportional to 
\begin{equation} \label{eq:kbotdef}
\bk_\bot = k_\psi \nabla \psi + k_\alpha \nabla \alpha \sim \frac{1}{\rho_i}
\end{equation}
is the dominant term. The parallel gradient is
\begin{equation} \label{eq:gradparphitborder}
\bun \cdot \nabla \phi^\tb_n = \sum_{k_\psi, k_\alpha} \bun \cdot \nabla \theta \frac{\partial \underline{\phi}^\tb_n}{\partial \theta} \exp ( i k_\psi \psi + i k_\alpha \alpha ) \sim \frac{\phi_n^\tb}{a}.
\end{equation}

Note that the orderings \eq{eq:phi0order}, \eq{eq:phinorder}, \eq{eq:gradphilworder} and \eq{eq:gradbotphitborder} imply
\begin{equation} \label{eq:gradphiorder}
\nabla \phi^\lw \sim \nabla \phi^\tb \sim \frac{T_e}{ea}.
\end{equation}
This size for the electric field is consistent with \eq{eq:Vestimate} because it implies that the $\bE \times \bB$ drift is of order $\rho_\ast v_{ti}$.

\subsection{Gyrokinetic variables} \label{sub:gkvariables}
Before expanding the distribution functions $f_s$ in $\rho_\ast \ll 1$, it is convenient to write the distribution functions in gyrokinetic variables. Gyrokinetics \cite{catto78, frieman82} separates the fast gyromotion time scale $\Omega_s^{-1}$ from the slower turbulent time $a/v_{ts}$ by expanding in the small parameter $\rho_s/a \ll 1$ while still allowing wavelengths comparable to the gyroradius. Here $\Omega_s = Z_s e B/m_s c$, $v_{ts} = \sqrt{2T_s/m_s}$ and $\rho_s = v_{ts}/\Omega_s$ are the gyrofrequency, thermal speed and gyroradius of species $s$. Gyrokinetics defines new phase space variables order by order in the small parameter $\rho_s/a \ll 1$ such that the Fokker-Planck equation \eq{eq:FPequation} written in these new coordinates satisfies certain properties (see section \ref{sec:gyrokinetics} and in particular, equations \eq{eq:Rdotorder}-\eq{eq:dotmu}). We can obtain the gyrokinetic variables $\{ \bR, u, \mu, \varphi \}$ up to second order in $\rho_s/a \ll 1$ following the recursive procedure in \cite{parra08} that can be applied to an electrostatic electric field $\bE = - \nabla \phi$ that satisfies assumptions \eq{eq:phi0order} - \eq{eq:gradphilworder}, \eq{eq:ddtphitborder}, \eq{eq:gradbotphitborder} and \eq{eq:gradparphitborder}. In this article, due to the expansion in $B_p/B \ll 1$ performed in sections \ref{sec:expansion} and \ref{sec:secondordertotal}, it is sufficient to know that the expansion can be carried out to second order. The details of the second order calculation are not needed \footnote{The complete expansion to second order in $\rho_s/a \ll 1$ is given in \cite{parra11a} following a formalism slightly different from the one used here.}. The gyrokinetic variables are the guiding center position
\begin{equation} \label{eq:Rdef}
\bR = \bR_g + \bR_2 + \ldots,
\end{equation}
the guiding center parallel velocity 
\begin{equation} \label{eq:udef}
u = v_{||} + u_1 + u_2 + \ldots,
\end{equation}
the magnetic moment 
\begin{equation} \label{eq:mudef}
\mu = \mu_0 + \mu_1 + \mu_2 + \ldots
\end{equation}
and the gyrokinetic gyrophase
\begin{equation} \label{eq:varphidef}
\varphi = \varphi_0 + \varphi_1 + \varphi_2 + \ldots
\end{equation}
Here
\begin{equation} \label{eq:Rgdef}
\bR_g = \boldr + \frac{1}{\Omega_s} \bv \times \bun
\end{equation}
is the lowest order Catto transformation \cite{catto78}, $v_{||} = \bv \cdot \bun$ is the particle velocity parallel to the magnetic field, $\bun(\boldr) = \bB/B$ is the unit vector in the direction of the magnetic field,
\begin{equation} \label{eq:mu0def}
\mu_0 = \frac{v_\bot^2}{2B}
\end{equation}
is the lowest order magnetic moment, $\bv_\bot = \bv - v_{||}\bun$ is the particle velocity perpendicular to the magnetic field,
\begin{equation} \label{eq:varphi0def}
\varphi_0 = \arctan \left ( \frac{\bv \cdot \eun_2}{\bv \cdot \eun_1} \right )
\end{equation}
is the lowest order gyrophase, and $\eun_1 (\boldr)$ and $\eun_2(\boldr)$ are two unit vectors perpendicular to each other and perpendicular to $\bun(\boldr)$ that satisfy $\eun_1 \times \eun_2 = \bun$. The quantities $u_1 \sim (\rho_s/a) v_{ts}$, $\mu_1 \sim (\rho_s/a) v_{ts}^2/B$ and $\varphi_1 \sim \rho_s/a$ are the first order corrections to the gyrokinetic variables, and the quanties $\bR_2 \sim (\rho_s/a)^2 a$, $u_2 \sim (\rho_s/a)^2 v_{ts}$, $\mu_2 \sim (\rho_s/a)^2 v_{ts}^2/B$ and $\varphi_2 \sim (\rho_s/a)^2$ are the second order corrections. For most of this article, we only need to know that $u_1$, $\mu_1$, $\varphi_1$, $\bR_2$, $u_2$, $\mu_2$ and $\varphi_2$ exist and can be calculated recursively. When calculating the quasineutrality equation, we will need the first order corrections to the parallel velocity and the magnetic moment, given by \cite{parra08}
\begin{eqnarray} \label{eq:u1def}
\fl u_1 = \frac{v_{||}}{\Omega_s} \bun \cdot \nabla \bun \cdot (\bv \times \bun) + \frac{1}{4\Omega_s} [ \bv_\bot ( \bv \times \bun ) +
( \bv \times \bun ) \bv_\bot ]:\nabla \bun + \frac{v_{\bot}^2}{2\Omega_s} \bun \cdot \nabla \times \bun
\end{eqnarray}
and
\begin{eqnarray} \label{eq:mu1def}
\fl \mu_1 = \frac{Z_s e \phiwig}{m_s B} - \frac{v_\bot^2}{2B^2\Omega_s} (\bv \times \bun)\cdot \nabla B - \frac{v_{||}^2}{B\Omega_s} \bun \cdot \nabla \bun \cdot (\bv \times \bun) \nonumber\\ - \frac{v_{||}}{4 B \Omega_s} [ \bv_\bot ( \bv \times \bun ) +
( \bv \times \bun ) \bv_\bot ]:\nabla \bun - \frac{v_{||}
v_{\bot}^2}{2B\Omega_s} \bun \cdot \nabla \times \bun.
\end{eqnarray}
Our double-dot convention is $\mathbf{a} \mathbf{c} : \matrixtop{\mathbf{M}} = \mathbf{c} \cdot \matrixtop{\mathbf{M}} \cdot \mathbf{a}$. The function $\phiwig$ is \footnote{The definitions of $\phiave$ and $\phiwig$ here are the definitions given in  \cite{parra11a, brizard07}, and they are slightly different from the definitions in \cite{parra08}. The difference is discussed in detail in \cite{parra09a} and it is $O(\rho_\ast^2 T_e/e)$.}
\begin{equation} \label{eq:phiwigdef}
\phiwig (\bR, \mu, \varphi, t) = \phi(\bR + \rhobf (\bR, \mu, \varphi), t) - \phiave (\bR, \mu, t),
\end{equation}
where
\begin{equation} \label{eq:phiavedef}
\phiave (\bR, \mu, t) = \frac{1}{2\pi} \int_0^{2\pi} \dd\varphi\, \phi (\bR + \rhobf (\bR, \mu, \varphi), t)
\end{equation}
is the average of $\phi(\boldr, t)$ around a circular gyro-orbit, and 
\begin{equation} \label{eq:rhobfdef}
\rhobf (\bR, \mu, \varphi) = \frac{m_s c}{Z_s e} \sqrt{\frac{2 \mu}{B(\bR)}} ( - \sin \varphi \, \eun_1 (\bR) + \cos \varphi \, \eun_2 (\bR))
\end{equation}
is the gyroradius.

The functions $\phiave$ and $\phiwig$ can be rewritten in more convenient forms. We show how to obtain these different forms of $\phiave$ and $\phiwig$ in \ref{app:phiavephiwig}. In particular, we obtain
\begin{equation}
\phiave (\bR, \mu, t) = \phiave^\lw + \phiave^\tb,
\end{equation}
where
\begin{equation} \label{eq:phiaveRlw}
\phiave^\lw = \phi_0 (\bR, t) + \phiave_1^\lw + \phiave^\lw_2 + O \left ( \rho_\ast^3 \frac{T_e}{e} \right )
\end{equation}
is the long wavelength component of $\phiave$, and
\begin{equation} \label{eq:phiaveRtb}
\fl \phiave^\tb = \sum_{k_\psi, k_\alpha} ( \underline{\phiave}_1^\tb + \underline{\phiave}^\tb_2 ) \exp( i k_\psi \psi (\bR) + i k_\alpha \alpha (\bR))+ O \left ( \rho_\ast^3 \frac{T_e}{e} \right )
\end{equation}
is the turbulent piece. The first order pieces $\phiave^\lw_1$ and $\underline{\phiave}^\tb_1$ are
\begin{equation} \label{eq:phiaveRlw1}
\phiave^\lw_1 = \phi_1^\lw (\bR, t)
\end{equation}
and
\begin{equation} \label{eq:underlinephiaveRtb1}
\underline{\phiave}^\tb_1 = J_0 (\Lambda_s) \underline{\phi}_1^\tb (k_\psi, k_\alpha, \psi(\bR), \theta(\bR), t),
\end{equation}
where $J_n$ is the $n$-th order Bessel function of the first kind,
\begin{equation} \label{eq:Lambdasdef}
\Lambda_s (\bR, \mu) = \frac{k_\bot (\bR) m_s c}{Z_s e} \sqrt{\frac{2\mu}{B(\bR)}}
\end{equation}
and $k_\bot (\bR)$ is the function
\begin{equation}
k_\bot (\boldr) = |\bk_\bot (\boldr)| = ( k_\psi^2 |\nabla \psi|^2 + 2 k_\psi k_\alpha \nabla \psi \cdot \nabla \alpha + k_\alpha^2 |\nabla \alpha|^2 )^{1/2}
\end{equation}
with $\boldr$ replaced by $\bR$. The second order pieces $\phiave^\lw_2$ and $\underline{\phiave}^\tb_2$ are given in \ref{app:phiavephiwig}, in equations \eq{eq:phiaveRlw2} and \eq{eq:underlinephiaveRtb2}.

\subsection{Distribution functions} \label{sub:fsordering}
Using the gyrokinetic variables $\{ \bR, u, \mu, \varphi \}$, the distribution function $f_s = f_s^\lw + f_s^\tb$ can be written as
\begin{eqnarray} \label{eq:flwexpansion}
\fl f_s^\lw (\bR, u, \mu, \varphi, t) = f_{Ms}(\psi(\bR), \theta(\bR) u, \mu, t) + f_{s1}^\lw (\psi(\bR), \theta(\bR), u, \mu, t) \nonumber\\+ f_{s2}^\lw (\psi(\bR), \theta(\bR), u, \mu, \varphi, t) + \ldots
\end{eqnarray}
and
\begin{equation} \label{eq:ftbexpansion}
\fl f_s^\tb (\bR, u, \mu, \varphi, t) = f_{s1}^\tb (\bR, u, \mu, t) + f_{s2}^\tb (\bR, u, \mu, \varphi, t) + \ldots
\end{equation}
The distribution function to lowest order only has a long wavelength component and it is Maxwellian because of assumption \eq{eq:collisiontauE},
\begin{equation} \label{eq:fMsdef}
\fl f_{Ms} = n_s (\psi(\bR), t) \left ( \frac{m_s}{2\pi T_s (\psi(\bR), t)} \right )^{3/2} \exp \left ( - \frac{m_s (u^2 + 2\mu B(\psi(\bR), \theta(\bR)))}{2 T_s(\psi(\bR), t)} \right ).
\end{equation}
Note that the density $n_s (\psi, t)$ and the temperature $T_s (\psi, t)$ are flux functions because of assumption \eq{eq:collisionmax}. The Maxwellian is stationary because we are assuming that the velocity is subsonic (see \eq{eq:Vestimate}). The long wavelength corrections to the distribution function $f_{s1}^\lw (\psi, \theta, u, \mu, t)$ and $f_{s2}^\lw (\psi, \theta, u, \mu, \varphi, t)$ are axisymmetric. The size of the corrections to the Maxwellian for ionic species ($s \neq e$) is
\begin{equation} \label{eq:fnorder}
\frac{f_{sn}^\lw}{f_{Ms}} \sim \frac{f_{sn}^\tb}{f_{Ms}} \sim \rho_\ast^n.
\end{equation}
The electron distribution function is an exception to this rule as shown below. Note that in \eq{eq:flwexpansion} and \eq{eq:ftbexpansion}, only the second order pieces $f_{s2}^\lw$ and $f_{s2}^\tb$ depend on the gyrophase $\varphi$. We will prove that this is the case in section \ref{sec:gyrokinetics}, although in section \ref{sec:secondordertotal} we are able to show that this gyrophase dependence is negligible in the $B_p/B \ll 1$ expansion.

The electron distribution function does not follow equation \eq{eq:fnorder} because $\sqrt{m_e/m_i} \ll 1$. The pieces of the distribution function even in $u$ are of order
\begin{equation} \label{eq:fenevenorder}
\frac{f_{en}^\lw (u) + f_{en}^\lw (-u)}{f_{Me}} \sim \frac{f_{en}^\tb(u) + f_{en}^\tb(-u)}{f_{Me}} \sim \rho_\ast^n.
\end{equation}
Equation \eq{eq:fenevenorder} is deduced from quasineutrality: the higher order corrections to the electron density, 
\begin{equation}
\int \dd^3v\, f_{en}^\lw \sim \frac{f_{en}^\lw (u) + f_{en}^\lw (-u)}{f_{Me}} n_e
\end{equation}
and
\begin{equation}
\int \dd^3v\, f_{en}^\tb \sim \frac{f_{en}^\tb(u) + f_{en}^\tb(-u)}{f_{Me}} n_e,
\end{equation}
have to be of the same order as the higher order corrections to the ion density, $\int \dd^3v\, f_{in}^\lw \sim \rho_\ast^n n_e$ and $\int \dd^3v\, f_{in}^\tb \sim \rho_\ast^n n_e$. The pieces of the distribution function odd in $u$ are
\begin{equation} \label{eq:fenoddorder}
\frac{f_{en}^\lw (u) - f_{en}^\lw (-u)}{f_{Me}} \sim \frac{f_{en}^\tb(u) - f_{en}^\tb(-u)}{f_{Me}} \sim \sqrt{\frac{m_e}{m_i}} \rho_\ast^n \ll \rho_\ast^n.
\end{equation}
The reason for this estimate is that the higher order electron flows, 
\begin{equation}
\int \dd^3v\, f_{en}^\lw \bv \sim \frac{f_{en}^\lw (u) - f_{en}^\lw (-u)}{f_{Me}} n_e v_{te}
\end{equation}
and
\begin{equation}
\int \dd^3v\, f_{en}^\tb \bv \sim \frac{f_{en}^\tb(u) - f_{en}^\tb(-u)}{f_{Me}} n_e v_{te},
\end{equation}
must be of the same order as the higher order corrections to the ion flow, $\int \dd^3v\, f_{in}^\lw \bv \sim \rho_\ast^n n_e v_{ti}$ and $\int \dd^3v\, f_{in}^\tb \bv \sim \rho_\ast^n n_e v_{ti}$, again due to quasineutrality. 

We assume that the time and length scales of the distribution functions and the electrostatic potential are similar. The long wavelength piece of the distribution function satisfies
\begin{equation} \label{eq:ddtflworder}
\frac{\partial}{\partial t} \ln f^\lw_{sn} \sim \rho_\ast^2 \frac{v_{ti}}{a}
\end{equation}
and
\begin{equation} \label{eq:gradflworder}
\nabla_\bR \ln f^\lw_{sn} \sim \frac{1}{a}.
\end{equation}
We write the turbulent piece of the distribution function as
\begin{equation} \label{eq:ftbscales}
\fl f^\tb_{sn} (\bR, u, \mu, \varphi, t) = \sum_{k_\psi, k_\alpha} \underline{f}^\tb_{sn} (k_\psi, k_\alpha, \psi(\bR), \theta(\bR), u, \mu, \varphi, t) \exp ( i k_\psi \psi (\bR) + i k_\alpha \alpha (\bR) ),
\end{equation}
where $k_\psi$ and $k_\alpha$ are ordered as in \eq{eq:kpsiorder} and \eq{eq:kalphaorder}. The dependence of $\underline{f}_{sn}^\tb$ on $\psi$ and $\theta$ represents the long wavelength of the turbulence along the magnetic field and the slow radial and poloidal variation of the turbulence fluctuations due to the spatially varying background profiles and magnetic field. Then,
\begin{equation}
\frac{\partial}{\partial \psi} \ln \underline{f}_{sn}^\tb \sim \frac{1}{aRB_p}
\end{equation}
and
\begin{equation}
\frac{\partial}{\partial \theta} \ln \underline{f}_{sn}^\tb \sim 1.
\end{equation}
The time derivative is ordered as
\begin{equation} \label{eq:ddtftborder}
\frac{\partial}{\partial t} \ln \underline{f}^\tb_{sn} \sim \frac{v_{ti}}{a},
\end{equation}
Finally, we order the velocity space derivatives as
\begin{equation}
\frac{\partial}{\partial u} \ln f_{sn}^\lw \sim \frac{\partial}{\partial u} \ln \underline{f}_{sn}^\tb \sim \frac{1}{v_{ts}},
\end{equation} 
\begin{equation}
\frac{\partial}{\partial \mu} \ln f_{sn}^\lw \sim \frac{\partial}{\partial \mu} \ln \underline{f}_{sn}^\tb \sim \frac{B}{v_{ts}^2}
\end{equation} 
and for $n \geq 2$
\begin{equation}
\frac{\partial}{\partial \varphi} \ln f_{sn}^\lw \sim \frac{\partial}{\partial \varphi} \ln \underline{f}_{sn}^\tb \sim 1.
\end{equation} 

From \eq{eq:ftbscales}, we find that the perpendicular gradient of the $n$-th order distribution function is
\begin{eqnarray} \label{eq:gradbotftborder}
\fl \nabla_{\bR\bot} f^\tb_{sn} (\bR, u, \mu, \varphi, t) = \nonumber\\ \sum_{k_\psi, k_\alpha} \left ( i \bk_\bot \underline{f}^\tb_{sn} + \nabla_\bR \psi \frac{\partial \underline{f}_{sn}^\tb}{\partial \psi} + \nabla_{\bR\bot} \theta \frac{\partial \underline{f}_{sn}^\tb}{\partial \theta} \right ) \exp ( i k_\psi \psi + i k_\alpha \alpha ) \sim \frac{\underline{f}_{sn}^\tb}{\rho_i},
\end{eqnarray}
where $\bk_\bot$ is defined in \eq{eq:kbotdef}, and the term proportional to $\bk_\bot$ in \eq{eq:gradbotftborder} is the largest term. The parallel gradient is
\begin{equation} \label{eq:gradparftborder}
\fl \bun (\bR) \cdot \nabla_\bR f^\tb_{sn} (\bR, u, \mu, \varphi, t) = \sum_{k_\psi, k_\alpha} \bun(\bR) \cdot \nabla_{\bR} \theta \frac{\partial \underline{f}_{sn}^\tb}{\partial \theta} \exp ( i k_\psi \psi + i k_\alpha \alpha ) \sim \frac{\underline{f}_{sn}^\tb}{a}.
\end{equation}

Assumptions \eq{eq:fnorder}, \eq{eq:gradflworder} and \eq{eq:gradbotftborder} give
\begin{equation} \label{eq:gradforder}
\nabla_\bR f^\lw_s \sim \nabla_\bR f^\tb_s \sim \frac{f_{Ms}}{a}.
\end{equation}

Since the distribution functions are functions of the gyrokinetic variables, we need to write them as functions of $\{\boldr, \bv\}$ to be able to integrate over velocity space to find densities, flows and pressures. We change variables in \ref{app:frv}. The long wavelength component of $f_s$ written in $\{\boldr, \bv \}$ is
\begin{eqnarray} \label{eq:fsaveT}
\fl [ f_s ]^\lw = f_{Ms} (\boldr, v_{||}, \mu_0, t) + [ f_s ]^\lw_1 + [ f_s ]^\lw_2 +O(\rho_\ast^3 f_{Ms}),
\end{eqnarray}
and the turbulent piece is
\begin{eqnarray} \label{eq:fswigT}
\fl [ f_s ]^\tb = \sum_{k_\psi, k_\alpha} ( \underline{[ f_s ]}^\tb_1 + \underline{[f_s]}^\tb_2) \exp ( i k_\psi \psi (\boldr) + i k_\alpha \alpha (\boldr) ) +O(\rho_\ast^3 f_{Ms}).
\end{eqnarray}
The square brackets denote that $f_s$ is written in $\{ \boldr, \bv \}$ coordinates instead of $\{ \bR, u, \mu, \varphi \}$ coordinates. Note that $f_s^\lw \neq [f_s]^\lw$ and $f_s^\tb \neq [f_s]^\tb$. The first order pieces are
\begin{eqnarray} \label{eq:fsaveT1}
[ f_s ]^\lw_1 = f_{s1}^\lw(\boldr, v_{||}, \mu_0, t) + \Delta f_{s1}^\lw
\end{eqnarray}
and
\begin{equation} \label{eq:underlinefswigT1}
\fl \underline{[f_s]}^\tb_1 = \underline{f}_{s1}^\tb (k_\psi, k_\alpha, \psi(\boldr), \theta(\boldr), v_{||}, \mu_0, t) \exp \left ( \frac{i\bk_\bot \cdot ( \bv \times \bun )}{\Omega_s} \right ) + \Delta \underline{f}_{s1}^\tb,
\end{equation}
where using \eq{eq:u1def} and \eq{eq:mu1lw},
\begin{eqnarray} \label{eq:DeltaFs1lwdef}
\fl \Delta f_{s1}^\lw = \frac{1}{\Omega_s} (\bv \times \bun) \cdot \left [ \frac{\nabla n_s}{n_s} + \frac{Z_s e \nabla \phi_0}{T_s} + \left ( \frac{m_s v^2}{2 T_s} - \frac{3}{2} \right ) \frac{\nabla T_s}{T_s} \right ] f_{Ms} \sim \frac{\rho_s}{a} f_{Ms}
\end{eqnarray}
and
\begin{eqnarray} \label{eq:DeltaunderlineFs1tbdef}
\fl \Delta \underline{f}_{s1}^\tb =  - \frac{Z_s e f_{Ms}}{T_s} \left [ 1 - J_0 (\lambda_s) \exp \left ( \frac{i \bk_\bot \cdot ( \bv \times \bun)}{\Omega_s}\right ) \right ] \underline{\phi}_1^\tb (k_\psi, k_\alpha, \psi(\boldr), \theta(\boldr), t).
\end{eqnarray}
Here
\begin{equation} \label{eq:lambdasdef}
\lambda_s (\boldr, v_\bot) = \frac{k_\bot (\boldr) v_\bot}{\Omega_s (\boldr)}.
\end{equation}
The second order pieces of the distribution function written in $\{\boldr, \bv\}$ coordinates are
\begin{eqnarray} \label{eq:fsaveT2}
[ f_s ]^\lw_2 = f_{s2}^\lw (\boldr, v_{||}, \mu_0, \varphi_0, t) + \Delta f^\lw_{s2}
\end{eqnarray}
and
\begin{equation} \label{eq:underlinefswigT2}
\fl \underline{[f_s]}^\tb_2 = \underline{f}_{s2}^\tb (k_\psi, k_\alpha, \psi(\boldr), \theta(\boldr), v_{||}, \mu_0, \varphi_0, t) \exp \left ( \frac{i\bk_\bot \cdot ( \bv \times \bun )}{\Omega_s} \right ) + \Delta \underline{f}_{s2}^\tb,
\end{equation}
where $\Delta f_{s2}^\lw$ and $\Delta \underline{f}_{s2}^\tb$ are given in \ref{app:frv}, in equations \eq{eq:DeltaFs2lwdef} and \eq{eq:DeltaunderlineFs2tbdef}. When we expand in $B_p/B \ll 1$ in section \ref{sec:secondordertotal}, we will see that the complicated functions $\Delta f_{s2}^\lw$ and $\Delta \underline{f}_{s2}^\tb$ simplify considerably. 

\section{Radial flux of toroidal angular momentum} \label{sec:radialflux}

To find the radial flux of toroidal angular momentum $\Pi$, we first manipulate expression \eq{eq:Pidefinition} using the Fokker-Planck equations \eq{eq:FPequation}. We follow the same procedure as in \cite{parra10a, parra11d}. The equation for $\Pi$ given here is different from the expression in \cite{parra10a, parra11d} because in this derivation we consider multiple ion species, interspecies collisions and the sources and sinks $Q_s$.

We want to find an expression for $\Pi$ that does not require high order pieces of the distribution functions and the potential. According to the estimate \eq{eq:Piestimate}, the distribution functions $f_s$ would have be known to $O(\rho_\ast^3 f_{Ms})$ to find the correct $\Pi$ using \eq{eq:Pidefinition}, and the contributions to $\Pi$ of $O(pR|\nabla \psi|)$, $O(\rho_\ast pR |\nabla \psi|)$ and $O(\rho_\ast^2 p R |\nabla \psi|)$ must vanish. Using moments of the Fokker-Planck equations \eq{eq:FPequation} as explained in \ref{app:Pisimplify}, we find that $\Pi$ can be written as
\begin{eqnarray} \label{eq:Pifinal}
\Pi = \Pi_{-1} + \Pi_0,
\end{eqnarray}
where
\begin{eqnarray} \label{eq:Piminusone}
\fl \Pi_{-1} = - \left \langle \left \langle \sum_{s \neq e} \sum_{k_\psi, k_\alpha} R m_s c i k_\alpha (\underline{\phi}_1^\tb)^\ast \int \dd^3v\, \underline{[f_s]}^\tb_1 (\bv \cdot \zun) \right \rangle_\psi \right \rangle_t \nonumber\\- \left \langle \sum_{s \neq e, s^\prime \neq e} \frac{R^2 m_s^2 c}{2Z_s e} \int \dd^3v\, C_{ss^\prime}^{(\ell)} \left [ [f_s]_1^\lw; [f_{s^\prime}]_1^\lw \right ] (\bv \cdot \zun)^2 \right \rangle_\psi
\end{eqnarray}
is the lowest order radial flux of toroidal angular momentum, formally of order $\rho_\ast^2 p R |\nabla \psi|$, and
\begin{eqnarray} \label{eq:Pizero}
\fl \Pi_0 = \frac{\partial}{\partial t} \left ( \sum_{s \neq e} \frac{\langle R^2 \rangle_\psi m_s c p_s}{2Z_s e} \right ) \nonumber\\ + \frac{1}{V^\prime} \frac{\partial}{\partial \psi} \Bigg [ V^\prime \Bigg ( - \left \langle \left \langle \sum_{s \neq e} \sum_{k_\psi, k_\alpha} \frac{R^2 m_s^2 c^2}{2 Z_s e} i k_\alpha (\underline{\phi}_1^\tb)^\ast \int \dd^3v\, \underline{[f_s]}_1^\tb (\bv \cdot \zun)^2 \right \rangle_\psi \right \rangle_t \nonumber\\ - \left \langle \sum_{s \neq e, s^\prime \neq e} \frac{R^3 m_s^3 c^2}{6Z_s^2 e^2} \int \dd^3v\, C_{ss^\prime}^{(\ell)} \left [ [f_s]_1^\lw ; [f_{s^\prime}]_1^\lw \right ] (\bv \cdot \zun)^3 \right \rangle_\psi \Bigg )  \Bigg ] \nonumber\\ - \left \langle \left \langle \sum_{s \neq e} \sum_{k_\psi, k_\alpha} R m_s c i k_\alpha (\underline{\phi}_1^\tb)^\ast \int \dd^3v\, \underline{[f_s]}_2^\tb (\bv \cdot \zun) \right \rangle_\psi \right \rangle_t \nonumber\\ - \left \langle \left \langle \sum_{s \neq e} \sum_{k_\psi, k_\alpha} R m_s c i k_\alpha (\underline{\phi}_2^\tb)^\ast \int \dd^3v\, \underline{[f_s]}_1^\tb (\bv \cdot \zun) \right \rangle_\psi \right \rangle_t \nonumber\\- \left \langle \sum_{s \neq e, s^\prime \neq e} \frac{R^2 m_s^2 c}{2Z_s e} \int \dd^3v\, C_{ss^\prime}^{(\ell)} \left [ [f_s]_2^\lw ; [f_{s^\prime}]_2^\lw\right ] (\bv \cdot \zun)^2 \right \rangle_\psi \nonumber\\- \left \langle \sum_{s \neq e, s^\prime \neq e} \frac{R^2 m_s^2 c}{2Z_s e} \int \dd^3v\, C_{ss^\prime} \left [ [f_s]_1^\lw, [f_{s^\prime}]_1^\lw \right ] (\bv \cdot \zun)^2 \right \rangle_\psi \nonumber\\- \left \langle \sum_{s \neq e, s^\prime \neq e} \frac{R^2 m_s^2 c}{2Z_s e} \int \dd^3v\, \left \langle \sum_{k_\psi, k_\alpha} C_{ss^\prime} \left [ (\underline{[f_s]}_1^\tb)^\ast, \underline{[f_{s^\prime}]}_1^\tb \right ] \right \rangle_t (\bv \cdot \zun)^2 \right \rangle_\psi \nonumber\\ - \sum_{s \neq e} \frac{\langle R^2 \rangle_\psi m_e c}{Z_s e} n_e \nu_{es} (T_e- T_i) - \left \langle \sum_{s \neq e} \frac{R^2 m_s^2 c}{2Z_s e} \int \dd^3v\, Q_s (\bv \cdot \zun)^2 \right \rangle_\psi
\end{eqnarray}
is the piece of $\Pi$ that is explicitly of order $\rho_\ast^3 p R |\nabla \psi|$. Here $z^\ast$ is the complex conjugate of $z$, 
\begin{equation}
\langle \ldots \rangle_t = \frac{1}{\Delta t} \int_{\Delta t} \dd t\, (\ldots) 
\end{equation}
is the average over the turbulent time scale, $p_s = n_s T_s$ is the pressure of species $s$,
\begin{eqnarray} \label{eq:Clineardef}
\fl C_{ss^\prime}^{(\ell)} [ g_s; g_{s^\prime}] = C_{ss^\prime} [ g_s, f_{Ms^\prime}] + C_{ss^\prime} [f_{Ms}, g_{s^\prime}] =\nonumber\\ \frac{\gamma_{ss^\prime}}{m_s} \nabla_v \cdot \Bigg [ \int \dd^3v^\prime\, f_{Ms} (\bv) f_{Ms^\prime} (\bv^\prime) \nabla_g \nabla_g g \cdot \Bigg ( \frac{1}{m_s} \nabla_v \left ( \frac{g_s (\bv)}{f_{Ms} (\bv)} \right ) \nonumber\\- \frac{1}{m_{s^\prime}} \nabla_{v^\prime} \left ( \frac{g_{s^\prime} (\bv^\prime)}{f_{Ms^\prime} (\bv^\prime)} \right ) \Bigg ) \Bigg ]
\end{eqnarray}
is the linearized collision operator for $s \neq e$ and $s^\prime \neq e$ (see \ref{app:Cse} and \ref{app:Ces}), and
\begin{equation} \label{eq:nuesdef}
\nu_{es} = \frac{4 \sqrt{2\pi}}{3} \frac{Z_s^2 e^4 n_s \ln \Lambda}{m_e^{1/2} T_e^{3/2}}
\end{equation}
is the collision frequency for collisions between electrons and species $s$. In the linearized collision operator \eq{eq:Clineardef}, we use a semicolon to separate the two arguments to emphasize that it is not a bilinear operator, but the sum of two linear operators: one linear on $g_s$ and another linear on $g_{s^\prime}$.

In \eq{eq:Pifinal} the $O(\rho_\ast^2 R p |\nabla \psi|)$ contribution $\Pi_{-1}$ vanishes for up-down symmetric tokamaks due to the symmetry described in \cite{parra11c, sugama11a} and revisited in section \ref{sec:symmetry}. Expression \eq{eq:Pifinal} only requires a distribution function correct to $O(\rho_\ast^2 f_{Ms})$ to find $\Pi$ to $O(\rho_\ast^3 R p |\nabla \psi|)$. Equation \eq{eq:Pifinal} is not the only way to write the radial flux of toroidal angular momentum.  A detailed explanation of why this form is convenient for $\nu_{ii} \lesssim v_{ti}/a$ is given in Appendix B of \cite{parra11d}. 

\section{Gyrokinetic equations} \label{sec:gyrokinetics}

In this section we discuss the gyrokinetic equations valid to second order in $\rho_\ast$ without performing the subsidiary expansion in $B_p/B \ll 1$. The gyrokinetic equations that we present here are different from the gyrokinetic equations in previous work \cite{parra10a, parra11d} for several reasons: we consider several ion species, the ion-electron collisions are included, we include the effect of sources of particles and energy, and we present a new treatment of the radial variation of the background gradients that does not require global gyrokinetic simulations.

The gyrokinetic equation for the distribution function $f_s (\bR, u, \mu, \varphi, t)$ is given by
\begin{equation} \label{eq:gkequation1}
\frac{\partial f_s}{\partial t} + \dot{\bR}_s \cdot \nabla_\bR f_s + \dot{u}_s \frac{\partial f_s}{\partial u} + \dot{\mu}_s \frac{\partial f_s}{\partial \mu} + \dot{\varphi}_s \frac{\partial f_s}{\partial \varphi} = \sum_{s^\prime} C_{ss^\prime} [ f_s, f_{s^\prime} ] + Q_s,
\end{equation}
where the time derivative is taken holding $\bR$, $u$, $\mu$ and $\varphi$ fixed, and the Vlasov operator applied to a function $h(\boldr, \bv, t)$ is denoted by
\begin{equation}
\dot{h} = \frac{\partial h}{\partial t} + \bv \cdot \nabla h + \frac{Z_s e}{m_s} \left ( - \nabla \phi + \frac{1}{c} \bv \times \bB \right ) \cdot \nabla_v h.
\end{equation}
The gyrokinetic variables $\bR$, $u$ and $\varphi$, given in \eq{eq:Rdef}, \eq{eq:udef} and \eq{eq:varphidef}, are chosen such that the coefficients $\dot{\bR}_s \sim v_{ts}$, $\dot{u}_s \sim v_{ts}^2/a$ and $\dot{\varphi}_s \sim \Omega_s$ do not depend on the gyrophase to a very high order in $\rho_\ast$ \cite{parra08}; in particular,
\begin{equation} \label{eq:Rdotorder}
\frac{\partial \dot{\bR}_s}{\partial \varphi} = O \left ( \rho_\ast^3 v_{ts} \right ),
\end{equation}
\begin{equation} \label{eq:udotorder}
\frac{\partial \dot{u}_s}{\partial \varphi} = O \left ( \rho_\ast^3 \frac{v_{ts}^2}{a} \right )
\end{equation}
and
\begin{equation} \label{eq:varphidotorder}
\frac{\partial \dot{\varphi}_s}{\partial \varphi} = O \left ( \rho_\ast^3 \Omega_s \right ).
\end{equation} 
The gyrokinetic variable $\mu$, given in \eq{eq:mudef}, is defined such that the coefficient $\dot{\mu}_s$ is zero to a very high order,
\begin{equation} \label{eq:dotmu}
\dot{\mu}_s = O \left ( \rho_\ast^3 \frac{v_{ts}^3}{Ba} \right ).
\end{equation}
The coefficients $\dot{\bR}_s$, $\dot{u}_s$ and $\dot{\varphi}_s$ are calculated to lowest order in \cite{parra08}. For our purposes, it is sufficient to know that
\begin{eqnarray} \label{eq:dotR}
\dot{\bR}_s = u \bun + \bv_{Ms} - \frac{c}{B} \nabla_\bR \phiave \times \bun + \dot{\bR}_{s2}
\end{eqnarray}
and
\begin{eqnarray} \label{eq:dotu}
\dot{u}_s = - \left ( \bun + \frac{u}{\Omega_s} \bun \times \kappabf \right ) \cdot \left ( \mu \nabla_\bR B + \frac{Z_s e}{m_s} \nabla_\bR \phiave \right ) + \dot{u}_{s2},
\end{eqnarray}
where
\begin{equation}
\bv_{Ms} = \frac{\mu}{\Omega_s} \bun \times \nabla_\bR B + \frac{u^2}{\Omega_s} \bun \times \kappabf
\end{equation}
are the $\nabla B$ and curvature drifts,
\begin{equation}
\kappabf = \bun \cdot \nabla_\bR \bun
\end{equation}
is the magnetic field line curvature, and the terms $\dot{\bR}_{s2}$ and $\dot{u}_{s2}$ are of order
\begin{equation} \label{eq:dotR2order}
\dot{\bR}_{s2} \sim \rho_\ast^2 v_{ts}
\end{equation}
and
\begin{equation} \label{eq:dotu2order}
\dot{u}_{s2} \sim \rho_\ast^2 \frac{v_{ts}^2}{a}.
\end{equation}
For the rest of the paper, we will only need to know that $\dot{\bR}_{s2}$ and $\dot{u}_{s2}$ can be calculated and that their size is given by \eq{eq:dotR2order} and \eq{eq:dotu2order}. 

The gyrophase dependent piece of $f_s$ is
\begin{equation}
\tilde{f}_s = f_s - \langle f_s \rangle.
\end{equation}
The triangular brackets denote the gyroaverage holding $\bR$, $u$, $\mu$ and $t$ fixed, i.e.
\begin{equation}
\langle h \rangle (\bR, u, \mu, t) = \frac{1}{2\pi} \int_0^{2\pi} \dd \varphi^\prime\, h (\bR, u, \mu, \varphi^\prime, t).
\end{equation}
From \eq{eq:gkequation1}, we find \cite{calvo12, parra08, brizard04}
\begin{equation}
\fl \tilde{f}_s \simeq - \frac{1}{\Omega_s} \sum_{s^\prime} \int^\varphi \dd\varphi^\prime\, ( C_{ss^\prime} [ f_s, f_{s^\prime} ] (\bR, u, \mu, \varphi^\prime, t) - \langle C_{ss^\prime} [ f_s, f_{s^\prime} ] \rangle (\bR, u, \mu, t) ).
\end{equation}
Using the fact that the lowest order distribution functions are Maxwellians with temperatures consistent with the mass of the species (see \eq{eq:collisiontauE} and \eq{eq:tequilibrationorder} and the discussion around them), the long wavelength and short wavelength pieces of the gyrophase dependent piece of the distribution function can be calculated, giving
\begin{equation} \label{eq:ftildelw}
\fl \tilde{f}^\lw_s \simeq \tilde{f}^\lw_{s2} = - \frac{1}{\Omega_s} \sum_{s^\prime} \int^\varphi \dd\varphi^\prime\, C_{ss^\prime}^{(\ell)} [ \Delta f_{s1}^\lw; \Delta f_{s^\prime1}^\lw ] \sim \frac{a \nu_{ii}}{v_{ti}} \left ( \frac{\rho_s}{a} \right )^2 f_{Ms} \lesssim \left ( \frac{\rho_s}{a} \right )^2 f_{Ms}
\end{equation}
and
\begin{eqnarray} \label{eq:ftildetb}
\fl \tilde{f}_{s}^\tb \simeq \tilde{f}_{s2}^\tb = \sum_{k_\psi, k_\alpha} \underline{\tilde{f}}_{s2}^\tb (k_\psi, k_\alpha, \psi(\bR), \theta(\bR), u, \mu, \varphi, t) \exp ( i k_\psi \psi (\bR) + i k_\alpha \alpha (\bR)) \nonumber\\ \sim \frac{a \nu_{ii}}{v_{ti}} \left ( \frac{\rho_s}{a} \right )^2 f_{Ms} \lesssim \left ( \frac{\rho_s}{a} \right )^2 f_{Ms},
\end{eqnarray}
where 
\begin{eqnarray} \label{eq:underlineftildetb}
\fl \underline{\tilde{f}}_{s2}^\tb = - \frac{1}{\Omega_s} \sum_{s^\prime} \int^\varphi \dd\varphi^\prime\, \Bigg ( C_{ss^\prime}^{(\ell)} \left [ \underline{[f_s]}_1^\tb ;  \underline{[f_{s^\prime}]}_1^\tb \right ] \exp( i \bk_\bot \cdot \rhobf ) \nonumber\\ - \left \langle C_{ss^\prime}^{(\ell)} \left [ \underline{[f_s]}_1^\tb ;  \underline{[f_{s^\prime}]}_1^\tb \right ] \exp( i \bk_\bot \cdot \rhobf ) \right \rangle \Bigg ),
\end{eqnarray}
and $\rhobf$ is defined in \eq{eq:rhobfdef}. Expression \eq{eq:underlineftildetb} is derived in \ref{app:gkcollision}. The linearized collision operators $C_{ss^\prime}^{(\ell)}$ are defined in \eq{eq:Clineardef}, and we have used \eq{eq:collisionorder1} and \eq{eq:collisionorder2} for the size of the collision frequencies to estimate the size of $\tilde{f}_s^\lw$ and $\tilde{f}_s^\tb$, given in \eq{eq:ftildelw} and \eq{eq:ftildetb}. These estimates led us to assume that the first order terms $f_{s1}^\lw$ and $f_{s1}^\tb$ did not depend on gyrophase in \eq{eq:flwexpansion} and \eq{eq:ftbexpansion}.

To find the equation for the gyroaveraged distribution function $\langle f_s \rangle (\bR, u, \mu, t)$, we gyroaverage \eq{eq:gkequation1}, and we use equations \eq{eq:ftildelw} and \eq{eq:ftildetb} for the gyrophase dependent part $\tilde{f}_s$ and equation \eq{eq:dotmu} for $\dot{\mu}_s$. The final result is an equation correct to order $(\rho_s/a)^2$,
\begin{equation} \label{eq:gkequation}
\fl \frac{\partial \langle f_s \rangle}{\partial t} + \dot{\bR}_s \cdot \nabla_\bR \langle f_s \rangle + \dot{u}_s \frac{\partial \langle f_s \rangle}{\partial u} = \sum_{s^\prime} \langle C_{ss^\prime} [ \langle f_s \rangle + \tilde{f}_s, \langle f_{s^\prime} \rangle + \tilde{f}_{s^\prime} ] \rangle + \langle Q_s \rangle.
\end{equation}
In addition to the gyrokinetic equation \eq{eq:gkequation}, we need to solve the quasineutrality equation \eq{eq:QNequation} to find the potential. Since $f_s (\bR, u, \mu, \varphi, t)$ is calculated as a function of the gyrokinetic variables, we need expressions \eq{eq:fsaveT1}, \eq{eq:underlinefswigT1}, \eq{eq:fsaveT2} and \eq{eq:underlinefswigT2} that give the distribution function $f_s$ written in $\{\boldr, \bv\}$ coordinates to order $(\rho_s/a)^2$. Equations \eq{eq:fsaveT1}, \eq{eq:underlinefswigT1}, \eq{eq:fsaveT2} and \eq{eq:underlinefswigT2} are also needed inside the collision operator in \eq{eq:gkequation} because the collision operator is more easily written in $\{ \boldr, \bv \}$ coordinates. 

With $\dot{\bR}_s$ and $\dot{u}_s$ given to the accuracy in \eq{eq:dotR} and \eq{eq:dotu}, and the expressions \eq{eq:fsaveT1}, \eq{eq:underlinefswigT1}, \eq{eq:fsaveT2} and \eq{eq:underlinefswigT2} for the distribution function $f_s$ written in $\{\boldr, \bv\}$ coordinates, the gyrokinetic equation and the quasineutrality equation are sufficiently accurate to give $f_{s1}^\lw$, $\phi_1^\lw$, $f_{s1}^\tb$, $\phi_1^\tb$, $f_{s2}^\lw$, $\phi_2^\lw$, $f_{s2}^\tb$ and $\phi_2^\tb$. These first and second order pieces are necessary to calculate the momentum flux $\Pi$ in \eq{eq:Pifinal}. The corrections $\bR_2$, $u_2$, $\mu_2$ and $\varphi_1$, and the terms $\dot{\bR}_{s2}$ and $\dot{u}_{s2}$ are then, in principle, needed. These higher order terms are calculated in \cite{parra11a}, but they are messy. We want to write simplified equations that are less complicated and easier to implement in existing codes. To do that, we will exploit the small parameter $B_p/B \ll 1$ in sections \ref{sec:expansion} and \ref{sec:secondordertotal}.

Before expanding in $B_p/B \ll 1$, we give and discuss the equations for $f_{s1}^\lw$, $\phi_1^\lw$, $f_{s1}^\tb$ and $\phi_1^\tb$ in subsections \ref{sub:lwfirstorder} and \ref{sub:tbfirstorder}. Finally, in subsections \ref{sub:lwsecondorder} and \ref{sub:tbsecondorder} we briefly discuss the second order equations.

\subsection{Long wavelength, first order equations} \label{sub:lwfirstorder}
Taking the first order long wavelength terms of equation \eq{eq:gkequation} (see, for example, \cite{calvo12}), we obtain
\begin{eqnarray} \label{eq:lwfirstordergk}
\fl \left ( u \bun \cdot \nabla_\bR \theta \frac{\partial}{\partial \theta} - \mu \bun \cdot \nabla_\bR B \frac{\partial}{\partial u} \right ) f_{s1}^\lw - \sum_{s^\prime} C_{ss^\prime}^{(\ell)} [ f_{s1}^\lw; f_{s^\prime 1}^\lw ] = - \bv_{Ms}\cdot \nabla_\bR \psi \Bigg [ \frac{\partial}{\partial \psi} \ln p_s \nonumber\\ + \frac{Z_s e}{T_s} \frac{\partial \phi_0}{\partial \psi} + \left ( \frac{m_s (u^2 + 2\mu B)}{2T_s} - \frac{5}{2} \right ) \frac{\partial}{\partial \psi} \ln T_s \Bigg ] f_{Ms} \nonumber\\ - \frac{Z_s e f_{Ms}}{T_s} u \bun \cdot \nabla_\bR \theta \frac{\partial \phi_1^\lw}{\partial \theta}.
\end{eqnarray}
We have used the decompositions \eq{eq:phiaveRlw} and \eq{eq:flwexpansion}. 

The linearized collision operator $C_{ss^\prime}^{(\ell)}$ in \eq{eq:Clineardef} is written more naturally in the $\{ \boldr, \bv \}$ coordinates, or, equivalently, in the $\{ \boldr, v_{||}, \mu_0, \varphi_0 \}$. We need to transform it to $\{ \bR, u, \mu, \varphi \}$ coordinates, but since to lowest order, $\bR \simeq \boldr$, $u \simeq v_{||}$, $\mu \simeq \mu_0$ and $\varphi \simeq \varphi_0$, to this order it is sufficient to replace $\{ \boldr, v_{||}, \mu_0, \varphi_0 \}$ by $\{ \bR, u, \mu, \varphi \}$. This is what is done in \eq{eq:lwfirstordergk}.

In equation \eq{eq:lwfirstordergk} for ionic species ($s\neq e$), the linearized ion-electron and impurity-electron collision operators, $C_{se}^{(\ell)}$, are negligible. We explain why in \ref{app:Cse}. If we want to consider the electron neoclassical particle flux, these terms cannot be neglected. The effect of the electron neoclassical particle flux is, however, usually small. In equation \eq{eq:lwfirstordergk} for the electrons ($s = e$), we use a simplified form for the linearized electron-ion and electron-impurity collision operator because $\sqrt{m_e/m_i} \sim \sqrt{m_e/m_z} \ll 1$ (see \ref{app:Ces}),
\begin{eqnarray} \label{eq:Cessimplify}
\fl C_{es^\prime}^{(\ell)} [ g_e; g_{s^\prime} ] \simeq \frac{4\sqrt{2\pi}}{3} \nu_{es^\prime} \left ( \frac{T_e}{m_e} \right )^{3/2} \nabla_v \cdot \left [ \nabla_v \nabla_v v \cdot \nabla_v \left ( g_e - \frac{m_e \bv \cdot \mathbf{U}_{s^\prime}}{T_e} f_{Me} \right ) \right ],
\end{eqnarray}
where $\nu_{es^\prime}$ is defined in \eq{eq:nuesdef}, and
\begin{equation} \label{eq:Usdef}
\mathbf{U}_{s^\prime} = \frac{1}{n_{s^\prime}} \int \dd^3v\, g_{s^\prime} \bv.
\end{equation}

The first order long wavelength electrostatic potential $\phi_1^\lw$ can be obtained from quasineutrality,
\begin{equation} \label{eq:lwfirstorderquasineutrality}
2 \pi \sum_{s \neq e} Z_s \int \dd v_{||} \, \dd \mu_0\, B f_{s1}^\lw (\boldr, v_{||}, \mu_0, t) = \frac{e\phi_1^\lw}{T_e} n_e,
\end{equation}
where we have used that the determinant of the Jacobian of the transformation between $\{ \bv \}$ and $\{ v_{||}, \mu_0, \varphi_0 \}$ is $B$, and we have taken into account that the electrons are adiabatic to lowest order (see \eq{eq:fe1evenadiabatic} below).

To estimate the size of $f_{s1}^\lw$ and compare it to our orderings \eq{eq:fnorder}, \eq{eq:fenevenorder} and \eq{eq:fenoddorder}, we define the function
\begin{eqnarray} \label{eq:gs1lwdef}
\fl g_{s1}^\lw = f_{s1}^\lw + \frac{I u}{\Omega_s} \left [ \frac{1}{p_s} \frac{\partial p_s}{\partial \psi} + \frac{Z_s e}{T_s} \frac{\partial \phi_0}{\partial \psi} + \left ( \frac{m_s (u^2 + 2\mu B)}{2T_s} - \frac{5}{2} \right ) \frac{1}{T_s} \frac{\partial T_s}{\partial \psi} \right ] f_{Ms} \nonumber\\ + \frac{Z_s e \phi_1^\lw}{T_s} f_{Ms},
\end{eqnarray}
with $I = RB_\zeta$ (see \eq{eq:Bdef}). The equation for $g_{s1}^\lw$ is obtained from the equation for $f_{s1}^\lw$, \eq{eq:lwfirstordergk}, by realizing that
\begin{equation} \label{eq:vMstrick}
\bv_{Ms} \cdot \nabla_\bR \psi = \left ( u \bun \cdot \nabla_\bR \theta \frac{\partial}{\partial \theta} - \mu \bun \cdot \nabla_\bR B \frac{\partial}{\partial u} \right ) \frac{I u}{\Omega_s}.
\end{equation}
With this expression, equation \eq{eq:lwfirstordergk} becomes
\begin{eqnarray} \label{eq:lwfirstordergkv2}
\left ( u \bun \cdot \nabla_\bR \theta \frac{\partial}{\partial \theta} - \mu \bun \cdot \nabla_\bR B \frac{\partial}{\partial u} \right ) g_{s1}^\lw - \sum_{s^\prime} C_{ss^\prime}^{(\ell)} [ f_{s1}^\lw; f_{s^\prime 1}^\lw ] = 0.
\end{eqnarray}
From equation \eq{eq:lwfirstordergkv2}, we see that \eq{eq:fnorder} is satisfied for ionic species ($s \neq e$). We also find that the lowest order piece of $f_{e1}^\lw$ that is even in $u$ is the Maxwell-Boltzmann response
\begin{equation} \label{eq:fe1evenadiabatic}
\frac{1}{2} ( f_{e1}^\lw (u) + f_{e1}^\lw (-u) ) = \frac{e\phi_1^\lw}{T_e} f_{Me} + O \left (\sqrt{\frac{m_e}{m_i}} \rho_\ast f_{Me} \right ),
\end{equation}
giving \eq{eq:fenevenorder}. Solving \eq{eq:lwfirstordergk} to next order shows that the piece of $f_{e1}^\lw$ odd in $u$ follows \eq{eq:fenoddorder}.

The solution to \eq{eq:lwfirstordergk} gives the long wavelength flow of the different ionic species, which in general has the form \cite{hinton76, helander02bk}
\begin{eqnarray} \label{eq:plasmaflowlw}
\fl n_s \bV_s^\lw = \int \dd^3v\, \langle f_s (\bR, u, \mu, \varphi, t) \rangle_\mathrm{T} \bv \simeq \int \dd^3 v\, \left ( f_{s1}^\lw (\boldr, v_{||}, \mu_0, t) + \Delta f_{s1}^\lw \right ) \bv \nonumber\\ = - \left ( c \frac{\partial \phi_0}{\partial \psi} + \frac{c}{Z_s e n_s} \frac{\partial p_s}{\partial \psi} \right ) n_s R \zun + K_s(\psi, t) \bB,
\end{eqnarray}
where we have used \eq{eq:fsaveT1}. The flux function $K_s(\psi, t)$ depends on the ion temperature gradient and the pressure gradients of the ions and the impurities. If the impurity density is very small, $K_i (\psi, t)$ only depends on the ion temperature gradient. The velocity $\bV_s^\lw$ is then completely determined except for the radial electric field $\partial \phi_0/\partial \psi$ (in particular, the poloidal component of the flow is completely determined due to collisional damping). Thus, the toroidal angular momentum equation \eq{eq:torangmom} can be understood as an equation for the lowest order radial electric field, $\partial \phi_0/\partial \psi$.

\subsection{Short wavelength, first order equations} \label{sub:tbfirstorder}
The turbulent pieces of the distribution functions are more easily given in terms of Fourier components (see \eq{eq:ftbscales}). Then, taking the short wavelength component of \eq{eq:gkequation} to first order in $\rho_\ast$, and Fourier analyzing it, we find
\begin{eqnarray} \label{eq:tbfirstordergk}
\fl \frac{\partial \underline{f}_{s1}^\tb}{\partial t} + \left ( u \bun \cdot \nabla_\bR \theta \frac{\partial}{\partial \theta} - \mu \bun \cdot \nabla_\bR B \frac{\partial}{\partial u} \right ) \underline{f}_{s1}^\tb + \left ( - i k_\alpha c \frac{\partial \phi_0}{\partial \psi} + i \bk_\bot \cdot \bv_{Ms} \right ) \underline{f}_{s1}^\tb \nonumber\\ - \sum_{s^\prime} C_{ss^\prime}^{GK} \left [ \underline{f}_{s1}^\tb; \underline{f}_{s^\prime 1}^\tb \right ] + \{ \underline{\phi}_1^\tb J_0 (\Lambda_s), \underline{f}_{s1}^\tb \} \nonumber\\ = - f_{Ms} \Bigg [ \frac{Z_s e}{T_s} \left ( u \bun \cdot \nabla_\bR + i \bk_\bot \cdot \bv_{Ms} \right ) \nonumber\\ + i k_\alpha c \Bigg ( \frac{\partial}{\partial \psi} \ln n_s  + \left ( \frac{m_s (u^2 + 2\mu B)}{2T_s} - \frac{3}{2} \right ) \frac{\partial}{\partial \psi} \ln T_s \Bigg ) \Bigg ] \underline{\phi}_1^\tb J_0 (\Lambda_s) \nonumber\\ + \sum_{s^\prime} C_{ss^\prime}^{GK} \left [ \frac{Z_s e \underline{\phi}_1^\tb}{T_s} J_0 (\Lambda_s) f_{Ms}; \frac{Z_{s^\prime} e \underline{\phi}_1^\tb}{T_{s^\prime}} J_0 (\Lambda_{s^\prime}) f_{Ms^\prime} \right ] .
\end{eqnarray}
Here we have used the decompositions \eq{eq:phiaveRtb} and \eq{eq:ftbexpansion}, and we have neglected the smallest terms in the formula \eq{eq:gradbotftborder} for $\nabla_{\bR\bot} f_{s1}^\tb$. The term
\begin{eqnarray} \label{eq:nonlinearity}
\fl \{ \underline{\phi}_1^\tb J_0 (\Lambda_s), \underline{f}_{s1}^\tb \} = c \sum_{k_\psi^\prime, k_\alpha^\prime} ( k_\psi^\prime k_\alpha - k_\alpha^\prime k_\psi) \underline{\phi}_1^\tb ( k_\psi^\prime, k_\alpha^\prime, \psi(\bR), \theta(\bR), t) J_0 (\Lambda_s^\prime) \nonumber\\ \times \underline{f}_{s1}^\tb ( k_\psi - k_\psi^\prime, k_\alpha - k_\alpha^\prime, \psi(\bR), \theta(\bR), u, \mu, t )
\end{eqnarray}
is the nonlinearity due to the turbulent $\bE \times \bB$ drift. Here $\Lambda_s^\prime = (k_\bot^\prime m_s c/Z_s e) \sqrt{2\mu/B}$ and $\bk_\bot^\prime = k_\psi^\prime \nabla_\bR \psi + k_\alpha^\prime \nabla_\bR \alpha$. The operator
\begin{eqnarray} \label{eq:gkcollision1}
\fl C_{ss^\prime}^{GK} [ \underline{f}_s; \underline{f}_{s^\prime} ] = \Bigg \langle C_{ss^\prime} ^{(\ell)} \left [ \underline{f}_s \exp \left ( \frac{i\bk_\bot \cdot (\bv \times \bun)}{\Omega_s} \right ); \underline{f}_{s^\prime} \exp \left ( \frac{i\bk_\bot \cdot (\bv \times \bun)}{\Omega_{s^\prime}} \right ) \right ] \nonumber\\ \times \exp ( i\bk_\bot \cdot \rhobf ) \Bigg \rangle
\end{eqnarray}
is the linearized gyrokinetic collision operator. It is discussed in detail in \ref{app:gkcollision}.

In equation \eq{eq:tbfirstordergk} for ionic species ($s\neq e$), the gyrokinetic ion-electron and impurity-electron collision operators, $C_{se}^{GK}$, are negligible. In equation \eq{eq:tbfirstordergk} for the electrons ($s = e$), the gyrokinetic electron-ion and electron-impurity collision operators are constructed as in \eq{eq:gkcollision1}, but with the simplified linearized collision operator \eq{eq:Cessimplify}.

Equation \eq{eq:tbfirstordergk} is solved in conjunction with the quasineutrality equation
\begin{eqnarray} \label{eq:tbfirstorderquasineutrality}
\fl 2 \pi \sum_s Z_s \int \dd v_{||}\, \dd \mu_0\, B J_0 (\lambda_s) \underline{f}_{s1}^\tb (k_\psi, k_\alpha, \psi(\boldr), \theta (\boldr), v_{||}, \mu_0, t) \nonumber\\ - \sum_{s} \frac{Z_s^2 e \underline{\phi}_1^\tb }{T_s} n_s ( 1 - \Gamma_0(b_s)) = 0,
\end{eqnarray}
where 
\begin{equation}
\Gamma_0 (b_s) = I_0 (b_s) \exp ( - b_s ),
\end{equation}
\begin{equation} \label{eq:bsdef}
b_s = \frac{k_\bot^2 T_s}{m_s \Omega_s^2},
\end{equation}
and $I_n$ is the $n$-th order modified Bessel function of the first kind. To find this equation, we have taken the first order, short wavelength component of the quasineutrality equation \eq{eq:QNequation}, and we have used \eq{eq:underlinefswigT1} to write the distribution function in terms of $\{\boldr, \bv\}$. In addition, we have employed that
\begin{equation}
\fl \int \dd^3v\, \exp \left ( \frac{i \bk_\bot \cdot (\bv \times \bun)}{\Omega_s} \right ) g (\boldr, v_{||}, \mu_0, t) = \int \dd^3v\, J_0 (\lambda_s) g (\boldr, v_{||}, \mu_0, t)
\end{equation}
for any function of phase space $g(\boldr, v_{||}, \mu_0, t)$ that is independent of $\varphi_0$, and that
\begin{equation}
\int_0^\infty \dd x\, x J_0^2 (x\sqrt{b})\exp\left ( - \frac{x^2}{2} \right ) = I_0 (b) \exp ( - b ).
\end{equation}

\subsection{Long wavelength, second order equations} \label{sub:lwsecondorder}
Taking the second order piece of the long wavelength component of \eq{eq:gkequation}, we find the equation for $\langle f_{s2}^\lw \rangle$ (the gyrophase dependent piece is given by \eq{eq:ftildelw}). At the end of this subsection, we will see that for momentum transport, we only need $\langle f_{s2}^\lw \rangle$ for ionic species ($s \neq e$),
\begin{eqnarray} \label{eq:lwsecondorderions}
\fl \left ( u \bun \cdot \nabla_\bR \theta \frac{\partial}{\partial \theta} - \mu \bun \cdot \nabla_\bR B \frac{\partial}{\partial u} \right ) \langle f_{s2}^\lw \rangle - \sum_{s^\prime \neq e} C_{ss^\prime}^{(\ell)} [ \langle f_{s2}^\lw \rangle; \langle f_{s^\prime 2}^\lw \rangle ] = - \frac{\partial f_{Ms}}{\partial t} \nonumber\\- \left ( \bv_{Ms} - \frac{c}{B} \nabla_\bR \phi_0 \times \bun \right ) \cdot \nabla_\bR f_{s1}^\lw + \Bigg [ \frac{Z_s e}{m_s} \bun \cdot \nabla_\bR \phi_1^\lw \nonumber\\+ \frac{u}{\Omega_s} (\bun \times \kappabf) \cdot \left ( \mu \nabla_\bR B + \frac{Z_s e}{m_s} \nabla_\bR \phi_0 \right ) \Bigg ] \frac{\partial f_{s1}^\lw}{\partial u} \nonumber\\ - \left ( \dot{\bR}_{s2}^\lw - \frac{c}{B} \nabla_\bR \phi_1^\lw \times \bun \right )\cdot \nabla_\bR \psi \Bigg [ \frac{\partial}{\partial \psi} \ln p_s \nonumber\\+ \left ( \frac{m_s ( u^2 + 2\mu B)}{2T_i} - \frac{5}{2} \right ) \frac{\partial}{\partial \psi} \ln T_i \Bigg ] f_{Ms} + \Bigg [ u \dot{u}_{s2}^\lw \nonumber\\ + \mu \dot{\bR}_{s2}^\lw \cdot \nabla_\bR B - \frac{Z_s e}{m_s} \bv_{Ms} \cdot \nabla_\bR \phi_1^\lw - \frac{Z_s e}{m_s} u \bun \cdot \nabla_\bR \phiave_2^\lw \Bigg ] \frac{m_s f_{Ms}}{T_i} \nonumber\\ - \left \langle \sum_{k_\psi, k_\alpha} \frac{c (\underline{\phi}_1^\tb)^\ast}{B} J_0 ( \Lambda_s ) i (\bk_\bot \times \bun) \cdot \left ( \nabla_\bR \psi \frac{\partial}{\partial \psi} + \nabla_\bR \theta \frac{\partial}{\partial \theta} \right ) \underline{f}_{s1}^\tb \right \rangle_t \nonumber\\ + \left \langle \sum_{k_\psi, k_\alpha} \frac{c (\underline{f}_{s1}^\tb)^\ast}{B} i (\bk_\bot \times \bun) \cdot \left ( \nabla_\bR \psi \frac{\partial}{\partial \psi} + \nabla_\bR \theta \frac{\partial}{\partial \theta} \right ) \underline{\phi}_1^\tb J_0 ( \Lambda_s )\right \rangle_t \nonumber\\ + \frac{Z_s e}{m_s} \left \langle \sum_{k_\psi, k_\alpha} \frac{\partial (\underline{f}_{s1}^\tb)^\ast}{\partial u} \left ( \bun \cdot \nabla_{\bR} \theta \frac{\partial}{\partial \theta} + \frac{u}{\Omega_s} i \bk_\bot \cdot (\bun \times \kappabf) \right ) \underline{\phi}_1^\tb J_0 (\Lambda_s) \right \rangle_t \nonumber\\ - \left \langle \sum_{k_\psi, k_\alpha} i \bk_\bot \cdot (\underline{\dot{\bR}}_{s2}^\tb)^\ast \underline{f}_{s1}^\tb \right \rangle_t + \sum_{s^\prime \neq e} \langle C_{ss^\prime,2}^\lw \rangle + C_{se} [ f_s, f_e ] + Q_s.
\end{eqnarray}
Here $\dot{\bR}_{s2}^\lw$ and $\dot{u}_{s2}^\lw$ are the long wavelength components of $\dot{\bR}_{s2}$ and $\dot{u}_{s2}$, and the underlined symbols $\underline{\dot{\bR}}^\tb_{s2} (k_\psi, k_\alpha)$ are the Fourier coefficients of the turbulent pieces of $\dot{\bR}_{s2}$ and $\dot{u}_{s2}$. The collisional piece $C_{ss^\prime, 2}^\lw$ is described in \ref{app:gkcollision}, and following \ref{app:Cse} the ion-electron and impurity-electron collision operators simplify to
\begin{eqnarray} \label{eq:Cseexpanded}
C_{se} [f_s, f_e] \simeq \frac{n_e m_e \nu_{es}}{n_s m_s} \left ( \frac{T_e}{T_i} - 1 \right ) \left ( \frac{m_s v^2}{T_i} - 3 \right ) f_{Ms}.
\end{eqnarray}
The frequency $\nu_{es}$ is defined in \eq{eq:nuesdef}.

The piece $\phi_2^\lw$ of the electrostatic potential enters in $\phiave_2^\lw$ (see \eq{eq:phiaveRlw2}), but it is unimportant for the momentum transport because the only effect that $\phi_2^\lw$ has on the distribution function is to add a Maxwell-Boltzmann response $(-Z_s e \phi_2^\lw/T_i) f_{Ms}$, and this Maxwell-Boltzmann response does not drive momentum flux, as is clear from \eq{eq:Pizero}, where $f_{s2}^\lw$ enters within a collision operator. For this reason we do not need to calculate $\phi_2^\lw$ and we do not need the second order, long wavelength quasineutrality equation. The second order piece $f_{e2}^\lw$ of the electron distribution function is then not needed for quasineutrality, and in addition, it does not enter in \eq{eq:Pizero}. As a result, we do not have to calculate it.

\subsection{Short wavelength, second order equations} \label{sub:tbsecondorder}
The equation for the gyroaveraged, second order, short wavelength piece of the distribution function is
\begin{eqnarray} \label{eq:tbsecondordergk}
\fl \frac{\partial \langle \underline{f}_{s2}^\tb \rangle}{\partial t} + \left ( u \bun \cdot \nabla_\bR \theta \frac{\partial}{\partial \theta} - \mu \bun \cdot \nabla_\bR B \frac{\partial}{\partial u} \right ) \langle \underline{f}_{s2}^\tb \rangle + \left ( - i k_\alpha c \frac{\partial \phi_0}{\partial \psi} + i \bk_\bot \cdot \bv_{Ms} \right ) \langle \underline{f}_{s2}^\tb \rangle \nonumber\\ - \sum_{s^\prime} C_{ss^\prime}^{GK} \left [ \langle \underline{f}_{s2}^\tb \rangle; \langle \underline{f}_{s^\prime2}^\tb \rangle \right ] + \{ \underline{\phiave}_2^\tb, \underline{f}_{s1}^\tb \} + \{ \underline{\phi}_1^\tb J_0 (\Lambda_s), \langle \underline{f}_{s2}^\tb \rangle \} = \nonumber\\ - f_{Ms} \Bigg [ \frac{Z_s e}{T_s} \left ( u \bun \cdot \nabla_\bR + i \bk_\bot \cdot \bv_{Ms} \right ) + i k_\alpha c \Bigg ( \frac{\partial}{\partial \psi} \ln n_s  \nonumber\\  + \left ( \frac{m_s (u^2 + 2\mu B)}{2T_s} - \frac{3}{2} \right ) \frac{\partial}{\partial \psi} \ln T_s \Bigg )\Bigg ] \underline{\phiave}_2^\tb - \underline{\dot{\bR}}_{s2}^\tb \cdot \nabla_\bR \psi \Bigg [ \frac{\partial}{\partial \psi} \ln n_s  \nonumber\\  + \left ( \frac{m_s (u^2 + 2\mu B)}{2T_s} - \frac{3}{2} \right ) \frac{\partial}{\partial \psi} \ln T_s \Bigg ] f_{Ms} + ( u \underline{\dot{u}}_{s2}^\tb \nonumber\\ + \mu \underline{\dot{\bR}}_{s2}^\tb \cdot \nabla_\bR B) \frac{m_s f_{Ms}}{T_s} + \Bigg [ \frac{c}{B} (i \bk_\bot \times \bun ) \cdot \nabla_\bR f_{s1}^\lw \nonumber\\ + \frac{Z_s e}{m_s} \frac{\partial f_{s1}^\lw}{\partial u} \Bigg ( \bun \cdot \nabla_\bR + \frac{u}{\Omega_s} i \bk_\bot \cdot (\bun \times \kappabf) \Bigg ) \Bigg ] \underline{\phi}_1^\tb J_0 (\Lambda_s) \nonumber\\ - f_{Ms} \Bigg [ \frac{Z_s e}{T_s} \bv_{Ms} \cdot \Bigg ( \nabla_\bR \psi \frac{\partial}{\partial \psi} + \nabla_\bR \theta \frac{\partial}{\partial \theta} \Bigg ) + \Bigg ( \frac{\partial}{\partial \psi} \ln n_s  \nonumber\\ + \left ( \frac{m_s (u^2 + 2\mu B)}{2T_s} - \frac{3}{2} \right ) \frac{\partial}{\partial \psi} \ln T_s \Bigg ) \frac{c}{B} (\nabla_\bR \psi \times \bun) \cdot \nabla_\bR \theta \frac{\partial}{\partial \theta} \Bigg ] \underline{\phi}_1^\tb J_0 (\Lambda_s) \nonumber \\ - \left ( - \frac{c}{B} \nabla_\bR \phi_0 \times \bun + \bv_{Ms} \right ) \cdot \left ( \nabla_\bR \psi \frac{\partial}{\partial \psi} + \nabla_\bR \theta \frac{\partial}{\partial \theta} \right ) \underline{f}_{s1}^\tb \nonumber\\ + \Bigg [ \frac{Z_s e}{m_s} \bun \cdot \nabla_\bR \phi_1^\lw + \frac{u}{\Omega_s} (\bun \times \kappabf) \cdot \left ( \mu \nabla_\bR B + \frac{Z_s e}{m_s} \nabla_\bR \phi_0 \right ) \Bigg ] \frac{\partial \underline{f}_{s1}^\tb}{\partial u} \nonumber\\ - i \bk_\bot \cdot \left ( \dot{\bR}_{s2}^\lw - \frac{c}{B} \nabla_\bR \phi_1^\lw \times \bun \right ) \underline{f}_{s1}^\tb \nonumber\\ + \sum_{k_\psi^\prime, k_\alpha^\prime} \Bigg [ \frac{c (\underline{\phi}_1^\tb)^\prime}{B} J_0 ( \Lambda_s^\prime ) i (\bk^\prime_\bot \times \bun) \cdot \left ( \nabla_\bR \psi \frac{\partial}{\partial \psi} + \nabla_\bR \theta \frac{\partial}{\partial \theta} \right ) (\underline{f}_{s1}^\tb)^{\prime\prime} \nonumber\\ - \frac{c (\underline{f}_{s1}^\tb)^\prime}{B} i (\bk_\bot^\prime \times \bun) \cdot \left ( \nabla_\bR \psi \frac{\partial}{\partial \psi} + \nabla_\bR \theta \frac{\partial}{\partial \theta} \right ) (\underline{\phi}_1^\tb)^{\prime\prime} J_0 ( \Lambda_s^{\prime\prime} ) \nonumber\\ + \frac{Z_s e}{m_s} \frac{\partial (\underline{f}_{s1}^\tb)^{\prime\prime}}{\partial u} \left ( \bun \cdot \nabla_{\bR} + \frac{u}{\Omega_s} i \bk_\bot^\prime \cdot (\bun \times \kappabf) \right ) (\underline{\phi}_1^\tb)^\prime J_0 (\Lambda^\prime_s) \nonumber\\ - i \bk_\bot^\prime \cdot (\underline{\dot{\bR}}_{s2}^\tb)^{\prime\prime} (\underline{f}_{s1}^\tb)^\prime \Bigg ] + \sum_{s^\prime} C_{ss^\prime}^{GK} \left [ \underline{\tilde{f}}_{s2}^\tb; \underline{\tilde{f}}_{s^\prime2}^\tb \right ] + \sum_{s^\prime} \langle \underline{C}_{ss^\prime,2}^\tb \rangle, 
\end{eqnarray}
where $\dot{\bR}_{s2}^\lw$ is the long wavelength piece of $\dot{\bR}_{s2}$, $\underline{\dot{\bR}}_{s2}^\tb (k_\psi, k_\alpha)$ and $\underline{\dot{u}}_{s2}^\tb(k_\psi, k_\alpha)$ are the Fourier coefficients of the turbulent pieces of $\dot{\bR}_{s2}$ and $\dot{u}_{s2}$, and the collisional contribution $C_{ss^\prime,2}^\tb$ is described in \ref{app:gkcollision}. From here on, a prime on a Fourier component such as $\underline{\phi}_1^\tb$ indicates that it depends on $k_\psi^\prime$ and $k_\alpha^\prime$, e.g., $(\underline{\phi}_1^\tb)^\prime = \underline{\phi}_1^\tb (k_\psi^\prime, k_\alpha^\prime, \psi(\bR), \theta(\bR), t)$, and two primes indicate that it depends on $k_\psi^{\prime\prime}$ and $k_\alpha^{\prime\prime}$, e.g., $(\underline{\phi}_1^\tb)^{\prime\prime} = \underline{\phi}_1^\tb (k_\psi^{\prime\prime}, k_\alpha^{\prime\prime}, \psi(\bR), \theta(\bR), t)$, where
\begin{equation}
k_\psi^{\prime\prime} = k_\psi - k_\psi^\prime,
\end{equation}
and
\begin{equation}
k_\alpha^{\prime\prime} = k_\alpha - k_\alpha^\prime.
\end{equation}

In equation \eq{eq:tbsecondordergk} for ionic species ($s\neq e$), the gyrokinetic ion-electron and impurity-electron collision operators, $C_{se}^{GK}$, are negligible. In equation \eq{eq:tbsecondordergk} for the electrons ($s = e$), the gyrokinetic electron-ion and electron-impurity collision operators and the collisional piece $C_{es, 2}^\tb$ must be simplified taking into account that $\sqrt{m_e/m_i} \sim \sqrt{m_e/m_z} \ll 1$ (see \ref{app:Ces}).

Equation \eq{eq:tbsecondordergk} is solved in conjunction with the quasineutrality equation
\begin{equation} \label{eq:tbsecondorderquasineutrality}
\fl 2\pi \sum_s Z_s \int \dd v_{||} \dd \mu_0\, B \left [ \left (\langle \underline{f}_{s2}^\tb \rangle + \underline{\tilde{f}}_{s2}^\tb \right ) \exp \left ( \frac{i\bk_\bot \cdot (\bv \times \bun)}{\Omega_s} \right ) + \Delta \underline{f}_{s2}^\tb \right ] = 0,
\end{equation}
where we have used \eq{eq:underlinefswigT2} to write $f_s$ as a function of $\{ \boldr, \bv \}$.

Importantly, the gyrophase dependent piece $\underline{\tilde{f}}_{s2}^\tb$ and the slow radial derivatives $\partial \underline{f}_{s1}^\tb/\partial \psi$ and $\partial \underline{\phi}_1^\tb/\partial \psi$ enter in both \eq{eq:tbsecondordergk} and \eq{eq:tbsecondorderquasineutrality}. The gyrophase dependent piece $\underline{\tilde{f}}_{s2}^\tb$ is given as a function of $\underline{f}_{s1}^\tb$ and $\underline{\phi}_1^\tb$ in \eq{eq:ftildetb}. The slow radial derivatives $\partial \underline{f}_{s1}^\tb/\partial \psi$ and $\partial \underline{\phi}_1^\tb/\partial \psi$ can be found by integrating in time the equations obtained from taking the radial derivative of \eq{eq:tbfirstordergk} and \eq{eq:tbfirstorderquasineutrality}. These equations are
\begin{eqnarray} \label{eq:radialderivativegk}
\fl \frac{\partial}{\partial t} \left ( \frac{\partial \underline{f}_{s1}^\tb}{\partial \psi} \right )+ \left ( u \bun \cdot \nabla_\bR \theta \frac{\partial}{\partial \theta} - \mu \bun \cdot \nabla_\bR B \frac{\partial}{\partial u} \right ) \frac{\partial \underline{f}_{s1}^\tb}{\partial \psi} + \left ( - i k_\alpha c \frac{\partial \phi_0}{\partial \psi} + i \bk_\bot \cdot \bv_{Ms} \right )  \frac{\partial \underline{f}_{s1}^\tb}{\partial \psi} \nonumber\\ - \sum_{s^\prime} C_{ss^\prime}^{GK} \left [  \frac{\partial \underline{f}_{s1}^\tb}{\partial \psi}; \frac{\partial \underline{f}_{s1}^\tb}{\partial \psi} \right ] + \left \{ \frac{\partial \underline{\phi}_1^\tb}{\partial \psi} J_0 (\Lambda_s), \underline{f}_{s1}^\tb \right \} + \left \{ \underline{\phi}_1^\tb J_0 (\Lambda_s), \frac{\partial \underline{f}_{s1}^\tb}{\partial \psi} \right \} \nonumber\\ = - f_{Ms} \Bigg [ \frac{Z_s e}{T_s} \left ( u \bun \cdot \nabla_\bR + i \bk_\bot \cdot \bv_{Ms} \right ) \nonumber\\ + i k_\alpha c \Bigg ( \frac{\partial}{\partial \psi} \ln n_s  + \left ( \frac{m_s (u^2 + 2\mu B)}{2T_s} - \frac{3}{2} \right ) \frac{\partial}{\partial \psi} \ln T_s \Bigg ) \Bigg ] \frac{\partial \underline{\phi}_1^\tb}{\partial \psi} J_0 (\Lambda_s) \nonumber\\ + \sum_{s^\prime} C_{ss^\prime}^{GK} \left [ \frac{Z_s e}{T_s} \frac{\partial \underline{\phi}_1^\tb}{\partial \psi} J_0 (\Lambda_s) f_{Ms}; \frac{Z_{s^\prime} e}{T_{s^\prime}}\frac{\partial \underline{\phi}_1^\tb}{\partial \psi} J_0 (\Lambda_{s^\prime}) f_{Ms^\prime} \right ] \nonumber\\ - u \frac{\partial}{\partial \psi} ( \bun \cdot \nabla_\bR \theta ) \frac{\partial \underline{f}_{s1}^\tb}{\partial \theta} + \mu \frac{\partial}{\partial \psi}( \bun \cdot \nabla_\bR B ) \frac{\partial \underline{f}_{s1}^\tb}{\partial u} \nonumber\\ - \underline{f}_{s1}^\tb \frac{\partial}{\partial \psi} \left ( -i  k_\alpha c \frac{\partial \phi_0}{\partial \psi} + i \bk_\bot \cdot \bv_{Ms} \right ) + \sum_{s^\prime} \frac{\partial C_{ss^\prime}^{GK}}{\partial \psi} \left [  \underline{f}_{s1}^\tb; \underline{f}_{s^\prime 1}^\tb \right ] \nonumber\\ - \left \{ \underline{\phi}_1^\tb \frac{\partial }{\partial \psi} (J_0 (\Lambda_s)), \underline{f}_{s1}^\tb \right \} - \frac{Z_s e f_{Ms}}{T_s} u \bun \cdot \nabla_\bR \theta \frac{\partial}{\partial \theta} \left [ \underline{\phi}_1^\tb \frac{\partial }{\partial \psi} (J_0 (\Lambda_s)) \right ] \nonumber\\- \frac{\partial }{\partial \psi} \left ( \frac{Z_s e f_{Ms}}{T_s} u \bun \cdot \nabla_\bR \theta \right ) \frac{\partial}{\partial \theta} ( \underline{\phi}_1^\tb J_0 (\Lambda_s) ) \nonumber\\- \underline{\phi}_1^\tb \frac{\partial }{\partial \psi} \Bigg [ \Bigg ( \frac{Z_s e}{T_s} i \bk_\bot \cdot \bv_{Ms}  J_0 (\Lambda_s) + i k_\alpha c J_0 (\Lambda_s) \Bigg ( \frac{\partial}{\partial \psi} \ln n_s \nonumber\\ + \left ( \frac{m_s (u^2 + 2\mu B)}{2T_s} - \frac{3}{2} \right ) \frac{\partial}{\partial \psi} \ln T_s \Bigg ) \Bigg ) f_{Ms} \Bigg ] \nonumber\\ + \sum_{s^\prime} \frac{\partial C_{ss^\prime}^{GK}}{\partial \psi} \left [ \frac{Z_s e \underline{\phi}_1^\tb}{T_s} J_0 (\Lambda_s) f_{Ms}; \frac{Z_{s^\prime} e \underline{\phi}_1^\tb}{T_{s^\prime}} J_0 (\Lambda_{s^\prime}) f_{Ms^\prime} \right ] \nonumber\\ + \sum_{s^\prime} C_{ss^\prime}^{GK} \left [  \underline{\phi}_1^\tb \frac{\partial}{\partial \psi} \left ( \frac{Z_s e}{T_s} J_0 (\Lambda_s) f_{Ms} \right ); \underline{\phi}_1^\tb \frac{\partial}{\partial \psi} \left (\frac{Z_{s^\prime} e}{T_{s^\prime}} J_0 (\Lambda_{s^\prime}) f_{Ms^\prime} \right ) \right ]
\end{eqnarray}
and
\begin{eqnarray} \label{eq:radialderivativequasineutrality}
\fl 2 \pi \sum_s Z_s \int \dd v_{||}\, \dd \mu_0\, B J_0 (\lambda_s) \frac{\partial \underline{f}_{s1}^\tb}{\partial \psi} - \sum_{s} \frac{Z_s^2 e }{T_s} n_s ( 1 - \Gamma_0(b_s)) \frac{\partial \underline{\phi}_1^\tb}{\partial \psi} \nonumber\\ = - 2 \pi \sum_s Z_s \int \dd v_{||}\, \dd \mu_0\, \underline{f}_{s1}^\tb \frac{\partial }{\partial \psi} (B J_0 (\lambda_s)) \nonumber\\ + \underline{\phi}_1^\tb \sum_{s} \frac{\partial }{\partial \psi} \left (\frac{Z_s^2 e }{T_s} n_s ( 1 - \Gamma_0(b_s)) \right ).
\end{eqnarray}
The symbol $\partial C_{ss^\prime}^{GK}/\partial \psi$ indicates that the operator $C_{ss^\prime}^{GK}$ has coefficients that depend on $\psi$, and those need to be differentiated. Equations \eq{eq:radialderivativegk} and \eq{eq:radialderivativequasineutrality} can be integrated in time at the same time as \eq{eq:tbfirstordergk} and \eq{eq:tbfirstorderquasineutrality}.

\section{Symmetry of the first order equations} \label{sec:symmetry}

In section \ref{sec:radialflux}, we have seen that the radial momentum flux has a piece, $\Pi_{-1}$ in \eq{eq:Piminusone}, that is formally of lower order in the expansion in $\rho_\ast$. This piece vanishes in up-down symmetric tokamaks. The reason is a symmetry of equations \eq{eq:lwfirstordergk}, \eq{eq:lwfirstorderquasineutrality}, \eq{eq:tbfirstordergk} and \eq{eq:tbfirstorderquasineutrality} \cite{parra11c, sugama11a}. In an up-down symmetric tokamak, under the transformation
\begin{equation} \label{eq:transformationsymmetry}
k_\psi \rightarrow - k_\psi, \; \theta \rightarrow - \theta, \; u \rightarrow - u,
\end{equation}
we find that \cite{parra11c}
\begin{equation} \label{eq:positivecoefficients}
\fl \bun \cdot \nabla_\bR \theta \rightarrow \bun \cdot \nabla_\bR \theta, \; B \rightarrow B, \; \bk_\bot \cdot \bv_{Ms} \rightarrow \bk_\bot \cdot \bv_{Ms}, \; k_\bot \rightarrow k_\bot,
\end{equation}
and that
\begin{equation} \label{eq:negativecoefficients}
\bv_{Ms} \cdot \nabla \psi \rightarrow - \bv_{Ms} \cdot \nabla \psi, \; \bk_\bot \cdot \nabla \psi \rightarrow - \bk_\bot \cdot \nabla \psi. 
\end{equation}
Then, the long wavelength pieces of the distribution functions and the potential, determined by \eq{eq:lwfirstordergk} and \eq{eq:lwfirstorderquasineutrality}, satisfy the symmetries
\begin{equation} \label{eq:symmetryfslw1}
f_{s1}^\lw (\psi(\bR), \theta(\bR), u, \mu, t) = - f_{s1}^\lw (\psi(\bR),- \theta(\bR), - u, \mu, t)
\end{equation}
and
\begin{equation} \label{eq:symmetryphilw1}
\phi_1^\lw (\psi(\boldr), \theta(\boldr), t) = - \phi_1^\lw (\psi(\boldr), - \theta(\boldr), t).
\end{equation}
To see the effect of this symmetry on $\Pi_{-1}$, we rewrite the second term in \eq{eq:Piminusone}. Using
\begin{equation} \label{eq:Rzetatrick}
R \zun = \frac{I}{B} \bun - \frac{1}{B} \bun \times \nabla \psi 
\end{equation}
that can be deduced from \eq{eq:Bdef}, and taking into account that $\Delta f_{s1}^\lw$ in \eq{eq:DeltaFs1lwdef} is even in $v_{||}$, we find that the second term in \eq{eq:Piminusone} is
\begin{eqnarray}
\fl - \left \langle \sum_{s \neq e, s^\prime \neq e} \frac{m_s^2 I^2 c}{2Z_s e B^2} \int \dd^3v\, C_{ss^\prime}^{(\ell)} [f_{s1}^\lw; f_{s^\prime 1}^\lw] v_{||}^2 \right \rangle_\psi \nonumber\\ - \left \langle \sum_{s \neq e, s^\prime \neq e} \frac{m_s^2 c}{2Z_s e B^2} \int \dd^3v\, C_{ss^\prime}^{(\ell)} [\Delta f_{s1}^\lw; \Delta f_{s^\prime 1}^\lw] [(\bv \times \bun) \cdot \nabla \psi]^2 \right \rangle_\psi = 0.
\end{eqnarray}
The first term in this equation vanishes because of the symmetry in \eq{eq:symmetryfslw1}, and the second term vanishes because it is proportional to the average over the gyrophase $\varphi_0$ of $\bv_\bot \bv_\bot \bv_\bot$.

To prove that the first term of $\Pi_{-1}$ in \eq{eq:Piminusone} also vanishes for up-down symmetric tokamaks, we rewrite it using that
\begin{eqnarray}
\fl \int \dd^3v\, \bv_\bot \exp \left ( \frac{i \bk_\bot \cdot (\bv \times \bun)}{\Omega_s} \right ) g (\boldr, v_{||}, \mu_0, t) = \nonumber\\ i \bun \times \bk_\bot \int \dd^3v\, \frac{v_\bot^2}{2\Omega_s} \frac{2 J_1 (\lambda_s)}{\lambda_s} g (\boldr, v_{||}, \mu_0, t)
\end{eqnarray}
for any function $g (\boldr, v_{||}, \mu_0, t)$ independent of gyrophase $\varphi_0$, and that
\begin{equation}
\fl \sqrt{b} \int_0^\infty \dd x\, x^2 J_0 (x\sqrt{b}) J_1(x \sqrt{b}) \exp \left ( - \frac{x^2}{2} \right ) = b (I_0 (b) - I_1(b)) \exp(-b).
\end{equation}
With these results and \eq{eq:Rzetatrick}, the  first term in \eq{eq:Piminusone} becomes
\begin{eqnarray} \label{eq:Piminusonetb}
\fl -\left \langle \left \langle \sum_{s \neq e} \sum_{k_\psi, k_\alpha} \frac{m_s I c}{B} i k_\alpha (\underline{\phi}_1^\tb)^\ast \int \dd^3v\, \underline{f}_{s1}^\tb J_0 (\lambda_s) v_{||} \right \rangle_\psi \right \rangle_t \nonumber\\ - \left \langle \left \langle \sum_{s \neq e} \sum_{k_\psi, k_\alpha} \frac{m_s c}{B} k_\alpha (\bk_\bot \cdot \nabla \psi) (\underline{\phi}_1^\tb)^\ast \int \dd^3v\, \underline{f}_{s1}^\tb \frac{v_\bot^2}{2\Omega_s} \frac{2 J_1 (\lambda_s)}{\lambda_s}\right \rangle_\psi \right \rangle_t \nonumber\\- \left \langle \left \langle \sum_{s \neq e} \sum_{k_\psi, k_\alpha} \frac{n_s m_s c^2}{B^2} k_\alpha (\bk_\bot \cdot \nabla \psi) |\underline{\phi}_1^\tb|^2 (\Gamma_0 (b_s) - \Gamma_1(b_s))\right \rangle_\psi \right \rangle_t,
\end{eqnarray}
where
\begin{equation}
\Gamma_1 (b_s) = I_1 (b_s) \exp( - b_s).
\end{equation}
The result in \eq{eq:Piminusonetb} vanishes due to the symmetry in \eq{eq:transformationsymmetry}, \eq{eq:positivecoefficients} and \eq{eq:negativecoefficients}. The reason is that in an up-down symmetric tokamak, for every solution to equations \eq{eq:tbfirstordergk} and \eq{eq:tbfirstorderquasineutrality},
\begin{equation} \label{eq:solutiontb1}
\fl \underline{f}_{s1}^\tb (k_\psi, k_\alpha, \psi (\bR), \theta(\bR), u, \mu, t), \; \underline{\phi}_1^\tb (k_\psi, k_\alpha, \psi(\boldr), \theta(\boldr), t),
\end{equation}
we can form another solution by using the symmetry in \eq{eq:transformationsymmetry}, \eq{eq:positivecoefficients} and \eq{eq:negativecoefficients}, i.e.,
\begin{equation} \label{eq:solutiontb2}
\fl - \underline{f}_{s1}^\tb (- k_\psi, k_\alpha, \psi (\bR), - \theta(\bR), - u, \mu, t), \; - \underline{\phi}_1^\tb (- k_\psi, k_\alpha, \psi(\boldr), - \theta(\boldr), t)
\end{equation}
is also a solution to equations \eq{eq:tbfirstordergk} and \eq{eq:tbfirstorderquasineutrality}. These two solutions have opposite momentum flux, that is, the quantity in \eq{eq:Piminusonetb} evaluated using the solution \eq{eq:solutiontb2} is equal in magnitude but opposite in sign to the quantity in \eq{eq:Piminusonetb} evaluated using the solution \eq{eq:solutiontb1}. These solutions differ only because they correspond to two different initial conditions. After time averaging, the flux cannot depend on the initial condition, so it must be that the momentum flux is equal to its negative, that is, it vanishes (see \cite{parra11c} for numerical evidence).

Therefore, in up-down symmetric tokamaks, we need to keep higher order terms, as explained in the introduction. For extreme up-down asymmetry, $\Pi_{-1}$ is the dominant contribution to the momentum flux \cite{camenen09b, camenen09c, ball14}.

\section{Expansion in $B_p/B \ll 1$} \label{sec:expansion}

The equations for the second order pieces of the distribution function and the electrostatic potential given in subsections \ref{sub:lwsecondorder} and \ref{sub:tbsecondorder} are not only difficult to implement, but the different physical effects in them are not apparent. To simplify these equations we assume that the poloidal component of the magnetic field is much smaller than the magnetic field itself, $B_p/B \ll 1$. In previous work \cite{parra10a, parra11d}, the expansion $B_p/B \ll 1$ was performed assuming that the turbulence characteristics did not depend strongly on $B_p/B \ll 1$. In this section we relax this assumption by allowing the turbulent eddy characteristic length to vary between the ion gyroradius and the ion poloidal gyroradius. Importantly, for turbulence with characteristic lengths much smaller than an ion poloidal gyroradius, the lowest order gyrokinetic equations satisfy a new nonlinear symmetry different from the one in \cite{parra11c, sugama11a}. This symmetry is discussed in paragraph~\ref{subsub:gyroradiusturbulence}.

In this section, we assume that $B_p/B = r/qR \ll 1$ and $r/R \sim 1$. This choice may seem to imply that we always take the safety factor $q$ much larger than unity, but this is not the case. Using this ordering, it is possible to take a subsidiary expansion in $r/R \ll 1$ in which $B_p/B \ll 1$ and $q \sim 1$. This subsidiary expansion is appropriate for the plasma around the magnetic axis.

In this section we study the implications that $B_p/B \ll 1$ has for the first order equations. The scaling of the long wavelength components with $B_p/B \ll 1$ is described in subsection \ref{sub:lwfirstorderBpB}, and the scaling of the turbulent pieces is given in subsection \ref{sub:tbfirstorderBpB}. We use the results of these two subsections to simplify the second order equations in section \ref{sec:secondordertotal}.

\subsection{First order, long wavelength pieces} \label{sub:lwfirstorderBpB}
The size of the long wavelength first order pieces $f_{s1}^\lw$ and $\phi_1^\lw$ scales linearly with $B/B_p$. To see this, note that the first term on the right side of \eq{eq:lwfirstordergk} determines the size of $f_{s1}^\lw$, and it is of order $(\rho_s/a) (v_{ts}/a) f_{Ms}$. This term is balanced by the first term on the left side of \eq{eq:lwfirstordergk}, of order $(B_p/B) (v_{ts}/a) f_{s1}^\lw$. By making these two terms comparable, we find
\begin{equation} \label{eq:fs1lworderBpB}
\frac{f_{s1}^\lw}{f_{Ms}} \sim \frac{B}{B_p} \rho_\ast \gg \rho_\ast
\end{equation}
for $s \neq e$. For electrons, the piece of $f_{e1}^\lw$ even in $u$ is of order 
\begin{equation} \label{eq:fe1lwevenorderBpB}
\frac{f_{e1}^\lw(u) + f_{e1}^\lw(-u)}{f_{Me}} \sim \frac{B}{B_p} \rho_\ast \gg \rho_\ast,
\end{equation}
as we will see shortly, whereas the odd piece is
\begin{equation} \label{eq:fe1lwoddorderBpB}
\frac{f_{e1}^\lw(u) - f_{e1}^\lw(-u)}{f_{Me}} \sim \frac{B}{B_p} \frac{\rho_e}{a} \gg \frac{\rho_e}{a}.
\end{equation}
The size of $\phi_1^\lw$ is obtained from quasineutrality equation \eq{eq:lwfirstorderquasineutrality},
\begin{equation} \label{eq:phi1lworderBpB}
\frac{e\phi_1^\lw}{T_e} \sim \frac{f_{s1}^\lw}{f_{Ms}} \sim \frac{B}{B_p} \rho_\ast \gg \rho_\ast.
\end{equation}
Since the piece of $f_{e1}^\lw$ even in $u$ is just the adiabatic response (see \eq{eq:fe1evenadiabatic}), this result leads to \eq{eq:fe1lwevenorderBpB}. For $B_p/B \ll 1$, the long wavelength first order pieces of the distribution function and the potential are large by $B/B_p \gg 1$ because their size is related to the width of the drift orbits, of order $(B/B_p) \rho_s \gg \rho_s$.

\subsection{First order, turbulent pieces} \label{sub:tbfirstorderBpB}
To study the scaling of turbulence with $B_p/B \ll 1$, we distinguish between two turbulent regimes with different perpendicular length scales. Before doing so, we give some basic balances that determine the other turbulent characteristics once the perpendicular length scale of the turbulence is known.

We need to determine the scaling of $f_{s1}^\tb$, $\phi_1^\tb$, $k_\psi$, $k_\alpha$ and $l_{||}$ with $B_p/B$ at the outer scale of the turbulence, i.e., at the scale where most of the free energy is contained. Here $l_{||}$ is a measure of the characteristic parallel eddy size and $B_p R/k_\alpha B$ and $(k_\psi R B_p)^{-1}$ are the eddy characteristic lengths in the perpendicular directions $\alpha$ and $\psi$. We use four simple assumptions to relate all these quantities to $k_\bot \sim k_\alpha B/RB_p$. These assumptions are
\begin{enumerate}

\item balance between different terms in the quasineutrality equation \eq{eq:tbfirstorderquasineutrality}, giving
\begin{equation} \label{eq:qnbalance}
\frac{f_{s1}^\tb}{f_{Ms}} \sim \frac{f_{e1}^\tb(u) + f_{e1}^\tb(-u)}{f_{Me}} \sim \frac{e
\phi_1^\tb}{T_e};
\end{equation}

\item balance between the nonlinear term and the drive term proportional to the density and temperature gradients in \eq{eq:tbfirstordergk},
\begin{equation} \label{eq:nonlinearomegastarbalance}
c k_\psi k_\alpha \phi_1^\mathrm{tb} \frac{f_{s1}^\mathrm{tb}}{f_{Ms}} \sim c k_\psi k_\alpha \phi_1^\mathrm{tb} \frac{f_{e1}^\mathrm{tb} (u) + f_{e1}^\tb(-u)}{f_{Me}} \sim \frac{c k_\alpha}{aRB_p} \phi_1^\mathrm{tb},
\end{equation}
where we have assumed $\nabla \ln n_s \sim \nabla \ln T_s \sim a^{-1}$;

\item critical balance between the parallel streaming and the nonlinear terms in \eq{eq:tbfirstordergk},
\begin{eqnarray} \label{eq:criticalbalance}
\fl \frac{v_{ts}}{l_{||}} \frac{f_{s1}^\mathrm{tb}}{f_{Ms}} \sim \frac{v_{te}}{l_{||}} \frac{f_{e1}^\mathrm{tb}(u) - f_{e1}^\mathrm{tb}(-u)}{f_{Me}} \sim c k_\psi k_\alpha \phi_1^\mathrm{tb} \frac{f_{s1}^\mathrm{tb}}{f_{Ms}} \nonumber\\ \sim c k_\psi k_\alpha \phi_1^\mathrm{tb} \frac{f_{e1}^\mathrm{tb} (u) + f_{e1}^\tb(-u)}{f_{Me}};
\end{eqnarray}

\item and isotropy in the perpendicular direction,
\begin{equation} \label{eq:isotropy}
k_\psi RB_p \sim \frac{k_\alpha B}{RB_p} \sim k_\bot.
\end{equation}

\end{enumerate} 
Note that we have distinguished between ionic species, $s \neq e$, and electrons, and that we have considered the symmetry in $u$ of the different terms in \eq{eq:tbfirstordergk} and \eq{eq:tbfirstorderquasineutrality} to obtain equations for the pieces of $f_{e1}^\tb$ odd and even in $u$. These assumptions were numerically checked and used to successfully predict the scaling of turbulent characteristics with the safety factor $q$ and the temperature gradient in \cite{barnes11b}.

From \eq{eq:qnbalance}, \eq{eq:nonlinearomegastarbalance} and \eq{eq:isotropy}, we find that
\begin{equation} \label{eq:ionquantitiesvskbot}
\frac{e\phi_1^\tb}{T_e} \sim \frac{f_{s1}^\tb}{f_{Ms}} \sim \frac{f_{e1}^\tb(u) + f_{e1}^\tb(-u)}{f_{Me}} \sim \frac{1}{k_\bot a}.
\end{equation}
From \eq{eq:qnbalance}, \eq{eq:nonlinearomegastarbalance}, \eq{eq:criticalbalance} and \eq{eq:isotropy}, we obtain
\begin{equation} \label{eq:Deltathetavskbot}
\frac{l_{||}}{a} \sim \frac{1}{k_\bot \rho_i} \gg \rho_\ast
\end{equation}
and
\begin{equation} \label{eq:fe1tboddvskbot}
\frac{f_{e1}^\tb(u) - f_{e1}^\tb(-u)}{f_{Me}} \sim \sqrt{\frac{m_e}{m_i}} \frac{1}{k_\bot a}.
\end{equation}
We only have to determine $k_\bot$ to have the scaling of all the interesting turbulent quantities with $B_p/B$. We consider two different cases: when the perpendicular length scale of the turbulence is of the order of the poloidal gyroradius, described in paragraph \ref{subsub:poloidalturbulence}, and when the turbulent characteristic length scale is the gyroradius, discussed in \ref{subsub:gyroradiusturbulence}.

\subsubsection{Poloidal gyroradius scale turbulence.} \label{subsub:poloidalturbulence}
In \cite{barnes11b} it was shown that for certain turbulent regimes, the parallel extent of the eddies is determined by the finite size of the tokamak, i.e.
\begin{equation}
l_{||} \sim \frac{B}{B_p} a.
\end{equation}
As a result, equation \eq{eq:Deltathetavskbot} gives
\begin{equation} \label{eq:kbotorderpol}
k_\bot \sim \frac{B_p}{B} \frac{1}{\rho_i},
\end{equation}
and equations \eq{eq:ionquantitiesvskbot} and \eq{eq:fe1tboddvskbot} give
\begin{equation} \label{eq:phi1tborderpol}
\frac{e \phi_1^\tb}{T_e} \sim \frac{B}{B_p} \rho_\ast,
\end{equation}
\begin{equation} \label{eq:fs1tborderpol}
\frac{f_{s1}^\tb}{f_{Ms}} \sim \frac{B}{B_p} \rho_\ast,
\end{equation}
\begin{equation} \label{eq:fe1tbevenorderpol}
\frac{f_{e1}^\tb (u) + f_{e1}^\tb(-u)}{f_{Me}} \sim \frac{B}{B_p} \rho_\ast
\end{equation}
and
\begin{equation} \label{eq:fe1tboddorderpol}
\frac{f_{e1}^\tb (u) - f_{e1}^\tb(-u)}{f_{Me}} \sim \frac{B}{B_p} \frac{\rho_e}{a}.
\end{equation}
Note that due to critical balance \eq{eq:criticalbalance} and the finite size of the tokamak in the parallel direction, we have obtained that the turbulent eddies can reach perpendicular scales of the order of the poloidal gyroradius $(B/B_p) \rho_i$, and as a consequence, the fluctuating pieces of the distribution function and the electrostatic potential are large by a factor of order $B/B_p \gg 1$. Bigger eddies cause bigger perturbations.

\subsubsection{Gyroradius scale turbulence.} \label{subsub:gyroradiusturbulence}

The turbulent regime studied in \cite{barnes11b} is observed when the instability driving the turbulence extends to very long wavelengths \cite{yoo15}. There are circumstances in which the instability does not exist for $k_\bot$ smaller than some cut off value \cite{yoo15}. In cases like this, assumption \eq{eq:kbotorderpol} is not correct and the scalings change. We expect $k_\bot$ to satisfy
\begin{equation} \label{eq:kbotrange}
\frac{B_p}{B} \ll k_\bot \rho_i \lesssim 1. 
\end{equation}
We will obtain equations valid for this range of $k_\bot$. We call this type of turbulence gyroradius scale turbulence because we keep finite gyroradius effects in the equations to be able to describe turbulence with $k_\bot \rho_i \sim 1$. 

The size of the turbulent pieces of the distribution function and the electrostatic potential are given in equations \eq{eq:ionquantitiesvskbot} and \eq{eq:fe1tboddvskbot}. Importantly, equations \eq{eq:Deltathetavskbot} and \eq{eq:kbotrange} lead to
\begin{equation}
\frac{B_p}{B} \lesssim \frac{B_p}{B} \frac{l_{||}}{a} \sim  \frac{B_p}{B} \frac{1}{k_\bot \rho_i} \ll 1.
\end{equation}
Thus, the turbulence has a very small characteristic parallel length scale compared to the connection length $(B/B_p) a$. This estimate for $l_{||}$ implies
\begin{equation} \label{eq:falsefastddtheta}
\frac{\partial}{\partial \theta} \ln \underline{\phi}_1^\tb \sim \frac{\partial}{\partial \theta} \ln \underline{f}_{s1}^\tb \sim \frac{B}{B_p} \frac{a}{l_{||}} \sim \frac{B}{B_p} k_\bot \rho_i \gg 1.
\end{equation}
The effect of this fast variation in $\theta$ on the turbulent momentum flux is affected by two factors. On the one hand, the fast variation in $\theta$ is due to the small parallel length scales of the turbulence, and as a result, it affects the parallel component of the gradient and not the poloidal component, as we show in \eq{eq:gradperpgyro} below. On the other hand, the fast variation in $\theta$ introduces a new symmetry that the lowest order gyrokinetic equations satisfy. 

When $l_{||} \ll (B/B_p) a$, the dependence of $\underline{f}_{s1}^\tb$ and $\underline{\phi}_1^\tb$ on the position along the magnetic field line has two different characteristic lengths: one due to the turbulence, and the other due to the tokamak size. We choose
\begin{equation} \label{eq:zetaminusalpha}
\int^\theta \dd \theta^\prime\, \frac{I (\psi) \mathcal{J} (\psi, \theta^\prime)}{[R(\psi, \theta^\prime)]^2} = \zeta - \alpha
\end{equation}
instead of $\theta$ to describe the fast variation in the parallel direction for reasons that will become clear shortly. Using this new variable to describe the fast variation along the magnetic field line, the functions $\underline{\phi}_1^\tb$ and $\underline{f}_{s1}^\tb$ can be written as
\begin{equation} \label{eq:underlinephi1tbgyro}
\fl \underline{\phi}_1^\tb (k_\psi, k_\alpha, \psi(\boldr), \theta(\boldr), t) = \sum_{k_\zeta} \underline{\check{\phi}}_1^\tb (k_\psi, \check{k}_\alpha, k_\zeta, \psi (\boldr), \theta (\boldr), t ) \exp ( i k_\zeta (\zeta (\boldr) - \alpha(\boldr)) )
\end{equation}
and 
\begin{eqnarray} \label{eq:underlinefs1tbgyro}
\fl \underline{f}_{s1}^\tb (k_\psi, k_\alpha, \psi(\bR), \theta(\bR), u, \mu, t) = \sum_{k_\zeta} \underline{\check{f}}_{s1}^\tb (k_\psi, \check{k}_\alpha, k_\zeta, \psi (\bR), \theta (\bR), u, \mu, t ) \nonumber\\ \times \exp ( i k_\zeta (\zeta (\bR) - \alpha (\bR)) ).
\end{eqnarray}
The functions $\underline{\check{\phi}}_1^\tb (k_\psi, \check{k}_\alpha, k_\zeta, \psi (\boldr), \theta (\boldr), t )$ and  $\underline{\check{f}}_{s1}^\tb (k_\psi, \check{k}_\alpha, k_\zeta, \psi (\bR), \theta (\bR), u, \mu, t )$ contain the slow dependence on $\theta$ due to the different characteristics of the turbulence at different poloidal locations, i.e., 
\begin{equation} \label{eq:truefastddtheta}
\frac{\partial}{\partial \theta} \ln \underline{\check{\phi}}_1^\tb \sim \frac{\partial}{\partial \theta} \ln \underline{\check{f}}_{s1}^\tb \sim 1,
\end{equation}
whereas the wavenumber 
\begin{equation}
k_\zeta \sim \frac{R}{l_{||}} 
\end{equation}
represents the large parallel gradients of the turbulence. Note that we are considering the functions $\underline{\check{\phi}}_1^\tb (k_\psi, \check{k}_\alpha, k_\zeta, \psi (\boldr), \theta (\boldr), t )$ and  $\underline{\check{f}}_{s1}^\tb (k_\psi, \check{k}_\alpha, k_\zeta, \psi (\bR), \theta (\bR), u, \mu, t )$ as dependent on
\begin{equation}
\check{k}_\alpha = k_\alpha - k_\zeta
\end{equation}
instead of $k_\alpha$ because the new symmetry of the equations will be clearer in $\check{k}_\alpha$. Substituting \eq{eq:underlinephi1tbgyro} and \eq{eq:underlinefs1tbgyro} into equations \eq{eq:phitbscales} and \eq{eq:ftbscales}, we find
\begin{equation} \label{eq:phi1tbgyro}
\fl \phi_1^\tb = \sum_{k_\psi, \check{k}_\alpha, k_\zeta} \underline{\check{\phi}}_1^\tb (k_\psi, \check{k}_\alpha, k_\zeta, \psi (\boldr), \theta (\boldr), t ) \exp ( i k_\psi \psi(\boldr) + i \check{k}_\alpha \alpha (\boldr) + i k_\zeta \zeta (\boldr) )
\end{equation}
and 
\begin{eqnarray} \label{eq:fs1tbgyro}
\fl f_{s1}^\tb = \sum_{k_\psi, \check{k}_\alpha, k_\zeta} \underline{\check{f}}_{s1}^\tb (k_\psi, \check{k}_\alpha, k_\zeta, \psi (\bR), \theta (\bR), u, \mu, t ) \nonumber\\ \times \exp ( i k_\psi \psi(\bR) + i \check{k}_\alpha \alpha (\bR) + i k_\zeta \zeta (\bR) )
\end{eqnarray}
The Fourier decompositions \eq{eq:phi1tbgyro} and \eq{eq:fs1tbgyro} are not meant to be implemented in a code; they are just a way to analyze turbulence with short characteristic parallel lengths. The form \eq{eq:underlinephi1tbgyro} for $\underline{\phi}_1^\tb$ that in turn leads to the decomposition \eq{eq:phi1tbgyro} is chosen such that the gradients of $\phi_1^\tb$ satisfy
\begin{eqnarray} \label{eq:gradpargyro}
\fl \bun \cdot \nabla \phi_1^\tb = \sum_{k_\psi, \check{k}_\alpha, k_\zeta} \Bigg ( \frac{i k_\zeta I}{R^2 B} \underline{\check{\phi}}_1^\tb + \bun \cdot \nabla \theta \frac{\partial  \underline{\check{\phi}}_1^\tb}{\partial \theta} \Bigg ) \exp ( i k_\psi \psi + i \check{k}_\alpha \alpha + i k_\zeta \zeta) = \nonumber\\  \sum_{k_\psi, \check{k}_\alpha, k_\zeta} \frac{i k_\zeta I}{R^2 B} \underline{\check{\phi}}_1^\tb  \exp ( i k_\psi \psi + i \check{k}_\alpha \alpha + i k_\zeta \zeta) + O \left ( \frac{B_p}{Ba} \phi_1^\tb \right ) \sim \frac{\phi_1^\tb}{l_{||}} 
\end{eqnarray}
and
\begin{eqnarray} \label{eq:gradperpgyro}
\fl \nabla_\bot \phi_1^\tb = \sum_{k_\psi, \check{k}_\alpha, k_\zeta} \Bigg ( i \check{\bk}_\bot \underline{\check{\phi}}_1^\tb + \nabla \psi \frac{\partial \underline{\check{\phi}}_1^\tb}{\partial \psi} + \nabla_\bot \theta \frac{\partial \underline{\check{\phi}}_1^\tb}{\partial \theta} - \frac{i k_\zeta \underline{\check{\phi}}_1^\tb}{R^2 B} \bun \times \nabla \psi  \Bigg )  \nonumber\\ \times \exp ( i k_\psi \psi + i \check{k}_\alpha \alpha + i k_\zeta \zeta) = \sum_{k_\psi, \check{k}_\alpha, k_\zeta} \Bigg ( i \check{\bk}_\bot  \underline{\check{\phi}}_1^\tb + \nabla \psi \frac{\partial \underline{\check{\phi}}_1^\tb}{\partial \psi} \nonumber\\ + \nabla_\bot \theta \frac{\partial \underline{\check{\phi}}_1^\tb}{\partial \theta}  \Bigg )  \exp ( i k_\psi \psi + i \check{k}_\alpha \alpha + i k_\zeta \zeta) + O \left ( \frac{B_p}{B l_{||}} \phi_1^\tb \right ) \sim k_\bot \phi_1^\tb,
\end{eqnarray}
where
\begin{equation} \label{eq:checkkbot}
\check{\bk}_\bot = k_\psi \nabla \psi + \check{k}_\alpha \nabla \alpha \simeq \bk_\bot.
\end{equation}
The form \eq{eq:underlinefs1tbgyro} for $\underline{f}_{s1}^\tb$ gives similar expressions for the gradients of $f_{s1}^\tb$. As mentioned in \eq{eq:zetaminusalpha}, we chose $\zeta - \alpha$ to describe the short parallel wavelength of the turbulence. The reason is clear in \eq{eq:gradperpgyro} where the term containing $k_\zeta$ is smaller than the terms containing the slow derivatives $\partial/\partial \psi$ and $\partial/\partial \theta$ by a factor of $(B_p/B)(a/l_{||}) \ll 1$ because $\nabla \zeta$ is very close to being parallel to $\bun$ when $B_p/B \ll 1$. Equations \eq{eq:gradpargyro} and \eq{eq:gradperpgyro} show that the correction to $\nabla_\bot$ due to $\partial/\partial \theta$ is of order $a^{-1}$ even when $\bun \cdot \nabla$ is of order $l_{||}^{-1} \gg (B_p/B) a^{-1}$. The new fast dependence on $\theta$ introduced in \eq{eq:underlinephi1tbgyro} and \eq{eq:underlinefs1tbgyro} is due to short parallel length scales. For this reason, it should not affect the poloidal gradient as much as it does the parallel gradient. In equation \eq{eq:gradperpgyro}, the large poloidal component of the gradient due to $k_\zeta$ has been absorbed into the lowest order perpendicular component of the gradient by defining the new wavevector $\check{\bk}_\bot$ in \eq{eq:checkkbot}.

The Fourier decompositions \eq{eq:phi1tbgyro} and \eq{eq:fs1tbgyro} not only give the right order of the correction to $\nabla_\bot$ due to $i k_\zeta$ and $\partial/\partial \theta$, but they show that there is a new symmetry in the system in addition to the symmetry described in section \ref{sec:symmetry}. Using \eq{eq:gradpargyro}, the parallel streaming term becomes
\begin{equation} \label{eq:parstreamgyro}
\fl u \bun \cdot \nabla_\bR f_{s1}^\tb- \mu \bun \cdot \nabla_\bR B \frac{\partial f_{s1}^\tb}{\partial u} \simeq \sum_{k_\psi, \check{k}_\alpha, k_\zeta} \frac{i k_\zeta I u}{R^2 B} \underline{\check{f}}_{s1}^\tb \exp ( i k_\psi \psi + i \check{k}_\alpha \alpha + i k_\zeta \zeta ),
\end{equation}
and equation \eq{eq:tbfirstordergk} for ionic species ($s \neq e$) simplifies to
\begin{eqnarray} \label{eq:tbfirstorderiongyro}
\fl \frac{\partial \underline{\check{f}}_{s1}^\tb}{\partial t} + \frac{i k_\zeta I u}{R^2 B} \underline{\check{f}}_{s1}^\tb + \left ( - i \check{k}_\alpha c \frac{\partial \phi_0}{\partial \psi} + i \check{\bk}_\bot \cdot \bv_{Ms} \right ) \underline{\check{f}}_{s1}^\tb - \sum_{s^\prime \neq e} C_{ss^\prime}^{GK} \left [ \underline{\check{f}}_{s1}^\tb; \underline{\check{f}}_{s^\prime 1}^\tb \right ]  \nonumber\\ + \{ \underline{\check{\phi}}_1^\tb J_0 (\check{\Lambda}_s), \underline{\check{f}}_{s1}^\tb \} = - f_{Ms} \Bigg [ \frac{Z_s e}{T_s} \left ( \frac{i k_\zeta I u}{R^2 B} + i \check{\bk}_\bot \cdot \bv_{Ms} \right ) \nonumber\\ + i \check{k}_\alpha c \Bigg ( \frac{\partial}{\partial \psi} \ln n_s  + \left ( \frac{m_s (u^2 + 2\mu B)}{2T_s} - \frac{3}{2} \right ) \frac{\partial}{\partial \psi} \ln T_s \Bigg ) \Bigg ] \underline{\check{\phi}}_1^\tb J_0 (\check{\Lambda}_s) \nonumber\\ + \sum_{s^\prime \neq e} C_{ss^\prime}^{GK} \left [ \frac{Z_s e \underline{\check{\phi}}_1^\tb}{T_s} J_0 (\check{\Lambda}_s) f_{Ms}; \frac{Z_{s^\prime} e \underline{\check{\phi}}_1^\tb}{T_{s^\prime}} J_0 (\check{\Lambda}_{s^\prime}) f_{Ms^\prime} \right ] .
\end{eqnarray}
Here $\check{\Lambda}_s$ and the collision operator $C_{ss^\prime}^{GK}$ are as defined in \eq{eq:Lambdasdef} and \eq{eq:gkcollision1}, but with $k_\alpha$ replaced by $\check{k}_\alpha$. The nonlinear term $\{ \underline{\check{\phi}}_1^\tb J_0 (\check{\Lambda}_s), \underline{\check{f}}_{s1}^\tb \}$ is slightly different from the one in \eq{eq:nonlinearity} because it has to be rewritten in terms of $\check{k}_\alpha$ and $k_\zeta$,
\begin{eqnarray} \label{eq:nonlinearlitygyro}
\fl \{ \underline{\check{\phi}}_1^\tb J_0 (\check{\Lambda}_s), \underline{\check{f}}_{s1}^\tb \} = c \sum_{k_\psi^\prime, \check{k}_\alpha^\prime, k_\zeta} ( k_\psi^\prime \check{k}_\alpha - \check{k}_\alpha^\prime k_\psi) \underline{\check{\phi}}_1^\tb ( k_\psi^\prime, \check{k}_\alpha^\prime, k_\zeta^\prime, \psi(\bR), \theta(\bR), t) J_0 (\check{\Lambda}_s^\prime) \nonumber\\ \times \underline{\check{f}}_{s1}^\tb ( k_\psi - k_\psi^\prime, \check{k}_\alpha - \check{k}_\alpha^\prime, k_\zeta - k_\zeta^\prime, \psi(\bR), \theta(\bR), u, \mu, t ).
\end{eqnarray}
Electrons need to be treated independently in this limit of short parallel lengths. We need to order $B_p/B$ with respect to $\sqrt{m_e/m_i}$. We assume 
\begin{equation} \label{eq:sqrtmassBpBorder}
k_\bot \rho_i \sqrt{\frac{m_e}{m_i}} \frac{B}{B_p} \ll 1
\end{equation}
because $\sqrt{m_e/m_i} (B/B_p)$ is usually around 0.1. Assumption \eq{eq:sqrtmassBpBorder} implies that the characteristic time scale of an electron orbit, $(B/B_p) (a/v_{te})$, is much shorter than the turbulent time scale, $(k_\bot \rho_i)^{-1} (a/v_{ti})$ (we estimate this time from the nonlinear term using \eq{eq:ionquantitiesvskbot}). By expanding the Fokker Planck equation \eq{eq:tbfirstordergk} for electrons ($s = e$) in $\sqrt{m_e/m_i} \ll 1$, and using the lowest order expression \eq{eq:parstreamgyro} for the parallel streaming term, one can show that for $k_\zeta \neq 0$,
\begin{equation} \label{eq:fe1tbsolutiongyrokzetaneq0}
\underline{\check{f}}_{e1}^\tb = \frac{e \underline{\check{\phi}}_1^\tb}{T_e} f_{Me}.
\end{equation}
For $k_\zeta = 0$, we use the next order correction to the parallel gradient in \eq{eq:gradpargyro} and assumption \eq{eq:sqrtmassBpBorder} to find that
\begin{equation} \label{eq:fe1tbsolutiongyrokzetaeq0}
\underline{\check{f}}_{e1}^\tb (k_\zeta = 0) = \underline{\check{h}}_{e1}^\tb + \frac{e \underline{\check{\phi}}_1^\tb (k_\zeta = 0)}{T_e} f_{Me},
\end{equation}
where $\underline{\check{h}}_{e1}^\tb (k_\psi, \check{k}_\alpha, \psi(\bR), \varepsilon, \mu, \sigma, t)$ is independent of $k_\zeta$, and it is constant along the lowest order particle trajectories that are the characteristics of the operator $u \bun \cdot \nabla_\bR - \mu \bun \cdot \nabla_\bR B (\partial/\partial u)$, i.e., $\underline{\check{h}}_{e1}^\tb (k_\psi, \check{k}_\alpha, \psi(\bR), \varepsilon, \mu, \sigma, t)$ is independent of $\theta$ when written as a function of $\psi$, the kinetic energy 
\begin{equation}
\varepsilon = \frac{u^2}{2} + \mu B(\bR),
\end{equation}
the magnetic moment $\mu$ and the sign of the parallel velocity $\sigma = u/|u|$. Finally, the function $\underline{\check{h}}_{e1}^\tb$ can be determined by going to next order in $k_\bot \rho_i \sqrt{m_e/m_i} (B/B_p)$ in equation \eq{eq:tbfirstordergk} and orbit averaging the equation. For a function $h$ of phase space, the orbit average is
\begin{equation} \label{eq:orbitavedef}
\fl \langle h \rangle_\tau = \left( \oint \frac{\dd \theta}{\sigma \sqrt{2 (\varepsilon - \mu B(\psi, \theta))} \bun \cdot \nabla_\bR \theta} \right )^{-1} \oint \frac{h (\psi, \theta, \alpha, \varepsilon, \mu, \sigma)}{\sigma \sqrt{2 (\varepsilon - \mu B(\psi, \theta))} \bun \cdot \nabla_\bR \theta} \dd \theta.
\end{equation}
Note that $h$ has been written as a function of $\psi$, $\theta$, $\alpha$, $\varepsilon$, $\mu$ and $\sigma$ before integrating over $\theta$. The same has been done with $u = \sigma \sqrt{2(\varepsilon - \mu B(\psi, \theta))}$. The integral over $\theta$ is between $0$ and $2\pi$ for passing particles, and between the two bounce points and summing over the two directions of velocity ($\sigma = +1$ and $\sigma = -1$) for trapped particles. After orbit averaging, equation \eq{eq:tbfirstordergk} for electrons ($s = e$) becomes
\begin{eqnarray} \label{eq:tbfirstorderelectrongyro}
\fl \left ( \frac{\partial}{\partial t} - i \check{k}_\alpha c \frac{\partial \phi_0}{\partial \psi} \right )\left ( \underline{\check{h}}_{e1}^\tb + \frac{e \langle \underline{\check{\phi}}_1^\tb (k_\zeta = 0) \rangle_\tau}{T_e} f_{Me} \right ) + i \langle \check{\bk}_\bot \cdot \bv_{Ms} \rangle_\tau \underline{\check{h}}_{e1}^\tb \nonumber\\  - \left \langle C_{ee}^{(\ell)} \left [ \underline{\check{h}}_{e1}^\tb \right ] \right \rangle_\tau - \sum_{s^\prime \neq e} \left \langle C_{es^\prime}^{(\ell)} \left [ \underline{\check{h}}_{e1}^\tb; \underline{[\check{f}_{s^\prime}]}_1^\tb (k_\zeta = 0) \right ] \right \rangle_\tau \nonumber\\ + \{ \langle \underline{\check{\phi}}_1^\tb (k_\zeta = 0) \rangle_\tau , \underline{\check{h}}_{e1}^\tb \} = - f_{Me} \, i \check{k}_\alpha c \Bigg [ \frac{\partial}{\partial \psi} \ln n_e \nonumber\\ + \left ( \frac{m_e (u^2 + 2\mu B)}{2T_e} - \frac{3}{2} \right ) \frac{\partial}{\partial \psi} \ln T_e \Bigg ] \langle \underline{\check{\phi}}_1^\tb (k_\zeta = 0) \rangle_\tau,
\end{eqnarray}
where there is no sum over $k_\zeta$ in the nonlinear term,
\begin{eqnarray} \label{eq:nonlinearlityelectrongyro}
\fl \{ \langle \underline{\check{\phi}}_1^\tb (k_\zeta = 0) \rangle_\tau , \underline{\check{h}}_{e1}^\tb \} = c \sum_{k_\psi^\prime, \check{k}_\alpha^\prime} ( k_\psi^\prime \check{k}_\alpha - \check{k}_\alpha^\prime k_\psi) \langle \underline{\check{\phi}}_1^\tb ( k_\psi^\prime, \check{k}_\alpha^\prime, k_\zeta = 0, \psi(\bR), \theta(\bR), t) \rangle_\tau \nonumber\\ \times \underline{\check{h}}_{e1}^\tb ( k_\psi - k_\psi^\prime, \check{k}_\alpha - \check{k}_\alpha^\prime, \psi(\bR), \varepsilon, \mu, \sigma, t ),
\end{eqnarray}
and $\underline{[\check{f}_{s^\prime}]}^\tb_1$ is defined analogously to $\underline{[f_{s^\prime}]}^\tb_1$ in \eq{eq:underlinefswigT1},
\begin{eqnarray}
\fl \underline{[\check{f}_{s^\prime}]}^\tb_1 = \Bigg ( \underline{\check{f}}_{s^\prime 1}^\tb + \frac{Z_{s^\prime} e \underline{\check{\phi}}_1^\tb}{T_i} J_0 (\check{\lambda}_{s^\prime}) f_{Ms^\prime} \Bigg )\exp \left ( \frac{i\check{\bk}_\bot \cdot ( \bv \times \bun )}{\Omega_{s^\prime}} \right ) - \frac{Z_{s^\prime} e \underline{\check{\phi}}_1^\tb}{T_i} f_{Ms^\prime}.
\end{eqnarray}
Here $\check{\lambda}_s$ is as defined in \eq{eq:lambdasdef}, but with $k_\alpha$ replaced by $\check{k}_\alpha$. Equations \eq{eq:tbfirstorderiongyro} and \eq{eq:tbfirstorderelectrongyro} have to be solved along with the quasineutrality equation 
\begin{equation} \label{eq:tbfirstorderqngyro}
2 \pi \sum_s Z_s \int \dd v_{||}\, \dd \mu_0\, B J_0 (\check{\lambda}_s) \underline{\check{f}}_{s1}^\tb - \sum_{s} \frac{Z_s^2 e \underline{\check{\phi}}_1^\tb }{T_s} n_s ( 1 - \Gamma_0(\check{b}_s)) = 0,
\end{equation}
that can be derived from \eq{eq:tbfirstorderquasineutrality}. Here $\check{b}_s$ is as defined in \eq{eq:bsdef}, but with $k_\alpha$ replaced by $\check{k}_\alpha$.

One of the consequences of the new equations \eq{eq:tbfirstorderiongyro} and \eq{eq:tbfirstorderelectrongyro}, which have to be solved along with the quasineutrality equation in \eq{eq:tbfirstorderqngyro}, is that the symmetry described in section \ref{sec:symmetry} is not unique. It can be split into two new symmetries, namely, if we have solutions $\underline{\check{f}}_{s1}^\tb (k_\psi, \check{k}_\alpha, k_\zeta, \psi(\bR), \theta(\bR), u, \mu, t)$ and $\underline{\check{\phi}}_1^\tb (k_\psi, \check{k}_\alpha, k_\zeta, \psi(\boldr), \theta(\boldr), t)$ to \eq{eq:tbfirstorderiongyro}, \eq{eq:tbfirstorderelectrongyro} and \eq{eq:tbfirstorderqngyro}, the transformation
\begin{equation} \label{eq:transformationsymmetrygyrovpar}
k_\zeta \rightarrow - k_\zeta, \; u \rightarrow - u
\end{equation}
applied to \eq{eq:tbfirstorderiongyro}, \eq{eq:tbfirstorderelectrongyro} and \eq{eq:tbfirstorderqngyro} gives that 
\begin{equation} \label{eq:symmetrygyrovpar}
\underline{\check{f}}_{s1}^\tb (k_\psi, \check{k}_\alpha, - k_\zeta, \psi(\bR), \theta(\bR), - u, \mu, t), \; \underline{\check{\phi}}_1^\tb (k_\psi, \check{k}_\alpha, - k_\zeta, \psi(\boldr), \theta(\boldr), t)
\end{equation}
are also solutions, and the transformation
\begin{equation} \label{eq:transformationsymmetrygyrokpsi}
k_\psi \rightarrow - k_\psi, \; \theta \rightarrow - \theta
\end{equation}
applied to the same equations gives that 
\begin{equation} \label{eq:symmetrygyrokpsi}
\fl - \underline{\check{f}}_{s1}^\tb (- k_\psi, \check{k}_\alpha, k_\zeta, \psi(\bR), - \theta(\bR), u, \mu, t), \; - \underline{\check{\phi}}_1^\tb (- k_\psi, \check{k}_\alpha, k_\zeta, \psi(\boldr), - \theta(\boldr), t)
\end{equation}
are solutions as well. The difference between the symmetry in \eq{eq:transformationsymmetrygyrovpar} and \eq{eq:symmetrygyrovpar}, and the symmetry in \eq{eq:solutiontb1} and \eq{eq:solutiontb2} is that we do not need to reverse the parameter $\theta$ that describes the slow dependence of the turbulence on the position within the flux surface. As a result, the symmetry in \eq{eq:transformationsymmetrygyrovpar} and \eq{eq:symmetrygyrovpar} does not depend on the flux surface being up-down symmetric. In the symmetry in \eq{eq:transformationsymmetrygyrokpsi} and \eq{eq:symmetrygyrokpsi}, we do not need to reverse $u$. These new symmetries are only valid to lowest order in $B_p/B \ll 1$, but they will affect our estimates for different intrinsic rotation generation mechanisms.

\section{Final second order equations and momentum flux} \label{sec:secondordertotal}

In this section, we simplify the equations presented in subsections \ref{sub:lwsecondorder} and \ref{sub:tbsecondorder} and the momentum flux given in section \ref{sec:radialflux} using the ordering assumptions described in section \ref{sec:expansion}. The equations given in this section are different from the ones in \cite{parra10a, parra11d} because we have included sources of particles and energy, we have considered several ion species, and the assumptions for the turbulence are different (i.e., we allow eddies of the order of the ion poloidal gyroradius, whereas in \cite{parra10a, parra11d} the turbulent eddies where assumed to be of the order of the ion gyroradius). The size estimates for the different terms are also different from the ones in \cite{parra10a, parra11d}. The new symmetry in \eq{eq:transformationsymmetrygyrovpar} and \eq{eq:symmetrygyrovpar} is partially responsible for these different estimates.

From the orderings given in subsection \ref{sub:tbfirstorderBpB}, the eddy turnover time (the size of the time derivative of the turbulent pieces) is 
\begin{equation} \label{eq:eddytimetotal}
\frac{\partial}{\partial t} \ln \phi^\tb \sim k_\bot \rho_i \frac{v_{ti}}{a}, 
\end{equation}
and we will see in subsection \ref{sub:lwsecondordertotal} that the transport time becomes
\begin{equation} \label{eq:tauEtbtotal}
\frac{\partial}{\partial t} \ln \phi^\lw \sim \frac{1}{\tau_E} \sim \frac{1}{k_\bot \rho_i} \rho_\ast^2 \frac{v_{ti}}{a}
\end{equation}
if turbulence dominates, and
\begin{equation} \label{eq:tauEnctotal}
\frac{\partial}{\partial t} \ln \phi^\lw \sim \frac{1}{\tau_E} \sim \frac{B^2}{B_p^2} \rho_\ast^2 \nu_{ii}
\end{equation}
if neoclassical transport dominates. Turbulent transport dominates for $a \nu_{ii}/v_{ti} \lesssim (k_\bot \rho_i)^{-1} (B_p/B)^2$. Due to \eq{eq:eddytimetotal}, \eq{eq:tauEtbtotal} and \eq{eq:tauEnctotal}, we need to adjust slightly our assumptions about sources, \eq{eq:Qordering}, collisionality, \eq{eq:collisionorder1}, and $\sqrt{m_e/m_i}$, \eq{eq:massorder}. We assume 
\begin{equation} \label{eq:Qorderingtotal}
Q_s \sim \frac{1}{k_\bot \rho_i} \rho_\ast^2 \frac{v_{ti}}{a} f_{Ms},
\end{equation}
where we have imposed that the sources must give a characteristic time comparable to the turbulent time scale \eq{eq:tauEtbtotal},
\begin{equation} \label{eq:collisionorder1total}
\frac{B}{B_p} \rho_\ast^2 \ll \frac{\nu_{ii} a}{v_{ti}} \sim \frac{\nu_{zi} a}{v_{tz}} \sim \frac{\nu_{ee} a}{v_{te}} \sim \frac{\nu_{ei} a}{v_{te}} \lesssim \frac{B_p^2}{B^2} \frac{1}{k_\bot \rho_i},
\end{equation}
where we have imposed $a \nu_{ii}/v_{ti} \lesssim (k_\bot \rho_i)^{-1} (B_p/B)^2$ to make turbulent transport dominate over neoclassical transport, and
\begin{equation} \label{eq:massordertotal}
\sqrt{\frac{m_e}{m_i}} \sim \sqrt{\frac{m_e}{m_z}} \sim \frac{1}{k_\bot \rho_i} \frac{v_{ti}}{a \nu_{ii}} \rho_\ast^2 \ll \frac{B_p}{B} \frac{1}{k_\bot \rho_i}.
\end{equation}
Note that the upper bound of \eq{eq:massordertotal} and the lower bound of \eq{eq:collisionorder1total} have been chosen to be consistent with each other and to imply that the electron characteristic orbit time, $(B/B_p)(a/v_{te})$, is much shorter than the eddy turnover time (see \eq{eq:sqrtmassBpBorder}).

To simplify the equations in subsections \ref{sub:lwsecondorder} and \ref{sub:tbsecondorder}, we use that according to \eq{eq:fs1lworderBpB} - \eq{eq:phi1lworderBpB}, \eq{eq:ionquantitiesvskbot} and \eq{eq:fe1tboddvskbot}, $\phi_1^\lw$ and $f_{s1}^\lw$ scale as $B/B_p$, becoming large for $B_p/B \ll 1$, and $\underline{\phi}_1^\tb$ and $\underline{f}_{s1}^\tb$ scale as $(k_\bot \rho_i)^{-1}$, becoming large for $k_\bot \rho_i \ll 1$, and comparable to $\phi_1^\lw$ and $f_{s1}^\lw$ for poloidal gyroradius scale turbulence, $k_\bot \rho_i \sim B_p/B$. Conversely the corrections $u_1$, $\mu_1$, $\varphi_1$, $\bR_2$, $u_2$ and $\mu_2$ to the gyrokinetic variables, and the corrections $\dot{\bR}_{s2}$ and $\dot{u}_{s2}$ to the drifts and the acceleration do not scale at all with $B_p$ or $k_\bot$. The reason why these pieces do not scale with $B_p$ is that the gyrokinetic expansion relies on the magnitude of the magnetic field being sufficiently large to make the gyroradius small compared to the characteristic length of the plasma, and as a result, the expansion is unaffected if the size of one of the components of the magnetic field is small. The size of $k_\bot$ does not matter as long as the size of the gradients of the potential are bounded as in \eq{eq:gradphiorder}. Thus, for $B_p/B \rightarrow 0$ or $k_\bot \rho_i \rightarrow 0$, the size of the corrections $u_1$, $\mu_1$, $\varphi_1$, $\bR_2$, $u_2$, $\mu_2$, $\dot{\bR}_{s2}$ and $\dot{u}_{s2}$ does not change. 

The pieces $\phi_1^\lw$ and $f_{s1}^\lw$ are the largest contributions unless $k_\bot \rho_i \sim B_p/B$, in which case $\underline{\phi}_1^\tb$ and $\underline{f}_{s1}^\tb$ are comparable. As a result, in general, terms that are quadratic in $\underline{\phi}_1^\tb$ and $\underline{f}_{s1}^\tb$ are only important when $k_\bot \rho_i \ll 1$, and we can neglect the finite gyroradius effects in them. In addition, the upper bound of \eq{eq:collisionorder1total} implies that collisions are small for $k_\bot \rho_i \gg B_p/B$, and consequently, when collisional terms are important, the finite gyroradius effects can be neglected, that is, in every collisional term we will neglect finite gyroradius corrections.

When $k_\bot \rho_i \gg B_p/B$, the turbulent pieces $f_{s1}^\tb$ and $\phi_1^\tb$ have characteristic parallel scale lengths small compared to the connection length, and we can use the forms \eq{eq:underlinephi1tbgyro} and \eq{eq:underlinefs1tbgyro} for the Fourier coefficients $\underline{f}_{s1}^\tb$ and $\underline{\phi}_1^\tb$. Then, the parallel and perpendicular gradients of $f_{s1}^\tb$ and $\phi_1^\tb$ are as given in \eq{eq:gradpargyro} and \eq{eq:gradperpgyro}, and the Fourier coefficients $\underline{\check{f}}_{s1}^\tb$ and $\underline{\check{\phi}}_1^\tb$ satisfy the symmetry in \eq{eq:transformationsymmetrygyrovpar} and \eq{eq:symmetrygyrovpar}. We will only use these properties of turbulence with $k_\bot \rho_i \gg B_p/B$ to estimate the size of different terms.

We give the equations for the long wavelength, second order pieces in subsection \ref{sub:lwsecondordertotal}, the equations for the turbulent, second order pieces in subsection \ref{sub:tbsecondordertotal}, and the formulas for the momentum flux in subsection \ref{sub:momentumfluxtotal}. To compare to derivations of gyrokinetic equations for sonic flows, it is convenient to have these equations in a frame rotating with speed $\Omega_{\zeta, E} = - c (\partial \phi_0/\partial \psi)$. We give these equations in \ref{app:equationsrotating}.

\subsection{Long wavelength, second order equations} \label{sub:lwsecondordertotal}
Applying the orderings discussed above to \eq{eq:lwsecondorderions}, we can neglect several terms. In particular, the terms proportional to $\dot{\bR}_{s2}$ and $\dot{u}_{s2}$ are small in $B_p/B \ll 1$. The function $\phiave_2^\lw$, defined in \eq{eq:phiaveRlw2}, becomes
\begin{equation}
\phiave_2^\lw \simeq \phi_2^\lw,
\end{equation} 
and $C_{ss^\prime,2}^\lw$, defined in \eq{eq:Cssprime2lw}, simplifies to
\begin{eqnarray}
C_{ss^\prime,2}^\lw \simeq C_{ss^\prime} \left [ f_{s1}^\lw, f_{s^\prime 1}^\lw \right ] + \left \langle \sum_{k_\psi, k_\alpha} C_{ss^\prime} \left [ (\underline{f}_{s1}^\tb)^\ast, \underline{f}_{s^\prime 1}^\tb \right ] \right \rangle_t.
\end{eqnarray}
To obtain this last expression, we have neglected the finite gyroradius corrections in the second term (see the discussion at the beginning of this section), and $\Delta f_{s2}^\lw \sim (B/B_p) \rho_\ast f_{Ms}$, defined in \eq{eq:DeltaFs2lwdef}, in the first term. With these approximations, we obtain
\begin{eqnarray} \label{eq:lwsecondorderionstotalv1}
\fl \left ( u \bun \cdot \nabla_\bR \theta \frac{\partial}{\partial \theta} - \mu \bun \cdot \nabla_\bR B \frac{\partial}{\partial u} \right ) \langle f_{s2}^\lw \rangle - \sum_{s^\prime \neq e} C_{ss^\prime}^{(\ell)} [ \langle f_{s2}^\lw \rangle; \langle f_{s^\prime 2}^\lw \rangle ] = - \frac{\partial f_{Ms}}{\partial t} \nonumber\\ - \frac{Z_s e f_{Ms}}{T_i} u \bun \cdot \nabla_\bR \phi_2^\lw + F_{s2, \nc}^\lw + F_{s2, \tb}^\lw + F_{s2, \Delta T}^\lw + F_{s2, Q}^\lw
\end{eqnarray}
for ionic species ($s \neq e$). We have divided the right side of \eq{eq:lwsecondorderionstotalv1} into different pieces with different physical origins,
\begin{eqnarray} \label{eq:Flwnctotal}
\fl F_{s2, \nc}^\lw = \left ( u \bun \cdot \nabla_\bR \theta \frac{\partial}{\partial \theta} - \mu \bun \cdot \nabla_\bR B \frac{\partial}{\partial u} \right ) \Delta h_{s2, \nc}^\lw \nonumber\\ + \bun \cdot \nabla_\bR \theta \frac{\partial}{\partial \psi} \left [ \frac{1}{\bun \cdot \nabla_\bR \theta}\frac{I u}{\Omega_s} \sum_{s^\prime \neq e} C_{ss^\prime}^{(\ell)} [ f_{s1}^\lw; f_{s^\prime 1}^\lw ] \right ] \nonumber\\ - \frac{\partial}{\partial u} \left [ \left ( \frac{c I}{B} \frac{\partial \phi_0}{\partial \psi} + \frac{ I \mu}{\Omega_s} \frac{\partial B}{\partial \psi} + \frac{Z_s e \phi_1^\lw}{m_s u} \right ) \sum_{s^\prime \neq e} C_{ss^\prime}^{(\ell)} [ f_{s1}^\lw; f_{s^\prime 1}^\lw ] \right ] \nonumber\\ + \sum_{s^\prime \neq e} C_{ss^\prime} \left [ f_{s1}^\lw, f_{s^\prime 1}^\lw \right ],
\end{eqnarray}
where $\Delta h_{s2, \nc}^\lw$ is defined in \eq{eq:Deltahs2nclwdef} (see discussion around \eq{eq:Flwnctotalv1} below),
\begin{eqnarray} \label{eq:Flwtbtotal}
\fl F_{s2, \tb}^\lw = - \left \langle \sum_{k_\psi, k_\alpha} \frac{c (\underline{\phi}_1^\tb)^\ast J_0 (\Lambda_s)}{B} i (\bk_\bot \times \bun) \cdot \left ( \nabla_\bR \psi \frac{\partial}{\partial \psi} + \nabla_\bR \theta \frac{\partial}{\partial \theta} \right ) \underline{f}_{s1}^\tb \right \rangle_t \nonumber\\ + \left \langle \sum_{k_\psi, k_\alpha} \frac{c (\underline{f}_{s1}^\tb)^\ast}{B} i (\bk_\bot \times \bun) \cdot \left ( \nabla_\bR \psi \frac{\partial}{\partial \psi} + \nabla_\bR \theta \frac{\partial}{\partial \theta} \right ) \underline{\phi}_1^\tb J_0 (\Lambda_s)\right \rangle_t \nonumber\\ + \frac{Z_s e}{m_s} \left \langle \sum_{k_\psi, k_\alpha} \frac{\partial (\underline{f}_{s1}^\tb)^\ast}{\partial u} \left ( \bun \cdot \nabla_{\bR} \theta \frac{\partial}{\partial \theta} + \frac{u}{\Omega_s} i \bk_\bot \cdot (\bun \times \kappabf) \right ) \underline{\phi}_1^\tb J_0 (\Lambda_s) \right \rangle_t \nonumber\\  + \sum_{s^\prime \neq e} \Bigg \langle \sum_{k_\psi, k_\alpha} C_{ss^\prime} \Bigg [ \underline{f}_{s1}^\tb, (\underline{f}_{s^\prime 1}^\tb)^\ast  \Bigg ] \Bigg \rangle_t,
\end{eqnarray}
\begin{eqnarray} \label{eq:FlwDeltaTtotal}
\fl F_{s2, \Delta T}^\lw = \frac{n_e m_e \nu_{es}}{n_s m_s} \left ( \frac{T_e}{T_i} - 1 \right ) \left ( \frac{m_s (u^2 + 2\mu B)}{T_i} - 3 \right ) f_{Ms} 
\end{eqnarray}
and
\begin{eqnarray} \label{eq:FlwQtotal}
\fl F_{s2, Q}^\lw = Q_s.
\end{eqnarray}
The subindex $\nc$ indicates that the piece is of neoclassical origin, the subindex $\tb$ that it is of turbulent origin, the subindex $\Delta T$ that it has to do with the temperature difference $T_i - T_e$, and $Q$ that it has to do with the sources $Q_s$. This separation of the right side of equation \eq{eq:lwsecondorderionstotalv1} into different pieces is based on the physical pictures given in section \ref{sec:pictures}. 

Two important comments about \eq{eq:lwsecondorderionstotalv1} are appropriate. First, expression \eq{eq:Flwnctotal} for $F_{s2, \nc}^\lw$ is the result of manipulating 
\begin{eqnarray} \label{eq:Flwnctotalv1}
\fl F_{s2, \nc}^\lw = - \left ( \bv_{Ms} - \frac{c}{B} \nabla_\bR \phi_0 \times \bun \right ) \cdot \nabla_\bR f_{s1}^\lw + \Bigg [ \frac{Z_s e}{m_s} \bun \cdot \nabla_\bR \phi_1^\lw \nonumber\\+ \frac{u}{\Omega_s} (\bun \times \kappabf) \cdot \left ( \mu \nabla_\bR B + \frac{Z_s e}{m_s} \nabla_\bR \phi_0 \right ) \Bigg ] \frac{\partial f_{s1}^\lw}{\partial u} \nonumber\\ + \frac{c}{B} (\nabla_\bR \phi_1^\lw \times \bun )\cdot \nabla_\bR \psi \Bigg [ \frac{\partial}{\partial \psi} \ln p_s \nonumber\\+ \left ( \frac{m_s ( u^2 + 2\mu B)}{2T_i} - \frac{5}{2} \right ) \frac{\partial}{\partial \psi} \ln T_i \Bigg ] f_{Ms} \nonumber\\  - \frac{Z_s e}{T_i} \bv_{Ms} \cdot \nabla_\bR \phi_1^\lw \, f_{Ms} + \sum_{s^\prime \neq e} C_{ss^\prime} \left [ f_{s1}^\lw, f_{s^\prime 1}^\lw \right ]
\end{eqnarray}
in the limit $B_p/B \ll 1$, as explained in \ref{app:Flwnc}. Second, the first three terms in $F_{s2, \tb}^\lw$ are the only terms quadratic in $\underline{\phi}_1^\tb$ and $\underline{f}_{s1}^\tb$ for which we need to keep finite gyroradius corrections. The reason is that even though $F_{s2, \tb}^\lw$ is small compared to $F_{s2, \nc}^\lw$ when $k_\bot \rho_i \gg B_p/B$ (in this case, $\underline{\phi}_1^\tb$ and $\underline{f}_{s1}^\tb$ are smaller than $\phi_1^\lw$ and $f_{s1}^\lw$), the piece of the distribution function $f_{s2, \tb}^\lw$ that $F_{s2, \tb}^\lw$ gives is large due to the upper bound that we have imposed on the collision frequency in \eq{eq:collisionorder1total} (see the estimates in table \ref{table:h2betatotal} and the discussion at the end of this subsection).

In equation \eq{eq:lwsecondorderionstotalv1}, one of the terms is
\begin{equation}
\frac{\partial f_{Ms}}{\partial t} = \left [ \frac{1}{n_s} \frac{\partial n_s}{\partial t} + \left ( \frac{m_s (u^2 + 2\mu B)}{2T_i} - \frac{3}{2} \right ) \frac{1}{T_i} \frac{\partial T_i}{\partial t} \right ] f_{Ms}.
\end{equation}
The equation for $\partial n_s/\partial t$ is obtained by integrating over velocity space and flux surface averaging equation \eq{eq:lwsecondorderionstotalv1} (see the discussion on solvability conditions in \cite{calvo12}), 
\begin{equation} \label{eq:dnsdttotal}
\frac{\partial n_s}{\partial t} = \left \langle 2\pi \int \dd u\, \dd \mu\, B ( F_{s2, \nc}^\lw + F_{s2, \tb}^\lw + F_{s2, Q}^\lw ) \right \rangle_\psi.
\end{equation} 
The equation for $\partial T_i/\partial t$ is found by multiplying \eq{eq:lwsecondorderionstotalv1} by $m_s(u^2 + 2\mu B)/3 - T_i$, integrating over velocity, flux surface averaging, and summing over ionic species, 
\begin{eqnarray} \label{eq:dTidttotal}
\fl \sum_{s \neq e} n_s \frac{\partial T_i}{\partial t} = \sum_{s \neq e} \Bigg \langle 2\pi \int \dd u\, \dd \mu\, B ( F_{s2, \nc}^\lw + F_{s2, \tb}^\lw + F_{s2, \Delta T}^\lw + F_{s2, Q}^\lw ) \nonumber\\ \times \left ( \frac{m_s ( u^2 + 2\mu B)}{3} - T_i \right ) \Bigg \rangle_\psi.
\end{eqnarray}
Note that the time derivatives $\partial n_s/\partial t$ and $\partial T_i/\partial t$ do not depend on $\langle f_{s2}^\lw \rangle$. 

Using the orderings for the turbulent fluctuations given in subsection \ref{sub:tbfirstorderBpB}, the turbulent terms $F_{s2, \tb}^\lw$, dominant for $a \nu_{ii}/v_{ti} \lesssim (k_\bot \rho_i)^{-1} (B_p/B)^2$, give
\begin{equation} \label{eq:dnsdttbestimate}
\frac{\partial n_s}{\partial t} \sim \frac{F_{s2, \tb}^\lw}{f_{Ms}} n_s \sim \frac{1}{k_\bot \rho_i} \rho_\ast^2 n_s \frac{v_{ti}}{a}
\end{equation}
and
\begin{equation} \label{eq:dTidttbestimate}
\frac{\partial T_i}{\partial t} \sim \frac{F_{s2, \tb}^\lw}{f_{Ms}} T_i \sim \frac{1}{k_\bot \rho_i} \rho_\ast^2 T_i \frac{v_{ti}}{a}.
\end{equation}
To obtain 
\begin{equation} \label{eq:Flwtbestimate}
\frac{F_{s2,\tb}^\lw}{f_{Ms}} \sim \frac{1}{k_\bot \rho_i} \rho_\ast^2 \frac{v_{ti}}{a}  
\end{equation}
for $k_\bot \rho_i \gg B_p/B$, we need to recall that for short parallel characteristic lengths, $\underline{\phi}_1^\tb$ and $\underline{f}_{s1}^\tb$ can be written in terms of the new Fourier coefficients $\underline{\check{\phi}}_1^\tb$ and $\underline{\check{f}}_{s1}^\tb$ defined in \eq{eq:underlinephi1tbgyro} and \eq{eq:underlinefs1tbgyro}. Using $\underline{\check{\phi}}_1^\tb$ and $\underline{\check{f}}_{s1}^\tb$ instead of $\underline{\phi}_1^\tb$ and $\underline{f}_{s1}^\tb$, the parallel gradient $\bun \cdot \nabla$ is of order $l_{||}^{-1} \gg (B_p/B) a^{-1}$, whereas the correction to the lowest order perpendicular gradient $i \check{\bk}_\bot$ is of order $a^{-1}$, that is, its size does not depend on $l_{||}$ (see \eq{eq:gradpargyro} and \eq{eq:gradperpgyro}). For example, according to \eq{eq:ionquantitiesvskbot} and \eq{eq:falsefastddtheta}, terms like the first term on the right side of \eq{eq:Flwtbtotal} seem to be of order $(B/B_p) \rho_\ast^2 f_{Ms} v_{ti}/a$. However, using $\underline{\check{\phi}}_1^\tb$ and $\underline{\check{f}}_{s1}^\tb$, defined in \eq{eq:underlinephi1tbgyro} and \eq{eq:underlinefs1tbgyro}, most of the large terms can be ignored. They become part of the lowest order equation, i.e., part of the wavevector $\check{\bk}_\bot$, and can then be ignored because they do not give momentum flux. The final form of the first term on the right side of \eq{eq:Flwtbtotal} becomes
\begin{eqnarray}
- \left \langle \sum_{k_\psi, \check{k}_\alpha, k_\zeta} \frac{c (\underline{\check{\phi}}_1^\tb)^\ast J_0 (\Lambda_s)}{B} i (\check{\bk}_\bot \times \bun) \cdot \left ( \nabla_\bR \psi \frac{\partial}{\partial \psi} + \nabla_\bR \theta \frac{\partial}{\partial \theta} \right ) \underline{\check{f}}_{s1}^\tb \right \rangle_t.
\end{eqnarray}
This term is of order $(k_\bot \rho_i)^{-1} \rho_\ast^2 f_{Ms} v_{ti}/a$ because of \eq{eq:truefastddtheta}.

Using the results for the long wavelength pieces in subsection \ref{sub:lwfirstorderBpB}, the neoclassical terms $F_{s2, \nc}^\lw$, dominant when $a \nu_{ii}/v_{ti} \gsim (k_\bot \rho_i)^{-1} (B_p/B)^2$, lead to
\begin{equation} \label{eq:dnsdtncestimate}
\frac{\partial n_s}{\partial t} \sim \frac{F_{s2, \nc}^\lw}{f_{Ms}} n_s \sim \frac{B^2}{B_p^2} \rho_\ast^2 n_s \nu_{ii}
\end{equation}
and
\begin{equation} \label{eq:dTidtncestimate}
\frac{\partial T_i}{\partial t} \sim \frac{F_{s2, \tb}^\nc}{f_{Ms}} T_i \sim \frac{B^2}{B_p^2} \rho_\ast^2 T_i \nu_{ii}.
\end{equation}
The estimates in \eq{eq:dnsdttbestimate}, \eq{eq:dTidttbestimate}, \eq{eq:dnsdtncestimate} and \eq{eq:dTidtncestimate} inspired the ordering in \eq{eq:tauEtbtotal} and \eq{eq:tauEnctotal}. 

We are going to rewrite \eq{eq:lwsecondorderionstotalv1} in a more convenient form. To do that, we first define the function
\begin{equation} \label{eq:h2lwdef}
h_{s2}^\lw = \langle f_{s2}^\lw \rangle + \frac{Z_s e \phi_2^\lw}{T_i} f_{Ms}.
\end{equation}
This is the only piece that matters for momentum transport because $f_{s2}^\lw$ enters as the argument of a linearized collision operator (see \eq{eq:Pizero}), and the Maxwell-Boltzmann response $ - (Z_s e \phi_2^\lw/T_i) f_{Ms}$ vanishes under the action of the linearized collision operator. Once \eq{eq:lwsecondorderionstotalv1} is written in terms of $h_{s2}^\lw$, we use that the equation is linear to split the function $h_{s2}^\lw$ into several pieces of different physical origin,
\begin{equation} \label{eq:splith2total}
h_{s2}^\lw = h_{s2, \nc}^\lw + h_{s2, \tb}^\lw + h_{s2, \Delta T}^\lw + h_{s2, Q}^\lw.
\end{equation}
The equations for these different pieces of $h_{s2}^\lw$ are
\begin{eqnarray} \label{eq:eqh2betatotal}
\fl \left ( u \bun \cdot \nabla_\bR \theta \frac{\partial}{\partial \theta} - \mu \bun \cdot \nabla_\bR B \frac{\partial}{\partial u} \right ) h_{s2, \beta}^\lw - \sum_{s^\prime \neq e} C_{ss^\prime}^{(\ell)} [  h_{s2, \beta}^\lw ; h_{s^\prime 2, \beta}^\lw ] = F_{s2, \beta}^\lw \nonumber\\ \fl - \frac{f_{Ms}}{n_s} \left \langle 2 \pi \int \dd u^\prime\, \dd \mu^\prime\, B F_{s2, \beta}^\lw (u^\prime, \mu^\prime) \right \rangle_\psi - f_{Ms} \left ( \frac{m_s ( u^2 + 2\mu B)}{2 T_i} - \frac{3}{2} \right )  \nonumber\\ \fl \times \left ( \sum_{s^{\prime\prime} \neq e} n_{s^{\prime\prime}} \right )^{-1} \sum_{s^\prime \neq e} \left \langle 2 \pi \int \dd u^\prime\, \dd \mu^\prime\, B F_{s^\prime 2, \beta}^\lw (u^\prime, \mu^\prime) \left ( \frac{m_{s^\prime} ( (u^\prime)^2 + 2\mu^\prime B)}{3 T_i} - 1 \right ) \right \rangle_\psi,
\end{eqnarray}
where $\beta = \nc, \tb, \Delta T, Q$. The only difference between the equations for the different pieces of $h_{s2}^\lw$ are the right side terms, completely determined by the $F_{s2,\beta}^\lw$ in \eq{eq:Flwnctotal}-\eq{eq:FlwQtotal}.

\begin{table}
\caption{Size of the pieces of the long wavelength, second order distribution function $h_{s2}^\lw$ for $B_p/B \lesssim k_\bot \rho_i \lesssim 1$.} \label{table:h2betatotal}
\begin{indented}
\item[]
\begin{tabular}{ l l }
\br $h_{s2, \beta}^\lw$ & Size  \\
\mr $h_{s2, \nc}^\lw$ & $(B/B_p)^2 \rho_\ast^2 f_{Ms}$  \\
$h_{s2, \tb}^\lw$ & $(k_\bot \rho_i)^{-1} (v_{ti}/ a\nu_{ii}) \rho_\ast^2 f_{Ms}$  \\
$h_{s2, \Delta T}^\lw$ & $\sqrt{m_e/m_i} (T_e/T_i - 1 ) f_{Ms}$  \\
$h_{s2, Q}^\lw$ & $Q_s/\nu_{ii}$  \\
\br
\end{tabular}
\end{indented}
\end{table}

The sizes of the different pieces of $h_{s2}^\lw$ are given in table \ref{table:h2betatotal}. The dominant piece in $h_{s2, \tb}^\lw$, $h_{s2, \Delta T}^\lw$ and $h_{s2, Q}^\lw$ is the orbit averaged piece, $\langle h_{s2, \beta} \rangle_\tau$, where the orbit average $\langle \ldots \rangle_\tau$ is defined in \eq{eq:orbitavedef}. Applying this average to \eq{eq:eqh2betatotal}, we find
\begin{eqnarray}
\fl - \sum_{s^\prime \neq e} \left \langle C_{ss^\prime}^{(\ell)} [ h_{s2, \beta}^\lw ; h_{s^\prime 2, \beta}^\lw ] \right \rangle_\tau = \left \langle F_{s2, \beta}^\lw \right \rangle_\tau - \frac{f_{Ms}}{n_s} \left \langle 2 \pi \int \dd u^\prime\ \dd \mu^\prime\, B F_{s2, \beta}^\lw (u^\prime, \mu^\prime) \right \rangle_\psi \nonumber\\ - f_{Ms} \left ( \frac{m_s ( u^2 + 2\mu B)}{2 T_i} - \frac{3}{2} \right ) \left ( \sum_{s^{\prime\prime} \neq e} n_{s^{\prime\prime}} \right )^{-1} \nonumber\\ \times \sum_{s^\prime \neq e} \left \langle 2 \pi \int \dd u^\prime\, \dd \mu^\prime\, B F_{s^\prime 2, \beta}^\lw (u^\prime, \mu^\prime) \left ( \frac{m_{s^\prime} ( (u^\prime)^2 + 2\mu^\prime B)}{3 T_i} - 1 \right ) \right \rangle_\psi.
\end{eqnarray}
Thus, for $h_{s2, \tb}^\lw$, $h_{s2, \Delta T}^\lw$ and $h_{s2, Q}^\lw$, the estimate is $h_{s2, \beta}^\lw \sim F_{s2, \beta}^\lw/\nu_{ii}$, with $F_{s2, \tb}^\lw \sim (k_\bot \rho_i)^{-1} \rho_\ast^2 f_{Ms} v_{ti}/a$, $F_{s2, \Delta T}^\lw \sim \nu_{ii} \sqrt{m_e/m_i} (T_e/T_i-1) f_{Ms}$ and $F_{s2, Q}^\lw \sim Q_s$ (the size of $F_{s2, \tb}^\lw$ is given in \eq{eq:Flwtbestimate}). According to \eq{eq:Flwnctotal}, $\langle F_{s2, \nc}^\lw \rangle_\tau \sim \nu_{ii} (B/B_p)^2 \rho_\ast^2 f_{Ms}$, giving $\langle h_{s2, \nc}^\lw \rangle_\tau\sim (B/B_p)^2 \rho_\ast^2 f_{Ms}$. The poloidally varying piece of $h_{s2, \nc}^\lw$, $h_{s2, \nc}^\lw - \langle h_{s2, \nc}^\lw \rangle_\tau$, must be comparable to $\Delta h_{s2, \nc}^\lw$, i.e. of order $(B/B_p)^2 \rho_\ast^2 f_{Ms}$. 

\subsection{Short wavelength, second order equations} \label{sub:tbsecondordertotal}

Using the orderings discussed at the beginning of this section, we can simplify equations \eq{eq:tbsecondordergk} and \eq{eq:tbsecondorderquasineutrality}. We can neglect the terms proportional to $\dot{\bR}_{s2}$ and $\dot{u}_{s2}$, and the piece $\underline{\tilde{f}}_{s2}^\tb$ defined in \eq{eq:underlineftildetb}. The function $\underline{\phiave}_2^\tb$, defined in \eq{eq:underlinephiaveRtb2}, is approximately
\begin{equation}
\underline{\phiave}_2^\tb \simeq \underline{\phi}_2^\tb J_0 (\Lambda_s),
\end{equation}
and the collisional piece $\underline{C}_{ss^\prime, 2}^\tb$, defined in \eq{eq:Cssprime2tb}, simplifies to
\begin{eqnarray}
\fl \underline{C}_{ss^\prime, 2}^\tb \simeq C_{ss^\prime} \left [ \underline{f}_{s1}^\tb, f_{s^\prime 1}^\lw \right ]+ C_{ss^\prime} \left [ f_{s1}^\lw, \underline{f}_{s^\prime 1}^\lw \right ] + \sum_{k_\psi^\prime, k_\alpha^\prime} C_{ss^\prime} \left [ (\underline{f}_{s1}^\tb)^\prime, (\underline{f}_{s^\prime 1}^\tb)^{\prime\prime} \right ],
\end{eqnarray}
where as explained at the beginning of this section, we neglect the finite gyroradius effects such as $\Delta \underline{f}_{s2}^\tb$, defined in \eq{eq:DeltaunderlineFs2tbdef}, because this is a collisional term. However, for $k_\bot \rho_i \sim 1$ the function $\Delta \underline{f}_{s2}^\tb$ is not negligible in terms that are not collisional. This function becomes, to lowest order,
\begin{equation} \label{eq:DeltaunderlineFs2tbapprox}
\Delta \underline{f}_{s2}^\tb \simeq \frac{Z_s e \underline{\phi}_1^\tb}{m_s B} \frac{\partial f_{s1}^\lw}{\partial \mu_0} \left (1 - J_0 (\lambda_s) \exp \left ( \frac{i \bk_\bot \cdot (\bv \times \bun)}{\Omega_s} \right ) \right ).
\end{equation}

Since equations \eq{eq:tbsecondordergk} and \eq{eq:tbsecondorderquasineutrality} are linear, we can split the functions $\langle \underline{f}_{s2}^\tb \rangle$ and $\underline{\phi}_{s2}^\tb$ into different pieces of different physical origin, namely
\begin{equation} \label{eq:splitf2tbtotal}
\langle \underline{f}_{s2}^\tb \rangle = \underline{f}_{s2, \nc}^\tb + \underline{f}_{s2, \grad}^\tb + \underline{f}_{s2, \acc}^\tb
\end{equation}
and
\begin{equation} \label{eq:splitphi2tbtotal}
\underline{\phi}_2^\tb = \underline{\phi}_{2, \nc}^\tb + \underline{\phi}_{2, \grad}^\tb + \underline{\phi}_{2, \acc}^\tb.
\end{equation}
The subindex $\nc$ indicates that the corrections are due to the effect of neoclassical flows on turbulent fluctuations, the subindex $\grad$ that they are due to the slow radial and poloidal variation of the turbulence characteristics, and the subindex $\acc$ indicates that the pieces of the distribution function and the potential have to do with the turbulent acceleration and deceleration of particles. This separation into different pieces is based on the physical pictures given in section \ref{sec:pictures}. The equations for the different pieces of the distribution function and the potential are
\begin{eqnarray} \label{eq:eqf2betatbtotal}
\fl \frac{\partial \underline{f}_{s2, \beta}^\tb}{\partial t} + \left ( u \bun \cdot \nabla_\bR \theta \frac{\partial}{\partial \theta} - \mu \bun \cdot \nabla_\bR B \frac{\partial}{\partial u} \right ) \underline{f}_{s2, \beta}^\tb + \left ( - i k_\alpha c \frac{\partial \phi_0}{\partial \psi} + i \bk_\bot \cdot \bv_{Ms} \right ) \underline{f}_{s2, \beta}^\tb \nonumber\\ - \sum_{s^\prime} C_{ss^\prime}^{(\ell)} \left [ \underline{f}_{s2, \beta}^\tb; \underline{f}_{s^\prime2, \beta}^\tb \right ] + \{ \underline{\phi}_{2, \beta}^\tb J_0 (\Lambda_s), \underline{f}_{s1}^\tb \} + \{ \underline{\phi}_1^\tb J_0 (\Lambda_s), \underline{f}_{s2, \beta}^\tb \} \nonumber\\ + f_{Ms} \Bigg [ \frac{Z_s e}{T_s} \left ( u \bun \cdot \nabla_\bR \theta \frac{\partial}{\partial \theta} + i \bk_\bot \cdot \bv_{Ms} \right ) \nonumber\\ + i k_\alpha c \Bigg ( \frac{\partial}{\partial \psi} \ln n_s  + \left ( \frac{m_s (u^2 + 2\mu B)}{2T_s} - \frac{3}{2} \right ) \frac{\partial}{\partial \psi} \ln T_s \Bigg )\Bigg ] \underline{\phi}_{2, \beta}^\tb J_0 (\Lambda_s) \nonumber\\ = \underline{F}_{s2, \beta}^\tb 
\end{eqnarray}
and
\begin{eqnarray} \label{eq:eqphi2betatbtotal}
\fl 2\pi \sum_s Z_s \int \dd v_{||} \dd \mu_0\, B \underline{f}_{s2, \beta}^\tb J_0 (\lambda_s) - \sum_{s} \frac{Z_s^2 e \underline{\phi}_{2, \beta}^\tb }{T_s} n_s ( 1 - \Gamma_0(b_s)) = \nonumber\\ - \sum_s Z_s \Delta \underline{n}_{s2,\beta}^\tb,
\end{eqnarray}
where $\beta = \nc, \grad, \acc$. The right sides of \eq{eq:eqf2betatbtotal} are
\begin{eqnarray} \label{eq:F2nctbtotal}
\fl \underline{F}_{s2, \nc}^\tb = \Bigg [ \frac{c}{B} (i \bk_\bot \times \bun ) \cdot \nabla_\bR f_{s1}^\lw + \frac{Z_s e}{m_s} \frac{\partial f_{s1}^\lw}{\partial u} \Bigg ( \bun \cdot \nabla_\bR \theta \frac{\partial}{\partial \theta} + \frac{u}{\Omega_s} i \bk_\bot \cdot (\bun \times \kappabf) \Bigg ) \Bigg ] \underline{\phi}_1^\tb J_0 (\Lambda_s) \nonumber\\ + \frac{c}{B} \left [ (\nabla_\bR \phi_0 \times \bun ) \cdot \nabla_\bR \theta \frac{\partial}{\partial \theta} + i \bk_\bot \cdot ( \nabla_\bR \phi_1^\lw \times \bun ) \right ] \underline{f}_{s1}^\tb \nonumber\\ + \left [ \frac{Z_s e}{m_s} \bun \cdot \nabla_\bR \phi_1^\lw + \frac{u}{\Omega_s} (\bun \times \kappabf) \cdot \left ( \mu \nabla_\bR B + \frac{Z_s e}{m_s} \nabla_\bR \phi_0 \right ) \right ] \frac{\partial \underline{f}_{s1}^\tb}{\partial u} \nonumber\\ + \sum_{s^\prime} \Bigg ( C_{ss^\prime} \left [ \underline{f}_{s1}^\tb, f_{s^\prime 1}^\lw \right ] + C_{ss^\prime} \left [ f_{s1}^\lw,  \underline{f}_{s^\prime 1}^\tb \right ] \nonumber\\ + \sum_{k_\psi^\prime, k_\alpha^\prime} C_{ss^\prime} \left [ (\underline{f}_{s1}^\tb)^\prime, (\underline{f}_{s^\prime 1}^\tb)^{\prime\prime} \right ]\Bigg ),
\end{eqnarray}
\begin{eqnarray} \label{eq:F2gradtbtotal}
\fl \underline{F}_{s2, \grad}^\tb = - f_{Ms} \Bigg [ \frac{Z_s e}{T_s} \bv_{Ms} \cdot \Bigg ( \nabla_\bR \psi \frac{\partial}{\partial \psi} + \nabla_\bR \theta \frac{\partial}{\partial \theta} \Bigg ) + \Bigg ( \frac{\partial}{\partial \psi} \ln n_s  \nonumber\\ + \left ( \frac{m_s (u^2 + 2\mu B)}{2T_s} - \frac{3}{2} \right ) \frac{\partial}{\partial \psi} \ln T_s \Bigg ) \frac{c}{B} (\nabla_\bR \psi \times \bun) \cdot \nabla_\bR \theta \frac{\partial}{\partial \theta} \Bigg ] \underline{\phi}_1^\tb \nonumber\\ -  \bv_{Ms} \cdot \left ( \nabla_\bR \psi \frac{\partial}{\partial \psi} + \nabla_\bR \theta \frac{\partial}{\partial \theta} \right ) \underline{f}_{s1}^\tb \nonumber\\ + \sum_{k_\psi^\prime, k_\alpha^\prime} \Bigg [ \frac{c (\underline{\phi}_1^\tb)^\prime}{B} i (\bk^\prime_\bot \times \bun) \cdot \left ( \nabla_\bR \psi \frac{\partial}{\partial \psi} + \nabla_\bR \theta \frac{\partial}{\partial \theta} \right ) (\underline{f}_{s1}^\tb)^{\prime\prime} \nonumber\\ - \frac{c (\underline{f}_{s1}^\tb)^\prime}{B} i (\bk_\bot^\prime \times \bun) \cdot \left ( \nabla_\bR \psi \frac{\partial}{\partial \psi} + \nabla_\bR \theta \frac{\partial}{\partial \theta} \right ) (\underline{\phi}_1^\tb)^{\prime\prime} \Bigg ]
\end{eqnarray}
and
\begin{eqnarray} \label{eq:F2acctbtotal}
\fl \underline{F}_{s2, \acc}^\tb = \sum_{k_\psi^\prime, k_\alpha^\prime} \frac{Z_s e}{m_s} \frac{\partial (\underline{f}_{s1}^\tb)^{\prime\prime}}{\partial u} \left ( \bun \cdot \nabla_{\bR} \theta \frac{\partial}{\partial \theta} + \frac{u}{\Omega_s} i \bk_\bot^\prime \cdot (\bun \times \kappabf) \right ) (\underline{\phi}_1^\tb)^\prime.
\end{eqnarray}
Here a prime on a Fourier coefficient indicates that it depends on $k_\psi^\prime$ and $k_\alpha^\prime$, and two primes that it depends on $k_\psi^{\prime\prime} = k_\psi - k_\psi^\prime$ and $k_\alpha^{\prime\prime} = k_\alpha - k_\alpha^\prime$. The right sides of \eq{eq:eqphi2betatbtotal} are
\begin{equation} \label{eq:Deltan2nctbtotal}
\Delta \underline{n}_{s2, \nc}^\tb = 2\pi \int \dd v_{||}\, \dd \mu_0\, (1 - J_0^2 (\lambda_s)) \frac{\partial f_{s1}^\lw}{\partial \mu_0}  \frac{Z_s \underline{\phi}_{1,\nc}^\tb}{m_s}
\end{equation}
and
\begin{equation}
\Delta \underline{n}_{s2, \beta}^\tb = 0
\end{equation}
for $\beta = \grad, \acc$. In the equations for ionic species ($s \neq e$), the ion-electron and impurity-electron collisions are negligible (see \ref{app:Cse}), and in the equations for electrons ($s = e$), the electron-ion and electron-impurity collision operators can be simplified using $\sqrt{m_e/m_i} \ll 1$, as shown in \ref{app:Ces}. To evaluate $\underline{F}_{s2, \grad}^\tb$, we need $\partial \underline{f}_{s1}^\tb/\partial \psi$ and $\partial \underline{\phi}_1^\tb/\partial \psi$. These derivatives can be calculated by neglecting terms of order $B_p/B \ll 1$ in equations \eq{eq:radialderivativegk} and \eq{eq:radialderivativequasineutrality} while assuming $k_\bot \rho_i \sim B_p/B$,
\begin{eqnarray} \label{eq:radialderivativegktotal}
\fl \frac{\partial}{\partial t} \left ( \frac{\partial \underline{f}_{s1}^\tb}{\partial \psi} \right )+ \left ( u \bun \cdot \nabla_\bR \theta \frac{\partial}{\partial \theta} - \mu \bun \cdot \nabla_\bR B \frac{\partial}{\partial u} \right ) \frac{\partial \underline{f}_{s1}^\tb}{\partial \psi} + \left ( - i k_\alpha c \frac{\partial \phi_0}{\partial \psi} + i \bk_\bot \cdot \bv_{Ms} \right )  \frac{\partial \underline{f}_{s1}^\tb}{\partial \psi} \nonumber\\ - \sum_{s^\prime} C_{ss^\prime}^{(\ell)} \left [  \frac{\partial \underline{f}_{s1}^\tb}{\partial \psi}; \frac{\partial \underline{f}_{s1}^\tb}{\partial \psi} \right ] + \left \{ \frac{\partial \underline{\phi}_1^\tb}{\partial \psi}, \underline{f}_{s1}^\tb \right \} + \left \{ \underline{\phi}_1^\tb , \frac{\partial \underline{f}_{s1}^\tb}{\partial \psi} \right \}\nonumber\\ = - f_{Ms} \Bigg [ \frac{Z_s e}{T_s} \left ( u \bun \cdot \nabla_\bR \theta \frac{\partial}{\partial \theta} + i \bk_\bot \cdot \bv_{Ms} \right ) \nonumber\\ + i k_\alpha c \Bigg ( \frac{\partial}{\partial \psi} \ln n_s  + \left ( \frac{m_s (u^2 + 2\mu B)}{2T_s} - \frac{3}{2} \right ) \frac{\partial}{\partial \psi} \ln T_s \Bigg ) \Bigg ] \frac{\partial \underline{\phi}_1^\tb}{\partial \psi} \nonumber\\ - u \frac{\partial}{\partial \psi} ( \bun \cdot \nabla_\bR \theta ) \frac{\partial \underline{f}_{s1}^\tb}{\partial \theta} + \mu \frac{\partial}{\partial \psi}( \bun \cdot \nabla_\bR B ) \frac{\partial \underline{f}_{s1}^\tb}{\partial u} \nonumber\\ - \underline{f}_{s1}^\tb \frac{\partial}{\partial \psi} \left ( -i  k_\alpha c \frac{\partial \phi_0}{\partial \psi} + i \bk_\bot \cdot \bv_{Ms} \right ) + \sum_{s^\prime} \frac{\partial C_{ss^\prime}^{(\ell)}}{\partial \psi} \left [  \underline{f}_{s1}^\tb; \underline{f}_{s1}^\tb \right ] \nonumber\\ - \frac{\partial }{\partial \psi} \left ( \frac{Z_s e f_{Ms}}{T_s} u \bun \cdot \nabla_\bR \theta \right ) \frac{\partial \underline{\phi}_1^\tb}{\partial \theta} \nonumber\\- \underline{\phi}_1^\tb \frac{\partial }{\partial \psi} \Bigg [ \frac{Z_s e f_{Ms}}{T_s} i \bk_\bot \cdot \bv_{Ms} + i k_\alpha c f_{Ms} \Bigg ( \frac{\partial}{\partial \psi} \ln n_s \nonumber\\ + \left ( \frac{m_s (u^2 + 2\mu B)}{2T_s} - \frac{3}{2} \right ) \frac{\partial}{\partial \psi} \ln T_s \Bigg )  \Bigg ]
\end{eqnarray}
and
\begin{eqnarray} \label{eq:radialderivativequasineutralitytotal}
\fl 2 \pi \sum_s Z_s \int \dd v_{||}\, \dd \mu_0\, B \frac{\partial \underline{f}_{s1}^\tb}{\partial \psi} = - 2 \pi \sum_s Z_s \int \dd v_{||}\, \dd \mu_0\, \underline{f}_{s1}^\tb \frac{\partial B}{\partial \psi}.
\end{eqnarray}
Note that we have kept finite gyroradius effects only in $\underline{F}_{s2, \nc}^\tb$, which is the piece that dominates for $k_\bot \rho_i \sim 1$ according to tables \ref{table:f2betatbtotal} and \ref{table:phi2betatbtotal} and the discussion in the next paragraph.

\begin{table}
\caption{Size of the pieces of the short wavelength, second order distribution function $\underline{f}_{s2}^\tb$ for $B_p/B \lesssim k_\bot \rho_i \lesssim 1$.}\label{table:f2betatbtotal}
\begin{indented}
\item[]
\begin{tabular}{ l l l l } 
\br $\underline{f}_{s2, \beta}^\tb$ & Size for $s\neq e$ & Size for $s = e$ \\
\mr $\underline{f}_{s2, \nc}^\tb$ & $(k_\bot \rho_i)^{-1} (B/B_p) \rho_\ast^2 f_{Ms}$ & Even in $u$:  $(k_\bot \rho_i)^{-1} (B/B_p) \rho_\ast^2 f_{Me}$ \\
& & Odd in $u$:  $(k_\bot \rho_i)^{-1} (B/B_p) \sqrt{m_e/m_i} \rho_\ast^2 f_{Me}$ \\
$\underline{f}_{s2, \grad}^\tb$ & $(k_\bot \rho_i)^{-2} \rho_\ast^2 f_{Ms}$ & Even in $u$: $(k_\bot \rho_i)^{-2} \rho_\ast^2 f_{Me}$ \\
& & Odd in $u$:  $(k_\bot \rho_i)^{-2} \sqrt{m_e/m_i} \rho_\ast^2 f_{Me}$ \\
$\underline{f}_{s2, \acc}^\tb$ & $(k_\bot \rho_i)^{-2} \rho_\ast^2 f_{Ms}$ & Even in $u$: $(k_\bot \rho_i)^{-2} \rho_\ast^2 f_{Me}$ \\
& & Odd in $u$:  $(k_\bot \rho_i)^{-2} \sqrt{m_e/m_i} \rho_\ast^2 f_{Me}$ \\
\br
\end{tabular}
\end{indented}
\end{table}

The sizes of the different pieces $\underline{f}_{s2,\beta}^\tb$ and $\underline{\phi}_{2, \beta}^\tb$ are given in tables \ref{table:f2betatbtotal} and \ref{table:phi2betatbtotal}. The size of $\underline{f}_{s2,\beta}^\tb$ for $s\neq e$ is deduced from the fact that the characteristic size of the left side of \eq{eq:eqf2betatbtotal} is $k_\bot \rho_i (v_{ti}/a) \underline{f}_{s2, \beta}^\tb$ according to \eq{eq:eddytimetotal}. Then, $\underline{f}_{s2, \beta}^\tb \sim (k_\bot \rho_i)^{-1} (a/v_{ti}) \underline{F}_{s2,\beta}^\tb$, and using \eq{eq:fs1lworderBpB}, \eq{eq:phi1lworderBpB}, \eq{eq:ionquantitiesvskbot} and \eq{eq:fe1tboddvskbot}, we find the second column of table \ref{table:f2betatbtotal}. Note that to obtain the size of $\underline{F}_{s2,\nc}^\tb$ and $\underline{F}_{s2, \grad}^\tb$ for $k_\bot \rho_i \gg B_p/B$, we have used the new Fourier coefficients $\underline{\check{\phi}}_1^\tb$ and $\underline{\check{f}}_{s1}^\tb$ in \eq{eq:underlinephi1tbgyro} and \eq{eq:underlinefs1tbgyro}. With these new coefficients, $\bun \cdot \nabla \sim l_{||}^{-1} \gg (B_p/B) a^{-1}$ whereas the correction to the lowest order perpendicular gradient $i \check{\bk}_\bot$ is of order $a^{-1}$, that is, it does not depend on $l_{||}$ (see \eq{eq:gradpargyro} and \eq{eq:gradperpgyro}). An example of how the new functions $\underline{\check{\phi}}_1^\tb$ and $\underline{\check{f}}_{s1}^\tb$ affect the estimates is given in the discussion around \eq{eq:Flwtbestimate}. From \eq{eq:eqphi2betatbtotal} we deduce $e \underline{\phi}_{2,\beta}^\tb/T_e \sim \underline{f}_{s2, \beta}^\tb/f_{Ms}$, leading to the results in table \ref{table:phi2betatbtotal}. The size of the electron distribution function is obtained in a similar manner. We only need to recall the discussion around equations \eq{eq:fenevenorder} and \eq{eq:fenoddorder} about the difference between pieces even and odd in $u$.

\begin{table}
\caption{Size of the pieces of the short wavelength, second order potential $\underline{\phi}_2^\tb$ for $B_p/B \lesssim k_\bot \rho_i \lesssim 1$.} \label{table:phi2betatbtotal}
\begin{indented}
\item[]
\begin{tabular}{ l l }
\br $\underline{\phi}_{2, \beta}^\tb$ & Size \\
\mr $\underline{\phi}_{2, \nc}^\tb$ & $(k_\bot \rho_i)^{-1}(B/B_p) \rho_\ast^2 T_e/e$  \\
$\underline{\phi}_{2, \grad}^\tb$ & $(k_\bot \rho_i)^{-2} \rho_\ast^2 T_e/e$ \\
$\underline{\phi}_{2, \acc}^\tb$ & $(k_\bot \rho_i)^{-2} \rho_\ast^2 T_e/e$ \\
\br
\end{tabular}
\end{indented}
\end{table}

\subsection{Momentum flux} \label{sub:momentumfluxtotal}
Using the orderings discussed at the beginning of this section, and the results of subsections \ref{sub:lwsecondordertotal} and \ref{sub:tbsecondordertotal}, we simplify the formula for the radial momentum flux $\Pi = \Pi_{-1} + \Pi_0$ given in \eq{eq:Pifinal}. The size of the pieces of the distribution function and the potential that contribute to the momentum flux is given in tables \ref{table:h2betatotal}, \ref{table:f2betatbtotal} and \ref{table:phi2betatbtotal}. The pieces $\tilde{f}_{s2}^\lw$ and $\underline{\tilde{f}}_{s2}^\tb$, given in \eq{eq:ftildelw} and \eq{eq:underlineftildetb}, and the pieces $\Delta f_{s1}^\lw$ and $\Delta f_{s2}^\lw$, defined in \eq{eq:DeltaFs1lwdef} and \eq{eq:DeltaFs2lwdef}, are all small in $B_p/B \ll 1$ compared to the other pieces of the distribution function, giving
\begin{equation}
[f_s]_n^\lw \simeq f_{sn}^\lw.
\end{equation}
The piece $\Delta \underline{f}_{s2}^\tb$, defined in \eq{eq:DeltaunderlineFs2tbdef} simplifies to the result in \eq{eq:DeltaunderlineFs2tbapprox}. Finally, to lowest order in $B_p/B \ll 1$,
\begin{equation}
R \bv \cdot \zun \simeq \frac{I v_{||}}{B}.
\end{equation}
This approximation is valid except for in \eq{eq:Pi0QFOWtotal}. We explain why below.

With these considerations, we find that the piece of the radial flux of toroidal angular momentum, $\Pi_{-1}$, is to lowest order in $B_p/B \ll 1$,
\begin{eqnarray}
\Pi_{-1} = \Pi_{-1, \ud}^\tb + \Pi_{-1,\ud}^\nc,
\end{eqnarray}
where
\begin{eqnarray} \label{eq:Piminusudtbtotal}
\Pi_{-1, \ud}^\tb = - \left \langle \left \langle \sum_{s \neq e} \sum_{k_\psi, k_\alpha} \frac{I m_s c }{B} i k_\alpha (\underline{\phi}_1^\tb)^\ast \int \dd^3v\, \underline{f}_{s1}^\tb J_0 (\lambda_s) v_{||} \right \rangle_\psi \right \rangle_t
\end{eqnarray}
and
\begin{eqnarray} \label{eq:Piminusudnctotal}
\Pi_{-1, \ud}^\nc = - \left \langle \sum_{s \neq e, s^\prime \neq e} \frac{I^2 m_s^2 c}{2Z_s e B^2} \int \dd^3v\, C_{ss^\prime}^{(\ell)} \left [ f_{s1}^\lw; f_{s^\prime 1}^\lw \right ] v_{||}^2 \right \rangle_\psi
\end{eqnarray}
are the lowest order turbulent and neoclassical momentum fluxes that only give a contribution when the tokamak is up-down asymmetric (see section~\ref{sec:symmetry}). Using the equations for $\partial n_s/\partial t$ and $\partial T_i/\partial t$, \eq{eq:dnsdttotal} and \eq{eq:dTidttotal}, and the split into different pieces of the distribution functions and the potential given in \eq{eq:splith2total}, \eq{eq:splitf2tbtotal} and \eq{eq:splitphi2tbtotal}, we obtain that $\Pi_0$ is
\begin{equation}
\Pi_0 = \Pi_0^\tb + \Pi_0^\nc + \Pi_0^\FOW,
\end{equation}
where
\begin{equation} \label{eq:Pi0tbtotal}
\Pi_0^\tb = \Pi^\tb_{0, \nc} + \Pi^\tb_{0, \grad} + \Pi^\tb_{0,\acc}
\end{equation}
is the turbulent momentum flux,
\begin{eqnarray} \label{eq:Pi0nctotal}
\fl \Pi_0^\nc = - \frac{1}{V^\prime} \frac{\partial}{\partial \psi} \left [ V^\prime \left \langle \sum_{s \neq e, s^\prime \neq e} \frac{I^3 m_s^3 c^2}{6Z_s^2 e^2 B^3} \int \dd^3v\, C_{ss^\prime}^{(\ell)} \left [ f_{s1}^\lw ; f_{s^\prime 1}^\lw \right ] v_{||}^3 \right \rangle_\psi \right ] \nonumber\\ - \left \langle \sum_{s \neq e, s^\prime \neq e} \frac{I^2 m_s^2 c}{2Z_s e B^2} \int \dd^3v\, C_{ss^\prime}^{(\ell)} \left [ h_{s2, \nc}^\lw ; h_{s^\prime 2, \nc}^\lw \right ] v_{||}^2 \right \rangle_\psi \nonumber\\- \left \langle \sum_{s \neq e, s^\prime \neq e} \frac{I^2 m_s^2 c}{2Z_s e B^2} \int \dd^3v\, C_{ss^\prime} \left [ f_{s1}^\lw, f_{s^\prime 1}^\lw \right ] v_{||}^2 \right \rangle_\psi \nonumber\\ + \left \langle \frac{I^2}{B^2} \right \rangle_\psi  \frac{c T_i}{2e} \left (\sum_{s^{\prime \prime} \neq e} n_{s^{\prime\prime}} \right )^{-1} \sum_{s \neq e, s^\prime \neq e} n_{s^\prime}  \Bigg \langle \int \dd^3v\, F_{s2, \nc}^\lw \Bigg [ \frac{m_s}{Z_s} \nonumber\\+ \frac{m_{s^\prime}}{Z_{s^\prime}} \left ( \frac{m_s v^2}{3T_i} -1 \right ) \Bigg ] \Bigg \rangle_\psi
\end{eqnarray}
is the neoclassical momentum flux, and
\begin{equation} \label{eq:Pi0FOWtotal}
\Pi_0^\FOW = \Pi_{0, \tb}^\FOW + \Pi_{0, \Delta T}^\FOW + \Pi_{0, Q}^\FOW
\end{equation}
is the momentum flux driven by finite orbit widths (see the discussion in subsection \ref{sub:picturesFOW}). The turbulent momentum flux $\Pi_0^\tb$ has been divided into three pieces, defined by
\begin{eqnarray} \label{eq:Pi0betatbtotal}
\fl \Pi_{0,\beta}^\tb = - \left \langle \left \langle \sum_{s \neq e} \sum_{k_\psi, k_\alpha} \frac{I m_s c}{B} i k_\alpha (\underline{\phi}_1^\tb)^\ast \int \dd^3v\, \underline{f}_{s2, \beta}^\tb J_0 (\lambda_s) v_{||} \right \rangle_\psi \right \rangle_t \nonumber\\ - \left \langle \left \langle \sum_{s \neq e} \sum_{k_\psi, k_\alpha} \frac{I m_s c}{B} i k_\alpha (\underline{\phi}_{2,\beta}^\tb)^\ast \int \dd^3v\, \underline{f}_{s1}^\tb J_0 (\lambda_s) v_{||} \right \rangle_\psi \right \rangle_t
\end{eqnarray}
for $\beta = \nc, \grad, \acc$. Similarly, the momentum flux due to finite orbit widths $\Pi_0^\FOW$ has been divided into three other pieces:
\begin{eqnarray} \label{eq:Pi0tbFOWtotal}
\fl \Pi_{0, \tb}^\FOW = - \frac{1}{V^\prime} \frac{\partial}{\partial \psi} \left [ V^\prime \left \langle \left \langle \sum_{s \neq e} \sum_{k_\psi, k_\alpha} \frac{I^2 m_s^2 c^2}{2 Z_s e B^2} i k_\alpha (\underline{\phi}_1^\tb)^\ast \int \dd^3v\, \underline{f}_{s1}^\tb v_{||}^2 \right \rangle_\psi \right \rangle_t  \right ] \nonumber\\- \left \langle \sum_{s \neq e, s^\prime \neq e} \frac{I^2 m_s^2 c}{2Z_s e B^2} \int \dd^3v\, C_{ss^\prime}^{(\ell)} \left [ h_{s2, \tb}^\lw ; h_{s^\prime 2, \tb}^\lw\right ] v_{||}^2 \right \rangle_\psi \nonumber\\- \Bigg \langle \sum_{s \neq e, s^\prime \neq e} \frac{I^2 m_s^2 c}{2Z_s eB^2} \int \dd^3v\, \Bigg \langle \sum_{k_\psi, k_\alpha} C_{ss^\prime} [ \underline{f}_{s1}^\tb, (\underline{f}_{s^\prime 1}^\tb)^\ast ] \Bigg \rangle_t v_{||}^2 \Bigg \rangle_\psi \nonumber\\ + \left \langle \frac{I^2}{B^2} \right \rangle_\psi  \frac{c T_i}{2e} \left (\sum_{s^{\prime \prime} \neq e} n_{s^{\prime\prime}} \right )^{-1} \sum_{s \neq e, s^\prime \neq e} n_{s^\prime}  \Bigg \langle \int \dd^3v\, F_{s2, \tb}^\lw \Bigg [ \frac{m_s}{Z_s} \nonumber\\+ \frac{m_{s^\prime}}{Z_{s^\prime}} \left ( \frac{m_s v^2}{3T_i} -1 \right ) \Bigg ] \Bigg \rangle_\psi,
\end{eqnarray}
\begin{eqnarray} \label{eq:Pi0DeltaTFOWtotal}
\fl \Pi_{0, \Delta T}^\FOW =- \left \langle \sum_{s \neq e, s^\prime \neq e} \frac{I^2 m_s^2 c}{2Z_s e B^2} \int \dd^3v\, C_{ss^\prime}^{(\ell)} \left [ h_{s2, \Delta T}^\lw ; h_{s^\prime 2, \Delta T}^\lw \right ] v_{||}^2 \right \rangle_\psi \nonumber\\ + \left \langle \frac{I^2}{B^2} \right \rangle_\psi  \frac{n_e m_e c}{e} ( T_e - T_i ) \left (\sum_{s^{\prime \prime} \neq e} n_{s^{\prime\prime}} \right )^{-1} \nonumber\\ \times \sum_{s \neq e, s^\prime \neq e} n_{s^\prime} \nu_{es} \left ( \frac{m_{s^\prime}}{Z_{s^\prime} m_s} - \frac{1}{Z_s} \right )
\end{eqnarray}
and
\begin{eqnarray} \label{eq:Pi0QFOWtotal}
\fl \Pi_{0, Q}^\FOW =- \left \langle \sum_{s \neq e, s^\prime \neq e} \frac{I^2 m_s^2 c}{2Z_s e B^2} \int \dd^3v\, C_{ss^\prime}^{(\ell)} \left [ h_{s2, Q}^\lw ; h_{s^\prime 2, Q}^\lw \right ] v_{||}^2 \right \rangle_\psi \nonumber\\ + \frac{c T_i}{2e} \left (\sum_{s^{\prime \prime} \neq e} n_{s^{\prime\prime}} \right )^{-1} \sum_{s \neq e, s^\prime \neq e} n_{s^\prime}  \Bigg \langle \int \dd^3v\, Q_s \Bigg [ \frac{\langle R^2 \rangle_\psi m_s}{Z_s} \nonumber\\+ \frac{\langle R^2 \rangle_\psi m_{s^\prime}}{Z_{s^\prime}} \left ( \frac{m_s v^2}{3T_i} -1 \right ) - \frac{R^2 m_s}{Z_s} \frac{m_s (\bv \cdot \zun)^2}{T_i}  \Bigg ] \Bigg \rangle_\psi.
\end{eqnarray}
Note that we have not used $R(\bv \cdot \zun) \simeq I v_{||}/B$ in \eq{eq:Pi0QFOWtotal} because $Q_s$ could have a large gyrophase dependent piece. This cannot happen in $h_{s2, Q}^\lw$ because the fast gyration of the particles would average it out. 

\begin{table}
\caption{Size of the different contribution to the momentum flux $\Pi$ for $B_p/B \lesssim k_\bot \rho_i \lesssim 1$. The estimates are normalized by $\Pi_N = \rho_\ast^3 p_i R |\nabla \psi|$.} \label{table:Pitotal}
\begin{indented}
\item[]
\begin{tabular}{ l l }
\br Momentum flux & Size/$\Pi_N$ \\
\mr $\Pi_{-1, \ud}^\tb$ & $\mathrm{max} \{ (k_\bot \rho_i)^{-2} (B_p/B), \sqrt{m_e/m_i} (B/B_p) \} \rho_\ast^{-1} \Delta_{\ud}$  \\
$\Pi_{-1, \ud}^\nc$ & $(B/B_p)^2 (a \nu_{ii}/v_{ti}) \rho_\ast^{-1} \Delta_{\ud}$  \\
$\Pi_{0, \nc}^\tb$ & $(k_\bot \rho_i)^{-1} (B/B_p)$ \\
$\Pi_{0, \grad}^\tb$ & $(k_\bot \rho_i)^{-2}$ \\
$\Pi_{0, \acc}^\tb$ & $(k_\bot \rho_i)^{-2}$ \\
$\Pi_{0}^\nc$ & $(B/B_p)^3 (a \nu_{ii}/v_{ti})$ \\
$\Pi_{0, \tb}^\FOW$ & $(k_\bot \rho_i)^{-1} (B/B_p)$ \\
$\Pi_{0, \Delta T}^\FOW$ & $(B/B_p) \sqrt{m_e/m_i} (T_e/T_i - 1) (a \nu_{ii}/v_{ti}) \rho_\ast^{-2}$ \\
$\Pi_{0, Q}^\FOW$ & $(B/B_p) (Q_s a/\rho_\ast^2 v_{ti} f_{Ms})$ \\
\br
\end{tabular}
\end{indented}
\end{table}

The size of the different pieces of $\Pi$ is given in table \ref{table:Pitotal}. The size of the different terms is deduced from the estimates given in tables \ref{table:h2betatotal}, \ref{table:f2betatbtotal} and \ref{table:phi2betatbtotal}. For $\Pi_{-1, \ud}^\tb$ and $\Pi_{-1,\ud}^\nc$, to indicate that they depend strongly on the up-down asymmetry of the flux surface, we have used the formal parameter $\Delta_{\ud}$ that measures how close the flux surface is to being up-down symmetric ($\Delta_{\ud} = 0$ for perfect up-down symmetry, and $\Delta_{\ud} \sim 1$ for extreme up-down asymmetry). In addition to up-down asymmetry, we have taken into account that turbulence with $k_\bot \rho_i \gg B_p/B$ also satisfies the symmetry described by equations \eq{eq:transformationsymmetrygyrovpar} and \eq{eq:symmetrygyrovpar} to lowest order in $(k_\bot \rho_i)^{-1} (B_p/B) \ll 1$ and $k_\bot \rho_i \sqrt{m_e/m_i} (B/B_p) \ll 1$ (see discussion around \eq{eq:sqrtmassBpBorder} for the parameter $k_\bot \rho_i \sqrt{m_e/m_i} (B/B_p) \ll 1$). Then, $\Pi_{-1,\ud}^\tb$ vanishes to lowest order in $(k_\bot \rho_i)^{-1} (B_p/B) \ll 1$ and $k_\bot \rho_i \sqrt{m_e/m_i} (B/B_p) \ll 1$ even for up-down asymmetric flux surfaces. In table \ref{table:Pitotal}, we have indicated that we need to choose the biggest of the two expansion parameters $(k_\bot \rho_i)^{-1} (B_p/B) \ll 1$ and $k_\bot \rho_i \sqrt{m_e/m_i} (B/B_p) \ll 1$.

The other contributions to the momentum flux do not depend on up-down asymmetry. The reason is that for an up-down symmetric flux surface, when we apply the symmetries discussed in section \ref{sec:symmetry} to equations \eq{eq:eqh2betatotal}, \eq{eq:eqf2betatbtotal} and \eq{eq:eqphi2betatbtotal}, we find that
\begin{equation}
h_{s2, \beta}^\lw ( \psi, \theta, u, \mu ) = h_{s2, \beta}^\lw ( \psi, - \theta, - u, \mu )
\end{equation}
and that equations \eq{eq:eqf2betatbtotal} and \eq{eq:eqphi2betatbtotal} are invariant under the transformation
\begin{equation}
k_\psi \rightarrow - k_\psi, \; \theta \rightarrow - \theta, \; u \rightarrow - u,\; \underline{f}_{s2, \beta}^\tb \rightarrow \underline{f}_{s2, \beta}^\tb, \; \underline{\phi}_{2, \beta}^\tb \rightarrow \underline{\phi}_{2, \beta}^\tb.
\end{equation}
Then, in principle, $\Pi_0^\tb$, $\Pi_0^\nc$ and $\Pi_0^\FOW$ do not vanish in up-down symmetric tokamaks, unlike $\Pi_{-1,\ud}^\tb$ and $\Pi_{-1, \ud}^\nc$.

Note that the equations presented in this subsection for $k_\bot \rho_i \sim 1$ (in this limit, $\Pi_{0, \grad}^\tb$, $\Pi_{0, \acc}^\tb$ and several terms in the equations for $h_{s2,\beta}^\lw$, $\underline{f}_{s2,\nc}^\tb$ and $\underline{\phi}_2^\tb$ are negligible) are the same ones derived in \cite{parra10a}, with the exception of the equations for electrons that were ignored. The electron equations were then introduced in \cite{parra11d}, although in this reference we obtained the equations in the frame rotating with velocity $\Omega_{\zeta, E} = - c (\partial \phi_0/\partial \psi)$. The equations in this rotating frame are given in \ref{app:equationsrotating}.

\section{Interpretation of the equations} \label{sec:pictures}

Using the momentum flux $\Pi$ calculated in subsection \ref{sub:momentumfluxtotal} and the conservation equation for toroidal angular momentum in \eq{eq:torangmom}, we can calculate the potential $\phi_0$ in the tokamak (the densities $n_s$ and the temperatures $T_i$ can be calculated using the particle and ion energy conservation equations, given in \eq{eq:dnsdttotal} and \eq{eq:dTidttotal}, and $T_e$ is determined by the electron energy conservation equation, not given in this article). The momentum flux $\Pi$ depends on the geometry of the flux surface and on first and second radial derivatives of the pressure, the potential $\phi_0$ and the temperature. Instead of working with the potential $\phi_0$, we use the plasma rotation. The plasma does not rotate rigidly because it has a parallel component of the velocity that is not exactly toroidal, given by the last term in \eq{eq:plasmaflowlw}. For this reason, we define the rotation of a flux surface $\Omega_\zeta$ as the rotation the flux surface should have to have the same total toroidal angular momentum, that is,
\begin{eqnarray}
\sum_s n_s m_s \langle R^2 \rangle_\psi \Omega_\zeta =  \sum_s \langle n_s m_s R \bV_s^\lw \cdot \zun \rangle_\psi.
\end{eqnarray}
Using \eq{eq:plasmaflowlw} we find
\begin{eqnarray} \label{eq:Omegadef}
\Omega_\zeta = - c \frac{\partial \phi_0}{\partial \psi} + \left ( \sum_{s^\prime \neq e} n_{s^\prime} m_{s^\prime} \right )^{-1} \sum_{s \neq e} \left ( - \frac{m_s c}{Z_s e} \frac{\partial p_s}{\partial \psi} + \frac{m_s K_s I}{\langle R^2 \rangle_\psi} \right ).
\end{eqnarray}
All the equations in this article can be written in terms of $\Omega_\zeta$ and $\partial \Omega_\zeta/\partial \psi$ if we use \eq{eq:Omegadef} to write $\partial \phi_0/\partial \psi$ as a function of $\Omega_\zeta$, $\partial n_s/\partial \psi$ and $\partial T_s/\partial \psi$. In particular, the conservation equation for toroidal angular momentum equation becomes
\begin{equation} \label{eq:torangmomOmega}
\frac{\partial}{\partial t} \left ( \sum_{s\neq e} n_s m_s \langle R^2 \rangle_\psi \Omega_\zeta \right ) = - \frac{1}{V^\prime} \frac{\partial}{\partial \psi} ( V^\prime \Pi ) + T_\zeta.
\end{equation}

In our ordering, the contributions $\Pi_{-1,\ud}^\tb$ and $\Pi_{-1,\ud}^\nc$ in \eq{eq:Piminusudtbtotal} and \eq{eq:Piminusudnctotal} do not depend on $\Omega_\zeta$ because they do not depend on $\partial \phi_0/\partial \psi$. The equations in the rotating frame where the radial electric field is zero, derived in \ref{app:equationsrotating}, show that $\Pi_{-1,\ud}^\tb$ and $\Pi_{-1,\ud}^\nc$ do not depend on $\partial \phi_0/\partial \psi$ because these pieces of the momentum flux do not change with the change of frame, i.e., they are the same even when the radial electric field is zero. The rotation frequency $\Omega_\zeta$ enters only in the higher order pieces $\Pi_0$, and in particular, in $\Pi_{0,\nc}^\tb$, $\Pi_{0, \grad}^\tb$ and $\Pi_0^\nc$. Since the equations for the second order pieces of the distribution function and the potential, given in subsections \ref{sub:lwsecondordertotal} and \ref{sub:tbsecondordertotal}, are linear, $\Omega_\zeta$ and $\partial \Omega_\zeta/\partial \psi$ only appear linearly in the inhomogeneous terms, and expression \eq{eq:Pi0betatbtotal} and \eq{eq:Pi0nctotal} for $\Pi_{0,\nc}^\tb$, $\Pi_{0, \grad}^\tb$ and $\Pi_0^\nc$ are linear in the second order pieces, the momentum flux must depend linearly on $\Omega_\zeta$ and $\partial \Omega_\zeta/\partial \psi$, i.e.
\begin{equation} \label{eq:PiOmega}
\Pi = - \chi_\zeta \frac{\partial \Omega_\zeta}{\partial \psi} + P_\zeta \Omega_\zeta + \Pi_\intr,
\end{equation}
where the intrinsic momentum flux $\Pi_\intr$ is the momentum flux for $\Omega_\zeta = 0$ and $\partial \Omega_\zeta/\partial \psi = 0$. The characteristic size of the momentum diffusivity $\chi_\zeta$ and the momentum convection $P_\zeta$ can be deduced from the sizes given in tables \ref{table:h2betatotal}, \ref{table:f2betatbtotal}, \ref{table:phi2betatbtotal} and \ref{table:Pitotal}. The momentum diffusivity $\chi_\zeta = \chi_\zeta^\tb + \chi_\zeta^\nc$ has two components: turbulent diffusivity and neoclassical diffusivity. The turbulent diffusivity is part of $\Pi_{0, \nc}^\tb$ and of $\Pi_{0, \grad}^\tb$, and it is of size
\begin{equation}
\chi_\zeta^\tb \sim \frac{1}{k_\bot a} \rho_i v_{ti} n_e m_i R^2 |\nabla \psi|^2.
\end{equation}
The neoclassical diffusivity is part of $\Pi_0^\nc$, and it is of size
\begin{equation}
\chi_\zeta^\nc \sim \frac{B^2}{B_p^2} \rho_i^2 \nu_{ii} n_e m_i R^2 |\nabla \psi|^2.
\end{equation}
The momentum convection $P_\zeta = \Gamma_\zeta^\tb + \Gamma_\zeta^\nc + P_\zeta^\tb$ has three components: the piece due to the turbulent particle transport $\Gamma_\zeta^\tb$, the piece due to the neoclassical particle transport $\Gamma_\zeta^\nc$, and the Coriolis pinch $P_\zeta^\tb$ described in \cite{peeters07}. The convection due to particle fluxes is 
\begin{eqnarray} \label{eq:Gammamtb}
\Gamma_\zeta^\tb = - \left \langle \left \langle \frac{I^2}{B^2} \sum_{s \neq e} \sum_{k_\psi, k_\alpha}  i k_\alpha m_s c (\underline{\phi}_1^\tb)^\ast \int \dd^3v\, \underline{f}_{s1}^\tb J_0 (\lambda_s) v_{||} \right \rangle_\psi \right \rangle_t
\end{eqnarray}
and
\begin{eqnarray} \label{eq:Gammamnc}
\Gamma_\zeta^\nc = - \left \langle \sum_{s \neq e, s^\prime \neq e} \frac{I^3 m_s^2 c}{Z_s e B^3} \int \dd^3v\, C_{ss^\prime}^{(\ell)} \left [ f_{s1}^\lw; f_{s^\prime 1}^\lw \right ] v_{||} \right \rangle_\psi.
\end{eqnarray}
Their size is given by
\begin{equation}
\Gamma_\zeta^\tb \sim \frac{1}{k_\bot a} \rho_\ast v_{ti} n_e m_i R^2 |\nabla \psi|
\end{equation}
and
\begin{equation}
\Gamma_\zeta^\nc \sim \frac{B^2}{B_p^2} \rho_\ast \rho_i \nu_{ii} n_e m_i R^2 |\nabla \psi|.
\end{equation}
The turbulent momentum pinch $P_\zeta^\tb$ is momentum flux that is proportional to $\Omega_\zeta$ and that is present even in the absence of particle flux \cite{peeters07}. Its characteristic size is
\begin{equation}
P_\zeta^\tb = P_\zeta - \Gamma_\zeta^\tb - \Gamma_\zeta^\nc \sim \frac{1}{k_\bot a} \rho_\ast v_{ti} n_e m_i R^2 |\nabla \psi|.
\end{equation}
Recall that $P_\zeta$, used in the equation above, is defined by \eq{eq:PiOmega}. It may be surprising that in a model for intrinsic rotation based on turbulent momentum redistribution, the dependence on the rotation $\Omega_\zeta$ is linear. The reason is that we have assumed that $R \Omega_\zeta \ll v_{ti}$ so that the rotation is too small to affect the turbulent diffusivity and convective flux. This assumption is incorrect for extreme up-down asymmetric configurations, or for large torques $T_\zeta$.

The steady state solution to equations \eq{eq:torangmomOmega} and \eq{eq:PiOmega} is
\begin{eqnarray} \label{eq:Omegasolution}
\fl \Omega_\zeta ( \psi ) = \Omega_\zeta ( \psi = \psi_a ) \exp \left ( \int_{\psi}^{\psi_a} \frac{P_\zeta(\psi^{\prime\prime})}{\chi_\zeta(\psi^{\prime\prime})} \dd \psi^{\prime\prime} \right ) \nonumber\\ - \int_\psi^{\psi_a} \frac{\Pi_\intr(\psi^{\prime})}{\chi_\zeta(\psi^{\prime})} \exp \left ( \int_{\psi}^{\psi^\prime} \frac{P_\zeta(\psi^{\prime\prime})}{\chi_\zeta(\psi^{\prime\prime})} \dd \psi^{\prime\prime} \right ) \dd \psi^{\prime} \nonumber\\ + \int_\psi^{\psi_a} \frac{\int_0^{\psi^\prime} V^\prime ( \psi^{\prime\prime\prime} ) T_\zeta ( \psi^{\prime\prime\prime} ) \dd \psi^{\prime\prime\prime}}{V^\prime (\psi^\prime) \chi_\zeta(\psi^{\prime})} \exp \left ( \int_{\psi}^{\psi^\prime} \frac{P_\zeta(\psi^{\prime\prime})}{\chi_\zeta(\psi^{\prime\prime})} \dd \psi^{\prime\prime} \right ) \dd \psi^{\prime}.
\end{eqnarray}
Here we have used the boundary conditions $\Pi = 0$ at the magnetic axis, $\psi = 0$, and $\Omega_\zeta = \Omega_\zeta (\psi = \psi_a)$ at the last closed flux surface, $\psi = \psi_a$. Equation \eq{eq:Omegasolution} shows that in the absence of external torque, $T_\zeta = 0$, the rotation does not depend on the size of $\Pi_\intr$, but on the relative size $\Pi_\intr/\chi_\zeta$. Since both $\Pi_\intr$ and $\chi_\zeta$ are proportional to the square of the turbulence amplitude, the final rotation does not depend on turbulence amplitude. Assuming that turbulence is the dominant mechanism for diffusion, i.e. $k_\bot \rho_i \lesssim (B_p/B)^2 (v_{ti}/a \nu_{ii})$, the momentum diffusivity is $\chi_\zeta \sim (k_\bot a)^{-1} \rho_i v_{ti} n_e m_i R^2 |\nabla \psi|^2$, and using table \ref{table:Pitotal}, we can calculate the size of the intrinsic rotation that the different mechanisms give,
\begin{equation}
\Delta \Omega_{\zeta, \intr} ( \psi ) = - \int_\psi^{\psi_a} \frac{\Pi_\intr(\psi^{\prime})}{\chi_\zeta(\psi^{\prime})} \exp \left ( \int_{\psi}^{\psi^\prime} \frac{P_\zeta(\psi^{\prime\prime})}{\chi_\zeta(\psi^{\prime\prime})} \dd \psi^{\prime\prime} \right ) \dd \psi^{\prime}.
\end{equation}
The different sizes of $\Delta \Omega_{\zeta,\intr}$ are given in table \ref{table:DeltaOmegatotal}. Ignoring up-down asymmetries, the size of the intrinsic rotation is $\Delta \Omega_{\zeta, \intr} \sim (B/B_p) \rho_\ast v_{ti}/R$, and it is mostly driven by $\Pi_{0, \nc}^\tb$ and the different pieces of $\Pi_0^\FOW$. Note that due to energy balance (see equation \eq{eq:dTidttotal}), 
\begin{equation}
\sqrt{\frac{m_e}{m_i}} \nu_{ii} \frac{T_e - T_i}{T_i} \sim \frac{Q_s}{f_{Ms}} \sim \frac{1}{\tau_E} \sim \frac{1}{k_\bot \rho_i} \rho_\ast^2 \frac{v_{ti}}{a},
\end{equation}
and the momentum fluxes $\Pi_{0, \Delta T}^\FOW$ and $\Pi_{0, Q}^\FOW$ will always be important. The pieces $\Pi_{0, \grad}^\tb$ and $\Pi_{0, \acc}^\tb$ can become important for turbulence with large eddies, $k_\bot \rho_i \sim B_p/B$. The estimate $\Delta \Omega_{\zeta,\intr} \sim (B/B_p) \rho_\ast v_{ti}/R$ was successfully checked with a large experimental database in \cite{parra12a}.

\begin{table}
\caption{Size of the intrinsic rotation $\Delta \Omega_{\zeta, \intr}$ driven by the different pieces of $\Pi$ for $B_p/B \lesssim k_\bot \rho_i \lesssim 1$.} \label{table:DeltaOmegatotal}
\begin{indented}
\item[]
\begin{tabular}{ l l }
\br Drive mechanism & Size of $\Delta \Omega_{\zeta, \intr}$ \\
\mr $\Pi_{-1, \ud}^\tb$ & $\mathrm{max} \{ (k_\bot \rho_i)^{-2} (B_p/B), \sqrt{m_e/m_i} (B/B_p) \} (k_\bot \rho_i)\Delta_{\ud} v_{ti}/R$  \\
$\Pi_{-1, \ud}^\nc$ & $(B/B_p)^2 (k_\bot \rho_i) (a \nu_{ii}/v_{ti}) \Delta_{\ud} v_{ti}/R$  \\
$\Pi_{0, \nc}^\tb$ & $(B/B_p) \rho_\ast v_{ti}/R$ \\
$\Pi_{0, \grad}^\tb$ & $(k_\bot \rho_i)^{-1} \rho_\ast v_{ti}/R$ \\
$\Pi_{0, \acc}^\tb$ & $(k_\bot \rho_i)^{-1} \rho_\ast v_{ti}/R$ \\
$\Pi_{0}^\nc$ & $(B/B_p)^3 (k_\bot \rho_i) (a \nu_{ii}/v_{ti}) \rho_\ast v_{ti}/R$ \\
$\Pi_{0, \tb}^\FOW$ & $(B/B_p) \rho_\ast v_{ti}/R$ \\
$\Pi_{0, \Delta T}^\FOW$ & $\sqrt{m_e/m_i} (T_e/T_i - 1) (a \nu_{ii}/v_{ti}) \rho_\ast^{-2} (k_\bot \rho_i) (B/B_p) \rho_\ast v_{ti}/R$ \\
$\Pi_{0, Q}^\FOW$ & $(Q_s a/\rho_\ast^2 v_{ti} f_{Ms}) (k_\bot \rho_i)(B/B_p) \rho_\ast v_{ti}/R$ \\
\br
\end{tabular}
\end{indented}
\end{table}

\begin{figure}

\begin{center}
\includegraphics[width = 8 cm]{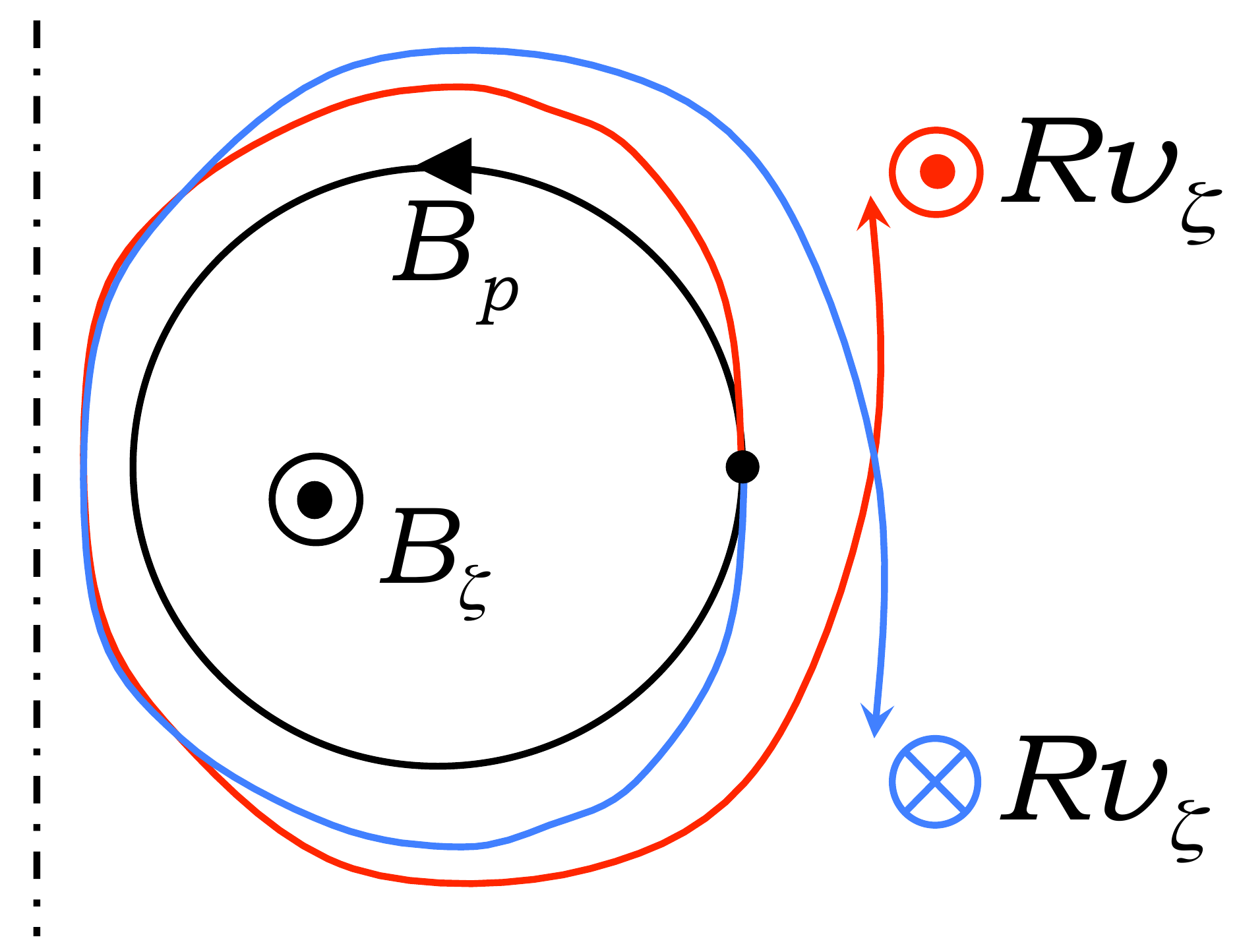}
\end{center}

\caption{\label{fig:symmetryslab} Sketch of two particles with opposite parallel velocity moving in an up-down symmetric tokamak. The axis of axisymmetry is represented by the dash-dot line, the black circle is a flux surface, the toroidal magnetic field is out of the page, and the poloidal magnetic field is pointing counter-clockwise. The trajectory of the particle with positive parallel velocity is plotted in red, and the trajectory of the particle with negative parallel velocity is given in blue.}
\end{figure}

The different drives in table \ref{table:DeltaOmegatotal} have different physical origin, but all of them drive intrinsic rotation because they break the symmetry of the lowest order kinetic equations, given in section \ref{sec:symmetry}. From here on we focus on the symmetry of the turbulence. To discuss the origin of the different drives of intrinsic rotation, we consider the equations with and without magnetic drift. The reason is that the magnetic drift is indispensable for several of the intrinsic rotation drives that we have deduced in section \ref{sec:secondordertotal}. We need to change the sign of $k_\psi$, $\underline{f}_{s1}^\tb$ and $\underline{\phi}_1^\tb$ in the symmetry described by equations \eq{eq:solutiontb1} and \eq{eq:solutiontb2} because tokamaks have a radial magnetic drift $\bv_{Ms} \cdot \nabla_\bR \psi$ and magnetic shear. If in equations \eq{eq:tbfirstordergk} and \eq{eq:tbfirstorderquasineutrality} we artificially make the magnetic shear and the radial component of the magnetic drift zero, there is a new symmetry of the equations in an up-down symmetric flux surface, given by
\begin{equation} \label{eq:symmetryslab}
\theta \rightarrow - \theta,\; u \rightarrow -u,\; \underline{f}_{s1}^\tb \rightarrow \underline{f}_{s1}^\tb, \; \underline{\phi}_1^\tb \rightarrow \underline{\phi}_1^\tb.
\end{equation}
Note that $k_\psi$, $\underline{f}_{s1}^\tb$ or $\underline{\phi}_1^\tb$ need not be reversed in this case. The partial symmetry \eq{eq:symmetryslab} can be understood by looking at two particles that move with opposite parallel velocities, as shown in Figure \ref{fig:symmetryslab}. Statistically, both particles will experience the same turbulence because the tokamak is up-down symmetric, and the turbulence in the upper half must be the same as in the lower half. Without loss of generality, we assume that for the magnitude of the parallel velocity of these two particles, $|u|$, the turbulent $\bE \times \bB$ drift pushes both particles outwards. Then, after a few eddy turnover times, both particles have moved radially outwards, as indicated in the figure, and as a result we have particle and energy turbulent transport. There is no net toroidal angular momentum flux because the toroidal projection of the parallel velocity of one of the particles cancels the projection of the other. Thus, if the background distribution function of particles is symmetric in $u$, there is no momentum flux. If the background distribution function is not symmetric in $u$, there are more particles with one sign of $u$ than particles with the other sign, and the picture in Figure \ref{fig:symmetryslab} just gives the momentum flux due to the turbulent particle flux, $\Gamma_\zeta^\tb \Omega_\zeta$, given in \eq{eq:Gammamtb}.

\begin{figure}

\begin{center}
\includegraphics[width = 10 cm]{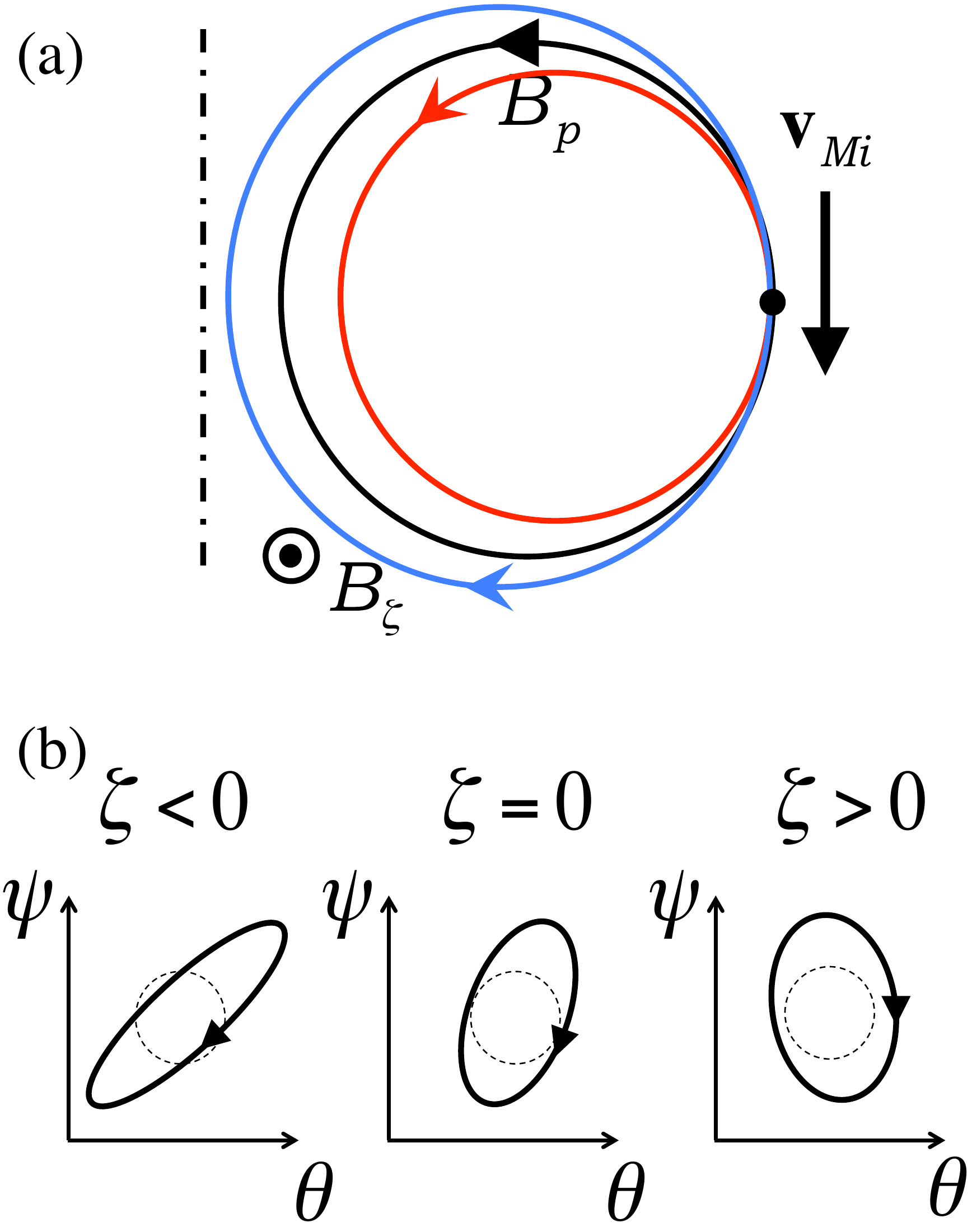}
\end{center}

\caption{\label{fig:symmetrykpsi} Sketch of the effect of (a) the radial magnetic drift $\bv_{Ms} \cdot \nabla_\bR \psi$ and (b) the magnetic shear on the symmetry of the turbulence. In (a) the drift orbits are sketched. The axis of axisymmetry is the dash-dot line, the toroidal magnetic field is out of the page, and the poloidal magnetic field is counter-clockwise. The orbit with positive parallel velocity is plotted in red, and the one with negative parallel velocity is plotted in blue. In (b) the cross section of a turbulent $\bE \times \bB$ eddy at different $\zeta = \mathrm{constant}$ planes is given. The direction of the $\bE \times \bB$ drift is indicated by the arrow, and the size of a typical gyroradius, plotted as a dashed circle, is given for comparison.}
\end{figure}

The partial symmetry \eq{eq:symmetryslab} is only valid when the magnetic shear and $\bv_{Ms} \cdot \nabla_\bR \psi$ are zero. When we have $\bv_{Ms} \cdot \nabla_\bR \psi \neq 0$, the particles follow drift orbits that break symmetry \eq{eq:symmetryslab}, as shown in Figure \ref{fig:symmetrykpsi}(a), where at the highlighted point, particles with positive parallel velocity move radially inwards, and particles with negative parallel velocity move radially outwards. These orbits make symmetry \eq{eq:symmetryslab} invalid, and we need to use the full symmetry in \eq{eq:solutiontb1} and \eq{eq:solutiontb2} where $k_\psi$, $\underline{f}_{s1}^\tb$ and $\underline{\phi}_1^\tb$ must reverse. The reversal in $k_\psi$ is necessary because it means that the particles experience the same turbulence fluctuations when they move radially outwards as they do when they move radially inwards, and hence, the asymmetry imposed by the particle orbits is not important. The turbulence is the same radially out and radially in because we have assumed that the turbulence characteristics vary very slowly in the radial direction, that is, the turbulent eddies are very small compared to characteristic length of the background density and temperature gradients. There is a next order asymmetry in the radial direction due to the slow derivatives $\partial \underline{f}_{s1}^\tb/\partial \psi$ and $\partial \underline{\phi}_1^\tb/\partial \psi$.

The magnetic shear also breaks the partial symmetry \eq{eq:symmetryslab}. The magnetic shear is included in our equations in the way that $\bk_\bot = k_\psi \nabla \psi + k_\alpha \nabla \alpha$ changes with $\theta$, and in particular, in the linear increase of $\nabla \alpha$ with $\theta$. The perpendicular wavevector $\bk_\bot$ breaks symmetry \eq{eq:symmetryslab} only in the terms that contain $i \bk_\bot \cdot \bv_{Ms}$, $J_0 (\Lambda_s)$ and $J_0 (\lambda_s)$ in \eq{eq:tbfirstordergk} and \eq{eq:tbfirstorderquasineutrality}. In the case of the terms with $i \bk_\bot \cdot \bv_{Ms}$, symmetry \eq{eq:symmetryslab} is only broken when $\bv_{Ms} \cdot \nabla_\bR \psi \neq 0$, that is, it is related to the drift orbits sketched in Figure \ref{fig:symmetrykpsi}(a). The finite gyroradius effects that give the terms with $J_0 (\Lambda_s)$ and $J_0 (\lambda_s)$ in \eq{eq:tbfirstordergk} and \eq{eq:tbfirstorderquasineutrality} are the only effects that the magnetic shear has by itself on the turbulence. The magnetic shear determines how the perpendicular structure of an eddy changes along the magnetic field. The effect of the magnetic shear is sketched in Figure \ref{fig:symmetrykpsi}(b), where an eddy with a given tilt at $\theta = 0$ and $\zeta = 0$ has different tilts at different $\zeta$ for positive magnetic shear $\partial q/\partial \psi > 0$. In the same figure, we compare the eddy width with the gyroradius of a characteristic particle. As the particle moves in its gyromotion, it averages over the eddy, so it is clear that the particle will experience a different $\bE \times \bB$ for $\zeta > 0$ than for $\zeta < 0$, and as a result, particles with opposite parallel velocities will not experience the same turbulence. The full symmetry given in \eq{eq:solutiontb1} and \eq{eq:solutiontb2} solves this problem because it indicates that eddies do not have a preferred tilt at $\theta = 0$, and hence, there will be eddies with a tilt opposite to the tilt depicted in Figure \ref{fig:symmetrykpsi}(b) that will compensate for the momentum flux driven by the eddy in this figure.

In this article, we have rigorously deduced the main symmetry breaking mechanisms for $B_p/B \ll 1$. The main mechanisms that break the symmetry of the turbulence and drive intrinsic rotation are: up-down asymmetry, neoclassical flows, the slow variation of the turbulence characteristics, the turbulent acceleration of the particles, and finite orbit widths. These different drives correspond to $\Pi_{-1,\ud}^\tb$, $\Pi_{0, \nc}^\tb$, $\Pi_{0, \grad}^\tb$, $\Pi_{0, \acc}^\tb$ and $\Pi_0^\FOW$, respectively. We discuss each one of these mechanisms below.

\subsection{Momentum flux driven by up-down asymmetry} \label{sub:picturesud}
It is obvious that an up-down asymmetric flux surface breaks the symmetry described in section \ref{sec:symmetry}. This effect has already been studied in \cite{camenen09b, camenen09c, ball14}. In our notation, this effect is included in $\Pi_{-1, \ud}^\tb$, and its size is given in tables \ref{table:Pitotal} and \ref{table:DeltaOmegatotal}. This drive is reduced when the turbulence has small perpendicular eddies, i.e. for large $k_\bot \rho_i$. The reason is that small eddies have shorter turn over times, typically of order $(k_\bot \rho_i)^{-1} (a/ v_{ti})$, and as a result, they do not have time to extend very far along the magnetic field lines because their typical parallel propagation velocity, $v_{ti}$, is not affected by the eddy size. Eddies that are small in the parallel direction are not able to sample a large portion of the flux surface, and as a result, are not affected by its up-down asymmetry.

\subsection{Momentum flux driven by neoclassical corrections} \label{sub:picturesnctb}
The effect of neoclassical corrections on turbulence was proposed as a drive for intrinsic rotation in \cite{parra10a, parra11d}, and it has been studied numerically in \cite{barnes13, lee14a, lee14b, lee14c}. The finite width of drift orbits depicted in Figure \ref{fig:symmetrykpsi}(a) drives neoclassical flows, poloidal perturbations to the density and temperature, and poloidal electric fields. All these effects break the symmetry of the turbulence even in up-down symmetric flux surfaces, and in our model this effect is included in $\Pi_{0, \nc}^\tb$. 

The poloidal electric field breaks the symmetry because it creates a potential $\phi_1^\lw$ that satisfies $\phi_1^\lw (\theta) = - \phi_1^\lw(-\theta)$, that is, it is odd in $\theta$. Due to this correction to the potential, particles with positive parallel velocity experience a different background potential than particles with negative parallel velocity, breaking the turbulence symmetry. Similar effects are obtained from poloidal perturbations to the background density and temperature.

The effect of neoclassical flows on momentum transport is more subtle. The momentum flux driven by an existing flow is not an intrinsic rotation drive, but neoclassical flows lead to intrinsic rotation. Consider a situation in which $\Omega_\zeta = 0$, where $\Omega_\zeta$ is defined in \eq{eq:Omegadef}. If for $\Omega_\zeta = 0$, $\Pi_{0, \nc}^\tb \neq 0$, then we can say that neoclassical flows drive intrinsic rotation. This can happen because $\Omega_\zeta = 0$ does not imply that each one of the contributions to $\Omega_\zeta$ given in \eq{eq:Omegadef} vanish, but that they cancel each other. The piece of $\Omega_\zeta$ proportional to the radial electric field $\partial \phi_0/\partial r$ modifies particle orbits and makes them precess toroidally, whereas the pressure and temperature gradient flows are due to the finite orbit width of the particles, but do not modify particle orbits. The pressure and temperature gradient drive flows with different poloidal dependence and direction. Due to these differences, the value of the turbulent diffusivity and pinch for each of the flows in \eq{eq:Omegadef} is different. As a result, even if these flows combine to give $\Omega_\zeta = 0$, the momentum flux they produce does not add to zero, and produces momentum redistribution and intrinsic rotation. The different turbulent diffusivities and pinches have been studied in \cite{lee14a, lee14b, lee14c}.

The size of these neoclassical corrections is related to the width of the drift orbits depicted in Figure \ref{fig:symmetrykpsi}(a). These orbits have a radial width of order the poloidal gyroradius, $(B/B_p) \rho_s$, and as a result, they drive intrinsic rotation that is small compared to the thermal speed by $(B/B_p) \rho_\ast$. 

For tokamaks with $B_p/B \ll 1$ and gyroradius scale turbulence in \cite{parra10a, parra11d}, the effect of neoclassical corrections on turbulence was predicted to be the most important effect for intrinsic rotation along with the finite orbit width effects described in subsection \ref{sub:picturesFOW}. This prediction has been confirmed in this article where we have allowed poloidal gyroradius scale turbulence in addition to gyroradius scale turbulence. Only when the turbulent eddies have large perpendicular scales can other effects such as the slow variation of the turbulence characteristics or the turbulent acceleration compete. We discuss these two important effects in subsections \ref{sub:picturesgradtb} and \ref{sub:picturesacctb}.

\subsection{Momentum flux driven by the variation of the turbulence characteristics} \label{sub:picturesgradtb}
The characteristics of the turbulence change radially and poloidally because the plasma density and temperature change with radius, and the magnetic field magnitude and direction change with poloidal angle. The length of variation of the turbulence characteristics is long compared to the characteristic eddy size, and it does not affect the turbulence to lowest order, but it is important for intrinsic rotation. This effect is included in our model via $\Pi_{0, \grad}^\tb$, and has been considered in detail in \cite{waltz11} and \cite{sung13}.

When we discussed the symmetry of the turbulence in section \ref{sec:pictures}, we mentioned that due to the drift orbits sketched in Figure \ref{fig:symmetrykpsi}(a), particles with positive parallel velocity move inwards, and particles with negative parallel velocity move outwards, but this asymmetry did not cause momentum transport because the turbulence was statistically the same radially in and radially out. This uniformity of the turbulence was only true to lowest order, and was based on the fact that turbulence eddies were small compared to the characteristic length of variation of the background density and temperature. By considering the next order slow radial variation of $\underline{f}_{s1}^\tb$ and $\underline{\phi}_1^\tb$, we break the symmetry. The ratio between characteristic size of the eddies and the background radial scale length is $(k_\bot a)^{-1} \ll 1$. Then, we expect the characteristic size of the momentum flux due to the slow radial variation of the turbulence characteristics to scale with $(k_\bot \rho_i)^{-1}$. If we compare $(k_\bot \rho_i)^{-1}$ to the characteristic size of $\Pi_{0, \nc}^\tb$, of order $(B/B_p) \rho_\ast$, we can see why only turbulence with large eddies, of the order of the poloidal gyroradius, can produce sufficient intrinsic momentum flux $\Pi_{0, \grad}^\tb$ to compete with $\Pi_{0, \nc}^\tb$.

The poloidal variation of the turbulence characteristics also matters for momentum transport. The reason is that perpendicular drifts such as the magnetic drift and the turbulent $\bE \times \bB$ drift move particles in the poloidal direction. This means that even in the absence of the radial magnetic drift $\bv_{Ms} \cdot \nabla_\bR \psi$ and hence, drift orbits, magnetic drifts can break the partial symmetry in \eq{eq:symmetryslab}. For example, a  downwards magnetic drift imposed on the situation presented in Figure \ref{fig:symmetryslab} breaks the symmetry because it means that positive parallel velocity particles move to the region $\theta > 0$ more slowly than particles with negative parallel velocity move to the region $\theta < 0$. Due to magnetic and turbulent $\bE \times \bB$ drifts, particles move in $\theta$, and as a result, they experience a slow gradient due to the poloidal variation of the turbulence characteristics. This change can be estimated by calculating the poloidal distance that a particle moves in an eddy turnover time, $(k_\bot \rho_i)^{-1}(a/ v_{ti})$. Given that the drifts are of order $\rho_\ast v_{ti}$, the particle moves poloidally a distance $k_\bot^{-1}$, and given that the characteristic length of variation in the poloidal direction is $a$, this effect is, as the radial variation, of order $(k_\bot a)^{-1} \ll 1$.

Note that magnetic shear is important for the symmetry of the turbulence, but we have not considered it in the physical picture presented here. The slow variation of the turbulence characteristics only matters for turbulence with large eddies, and in this case the finite gyroradius effects become unimportant. Without finite gyroradius effects, the magnetic shear only enters in a term proportional to $\bv_{Ms} \cdot \nabla_\bR \psi$ (see the discussion following equation \eq{eq:symmetryslab}). Thus, for $B_p/B \ll 1$ the magnetic shear does not drive intrinsic rotation by itself, but in conjunction with the radial magnetic drift $\bv_{Ms} \cdot \nabla_\bR \psi$. A consequence of this result is that $\Pi_{0, \grad}^\tb$ must depend mainly on the magnetic drift: on the radial component $\bv_{Ms} \cdot \nabla_\bR \psi$ due to the radial variation of the turbulence characteristics, and on the poloidal component $\bv_{Ms} \cdot \nabla_\bR \theta$ due to the poloidal variation.

\subsection{Momentum flux driven by turbulent acceleration} \label{sub:picturesacctb}
\begin{figure}
\begin{center}
\includegraphics[width = 10 cm]{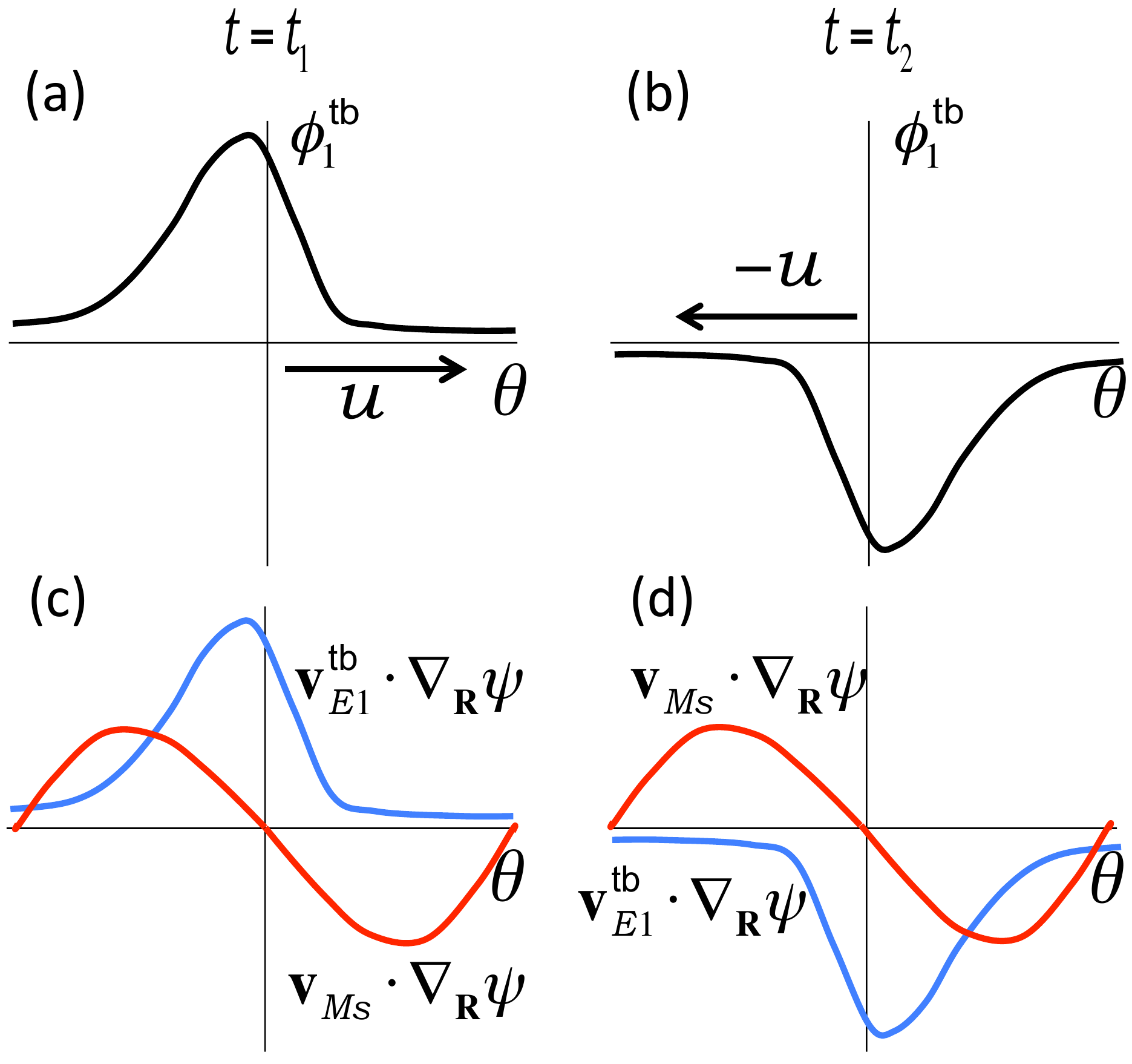}
\end{center}

\caption{\label{fig:nonlinearsymmetry} The first line of figures are sketches of the turbulent piece of the potential on the magnetic field line with $\psi = \psi_0$ and $\alpha = \alpha_0$ at times (a) $t = t_1$ and (b) $t = t_2$. The flux surface is symmetric with respect to $\theta = 0$. The second line of figures are sketches of the turbulent radial $\bE \times \bB$ drift, $\bv_{E1}^\tb \cdot \nabla_\bR \psi$, (blue) and the radial magnetic drift, $\bv_{Ms} \cdot \nabla_\bR \psi$, (red) on the magnetic field line with $\psi = \psi_0$ and $\alpha = \alpha_0$ at times (c) $t = t_1$ and (d) $t = t_2$.}
\end{figure}

In an eddy turnover time, $(k_\bot \rho_i)^{-1} (a/ v_{ti})$, the particle is accelerated by a parallel electric field $\rho_\ast T_e /e a$, giving a change in the parallel velocity of order $(k_\bot a)^{-1} v_{ti} \ll v_{ti}$. This small change in the parallel velocity can break the symmetry of the turbulence if there is a finite radial magnetic drift $\bv_{Ms} \cdot \nabla_\bR \psi$. If we force the radial component of the magnetic drift $\bv_{Ms} \cdot \nabla_\bR \psi$ to be zero, the piece of the momentum flux due to the turbulent acceleration, $\Pi_{0, \acc}^\tb$, vanishes because the partial symmetry \eq{eq:symmetryslab} is satisfied. Then, $\Pi_{0, \acc}^\tb$ must depend strongly on the radial component of the magnetic drift, $\bv_{Ms} \cdot \nabla_\bR \psi$. To see why the radial magnetic drift is necessary, we need to explain why $\underline{f}_{s1}^\tb$ and $\underline{\phi}_1^\tb$ change sign in the full symmetry in \eq{eq:solutiontb1} and \eq{eq:solutiontb2}. This change of sign is needed because the radial component of the $\bE \times \bB$ drift must change sign, as we proceed to explain. We focus on the turbulence around the magnetic field line defined by $\psi = \psi_0$ and $\alpha = \alpha_0$. The idea behind the full symmetry in \eq{eq:solutiontb1} and \eq{eq:solutiontb2} is that if at time $t = t_1$ we have a turbulent potential configuration $\phi_1^\tb (\psi - \psi_0, \alpha - \alpha_0, \theta, t_1)$ like the one sketched in figure \ref{fig:nonlinearsymmetry}(a), at another time $t = t_2$ we find a potential $\phi_1^\tb (\psi - \psi_0, \alpha - \alpha_0, \theta, t_2)$ such that to lowest order, particles with parallel velocity $- u$ at $-\theta$ and $t = t_2$ cancel the contribution to the momentum flux due to particles with parallel velocity $u$ at $\theta$ and $t = t_1$. To determine the relation between $\phi_1^\tb (\psi - \psi_0, \alpha - \alpha_0, \theta, t_1)$ and $\phi_1^\tb (\psi - \psi_0, \alpha - \alpha_0, \theta, t_2)$ we must consider the radial component of the magnetic and $\bE \times \bB$ drifts. The radial components of the turbulent $\bE \times \bB$ drift and the magnetic drift, $\bv_{E1}^\tb \cdot \nabla_\bR \psi = c (\partial \phi_1^\tb/\partial \alpha)$ and $\bv_{Ms} \cdot \nabla_\bR \psi$, at $t = t_1$ are sketched in figure \ref{fig:nonlinearsymmetry}(c). A particle with positive parallel velocity $u$ leaving from $\theta = 0$ (sketched as an arrow to the right in figure \ref{fig:nonlinearsymmetry}(a)) sees then a positive radial $\bE \times \bB$ and a negative radial magnetic drift. At time $t = t_2$, a particle with parallel velocity $-u$ leaving from $\theta = 0$ (sketched as an arrow to the left in figure \ref{fig:nonlinearsymmetry}(b)) must feel the same radial drift to give a contribution to the momentum flux that exactly opposes the momentum flux due to a particle with parallel velocity $u$ at $\theta = 0$ and $t = t_1$. Since the radial magnetic drift $\bv_{Ms} \cdot \nabla_\bR \psi$ is odd in $\theta$, the radial $\bE \times \bB$ at $t = t_2$ must be related to the radial $\bE \times \bB$ at $t = t_1$ by $(\bv_{E1}^\tb \cdot \nabla_\bR \psi) (\theta, t = t_2) = - ( \bv_{E1}^\tb \cdot \nabla_\bR \psi) (-\theta, t=t_1)$. When this is satisfied, the particle with positive parallel velocity $u$ leaving from $\theta = 0$ at $t = t_1$ and the particle with parallel velocity $-u$ leaving from $\theta = 0$ at $t = t_2$ feel total radial drifts of the same magnitude. The fact that the radial drifts have opposite signs at $t = t_1$ and $t = t_2$ is not important because the turbulence is uniform in the radial direction to lowest order. Then, as a consequence of having a radial magnetic drift $\bv_{Ms} \cdot \nabla_\bR \psi$, the radial $\bE \times \bB$ drift at $t = t_2$ must be the one sketched in figure \ref{fig:nonlinearsymmetry}(d). The corresponding turbulent piece of the potential is $\phi_1^\tb ( \psi - \psi_0, \alpha - \alpha_0, \theta, t_2) = - \phi_1^\tb ( - (\psi - \psi_0), \alpha - \alpha_0, - \theta, t_1)$, and it is sketched in figure \ref{fig:nonlinearsymmetry}(b).

The symmetry between the potentials in figures \ref{fig:nonlinearsymmetry}(a) and \ref{fig:nonlinearsymmetry}(b) is broken by the turbulent parallel acceleration. In the potential in figure \ref{fig:nonlinearsymmetry}(a), a particle with positive parallel velocity $u$ leaving from $\theta = 0$ (sketched in the figure as an arrow to the right) is accelerated by the potential, whereas a particle with parallel velocity $-u$ leaving from $\theta = 0$ in figure \ref{fig:nonlinearsymmetry}(b) (sketched as an arrow to the left) is decelerated. As a result, a particle with positive parallel velocity $u$ at $t = t_1$ spends more time in the region $\theta > 0$ than a particle with parallel velocity $-u$ at $t = t_2$ spends in the region $\theta < 0$, and the contributions to momentum flux from these two particles do not cancel exactly, giving $\Pi_{0,\acc}^\tb \neq 0$. The size of $\Pi_{0,\acc}^\tb$ can be estimated from the typical change in the parallel velocity due to the turbulent acceleration, $(k_\bot a)^{-1} v_{ti}$, giving intrinsic rotation speeds small in $(k_\bot a)^{-1} \ll 1$. This effect can only be comparable to the effect of the neoclassical corrections for turbulent eddies of the order of the poloidal gyroradius.

\subsection{Momentum flux driven by finite orbit widths} \label{sub:picturesFOW}
\begin{figure}

\begin{center}
\includegraphics[width = 10cm]{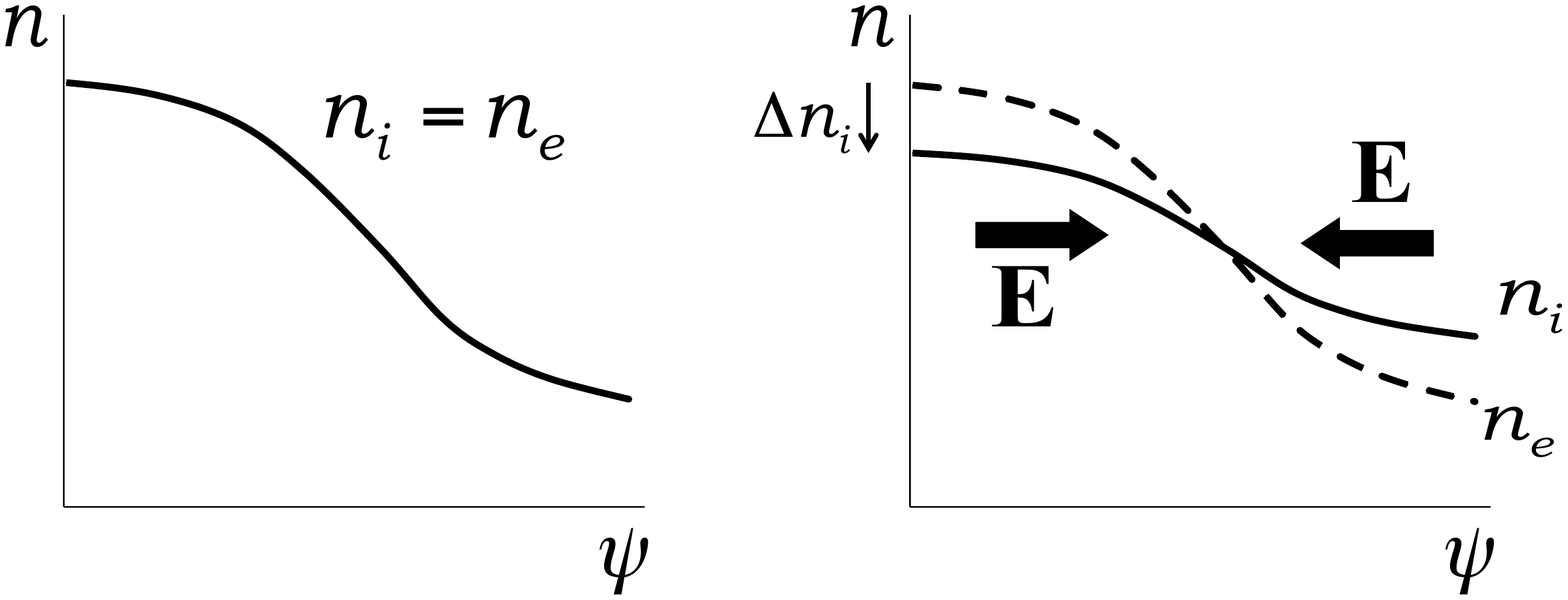}
\end{center}

\caption{\label{fig:PiFOW} Sketch of the effect of finite orbit widths on quasineutrality, and consequently, on rotation. The left graph gives the ion and electron density profile before increasing the average particle orbit width. The right graph gives the ion density (solid line) and the electron density (dashed line) after the average particle orbit width has been increased, and the radial electric field $\bE$ set up due to the charge difference.}
\end{figure}

Due to the presence of drift orbits such as the ones sketched in figure \ref{fig:symmetrykpsi}(a), the parallel velocity and the radial position are correlated. Then, fluctuations in the parallel velocity will be correlated to fluctuations in the radial position and any mechanism that changes the parallel velocity of particles will induce momentum flux. This effect is captured by the finite orbit width momentum flux $\Pi_0^\FOW$. This momentum flux is better understood using quasineutrality. The momentum flux $\Pi_0^\FOW$ is the result of a local increase or decrease of the average energy of the plasma at a radial position that can be driven by turbulence, the collisional transfer of energy between electrons and ions, or sources, leading to the momentum fluxes $\Pi_{0, \tb}^\FOW$, $\Pi_{0, \Delta T}^\FOW$ and $\Pi_{0, Q}^\FOW$. When the plasma average energy is increasing at a given radial position, the average width of the particle orbits increases. The easiest way to calculate this increase is to use the canonical angular momentum,
\begin{equation}
\psi_\ast = \psi - \frac{m_s c}{Z_s e} R (\bv \cdot \zun),
\end{equation}
which must be conserved unless momentum is injected. To give a simple physical picture, we take a plasma formed by one single ion species and electrons in which we increase the average energy at a flux surface by heating, but we do not inject toroidal angular momentum, i.e., $\overline{R \bv\cdot\zun}$ does not change in time. Here $\overline{(\ldots)}$ is an average over particles and over the flux surface. An increase in energy gives an increase in the average value $\overline{R^2 (\bv\cdot\zun)^2}$ and as a consequence, in the average width of the orbits $\overline{(\Delta \psi)^2}$. Before we heat up the plasma to increase $\overline{R^2 (\bv\cdot\zun)^2}$, the plasma is quasineutral. After $\overline{R^2 (\bv\cdot\zun)^2}$ has increased, the ion density changes because of the average increase in particle orbits. The difference between the ion density before and after the increase in $\overline{R^2 (\bv\cdot\zun)^2}$ is sketched in Figure \ref{fig:PiFOW}, and can be estimated by Taylor expanding the ion density $n_i (\psi)$ around $\psi_\ast$ to find
\begin{equation}
n_i (\psi) \simeq n_i (\psi_\ast) + \frac{m_i c}{Z_i e} R (\bv \cdot \zun) \frac{\partial n_i}{\partial \psi_\ast} + \frac{m_i^2 c^2}{2Z_i^2 e^2} R^2 (\bv \cdot \zun)^2 \frac{\partial^2 n_i}{\partial \psi_\ast^2}.
\end{equation}
Averaging over this equation and substracting the density before increasing $\overline{R^2 (\bv\cdot\zun)^2}$ from the density after increasing it, we obtain the change in ion density
\begin{equation}
\Delta n_i = \frac{m_i^2 c^2}{2Z_i^2 e^2} \Delta [ \overline{R^2 (\bv \cdot \zun)^2} ] \frac{\partial^2 n_i}{\partial \psi^2},
\end{equation}
where $\Delta [ \overline{R^2 (\bv \cdot \zun)^2} ]$ is the increase in $\overline{R^2 (\bv \cdot \zun)^2}$ (recall that we are assuming $\Delta [ \overline{R \bv \cdot \zun} ] = 0$). The electrons have orbits with a much smaller radial width, so the equivalent change in electron density, $\Delta n_e$, is negligible, and as a result, we have a charge imbalance $Z \Delta n_i$. A radial electric field is set up to compensate this charge imbalance. Considering the polarization of the plasma due to ion drift orbits, of width $(B/B_p) \rho_i$, we find that the radial electric field $\nabla \psi ( \partial \phi_0/\partial \psi )$ produces a polarization ion density \cite{rosenbluth98, hinton99} of order
\begin{equation}
n_{i, p} \sim - \frac{n_i m_i R^2 c^2}{Z_i e}\frac{\partial^2 \phi_0}{\partial \psi^2}.
\end{equation}
By balancing $n_{i,p}$ with $\Delta n_i$ and taking a time derivative we find
\begin{equation}
- \frac{\partial}{\partial t} \left ( n_i m_i R^2 c \frac{\partial \phi_0}{\partial \psi} \right ) \sim \frac{\partial}{\partial \psi} \left [ \frac{m_i c}{2Z_i e} \frac{\partial}{\partial t} ( n_i m_i \overline{R^2 (\bv \cdot \zun)^2}) \right ].
\end{equation}
By identifying the time derivative of $c(\partial \phi_0/\partial \psi)$ with the time derivative of $\Omega_\zeta$, we can identify 
\begin{equation} \label{eq:roughPiFOW}
\Pi_0^\FOW \sim - \frac{m_i c}{2Z_i e} \frac{\partial}{\partial t} ( n_i m_i \overline{R^2 (\bv \cdot \zun)^2})
\end{equation}
with a piece of the momentum flux. Here the time derivative of $n_i m_i \overline{R^2 (\bv \cdot \zun)^2}$ can be due to turbulence, collisional energy exchange between ions and electrons, or energy injection. This is not meant to be a rigorous derivation, but just a physical picture to indicate how finite orbit widths affect momentum transport. Note that expressions \eq{eq:Pi0tbFOWtotal}, \eq{eq:Pi0DeltaTFOWtotal} and \eq{eq:Pi0QFOWtotal} are more complicated than \eq{eq:roughPiFOW}. The more complex formulas \eq{eq:Pi0tbFOWtotal}, \eq{eq:Pi0DeltaTFOWtotal} and \eq{eq:Pi0QFOWtotal} are the result of being more careful with the derivation and taking into account that $\Omega_\zeta$ is not only $-c(\partial \phi_0/\partial \psi)$.

\section{Conclusions} \label{sec:conclusion}

We have obtained the equations for the evolution of intrinsic rotation in conventional tokamaks by exploiting the smallness of the poloidal magnetic field with respect to the total magnetic field, $B_p/B \ll 1$. The momentum redistribution that leads to intrinsic rotation is most difficult to calculate in an up-down symmetric tokamak, in which the momentum flux is zero to lowest order in $\rho_\ast$ due to a symmetry of the turbulent fluctuations. In this case, the turbulent momentum flux, and consequently the distribution function and the electrostatic potential, are needed to a higher order in $\rho_\ast$ than is usual in gyrokinetics. The full higher order equations are difficult to implement in a code, and have been simplified by expanding in $B_p/B \ll 1$. This expansion had been attempted before in \cite{parra10a, parra11d}, but in these articles, the turbulence was assumed to have characteristic perpendicular lengths of the order of the ion gyroradius. In this article, we have allowed turbulence with characteristic lengths of the order of the poloidal gyroradius, that is, with perpendicular wavelengths of the order of $(B/B_p) \rho_i$. The final equations for the higher order pieces are given in section \ref{sec:secondordertotal}. In \cite{parra10a, parra11d}, for gyroradius scale turbulence, we found that the momentum flux was mostly driven by the effect of the neoclassical corrections on the turbulence, and by finite orbit widths. This conclusion is confirmed here, but in addition, we find that for poloidal gyroradius scale turbulence, we have some new intrinsic rotation drives: the slow variation of the turbulence characteristics, and turbulent acceleration. These effects must be retained for ion poloidal gyroradius turbulence. In section \ref{sec:pictures}, all these drives are discussed in detail.

The equations of section \ref{sec:secondordertotal} are written for flux tube simulations. Unlike previous flux tube gyrokinetic equations, they capture the effect of the slow radial variation of the density and temperature gradients. This effect is contained in the second order piece $\underline{f}_{s2, \grad}^\tb$ that depends on the slow radial derivatives $\partial \underline{f}_{s1}^\tb/\partial \psi$ and $\partial \underline{\phi}_1^\tb/\partial \psi$. These derivatives can be calculated by integrating equations \eq{eq:radialderivativegktotal} and \eq{eq:radialderivativequasineutralitytotal}.

\ack{The authors would like to thank Peter J. Catto, Ivan Calvo, Jungpyo Lee, Greg Hammett and John Krommes for many helpful discussions. This work has been carried out within the framework of the EUROfusion Consortium and has received funding from the European Union's Horizon 2020 research and innovation programme under grant agreement number 633053. The views and opinions expressed herein do not necessarily reflect those of the European Commission. This research was supported in part by the RCUK Energy Programme (grant number EP/I501045) and by US Department of Energy Grant No. DE-SC008435.}

\appendix

\section{Calculation of $\phiave$ and $\phiwig$} \label{app:phiavephiwig}

We can evaluate part of $\phiave$, defined in \eq{eq:phiavedef}, by Taylor expanding the long wavelength pieces of $\phi(\boldr, t)$ around $\bR$ to find
\begin{eqnarray} \label{eq:philwexpand}
\fl \phi^\lw (\boldr, t) = \phi_0 (\bR, t) + \phi_1^\lw (\bR, t) + \rhobf \cdot \nabla_\bR \phi_0 (\bR, t) + \phi_2^\lw (\bR, t) + \rhobf \cdot \nabla_\bR \phi_1^\lw (\bR, t) \nonumber\\ + \frac{1}{2} \rhobf \rhobf : \nabla_\bR \nabla_\bR \phi_0 (\bR, t) + O\left (\rho_\ast^3 \frac{T_e}{e} \right ).
\end{eqnarray}
Then, it is clear that $\phiave^\lw = (2\pi)^{-1} \int_0^{2\pi} \dd \varphi\, \phi^\lw$ gives \eq{eq:phiaveRlw} with 
\begin{equation} \label{eq:phiaveRlw2}
\phiave_2^\lw = \phi_2^\lw + \frac{\mu B}{2 \Omega_s^2} (\matI - \bun \bun) : \nabla_\bR \nabla_\bR \phi_0,
\end{equation}
where we have used $\langle \rhobf \rhobf \rangle = (\mu B/\Omega_s^2) (\matI - \bun \bun)$. Note that the arguments of the functions $\phi_0$, $\phi_1^\lw$ and $\phi_2^\lw$ are given by $\phi_0 = \phi_0(\bR, t)$, $\phi_1^\lw = \phi_1^\lw(\bR, t)$ and $\phi_2^\lw = \phi_2^\lw (\bR, t)$.

Similarly, we can Taylor expand $\phi^\tb (\boldr, t)$ in \eq{eq:phitbscales} to find
\begin{eqnarray} \label{eq:phitbRumuvartheta}
\fl \phi^\tb (\boldr, t) = \sum_{k_\psi, k_\alpha} \Bigg ( \underline{\phi}_1^\tb + \underline{\phi}_2^\tb + \frac{i}{2} \rhobf \rhobf : \nabla_\bR \bk_\bot\,\underline{\phi}_1^\tb  + \rhobf \cdot \nabla_\bR \underline{\phi}_1^\tb \Bigg ) \exp( i \bk_\bot \cdot \rhobf ) \nonumber\\ \times \exp ( i k_\psi \psi (\bR) +  i k_\alpha \alpha (\bR) ) + O\left (\rho_\ast^3 \frac{T_e}{e} \right ),
\end{eqnarray}
where the arguments of the functions $\underline{\phi}_1^\tb$ and $\underline{\phi}_2^\tb$ are given by $\underline{\phi}_1^\tb = \underline{\phi}_1^\tb (k_\psi, k_\alpha, \psi(\bR), \theta(\bR), t)$ and $\underline{\phi}_2^\tb = \underline{\phi}_2^\tb (k_\psi, k_\alpha, \psi(\bR), \theta(\bR), t)$. The gradients $\nabla_\bR \bk_\bot$ and $\nabla_\bR \underline{\phi}^\tb_1$ indicate derivatives with respect to the slow variables $\psi(\bR)$ and $\theta(\bR)$. According to \eq{eq:phitbRumuvartheta}, $\phiave^\tb = (2\pi)^{-1} \int_0^{2\pi} \dd \varphi \, \phi^\tb$ is given by \eq{eq:phiaveRtb} with
\begin{eqnarray} \label{eq:underlinephiaveRtb2}
\fl \underline{\phiave}^\tb_2 = J_0 (\Lambda_s) \underline{\phi}_2^\tb + \frac{2J_1(\Lambda_s)}{\Lambda_s}\frac{i \mu B}{2 \Omega_s^2} \left ( \matI - \bun \bun \right ) : \nabla_\bR \bk_\bot\, \underline{\phi}_1^\tb  \nonumber\\ - G(\Lambda_s) \frac{i k_\bot \mu^2 B^2 }{4\Omega_s^4 }  \bk_\bot \cdot \nabla_\bR k_\bot\, \underline{\phi}_1^\tb + \frac{2 J_1 (\Lambda_s)}{\Lambda_s} \frac{i \mu B}{\Omega_s^2}  \bk_\bot \cdot \nabla_\bR \underline{\phi}_1^\tb,
\end{eqnarray}
where
\begin{equation}
G(\Lambda_s) = \frac{8 J_1 (\Lambda_s) - 4\Lambda_s (J_0 (\Lambda_s) - J_2 (\Lambda_s))}{\Lambda_s^3}
\end{equation}
is defined such that $G \rightarrow 1$ for $\Lambda_s \rightarrow 0$, and $\Lambda_s$ is defined in \eq{eq:Lambdasdef}. Here we have used
\begin{equation}
\frac{1}{2\pi} \int_0^{2\pi} \dd \varphi\, \exp( i\bk_\bot \cdot \rhobf ) = J_0 (\Lambda_s),
\end{equation}
\begin{equation}
\frac{1}{2\pi} \int_0^{2\pi} \dd \varphi\, \rhobf \exp( i\bk_\bot \cdot \rhobf ) = \frac{2 J_1 (\Lambda_s)}{\Lambda_s} \frac{i \bk_\bot \mu B}{\Omega_s^2}
\end{equation}
and
\begin{equation}
\fl \frac{1}{2\pi} \int_0^{2\pi} \dd \varphi\, \rhobf \rhobf \exp( i\bk_\bot \cdot \rhobf ) = \frac{2J_1(\Lambda_s)}{\Lambda_s}\frac{\mu B}{\Omega_s^2} \left ( \matI - \bun \bun \right ) - G(\Lambda_s) \frac{\mu^2 B^2 }{2\Omega_s^4 } \bk_\bot \bk_\bot.
\end{equation}

The function $\phiwig^\lw$ will be needed in $\{\boldr, \bv \}$ coordinates. We Taylor expand \eq{eq:phiaveRlw} around $\boldr$ and $\mu_0$, and we subtract the result from $\phi^\lw$ to find
\begin{eqnarray} \label{eq:phiwiglwrv}
\fl \phiwig^\lw (\boldr, \bv, t) = \phi^\lw - \phiave^\lw = - \frac{1}{\Omega_s} (\bv \times \bun) \cdot \nabla \phi_0 - \frac{1}{\Omega_s} (\bv \times \bun) \cdot \nabla \phi_1^\lw \nonumber\\ - \frac{1}{2\Omega_s^2} \Bigg [ (\bv \times \bun) (\bv \times \bun) + \frac{v_\bot^2}{2} (\matI - \bun \bun) \Bigg ] : \nabla \nabla \phi_0 - \bR_2 \cdot \nabla \phi_0 \nonumber\\ + O\left (\rho_\ast^3 \frac{T_e}{e} \right),
\end{eqnarray}
where the arguments of the functions $\phi_0$, $\phi_1^\lw$ and $\phi_2^\lw$ are given by $\phi_0 = \phi_0 (\boldr, t)$, $\phi_1^\lw = \phi_1^\lw (\boldr, t)$ and $\phi_2^\lw = \phi_2^\lw (\boldr, t)$. We need $\phiwig^\tb$ in $\{\boldr, \bv \}$ coordinates as well. We Taylor expand \eq{eq:phiaveRtb} around $\boldr$ and $\mu_0$, and then we subtract the result from $\phi^\tb$ to find
\begin{eqnarray} \label{eq:phiwigtbrv}
\fl \phiwig^\tb (\boldr, \bv, t) = \phi^\tb - \phiave^\tb = \sum_{k_\psi, k_\alpha} \Bigg \{ \left (1 - J_0 (\lambda_s) \exp \left ( \frac{i \bk_\bot \cdot (\bv \times \bun)}{\Omega_s} \right ) \right ) (\underline{\phi}_1^\tb + \underline{\phi}_2^\tb) \nonumber\\ + \Bigg [ \left ( \frac{v_\bot  \sqrt{B}}{\Omega_s^2} (\bv \times \bun) \cdot \nabla \left ( \frac{k_\bot}{\sqrt{B}} \right ) + \frac{k_\bot m_s c \mu_1}{Z_s e v_\bot} \right ) J_1 (\lambda_s) \underline{\phi}_1^\tb \nonumber\\ - i J_0 (\lambda_s) \left ( \bk_\bot \cdot  \bR_2 + \frac{1}{2\Omega_s^2} (\bv \times \bun) (\bv \times \bun) : \nabla \bk_\bot \right ) \underline{\phi}_1^\tb \nonumber\\ - \frac{J_0(\lambda_s)}{\Omega_s} (\bv \times \bun) \cdot \nabla \underline{\phi}_1^\tb - \frac{2J_1(\lambda_s)}{\lambda_s}\frac{i v_\bot^2}{4 \Omega_s^2} \left ( \matI - \bun \bun \right ) : \nabla \bk_\bot\, \underline{\phi}_1^\tb \nonumber\\ + G(\lambda_s) \frac{i k_\bot v_\bot^4 }{16\Omega_s^4 }  \bk_\bot \cdot \nabla k_\bot\, \underline{\phi}_1^\tb -\frac{2 J_1 (\lambda_s)}{\lambda_s} \frac{i v_\bot^2}{2\Omega_s^2} \bk_\bot \cdot \nabla \underline{\phi}_1^\tb \Bigg ] \nonumber \\ \times  \exp \left ( \frac{i \bk_\bot \cdot (\bv \times \bun)}{\Omega_s} \right ) \Bigg \} \exp ( i k_\psi \psi (\boldr) +  i k_\alpha \alpha (\boldr) ) + O\left (\rho_\ast^3 \frac{T_e}{e} \right ),
\end{eqnarray}
where we have Taylor expanded $\Lambda_s (\bR, \mu)$ around $\boldr$ and $\mu_0$ to obtain
\begin{eqnarray}
\fl J_0 (\Lambda_s) \simeq J_0 (\lambda_s) - J_1 (\lambda_s) \left ( \frac{1}{\Omega_s} (\bv \times \bun) \cdot \nabla \lambda_s (\boldr, \mu_0) + \mu_1 \frac{\partial \lambda_s}{\partial \mu_0} (\boldr, \mu_0) \right ) \nonumber\\ =J_0 (\lambda_s) - J_1 (\lambda_s)\left ( \frac{v_\bot \sqrt{B}}{\Omega_s^2} (\bv \times \bun) \cdot \nabla \left ( \frac{k_\bot}{\sqrt{B}} \right ) + \frac{k_\bot m_s c \mu_1}{Z_s e v_\bot} \right ).
\end{eqnarray}
The arguments of the functions $\underline{\phi}_1^\tb$ and $\underline{\phi}_2^\tb$ are given by $\underline{\phi}_1^\tb = \underline{\phi}_1^\tb (k_\psi, k_\alpha, \psi(\boldr), \theta(\boldr), t)$ and $\underline{\phi}_2^\tb = \underline{\phi}_2^\tb (k_\psi, k_\alpha, \psi(\boldr), \theta(\boldr), t)$.

From \eq{eq:phiwiglwrv} and \eq{eq:phiwigtbrv}, we can deduce the long wavelength and short wavelength components of $\phiwig$,
\begin{equation}
\phiwig = [ \phiwig ]^\lw + [\phiwig]^\tb.
\end{equation}
The long wavelength component of $\phiwig$ does not coincide with $\phiwig^\lw$ because $\bR_2$ and $\mu_1$ depend on the electrostatic potential and hence have turbulent components. We find that the long wavelength piece is
\begin{equation} \label{eq:phiwigrlw}
[ \phiwig ]^\lw = [ \phiwig ]^\lw_1 + [ \phiwig ]^\lw_2,
\end{equation}
with
\begin{equation} \label{eq:phiwigrlw1}
[ \phiwig ]^\lw_1 = - \frac{1}{\Omega_s} (\bv \times \bun) \cdot \nabla \phi_0
\end{equation}
and
\begin{eqnarray} \label{eq:phiwigrlw2}
\fl [\phiwig]^\lw_2 = - \frac{1}{\Omega_s} (\bv \times \bun) \cdot \nabla \phi_1^\lw - \frac{1}{2\Omega_s^2} \Bigg [ (\bv \times \bun) (\bv \times \bun) + \frac{v_\bot^2}{2} (\matI - \bun \bun) \Bigg ] : \nabla \nabla \phi_0 \nonumber\\- \bR_2^\lw \cdot \nabla \phi_0 + \sum_{k_\psi, k_\alpha} \Bigg [ \Bigg ( \frac{k_\bot m_s c J_1 (\lambda_s)}{Z_s e v_\bot} (\underline{\mu}_1^\tb)^\ast - i \bk_\bot \cdot (\underline{\bR}_2^\tb)^\ast J_0 (\lambda_s) \Bigg ) \nonumber\\ \times \exp \left ( \frac{i \bk_\bot \cdot (\bv \times \bun)}{\Omega_s} \right ) \Bigg ] \underline{\phi}_1^\tb.
\end{eqnarray}
Here $\bR_2^\lw$ is the long wavelength piece of $\bR_2$, and $\underline{\mu}_1^\tb$ and $\underline{\bR}_2^\tb$ are the Fourier coefficients of the turbulent pieces of $\mu_1$ and $\bR_2$. The function $\phiwig (\boldr, \bv, t)$ also has short wavelength components,
\begin{equation} \label{eq:phiwigrtb}
[ \phiwig ]^\tb = \sum_{k_\psi, k_\alpha} ( \underline{[ \phiwig ]}^\tb_1 + \underline{[ \phiwig ]}^\tb_2) \exp ( i k_\psi \psi (\boldr) + i k_\alpha \alpha (\boldr)),
\end{equation}
with
\begin{equation} \label{eq:underlinephiwigrtb1}
\underline{[ \phiwig ]}^\tb_1 =  \left (1 - J_0 (\lambda_s) \exp \left ( \frac{i \bk_\bot \cdot (\bv \times \bun)}{\Omega_s} \right ) \right ) \underline{\phi}_1^\tb
\end{equation}
and
\begin{eqnarray} \label{eq:underlinephiwigrtb2}
\fl \underline{[\phiwig]}^\tb_2 = \left (1 - J_0 (\lambda_s) \exp \left ( \frac{i \bk_\bot \cdot (\bv \times \bun)}{\Omega_s} \right ) \right ) \underline{\phi}_2^\tb - \underline{\bR}_2^\tb \cdot \nabla \phi_0 \nonumber\\ + \Bigg [ \left ( \frac{v_\bot  \sqrt{B}}{\Omega_s^2} (\bv \times \bun) \cdot \nabla \left ( \frac{k_\bot}{\sqrt{B}} \right ) + \frac{k_\bot m_s c \mu_1^\lw}{Z_s e v_\bot} \right ) J_1 (\lambda_s) \underline{\phi}_1^\tb \nonumber\\ - i J_0 (\lambda_s) \left ( \bk_\bot \cdot  \bR_2^\lw + \frac{1}{2\Omega_s^2} (\bv \times \bun) (\bv \times \bun) : \nabla \bk_\bot \right ) \underline{\phi}_1^\tb \nonumber\\ - \frac{J_0(\lambda_s)}{\Omega_s} (\bv \times \bun) \cdot \nabla \underline{\phi}_1^\tb - \frac{2J_1(\lambda_s)}{\lambda_s}\frac{i v_\bot^2}{4 \Omega_s^2} \left ( \matI - \bun \bun \right ) : \nabla \bk_\bot\, \underline{\phi}_1^\tb \nonumber\\ + G(\lambda_s) \frac{i k_\bot v_\bot^4 }{16\Omega_s^4 }  \bk_\bot \cdot \nabla k_\bot\, \underline{\phi}_1^\tb -\frac{2 J_1 (\lambda_s)}{\lambda_s} \frac{i v_\bot^2}{2\Omega_s^2} \bk_\bot \cdot \nabla \underline{\phi}_1^\tb \Bigg ] \nonumber\\ \times \exp \left ( \frac{i \bk_\bot \cdot (\bv \times \bun)}{\Omega_s} \right ) + \sum_{k_\psi^\prime, k_\alpha^\prime} \Bigg [ \frac{k_\bot^\prime m_s c J_1 (\lambda_s^\prime)}{Z_s e v_\bot} (\underline{\mu}_1^\tb)^{\prime\prime} \nonumber\\ - i \bk_\bot^\prime \cdot (\underline{\bR}_2^\tb)^{\prime\prime} J_0 (\lambda_s^\prime) \Bigg ] \exp \left ( \frac{i \bk_\bot^\prime \cdot (\bv \times \bun)}{\Omega_s} \right ) (\underline{\phi}_1^\tb)^\prime,
\end{eqnarray}
where the prime $^\prime$ indicates that the term depends on $k_\psi^\prime$ and $k_\alpha^\prime$, and the double prime $^{\prime\prime}$ that it depends on $k_\psi^{\prime\prime} = k_\psi - k_\psi^\prime$ and $k_\alpha^{\prime\prime} = k_\alpha - k_\alpha^\prime$. Here $\mu_1^\lw$ is the long wavelength piece of $\mu_1$.

\section{Distribution functions written in $\{\boldr, \bv \}$ variables} \label{app:frv}

To express the distribution functions as functions of $\{\boldr, \bv\}$, we Taylor expand them around $\boldr$, $v_{||}$, $\mu_0$ and $\varphi_0$. We find
\begin{eqnarray} \label{eq:flwrv}
\fl f_s^\lw = f_{Ms} + f_{s1}^\lw + \frac{1}{\Omega_s} (\bv \times \bun) \cdot \nabla f_{Ms} + u_1 \frac{\partial f_{Ms}}{\partial v_{||}} + \mu_1 \frac{\partial f_{Ms}}{\partial \mu_0} + f_{s2}^\lw \nonumber\\+ \frac{1}{2\Omega_s^2} (\bv\times\bun)(\bv\times\bun) :\nabla \nabla f_{Ms} + \frac{u_1^2}{2} \frac{\partial^2 f_{Ms}}{\partial v_{||}^2} + \frac{\mu_1^2}{2} \frac{\partial^2 f_{Ms}}{\partial \mu_0^2}\nonumber\\ + \frac{u_1}{\Omega_s} (\bv \times \bun) \cdot \nabla \left ( \frac{\partial f_{Ms}}{\partial v_{||}} \right ) + \frac{\mu_1}{\Omega_s} (\bv \times \bun) \cdot \nabla \left ( \frac{\partial f_{Ms}}{\partial \mu_0} \right ) + u_1 \mu_1 \frac{\partial^2 f_{Ms}}{\partial v_{||} \partial \mu_0} \nonumber\\+ \bR_2 \cdot \nabla f_{Ms} + u_2 \frac{\partial f_{Ms}}{\partial v_{||}} + \mu_2 \frac{\partial f_{Ms}}{\partial \mu_0} + \frac{1}{\Omega_s} (\bv \times \bun) \cdot \nabla f_{s1}^\lw + u_1 \frac{\partial f_{s1}^\lw}{\partial v_{||}} \nonumber\\+ \mu_1 \frac{\partial f_{s1}^\lw}{\partial \mu_0} +O(\rho_\ast^3 f_{Ms})
\end{eqnarray}
and
\begin{eqnarray} \label{eq:ftbrv}
\fl f_s^\tb = \sum_{k_\psi, k_\alpha} \Bigg [ \underline{f}_{s1}^\tb + \underline{f}_{s2}^\tb + i \left ( \bk_\bot \cdot  \bR_2 + \frac{1}{2\Omega_s^2} (\bv \times \bun) (\bv \times \bun) : \nabla \bk_\bot \right ) \underline{f}_{s1}^\lw \nonumber\\ + \frac{1}{\Omega_s} (\bv \times \bun) \cdot \nabla \underline{f}_{s1}^\tb + u_1 \frac{\partial \underline{f}_{s1}^\tb}{\partial v_{||}}  + \mu_1 \frac{\partial \underline{f}_{s1}^\tb}{\partial \mu_0} \Bigg ] \nonumber\\ \times \exp \left ( \frac{i \bk_\bot \cdot ( \bv \times \bun)}{\Omega_s} \right ) \exp ( i k_\psi \psi (\boldr) + i k_\alpha \alpha (\boldr) ) + O(\rho_\ast^3 f_{Ms}),
\end{eqnarray}
where the arguments of the functions $f_{Ms}$, $f_{s1}^\lw$, $f_{s2}^\lw$, $\underline{f}_{s1}^\tb$ and $\underline{f}_{s2}^\tb$ are given by $f_{Ms} (\boldr, v_{||}, \mu_0, t)$, $f_{s1}^\lw (\boldr, v_{||}, \mu_0, t)$, $f_{s2}^\lw(\boldr, v_{||}, \mu_0, \varphi_0, t)$, $\underline{f}_{s1}^\tb(k_\psi, k_\alpha, \psi(\boldr), \theta(\boldr), v_{||}, \mu_0, t)$ and $\underline{f}_{s2}^\tb(k_\psi, k_\alpha, \psi(\boldr), \theta(\boldr), v_{||}, \mu_0, \varphi_0, t)$.

It is useful to split the functions $f_s (\boldr, \bv, t)$ into long wavelength and turbulent pieces in the $\{\boldr, \bv \}$ variables. The long wavelength and the turbulent components of $f_s$ written in $\{\boldr, \bv \}$ are given in \eq{eq:fsaveT} and \eq{eq:fswigT}, respectively. To calculate the first order pieces of \eq{eq:fsaveT} and \eq{eq:fswigT}, given in \eq{eq:fsaveT1} and \eq{eq:underlinefswigT1}, we have used the definitions of $u_1$ and $\mu_1$ in \eq{eq:u1def} and \eq{eq:mu1def}, and we have employed \eq{eq:phiwigrlw} and \eq{eq:phiwigrtb} to rewrite $\phiwig$ in a more convenient form. Note that 
\begin{equation}
u_1 = u_1^\lw, 
\end{equation}
and that according to \eq{eq:mu1def}, \eq{eq:phiwigrlw1} and \eq{eq:underlinephiwigrtb1}, $\mu_1^\lw$ and $\underline{\mu}_1^\tb$ are given by
\begin{eqnarray} \label{eq:mu1lw}
\fl \mu_1^\lw = - \frac{c}{B^2} (\bv \times \bun) \cdot \nabla \phi_0 - \frac{v_\bot^2}{2B^2\Omega_s} (\bv \times \bun)\cdot \nabla B - \frac{v_{||}^2}{B\Omega_s} \bun \cdot \nabla \bun \cdot (\bv \times \bun) \nonumber\\ - \frac{v_{||}}{4 B \Omega_s} [ \bv_\bot ( \bv \times \bun ) + ( \bv \times \bun ) \bv_\bot ]:\nabla \bun - \frac{v_{||} v_{\bot}^2}{2B\Omega_s} \bun \cdot \nabla \times \bun.
\end{eqnarray}
and
\begin{equation} \label{eq:mu1tb}
\underline{\mu}_1^\tb = \frac{Z_s e}{m_s B} \left (1 - J_0 (\lambda_s) \exp \left ( \frac{i \bk_\bot \cdot (\bv \times \bun)}{\Omega_s} \right ) \right ) \underline{\phi}_1^\tb.
\end{equation}

To obtain the second order pieces in \eq{eq:fsaveT2} and \eq{eq:underlinefswigT2}, we have employed again \eq{eq:phiwigrlw} and \eq{eq:phiwigrtb} to write $\phiwig$ explicitly. To write \eq{eq:fsaveT2} and \eq{eq:underlinefswigT2} in a compact form, we have defined the functions
\begin{eqnarray} \label{eq:DeltaFs2lwdef}
\fl \Delta f_{s2}^\lw = \frac{Z_s e f_{Ms}}{T_s} \Bigg [  \frac{1}{\Omega_s} (\bv \times \bun) \cdot \nabla \phi_1^\lw + \frac{1}{2\Omega_s^2} \Bigg ( (\bv \times \bun) (\bv \times \bun) + \frac{v_\bot^2}{2} (\matI - \bun \bun) \Bigg ) : \nabla \nabla \phi_0 \nonumber\\ + \bR^\lw_2 \cdot \nabla \phi_0 \Bigg ] + \frac{1}{2\Omega_s^2} (\bv\times\bun)(\bv\times\bun) :\nabla \nabla f_{Ms} + \frac{u_1^2}{2} \frac{\partial^2 f_{Ms}}{\partial v_{||}^2} \nonumber\\+ \frac{(\mu_1^\lw)^2}{2} \frac{\partial^2 f_{Ms}}{\partial \mu_0^2} + \frac{u_1}{\Omega_s} (\bv \times \bun) \cdot \nabla \left ( \frac{\partial f_{Ms}}{\partial v_{||}} \right ) + \frac{\mu_1^\lw}{\Omega_s} (\bv \times \bun) \cdot \nabla \left ( \frac{\partial f_{Ms}}{\partial \mu_0} \right ) \nonumber\\+ u_1 \mu_1^\lw \frac{\partial^2 f_{Ms}}{\partial v_{||} \partial \mu_0} + \bR_2^\lw \cdot \nabla f_{Ms} + u_2^\lw \frac{\partial f_{Ms}}{\partial v_{||}}+ \mu_2^\lw \frac{\partial f_{Ms}}{\partial \mu_0}  \nonumber\\+ \frac{1}{\Omega_s} (\bv \times \bun) \cdot \nabla f_{s1}^\lw + u_1 \frac{\partial f_{s1}^\lw}{\partial v_{||}} + \mu_1^\lw \frac{\partial f_{s1}^\lw}{\partial \mu_0} \nonumber\\ + \sum_{k_\psi, k_\alpha} \Bigg \{ \frac{1}{2} | \underline{\mu}_1^\tb |^2 \frac{\partial^2 f_{Ms}}{\partial \mu_0^2} + \Bigg [ \Bigg ( i \bk_\bot \cdot (\underline{\bR}_2^\tb)^\ast + (\underline{\mu}_1^\tb)^\ast \frac{\partial}{\partial \mu_0} \Bigg ) \underline{f}_{s1}^\tb \nonumber\\ - \frac{Z_s e f_{Ms}}{T_s} \Bigg ( \frac{k_\bot m_s c  J_1 (\lambda_s)}{Z_s e v_\bot} (\underline{\mu}_1^\tb)^\ast - i \bk_\bot \cdot (\underline{\bR}_2^\tb)^\ast J_0 (\lambda_s) \Bigg )  \underline{\phi}_1^\tb \Bigg ] \nonumber\\ \times \exp \left ( \frac{i \bk_\bot \cdot (\bv \times \bun)}{\Omega_s} \right ) \Bigg \}
\end{eqnarray}
and
\begin{eqnarray} \label{eq:DeltaunderlineFs2tbdef}
\fl \Delta \underline{f}_{s2}^\tb = \mu_1^\lw \underline{\mu}_1^\tb \frac{\partial^2 f_{Ms}}{\partial \mu_0^2} + \frac{\underline{\mu}_1^\tb}{\Omega_s} (\bv \times \bun) \cdot \nabla \left ( \frac{\partial f_{Ms}}{\partial \mu_0} \right ) + u_1  \underline{\mu}_1^\tb \frac{\partial^2 f_{Ms}}{\partial v_{||} \partial \mu_0} + \underline{\bR}_2^\tb \cdot \nabla f_{Ms} \nonumber\\ + \underline{u}_2^\tb \frac{\partial f_{Ms}}{\partial v_{||}} + \underline{\mu}_2^\tb \frac{\partial f_{Ms}}{\partial \mu_0} + \underline{\mu}_1^\tb \frac{\partial f_{s1}^\lw}{\partial \mu_0} \nonumber\\+ \Bigg [  i \left ( \bk_\bot \cdot \bR_2^\lw + \frac{1}{2\Omega_s^2} (\bv \times \bun) (\bv \times \bun) : \nabla \bk_\bot \right ) \underline{f}_{s1}^\tb \nonumber\\ + \frac{1}{\Omega_s} (\bv \times \bun) \cdot \nabla \underline{f}_{s1}^\tb + u_1 \frac{\partial \underline{f}_{s1}^\tb}{\partial v_{||}} + \mu_1^\lw \frac{\partial \underline{f}_{s1}^\tb}{\partial \mu_0} \Bigg ] \exp \left ( \frac{i \bk_\bot \cdot (\bv \times \bun)}{\Omega_s} \right )  \nonumber\\ - \frac{Z_s e}{T_s} \Bigg \{ \left (1 - J_0 (\lambda_s) \exp \left ( \frac{i \bk_\bot \cdot (\bv \times \bun)}{\Omega_s} \right ) \right ) \underline{\phi}_2^\tb - \underline{\bR}_2^\tb \cdot \nabla \phi_0 \nonumber\\ + \Bigg [ \left ( \frac{v_\bot  \sqrt{B}}{\Omega_s^2} (\bv \times \bun) \cdot \nabla \left ( \frac{k_\bot}{\sqrt{B}} \right ) + \frac{k_\bot m_s c \mu_1^\lw}{Z_s e v_\bot} \right ) J_1 (\lambda_s) \underline{\phi}_1^\tb \nonumber\\ - i J_0 (\lambda_s) \left ( \bk_\bot \cdot  \bR_2^\lw + \frac{1}{2\Omega_s^2} (\bv \times \bun) (\bv \times \bun) : \nabla \bk_\bot \right ) \underline{\phi}_1^\tb \nonumber\\ - \frac{J_0(\lambda_s)}{\Omega_s} (\bv \times \bun) \cdot \nabla \underline{\phi}_1^\tb - \frac{2J_1(\lambda_s)}{\lambda_s}\frac{i v_\bot^2}{4 \Omega_s^2} \left ( \matI - \bun \bun \right ) : \nabla \bk_\bot\, \underline{\phi}_1^\tb \nonumber\\ + G(\lambda_s) \frac{i k_\bot v_\bot^4 }{16\Omega_s^4 }  \bk_\bot \cdot \nabla k_\bot\, \underline{\phi}_1^\tb -\frac{2 J_1 (\lambda_s)}{\lambda_s} \frac{i v_\bot^2}{2\Omega_s^2} \bk_\bot \cdot \nabla \underline{\phi}_1^\tb \Bigg ] \nonumber\\ \times \exp \left ( \frac{i \bk_\bot \cdot (\bv \times \bun)}{\Omega_s} \right )\Bigg \}f_{Ms} + \sum_{k_\psi^\prime, k_\alpha^\prime} \Bigg \{ \frac{1}{2}  (\underline{\mu}_1^\tb)^{\prime\prime} (\underline{\mu}_1^\tb)^\prime \frac{\partial^2 f_{Ms}}{\partial \mu_0^2}   \nonumber\\ + \Bigg [ \Bigg ( i \bk_\bot^\prime \cdot (\underline{\bR}_2^\tb)^{\prime\prime} + (\underline{\mu}_1^\tb)^{\prime\prime} \frac{\partial}{\partial \mu_0} \Bigg ) (\underline{f}_{s1}^\tb)^\prime \nonumber\\ - \frac{Z_s e f_{Ms}}{T_s} \Bigg ( \frac{k_\bot^\prime m_s c  J_1 (\lambda_s^\prime)}{Z_s e v_\bot} (\underline{\mu}_1^\tb)^{\prime\prime} - i \bk_\bot^\prime \cdot (\underline{\bR}_2^\tb)^{\prime\prime} J_0 (\lambda_s^\prime) \Bigg )  (\underline{\phi}_1^\tb)^\prime \Bigg ] \nonumber\\ \times \exp \left ( \frac{i \bk_\bot^\prime \cdot (\bv \times \bun)}{\Omega_s} \right ) \Bigg \}.
\end{eqnarray}
Here $u_2^\lw$ and $\mu_2^\lw$ are the long wavelength pieces of $u_2$ and $\mu_2$, and $\underline{u}_2^\tb$ and $\underline{\mu}_2^\tb$ are the Fourier coefficients of the turbulent pieces of $u_2$ and $\mu_2$. The prime $^\prime$ indicates that the term depends on $k_\psi^\prime$ and $k_\alpha^\prime$, and the double prime $^{\prime\prime}$ that it depends on $k_\psi^{\prime\prime} = k_\psi - k_\psi^\prime$ and $k_\alpha^{\prime\prime} = k_\alpha - k_\alpha^\prime$.
 
\section{Derivation of the form of $\Pi$ in \eq{eq:Pifinal}} \label{app:Pisimplify}

We proceed to take moments of \eq{eq:FPequation} to find expression \eq{eq:Pifinal}. First, we take the moment $(m_s^2 c/2 Z_s e) R^2 (\bv\cdot \zun)^2$ to find
\begin{eqnarray} \label{eq:Pisimplify1}
\fl R m_s \int \dd^3v\, f_s (\bv \cdot \zun) \left ( \bv \cdot \nabla \psi \right ) = \frac{\partial}{\partial t} \left [ \frac{R^2 m_s^2 c}{2Z_s e} \int \dd^3v\, f_s (\bv \cdot \zun)^2  \right ] \nonumber\\+ \nabla \cdot \left [ \frac{R^2 m_s^2 c}{2Z_s e} \int \dd^3v\, f_s \bv (\bv \cdot \zun)^2  \right ] + R m_s c \frac{\partial \phi}{\partial \zeta} \int \dd^3v\, f_s (\bv \cdot \zun) \nonumber\\- \frac{R^2 m_s^2 c}{2Z_s e} \int \dd^3v\, \sum_{s^\prime} C_{ss^\prime} [f_s, f_{s^\prime}] (\bv \cdot \zun)^2 - \frac{R^2 m_s^2 c}{2Z_s e} \int \dd^3v\, Q_s (\bv \cdot \zun)^2,
\end{eqnarray}
where we have used that $R\bB \times \zun = \nabla \psi$ and that $\nabla (R\zun) = \nabla R \zun - \zun \nabla R$ is an antisymmetric tensor. The term on the left side of \eq{eq:Pisimplify1} comes from integrating by parts $(m_s^2 c/2 Z_s e) R^2 (\bv \cdot \zun)^2 (Z_s e/m_s c)(\bv \times \bB) \cdot \nabla_v f_s$. Flux surface averaging \eq{eq:Pisimplify1} (see \eq{eq:transpfluxaverage}), summing over species and coarse grain averaging, we find a new expression for $\Pi$,
\begin{eqnarray} \label{eq:Pisimplify2}
\fl \Pi = \frac{\partial}{\partial t} \left \langle \left \langle \sum_s \frac{R^2 m_s^2 c}{2Z_s e} \int \dd^3v\, f_s (\bv \cdot \zun)^2 \right \rangle_\psi \right \rangle_\mathrm{T} \nonumber\\+ \frac{1}{V^\prime} \frac{\partial}{\partial \psi} \left [ V^\prime \left \langle \left \langle \sum_s \frac{R^2 m_s^2 c}{2Z_s e} \int \dd^3v\, f_s (\bv \cdot \zun)^2 \left ( \bv \cdot \nabla \psi \right ) \right \rangle_\psi \right \rangle_\mathrm{T}  \right ] \nonumber\\ + \left \langle \left \langle \sum_s R m_s c \frac{\partial \phi}{\partial \zeta} \int \dd^3v\, f_s (\bv \cdot \zun) \right \rangle_\psi \right \rangle_\mathrm{T} \nonumber\\- \left \langle \left \langle \sum_{s, s^\prime} \frac{R^2 m_s^2 c}{2Z_s e} \int \dd^3v\, C_{ss^\prime} [f_s, f_{s^\prime}] (\bv \cdot \zun)^2 \right \rangle_\psi \right \rangle_\mathrm{T} \nonumber\\- \left \langle \left \langle \sum_s \frac{R^2 m_s^2 c}{2Z_s e} \int \dd^3v\, Q_s (\bv \cdot \zun)^2 \right \rangle_\psi \right \rangle_\mathrm{T}.
\end{eqnarray}

We apply the ordering assumptions in section \ref{sec:ordering} to \eq{eq:Pisimplify2}. Here we assume that $R \sim a$ and $B \sim B_p$. The expansion in $B_p/B \ll 1$ is treated as a subsidiary expansion in sections \ref{sec:expansion} and \ref{sec:secondordertotal}. From \eq{eq:philwexpansion}, \eq{eq:phinorder}, \eq{eq:gradbotphitborder} and \eq{eq:gradparphitborder}, we find that 
\begin{equation} \label{eq:dphidzetaestimate}
c \frac{\partial \phi}{\partial \zeta} = c R\zun \cdot \nabla \phi^\tb \simeq - \frac{c}{B} \nabla_\bot \phi^\tb \cdot (\bun \times \nabla \psi) \sim \rho_\ast v_{ti} |\nabla \psi|,
\end{equation}
where we have neglected $\bun \cdot \nabla \phi^\tb$ because of \eq{eq:gradparphitborder}, and we have used \eq{eq:Rzetatrick}. To simplify the terms with collision operators, we use that $f_s$ is to lowest order a Maxwellian with a temperature consistent with the rest of the species and its mass (see \eq{eq:collisiontauE} and \eq{eq:tequilibrationorder} and the discussion around them), and that we have ordered the collision frequencies according to \eq{eq:collisionorder1} and \eq{eq:collisionorder2}, giving
\begin{eqnarray} \label{eq:Cssprimeexpanded}
\fl \langle C_{ss^\prime} [f_s, f_{s^\prime}] \rangle_\mathrm{T} = C_{ss^\prime}^{(\ell)} \left [ [f_s]_1^\lw ; [f_{s^\prime}]_1^\lw \right ] + C_{ss^\prime}^{(\ell)} \left [ [f_s]_2^\lw ; [f_{s^\prime}]_2^\lw \right ] + C_{ss^\prime} \left [ [f_s^\lw]_1, [f_{s^\prime}]_1^\lw \right ] \nonumber\\+ \left \langle \sum_{k_\psi, k_\alpha} C_{ss^\prime} \left [ (\underline{[f_s]}_1^\tb)^\ast, \underline{[f_{s^\prime}]}_1^\tb \right ]  \right \rangle_t + O \left ( \frac{v_{ts}}{a} \rho_\ast^3 f_{Ms} \right ).
\end{eqnarray}
for $s^\prime \neq e$. The pieces of the distribution function $[f_s]_1^\lw$, $\underline{[f_s]}_1^\tb$ and $[f_s]_2^\lw$ are defined in \eq{eq:fsaveT1}, \eq{eq:underlinefswigT1} and \eq{eq:fsaveT2}. The ion-electron and impurity-electron collision operators are an exception to \eq{eq:Cssprimeexpanded} due to our assumption \eq{eq:massorder}. Following \ref{app:Cse}, the ion-electron and impurity-electron collision operators simplify to \eq{eq:Cseexpanded}. Finally, after coarse grain averaging, the time derivatives are of order of the inverse of the transport time scale, giving
\begin{equation} \label{eq:transporttime}
\frac{\partial}{\partial t} \sim \frac{1}{\tau_E} \sim \rho_\ast^2 \frac{v_{ti}}{a}.
\end{equation}
Then, to find the contribution of the first term of the right side of \eq{eq:Pisimplify2} to order $\rho_\ast^3 p R |\nabla \psi|$, it is sufficient to use the Maxwellian, $f_s \simeq f_{Ms}$. For the third and fourth terms on the right side of \eq{eq:Pisimplify2}, we need the distribution functions to second order. With these considerations, we obtain
\begin{eqnarray} \label{eq:Pisimplify2v2}
\fl \Pi = \Pi_{-1} + \frac{\partial}{\partial t} \left ( \sum_{s \neq e} \frac{\langle R^2 \rangle_\psi m_s c p_s}{2Z_s e} \right ) \nonumber\\+ \frac{1}{V^\prime} \frac{\partial}{\partial \psi} \left [ V^\prime \left \langle \left \langle \sum_{s\neq e} \frac{R^2 m_s^2 c}{2Z_s e} \int \dd^3v\, f_s (\bv \cdot \zun)^2 \left ( \bv \cdot \nabla \psi  \right ) \right \rangle_\psi \right \rangle_\mathrm{T}  \right ] \nonumber\\ - \left \langle \left \langle \sum_{s \neq e} \sum_{k_\psi, k_\alpha} R m_s c i k_\alpha (\underline{\phi}_1^\tb)^\ast \int \dd^3v\, \underline{[f_s]}_2^\tb (\bv \cdot \zun) \right \rangle_\psi \right \rangle_t \nonumber\\ - \left \langle \left \langle \sum_{s \neq e} \sum_{k_\psi, k_\alpha} R m_s c i k_\alpha (\underline{\phi}_2^\tb)^\ast \int \dd^3v\, \underline{[f_s]}_1^\tb (\bv \cdot \zun) \right \rangle_\psi \right \rangle_t \nonumber\\- \left \langle \sum_{s \neq e, s^\prime \neq e} \frac{R^2 m_s^2 c}{2Z_s e} \int \dd^3v\, C_{ss^\prime}^{(\ell)} \left [ [f_s]_2^\lw ; [f_{s^\prime}]_2^\lw\right ] (\bv \cdot \zun)^2 \right \rangle_\psi \nonumber\\- \left \langle \sum_{s \neq e, s^\prime \neq e} \frac{R^2 m_s^2 c}{2Z_s e} \int \dd^3v\, C_{ss^\prime} \left [ [f_s]_1^\lw , [f_{s^\prime}]_1^\lw \right ] (\bv \cdot \zun)^2 \right \rangle_\psi \nonumber\\- \left \langle \sum_{s \neq e, s^\prime \neq e} \frac{R^2 m_s^2 c}{2Z_s e} \int \dd^3v\, \left \langle \sum_{k_\psi, k_\alpha} C_{ss^\prime} \left [(\underline{[f_s]}_1^\tb)^\ast, \underline{[f_{s^\prime}]}_1^\tb\right ] \right \rangle_t (\bv \cdot \zun)^2 \right \rangle_\psi \nonumber\\ - \sum_{s \neq e} \frac{\langle R^2 \rangle_\psi m_e c}{Z_s e} n_e \nu_{es} (T_e- T_s) - \left \langle \sum_{s \neq e} \frac{R^2 m_s^2 c}{2Z_s e} \int \dd^3v\, Q_s (\bv \cdot \zun)^2 \right \rangle_\psi,
\end{eqnarray}
where we have used \eq{eq:Cseexpanded} to write $\int \dd^3v\, C_{se} [ f_s, f_e ] (\bv \cdot \zun)^2 \simeq 2 n_e m_e \nu_{es} (T_e - T_i)/m_s^2$ for $s \neq e$, and we have dropped many of the terms having to do with electrons because $\sqrt{m_e/m_i} \ll 1$ (recall \eq{eq:fenevenorder} and \eq{eq:fenoddorder}). The piece $\Pi_{-1}$ is defined in \eq{eq:Piminusone}.

The third term on the right side of \eq{eq:Pisimplify2v2} can be written in a more convenient form employing another moment of equation \eq{eq:FPequation}. Multiplying equation \eq{eq:FPequation} by $(m_s^3 c^2/6 Z^2 e^2) R^3 (\bv \cdot \zun)^3$, integrating over velocity space, flux surface averaging, coarse grain averaging and summing over species, we find
\begin{eqnarray} \label{eq:Pisimplify3}
\fl  \left \langle \left \langle \sum_s \frac{R^2 m_s^2 c}{2Z_s e} \int \dd^3v\, f_s (\bv \cdot \zun)^2 \left ( \bv \cdot \nabla \psi \right ) \right \rangle_\psi \right \rangle_\mathrm{T}  \nonumber\\ = \frac{\partial}{\partial t} \left [ \left \langle \left \langle \sum_s \frac{R^3 m_s^3 c^2}{6Z_s^2 e^2} \int \dd^3v\, f_s (\bv \cdot \zun)^3 \right \rangle_\psi \right \rangle_\mathrm{T}  \right ] \nonumber\\+ \frac{1}{V^\prime} \frac{\partial}{\partial \psi} \left [ V^\prime \left \langle \left \langle \sum_s \frac{R^3 m_s^3 c^2}{6Z_s^2 e^2} \int \dd^3v\, f_s (\bv \cdot \zun)^3 \left ( \bv \cdot \nabla \psi \right ) \right \rangle_\psi \right \rangle_\mathrm{T}  \right ] \nonumber\\ + \left \langle \left \langle \sum_s \frac{R^2 m_s^2 c^2}{2 Z_s e} \frac{\partial \phi}{\partial \zeta} \int \dd^3v\, f_s (\bv \cdot \zun)^2 \right \rangle_\psi \right \rangle_\mathrm{T} \nonumber\\- \left \langle \left \langle \sum_{s, s^\prime} \frac{R^3 m_s^3 c^2}{6Z_s^2 e^2} \int \dd^3v\, C_{ss^\prime} [f_s, f_{s^\prime}] (\bv \cdot \zun)^3 \right \rangle_\psi \right \rangle_\mathrm{T} \nonumber\\- \left \langle \left \langle \sum_s \frac{R^3 m_s^3 c^2}{6Z_s^2 e^2} \int \dd^3v\, Q_s (\bv \cdot \zun)^3 \right \rangle_\psi \right \rangle_\mathrm{T}.
\end{eqnarray}
Using \eq{eq:transporttime} and \eq{eq:Qordering}, we see that the first and fifth terms on the right side of \eq{eq:Pisimplify3} are negligible. Finally, using \eq{eq:dphidzetaestimate}, \eq{eq:Cssprimeexpanded} and \eq{eq:Cseexpanded}, we find that only first order corrections are needed to find the third and fourth terms up to order $\rho_\ast^3 p a R |\nabla \psi|^2$. Then,
\begin{eqnarray} \label{eq:Pisimplify5}
\fl  \left \langle \left \langle \sum_s \frac{R^2 m_s^2 c}{2Z_s e} \int \dd^3v\, f_s (\bv \cdot \zun)^2 \left ( \bv \cdot \nabla \psi \right ) \right \rangle_\psi \right \rangle_\mathrm{T}  \nonumber\\ = - \left \langle \left \langle \sum_{s \neq e} \sum_{k_\psi, k_\alpha} \frac{R^2 m_s^2 c^2}{2 Z_s e} i k_\alpha (\underline{\phi}_1^\tb)^\ast \int \dd^3v\, \underline{[f_s]}_1^\tb (\bv \cdot \zun)^2 \right \rangle_\psi \right \rangle_t \nonumber\\- \left \langle \sum_{s \neq e, s^\prime \neq e} \frac{R^3 m_s^3 c^2}{6Z_s^2 e^2} \int \dd^3v\, C_{ss^\prime}^{(\ell)} \left [ [f_s]_1^\lw ; [f_{s^\prime}]_1^\lw \right ] (\bv \cdot \zun)^3 \right \rangle_\psi.
\end{eqnarray}
Here we have also neglected the second term on the right side of \eq{eq:Pisimplify3} because it is negligible compared to $\rho_\ast^3 p a R |\nabla \psi|^2$, the size needed for the left side of \eq{eq:Pisimplify3} to give a contribution comparable to the other terms in \eq{eq:Pisimplify2v2}. Indeed, multiplying \eq{eq:FPequation} by $(m_s^4 c^3/24 Z_s^3 e^3) R^4 (\bv \cdot \zun)^4$, integrating over velocity space, flux surface averaging, coarse grain averaging and summing over species, we find
\begin{eqnarray} \label{eq:Pisimplify4}
\fl  \left \langle \left \langle \sum_s \frac{R^3 m_s^3 c^2}{6Z_s^2 e^2} \int \dd^3v\, f_s (\bv \cdot \zun)^3 \left ( \bv \cdot \nabla \psi \right ) \right \rangle_\psi \right \rangle_\mathrm{T}  \nonumber\\ = \frac{\partial}{\partial t} \left [ \left \langle \left \langle \sum_s \frac{R^4 m_s^4 c^3}{24Z_s^3 e^3} \int \dd^3v\, f_s (\bv \cdot \zun)^4 \right \rangle_\psi \right \rangle_\mathrm{T}  \right ] \nonumber\\+ \frac{1}{V^\prime} \frac{\partial}{\partial \psi} \left [ V^\prime \left \langle \left \langle \sum_s \frac{R^4 m_s^4 c^3}{24Z_s^3 e^3} \int \dd^3v\, f_s (\bv \cdot \zun)^4 \left ( \bv \cdot \nabla \psi \right ) \right \rangle_\psi \right \rangle_\mathrm{T}  \right ] \nonumber\\ + \left \langle \left \langle \sum_s \frac{R^3 m_s^3 c^3}{6 Z_s^2 e^2} \frac{\partial \phi}{\partial \zeta} \int \dd^3v\, f_s (\bv \cdot \zun)^3 \right \rangle_\psi \right \rangle_\mathrm{T} \nonumber\\- \left \langle \left \langle \sum_{s, s^\prime} \frac{R^4 m_s^4 c^3}{24Z_s^3 e^3} \int \dd^3v\, C_{ss^\prime} [f_s, f_{s^\prime}] (\bv \cdot \zun)^4 \right \rangle_\psi \right \rangle_\mathrm{T} \nonumber\\- \left \langle \left \langle \sum_s \frac{R^4 m_s^4 c^3}{24Z_s^3 e^3} \int \dd^3v\, Q_s (\bv \cdot \zun)^4 \right \rangle_\psi \right \rangle_\mathrm{T}.
\end{eqnarray}
Using \eq{eq:transporttime} and \eq{eq:Qordering}, we see that the first and fifth terms on the right side of \eq{eq:Pisimplify3} are negligible. Using \eq{eq:dphidzetaestimate}, \eq{eq:Cssprimeexpanded} and \eq{eq:Cseexpanded}, we find that the third and fourth terms are also negligible. The second term in \eq{eq:Pisimplify4} vanishes to lowest order because $f_s \simeq f_{Ms}$ is sufficient to evaluate this term to order $\rho_\ast^3 p a^2 R |\nabla \psi|^3$.

Substituting \eq{eq:Pisimplify5} into \eq{eq:Pisimplify2v2}, we finally obtain \eq{eq:Pifinal}.

\section{Ion-electron and impurity-electron collision operators} \label{app:Cse}

In this Appendix, we simplify the operator for collisions between species $s$ and electrons,
\begin{eqnarray} \label{eq:Cse1}
\fl C_{se} [ f_s, f_e ] = \frac{\gamma_{es}}{m_s} \nabla_v \cdot \left [ \int \dd^3 v^\prime\, \nabla_w \nabla_w w \cdot \left ( \frac{f_e(\bv^\prime)}{m_s} \nabla_v f_s (\bv) - \frac{f_s (\bv)}{m_e} \nabla_{v^\prime} f_e (\bv^\prime) \right ) \right ] ,
\end{eqnarray}
using assumptions \eq{eq:massorder}, \eq{eq:fenevenorder} and \eq{eq:fenoddorder}. We expand \eq{eq:Cse1} to first order in $\sqrt{m_e/m_i}$. Since $\bv^\prime \sim v_{te} \gg \bv \sim v_{ts}$,
\begin{equation} \label{eq:gradgradgexpansionse}
\nabla_w \nabla_w w = \nabla_{v^\prime} \nabla_{v^\prime} v^\prime - \bv \cdot \nabla_{v^\prime} \nabla_{v^\prime} \nabla_{v^\prime} v^\prime + O \left ( \frac{m_e}{m_s} \frac{1}{v_{te}} \right ).
\end{equation}
Using this result, $f_s \simeq f_{Ms}$ and the notation $g_e = f_e - f_{Me} \ll f_{Me}$, equation \eq{eq:Cse1} simplifies to
\begin{eqnarray} \label{eq:Cse2}
\fl C_{se} [ f_s, f_e ] \simeq \frac{\gamma_{es}}{m_s} \nabla_v \cdot \Bigg [ \int \dd^3 v^\prime\, \nabla_{v^\prime} \nabla_{v^\prime} v^\prime \cdot \left ( \frac{f_{Me} (\bv^\prime)}{m_s} \nabla_v f_{Ms} (\bv) - \frac{f_{Ms} (\bv)}{m_e} \nabla_{v^\prime} g_e (\bv^\prime) \right ) \nonumber \\ + \int \dd^3 v^\prime\,  \frac{f_{Ms} (\bv)}{m_e} \bv \cdot \nabla_{v^\prime} \nabla_{v^\prime} \nabla_{v^\prime} v^\prime \cdot \nabla_{v^\prime} f_{Me} (\bv^\prime) \Bigg ],
\end{eqnarray}
where we have already employed
\begin{equation} \label{eq:nablavfMs}
\nabla_v f_{Ms} (\bv) = - \frac{m_s \bv}{T_s} f_{Ms} (\bv)
\end{equation}
and
\begin{equation} \label{eq:vnablavnablavv}
\bv^\prime \cdot \nabla_{v^\prime} \nabla_{v^\prime} v^\prime = 0
\end{equation}
to cancel one term. Using \eq{eq:nablavfMs} and
\begin{equation} \label{eq:vnablavnablavnablavv}
\bv^\prime \cdot \nabla_{v^\prime} \nabla_{v^\prime} \nabla_{v^\prime} v^\prime = \nabla_{v^\prime} ( \bv^\prime \cdot \nabla_{v^\prime} \nabla_{v^\prime} v^\prime) - \nabla_{v^\prime} \nabla_{v^\prime} v^\prime = - \nabla_{v^\prime} \nabla_{v^\prime} v^\prime,
\end{equation} 
and integrating by parts the term that contains $g_e$, we obtain
\begin{eqnarray} \label{eq:Cse3}
\fl C_{se} [ f_s, f_e ] \simeq \frac{\gamma_{es}}{m_s} \nabla_v \cdot \Bigg [ f_{Ms} (\bv) \left ( \frac{1}{T_e} - \frac{1}{T_i} \right ) \bv \cdot \int \dd^3 v^\prime\, \nabla_{v^\prime} \nabla_{v^\prime} v^\prime f_{Me} (\bv^\prime) \nonumber\\ + \frac{f_{Ms} (\bv)}{m_e} \int \dd^3 v^\prime\, \nabla_{v^\prime} \nabla_{v^\prime}^2 v^\prime g_e (\bv^\prime) \Bigg ].
\end{eqnarray}
Using 
\begin{equation} \label{eq:nablavlaplacianvv}
\nabla_{v^\prime} \nabla_{v^\prime}^2 v^\prime = - \frac{2 \bv^\prime}{(v^\prime)^3} 
\end{equation}
and $\int \dd^3 v^\prime\, \nabla_{v^\prime} \nabla_{v^\prime} v^\prime f_{Me} (v^\prime) = (2\sqrt{2}/ 3 \sqrt{\pi}) n_e \sqrt{m_e/T_e} \matI$, we find
\begin{eqnarray} \label{eq:Cse4}
\fl C_{se} [ f_s, f_e ] \simeq \left [ \frac{n_e m_e \nu_{es}}{n_s m_s} \left ( \frac{T_e}{T_s} - 1 \right ) \left ( \frac{m_s v^2}{T_i} - 3 \right ) + \frac{2\gamma_{es}}{m_e T_i} \bv \cdot \int \dd^3 v^\prime\, g_e (\bv^\prime) \frac{\bv^\prime}{(v^\prime)^3} \right ] f_{Ms}.
\end{eqnarray}
According to \eq{eq:fenoddorder}, \eq{eq:DeltaFs1lwdef} and \eq{eq:DeltaunderlineFs1tbdef}, the largest pieces of $g_e = f_e - f_{Me}$ that are either odd in the parallel velocity or are gyrophase dependent are of order $(\rho_e/a) f_{Me}$. Then, the last term in \eq{eq:Cse4} is of order $\nu_{ei} (m_e/m_i) \rho_\ast f_{Ms}$, being much smaller than the first, of order $\nu_{ei} (m_e/m_i) f_{Ms}$. By neglecting the second term, we finally obtain \eq{eq:Cseexpanded}. 

Note that the pieces of $C_{se}$ given in \eq{eq:Cseexpanded} are of order $\nu_{ei} (m_e/m_i) f_{Ms} \sim f_{Ms}/\tau_E \sim \rho_\ast^2 f_{Ms} v_{ti}/a$ (recall \eq{eq:tequilibrationorder}), and hence we need not keep higher order terms. 

\section{Electron-ion and electron-impurity collision operators} \label{app:Ces}

In this Appendix, we simplify the operator for collisions between electrons and species $s$,
\begin{eqnarray} \label{eq:Ces1}
\fl C_{es} [ f_e, f_s ] = \frac{\gamma_{es}}{m_e} \nabla_v \cdot \left [ \int \dd^3 v^\prime\, \nabla_w \nabla_w w \cdot \left ( \frac{f_s(\bv^\prime)}{m_e} \nabla_v f_e (\bv) - \frac{f_e (\bv)}{m_s} \nabla_{v^\prime} f_s (\bv^\prime) \right ) \right ].
\end{eqnarray}
Defining $g_e = f_e - f_{Me}$ and $g_s = f_s - f_{Ms}$, we find that
\begin{eqnarray} \label{eq:Ces2}
\fl C_{es} [ f_e, f_s ] = C_{es} [ f_{Me}, f_{Ms} ] + C_{es}^{(\ell)} [ g_e; g_s ] + C_{es} [ g_e, g_s ],
\end{eqnarray}
where
\begin{eqnarray} \label{eq:Ceslinear1}
\fl C_{es}^{(\ell)} [ g_e; g_s ] = C_{es} [ g_e, f_{Ms} ] + C_{es} [ f_{Me}, g_s ] \nonumber\\= \frac{\gamma_{es}}{m_e} \nabla_v \cdot \Bigg [ \int \dd^3v^\prime \nabla_w \nabla_w w \cdot \Bigg ( \frac{f_{Ms} (\bv^\prime)}{m_e} \nabla_v g_e (\bv) - \frac{ \bv}{T_e} f_{Me} (\bv) g_s  (\bv^\prime) \nonumber\\ - \frac{f_{Me} (\bv)}{m_s} \nabla_{v^\prime} g_s (\bv^\prime) + \frac{\bv^\prime}{T_i} g_e (\bv) f_{Ms} (\bv^\prime )\Bigg ) \Bigg ].
\end{eqnarray}
To obtain \eq{eq:Ceslinear1} we have used \eq{eq:nablavfMs}. The linearized collision operator $C_{es}^{(\ell)} [g_e; g_s]$ cannot be obtained from the general formula \eq{eq:Clineardef} for the linearlized collision operator between species $s$ and $s^\prime$ because to deduce \eq{eq:Clineardef} we assumed that $T_s = T_{s^\prime}$. We proceed to simplify each one of the terms in \eq{eq:Ces2} independently.

The collision operator applied on the Maxwellians can be simplified by realizing that $\bv \sim v_{te} \gg \bv^\prime \sim v_{ts}$. Then,
\begin{equation} \label{eq:gradgradgexpansiones}
\fl \nabla_w \nabla_w w = \nabla_v \nabla_v v - \bv^\prime \cdot \nabla_v \nabla_v \nabla_v v + \frac{1}{2} \bv^\prime \bv^\prime : \nabla_v \nabla_v \nabla_v \nabla_v v + O \left [ \left ( \frac{m_e}{m_s} \right )^{3/2} \frac{1}{v_{te}} \right ],
\end{equation}
Using this result, we find
\begin{eqnarray} \label{eq:CesMaxwellians1}
\fl C_{es} [ f_{Me}, f_{Ms} ] \simeq \frac{\gamma_{es}}{m_e} \nabla_v \cdot \Bigg [ f_{Me} (\bv) \int \dd^3 v^\prime\, f_{Ms}(\bv^\prime) \Bigg ( - \frac{1}{2T_e} \bv^\prime \bv^\prime : \nabla_v \nabla_v \nabla_v \nabla_v v \cdot \bv \nonumber\\ - \frac{1}{T_i} \bv^\prime \bv^\prime: \nabla_v \nabla_v \nabla_v v \Bigg ) \Bigg ],
\end{eqnarray}
where we have employed \eq{eq:nablavfMs}, \eq{eq:vnablavnablavv} and 
\begin{equation} \label{eq:intfMsv}
\int \dd^3 v^\prime\, f_{Ms} (\bv^\prime) \bv^\prime = 0 
\end{equation}
to cancel several terms. Using \eq{eq:vnablavnablavnablavv}, we find 
\begin{equation}
\nabla_v \nabla_v \nabla_v \nabla_v v \cdot \bv = \nabla_v ( \nabla_v \nabla_v \nabla_v v \cdot \bv) - \nabla_v \nabla_v \nabla_v v = - 2 \nabla_v \nabla_v \nabla_v v.
\end{equation}
Employing this result, \eq{eq:nablavlaplacianvv}, $\nabla_v^2 \nabla_v^2 v = - 8\pi \delta(\bv)$, with $\delta(\bv)$ the Dirac delta function, and $\int \dd^3v^\prime\, f_{Ms} (\bv^\prime) \bv^\prime \bv^\prime = (n_s T_i/m_s) \matI$, equation \eq{eq:CesMaxwellians1} finally becomes
\begin{eqnarray} \label{eq:CesMaxwellians}
\fl C_{es} [ f_{Me}, f_{Ms} ] \simeq \frac{3 \sqrt{2\pi}}{2} \frac{m_e \nu_{es}}{m_s} \left ( \frac{T_e}{m_e v^2} \right )^{3/2} \Bigg ( \frac{T_i}{T_e} - 1 \Bigg ) \Bigg ( \frac{m_e v^2}{T_e} - 4 \pi v^3 \delta (\bv)\Bigg ) f_{Me},
\end{eqnarray}
where the frequency $\nu_{es}$ is defined in \eq{eq:nuesdef}. Note that $C_{es} [ f_{Me}, f_{Ms} ]$ is of order $\nu_{ei} (m_e/m_i) f_{Me} \sim f_{Me}/\tau_E \sim \rho_\ast^2 f_{Me} v_{ti}/L$ (recall \eq{eq:tequilibrationorder}), and hence it is negligible except for the very high order piece $f_{e2}^\lw$, which is not relevant for momentum transport as we explain in subsection \ref{sub:lwsecondorder}.

The linearized collision operator for collisions between electrons and ions \eq{eq:Ceslinear1} can be simplified using the two lowest order terms in the expansion \eq{eq:gradgradgexpansiones}, $\nabla_w \nabla_w w \simeq \nabla_v \nabla_v v - \bv^\prime \cdot \nabla_v \nabla_v \nabla_v v$. It may seem that in fact it is sufficient with $\nabla_w \nabla_w w \simeq \nabla_v \nabla_v v$ because to lowest order in $\sqrt{m_e/m_s} \ll 1$, equation \eq{eq:Ceslinear1} is formally
\begin{eqnarray} \label{eq:Ceslinear2}
\fl C_{es}^{(\ell)} [ g_e; g_s]  \simeq \frac{\gamma_{es}}{m_e^2} \nabla_v \cdot \Bigg ( \int \dd^3v^\prime\, f_{Ms} (\bv^\prime) \nabla_v \nabla_v v \cdot \nabla_v g_e (\bv) \Bigg ),
\end{eqnarray}
where we have neglected the third and fourth term in \eq{eq:Ceslinear1} as small in $\sqrt{m_e/m_s}$, and we have used \eq{eq:vnablavnablavv} to eliminate the second term. However, there are pieces of $g_e/f_{Me}$ that are small by $\sqrt{m_e/m_s}$ compared to $g_s/f_{Ms}$ (see the estimate for the piece of $g_e$ odd in $u$ given in \eq{eq:fenoddorder} and compare it to \eq{eq:fnorder}). To correctly determine these pieces, we have to go to higher order in $\sqrt{m_e/m_s}$ and use $\nabla_w \nabla_w w \simeq \nabla_v \nabla_v v - \bv^\prime \cdot \nabla_v \nabla_v \nabla_v v$. Then, equation \eq{eq:Ceslinear1} becomes
\begin{eqnarray} \label{eq:Ceslinear3}
\fl C_{es}^{(\ell)} [ g_e; g_s] = \frac{\gamma_{es}}{m_e} \nabla_v \cdot \Bigg ( \frac{n_s}{m_e}  \nabla_v \nabla_v v \cdot \nabla_v g_e (\bv) \nonumber\\ + \frac{f_{Me} (\bv)}{T_e} \bv \cdot \nabla_v \nabla_v \nabla_v v \cdot\int \dd^3v^\prime\,  g_s  (\bv^\prime) \bv^\prime \Bigg),
\end{eqnarray}
where we have used \eq{eq:vnablavnablavv}, \eq{eq:intfMsv}, $\int \dd^3v^\prime\, f_{Ms} (\bv^\prime) = n_s$ and
\begin{equation} \label{eq:intnablavgs}
\int \dd^3v^\prime \nabla_{v^\prime} g_s (\bv^\prime) = 0 
\end{equation}
to simplify some terms. Using \eq{eq:vnablavnablavnablavv} and $f_{Me} (\bv) \nabla_v \nabla_v v = \nabla_v \nabla_v v \cdot \nabla_v ( \bv f_{Me}(\bv) )$, we convert \eq{eq:Ceslinear3} into \eq{eq:Cessimplify}.

Finally the nonlinear piece $C_{es} [ g_e, g_s ]$ can be simplified using the two lowest order terms in the expansion \eq{eq:gradgradgexpansiones}, $\nabla_w \nabla_w w \simeq \nabla_v \nabla_v v - \bv^\prime \cdot \nabla_v \nabla_v \nabla_v v$. The second term in the approximation for $\nabla_w \nabla_w w$, of higher order in $\sqrt{m_e/m_i}$, is needed for the same reason that it was needed in the linearized collision operator $C_{es}^{(\ell)} [ g_e; g_s]$: there are pieces of $g_e$ that are small by $\sqrt{m_e/m_s}$ (see the estimate for the piece of $g_e$ odd in $u$ in \eq{eq:fenoddorder} and compare it to \eq{eq:fnorder}). Using $\nabla_w \nabla_w w \simeq \nabla_v \nabla_v v - \bv^\prime \cdot \nabla_v \nabla_v \nabla_v v$, we find
\begin{eqnarray} \label{eq:Cesnonlinear1}
\fl C_{es} [ g_e, g_s ] \simeq \frac{\gamma_{es}}{m_e} \nabla_v \cdot \Bigg [ \int \dd^3 v^\prime\, \Bigg ( \frac{g_s(\bv^\prime)}{m_e} \nabla_v \nabla_v v \cdot \nabla_v g_e (\bv) \nonumber\\ - \frac{g_s(\bv^\prime)}{m_e} \bv^\prime \cdot \nabla_v \nabla_v \nabla_v v \cdot \nabla_v g_e (\bv) - \frac{g_e (\bv)}{m_s} \nabla_v \nabla_v v \cdot \nabla_{v^\prime} g_s (\bv^\prime) \Bigg ) \Bigg ].
\end{eqnarray}
Employing \eq{eq:intnablavgs} and
\begin{equation}
\bv^\prime \cdot \nabla_v \nabla_v \nabla_v v = - \frac{1}{v^3}  (\bv \bv^\prime + \bv^\prime \bv) - \frac{\bv \cdot \bv^\prime}{v^5} (v^2 \matI - 3 \bv \bv),
\end{equation}
we finally obtain
\begin{eqnarray} \label{eq:Cesnonlinear}
\fl C_{es} [ g_e, g_s ] \simeq \frac{\gamma_{es} n_s}{m_e^2} \nabla_v \cdot \Bigg [ \Bigg ( \frac{\int \dd^3 v^\prime\, g_s(\bv^\prime)}{n_s} \frac{v^2 \matI - \bv\bv}{v^3} \nonumber\\ + \frac{\bv \mathbf{U}_s + \mathbf{U}_s \bv}{v^3} + \frac{\bv \cdot \mathbf{U}_s}{v^5} (v^2 \matI - 3 \bv \bv) \Bigg ) \cdot \nabla_v g_e (\bv)  \Bigg ],
\end{eqnarray}
where $\mathbf{U}_s$ is defined in \eq{eq:Usdef}.

\section{Collision operators in the gyrokinetic equation} \label{app:gkcollision}
Collision operators are more naturally applied to distribution functions written in $\{\boldr, \bv\}$ coordinates. We use \eq{eq:fsaveT} and \eq{eq:fswigT} to write the distribution function in the $\{ \boldr, \bv \}$ variables, finding
\begin{eqnarray} \label{eq:collisionrv}
\fl C_{ss^\prime} [ f_s, f_{s^\prime} ] = C_{ss^\prime}^{(\ell)} \left [ [f_s]_1^\lw; [f_{s^\prime}]_1^\lw \right ] + C_{ss^\prime}^{(\ell)} \left [ [f_s]_2^\lw; [f_{s^\prime}]_2^\lw \right ] + C_{ss^\prime} \left [ [f_s]_1^\lw, [f_{s^\prime}]_1^\lw \right ] \nonumber \\+ C_{ss^\prime} [ f_{Ms}, f_{Ms^\prime} ] + \sum_{k_\psi, k_\alpha} \Bigg ( C_{ss^\prime}^{(\ell)} \left [ \underline{[f_s]}_1^\tb; \underline{[f_{s^\prime}]}_1^\tb \right ] + C_{ss^\prime}^{(\ell)} \left [ \underline{[f_s]}_2^\tb; \underline{[f_{s^\prime}]}_2^\tb \right ] \nonumber\\+ C_{ss^\prime} \left [ \underline{[f_s]}_1^\tb, [f_{s^\prime}]_1^\lw \right ]+ C_{ss^\prime} \left [ [f_s]_1^\lw, \underline{[f_{s^\prime}]}_1^\tb \right ] \nonumber\\+ \sum_{k_\psi^\prime, k_\alpha^\prime} C_{ss^\prime} \left [ (\underline{[f_s]}_1^\tb)^\prime, (\underline{[f_{s^\prime}]}_1^\tb)^{\prime\prime} \right ] \Bigg ) \exp ( i k_\psi \psi(\boldr) + i k_\alpha \alpha (\boldr)).
\end{eqnarray}
Here a prime on a Fourier coefficient indicates that it depends on $k_\psi^\prime$ and $k_\alpha^\prime$, and two primes that it depends on $k_\psi^{\prime\prime} = k_\psi - k_\psi^\prime$ and $k_\alpha^{\prime\prime} = k_\alpha - k_\alpha^\prime$. We have kept $C_{ss^\prime} [ f_{Ms}, f_{Ms^\prime} ]$ because it can be non-zero for ion-electron, impurity-electron, electron-ion and electron-impurity collisions (see \ref{app:Cse} and \ref{app:Ces}).

Equation \eq{eq:collisionrv} is written in $\{\boldr, \bv\}$ variables, and we need to rewrite it in $\{\bR, u, \mu, \varphi \}$ variables. We first rewrite $C_{ss^\prime} [ f_s, f_{s^\prime} ] (\boldr, \bv)$ as a function of the variables $\{ \boldr, v_{||}, \mu_0, \varphi_0 \}$, that is, $C_{ss^\prime} [ f_s, f_{s^\prime} ] (\boldr, \bv) = C_{ss^\prime} [ f_s, f_{s^\prime} ] (\boldr, v_{||}, \mu_0, \varphi_0 )$. We then invert relations \eq{eq:Rdef}-\eq{eq:varphidef} to find 
\begin{equation} \label{eq:rRumuvarphi}
\boldr(\bR, u, \mu, \varphi, t) \simeq \bR + \rhobf (\bR, \mu, \varphi) + \boldr_2 (\bR, u, \mu, \varphi, t),
\end{equation}
\begin{equation} \label{eq:vparRumuvarphi}
v_{||} (\bR, u, \mu, \varphi, t) \simeq u + v_{||, 1}(\bR, u, \mu, \varphi, t), 
\end{equation}
\begin{equation}  \label{eq:mu0Rumuvarphi}
\mu_0 (\bR, u, \mu, \varphi, t) \simeq \mu + \mu_{0, 1}(\bR, u, \mu, \varphi, t)
\end{equation} 
and 
\begin{equation}  \label{eq:varphi0Rumuvarphi}
\varphi_0 (\bR, u, \mu, \varphi, t) \simeq \varphi + \varphi_{0,1}(\bR, u, \mu, \varphi, t).
\end{equation}
The corrections $\boldr_2$, $v_{||,1}$, $\mu_{0,1}$ and $\varphi_{0,1}$ are of order $(\rho_s/a)^2 a$, $(\rho_s/a) v_{ts}$, $(\rho_s/a) v_{ts}^2/B$ and $\rho_s/a$, respectively. Importantly, these corrections remain of this order when the subsidiary expansion $B_p/B \ll 1$ is performed in sections \ref{sec:expansion} and \ref{sec:secondordertotal} because the corrections $\bR_2$, $u_1$, $\mu_1$ and $\varphi_1$ in \eq{eq:Rdef}, \eq{eq:udef}, \eq{eq:mudef} and \eq{eq:varphidef} do not scale with $B_p/B$ in any particular way.

Using \eq{eq:rRumuvarphi}, \eq{eq:vparRumuvarphi}, \eq{eq:mu0Rumuvarphi} and \eq{eq:varphi0Rumuvarphi}, we Taylor expand $C_{ss^\prime} [ f_s, f_{s^\prime} ] (\boldr, v_{||}, \mu_0, \varphi_0)$ around $\bR$, $u$, $\mu$ and $\varphi$. We split the result into a long wavelength piece and a turbulent piece,
\begin{equation}
C_{ss^\prime} [ f_s, f_{s^\prime} ] = C_{ss^\prime}^\lw + \sum_{k_\psi, k_\alpha} \underline{C}_{ss^\prime}^\tb \exp ( i k_\psi \psi (\bR) + i k_\alpha \alpha (\bR) ).
\end{equation}
The long wavelength piece is
\begin{eqnarray} \label{eq:Cssprimelw}
C_{ss^\prime}^\lw = C_{ss^\prime}^{(\ell)} \left [ [f_s]_1^\lw; [f_{s^\prime}]_1^\lw \right ] + C_{ss^\prime}^{(\ell)} \left [ f_{s2}^\lw; f_{s2}^\lw \right ] + C_{ss^\prime} [ f_{Ms}, f_{Ms^\prime} ] + C_{ss^\prime, 2}^\lw,
\end{eqnarray}
where
\begin{eqnarray} \label{eq:Cssprime2lw}
\fl C_{ss^\prime,2}^\lw = \left ( \rhobf \cdot \nabla_\bR + v_{||,1} \frac{\partial}{\partial u} + \mu_{0,1}^\lw \frac{\partial}{\partial \mu} + \varphi_{0,1}^\lw \frac{\partial}{\partial \varphi} \right ) \left ( C_{ss^\prime}^{(\ell)} \left [ [f_s]_1^\lw; [f_{s^\prime}]_1^\lw \right ] \right ) \nonumber\\+ C_{ss^\prime}^{(\ell)} \left [ \Delta f_{s2}^\lw; \Delta f_{s^\prime 2}^\lw \right ] + C_{ss^\prime} \left [ [f_s]_1^\lw, [f_{s^\prime}]_1^\lw \right ] \nonumber \\ + \Bigg \langle  \sum_{k_\psi, k_\alpha} \Bigg [ C_{ss^\prime} \left [ (\underline{[f_s]}_1^\tb)^\ast, \underline{[f_{s^\prime}]}_1^\tb \right ] + \Bigg ( i \bk_\bot \cdot (\underline{\boldr}_2^\tb)^\ast + (\underline{\mu}_{0,1}^\tb)^\ast \frac{\partial}{\partial \mu} \nonumber\\ + (\underline{\varphi}_{0,1}^\tb)^\ast \frac{\partial}{\partial \varphi} \Bigg ) \left ( C_{ss^\prime}^{(\ell)} \left [ \underline{[f_s]}_1^\tb; \underline{[f_{s^\prime}]}_1^\tb \right ] \right ) \exp ( i \bk_\bot \cdot \rhobf ) \Bigg ] \Bigg \rangle_t.
\end{eqnarray}
From the lowest order pieces of \eq{eq:Cssprimelw}, we obtain the gyrophase dependent piece $\tilde{f}_{s2}^\lw$ in \eq{eq:ftildelw} and the collision operator in \eq{eq:lwfirstordergk}. The gyroaverage of the second order pieces of \eq{eq:Cssprimelw} give the collisional contributions to \eq{eq:lwsecondorderions}.

The short wavelength piece of the collision operator is
\begin{eqnarray} \label{eq:Cssprimetb}
\fl \underline{C}_{ss^\prime}^\tb = C_{ss^\prime}^{(\ell)} \left [ \underline{[f_s]}_1^\tb; \underline{[f_{s^\prime}]}_1^\tb \right ] \exp ( i \bk_\bot \cdot \rhobf ) + \underline{C}_{ss^\prime, 2}^\tb \nonumber\\+ C_{ss^\prime}^{(\ell)} \left [ \underline{f}_{s2}^\tb \exp \left ( \frac{i \bk_\bot \cdot (\bv \times \bun)}{\Omega_s} \right ); \underline{f}_{s^\prime2}^\tb \exp \left ( \frac{i \bk_\bot \cdot (\bv \times \bun)}{\Omega_{s^\prime}} \right ) \right ] \nonumber\\ \times \exp ( i \bk_\bot \cdot \rhobf ),
\end{eqnarray}
where
\begin{eqnarray} \label{eq:Cssprime2tb}
\fl \underline{C}_{ss^\prime, 2}^\tb = \Bigg [ C_{ss^\prime}^{(\ell)} \left [ \Delta \underline{f}_{s2}^\tb; \Delta \underline{f}_{s^\prime 2}^\tb \right ] + C_{ss^\prime} \left [ \underline{[f_s]}_1^\tb, [f_{s^\prime}]_1^\lw \right ]+ C_{ss^\prime} \left [ [f_s]_1^\lw, \underline{[f_{s^\prime}]}_1^\tb \right ] \nonumber\\ + \Bigg ( i \bk_\bot \cdot \boldr_2^\lw + \frac{i}{2} \rhobf \rhobf : \nabla_\bR \bk_\bot + \rhobf \cdot \nabla_\bR \psi \frac{\partial}{\partial \psi} + \rhobf \cdot \nabla_\bR \theta \frac{\partial}{\partial \theta} \nonumber\\ + v_{||,1} \frac{\partial}{\partial u} + \mu_{0,1}^\lw \frac{\partial}{\partial \mu} + \varphi_{0,1}^\lw \frac{\partial}{\partial \varphi} \Bigg ) \left ( C_{ss^\prime}^{(\ell)} \left [ [f_s]_1^\lw; [f_{s^\prime}]_1^\lw \right ] \right ) \Bigg ] \exp ( i \bk_\bot \cdot \rhobf ) \nonumber\\ + \sum_{k_\psi^\prime, k_\alpha^\prime} \Bigg [ C_{ss^\prime} \left [ (\underline{[f_s]}_1^\tb)^\prime, (\underline{[f_{s^\prime}]}_1^\tb)^{\prime\prime} \right ]  \exp ( i \bk_\bot \cdot \rhobf ) + \Bigg ( i \bk_\bot^\prime \cdot (\underline{\boldr}_2^\tb)^{\prime\prime} \nonumber\\ + (\underline{\mu}_{0,1}^\tb)^{\prime\prime} \frac{\partial}{\partial \mu} + (\underline{\varphi}_{0,1}^\tb)^{\prime\prime} \frac{\partial}{\partial \varphi} \Bigg ) \left ( C_{ss^\prime}^{(\ell)} \left [ (\underline{[f_s]}_1^\tb)^\prime; (\underline{[f_{s^\prime}]}_1^\tb)^\prime \right ] \right ) \exp ( i \bk_\bot^\prime \cdot \rhobf )\Bigg ].
\end{eqnarray}
Here a prime on a Fourier coefficient indicates that it depends on $k_\psi^\prime$ and $k_\alpha^\prime$, and two primes that it depends on $k_\psi^{\prime\prime} = k_\psi - k_\psi^\prime$ and $k_\alpha^{\prime\prime} = k_\alpha - k_\alpha^\prime$. The lowest order pieces of \eq{eq:Cssprimetb} determine the gyrophase dependent piece $\underline{\tilde{f}}_{s2}^\tb$ in \eq{eq:underlineftildetb} and the collisional terms in \eq{eq:tbfirstordergk}. The collisional terms in \eq{eq:tbsecondordergk} come from the second order pieces of \eq{eq:Cssprimetb}.

\section{Derivation of equation \eq{eq:Flwnctotal}} \label{app:Flwnc}
In this Appendix, we derive \eq{eq:Flwnctotal} from \eq{eq:Flwnctotalv1} following the procedure in Appendix H of \cite{calvo12}. To rewrite \eq{eq:Flwnctotalv1}, we use the definition of $g_{s1}^\lw$ in \eq{eq:gs1lwdef} and the relations
\begin{eqnarray} \label{eq:Flwncg1lw}
\fl - \left [ \left ( \bv_{Ms} - \frac{c}{B} \nabla_\bR \phi_0 \times \bun \right ) \cdot \nabla_\bR - \frac{u}{\Omega_s} (\bun \times \kappabf) \cdot \left ( \mu \nabla_\bR B + \frac{Z_s e}{m_s} \nabla_\bR \phi_0 \right ) \frac{\partial}{\partial u} \right ] g_{s1}^\lw \nonumber\\ = \left ( u \bun \cdot \nabla_\bR - \mu \bun \cdot \nabla_\bR B \frac{\partial}{\partial u} \right ) \left [ \frac{I}{\Omega_s} \left ( \frac{Z_s e}{m_s} \frac{\partial \phi_0}{\partial \psi} + \mu \frac{\partial B}{\partial \psi} \right ) \frac{\partial g_{s1}^\lw}{\partial u} - \frac{Iu}{\Omega_s} \frac{\partial g_{s1}^\lw}{\partial \psi} \right ] \nonumber\\ + \bun \cdot \nabla_\bR \theta \frac{\partial}{\partial \psi} \left [ \frac{1}{\bun \cdot \nabla_\bR \theta}\frac{I u}{\Omega_s} \left ( u \bun \cdot \nabla_\bR - \mu \bun \cdot \nabla_\bR B \frac{\partial}{\partial u} \right ) g_{s1}^\lw \right ] \nonumber\\ - \frac{\partial}{\partial u} \left [ \frac{I}{\Omega_s} \left ( \frac{Z_s e}{m_s} \frac{\partial \phi_0}{\partial \psi} + \mu \frac{\partial B}{\partial \psi} \right ) \left ( u \bun \cdot \nabla_\bR - \mu \bun \cdot \nabla_\bR B \frac{\partial}{\partial u} \right ) g_{s1}^\lw \right ] \nonumber\\ + \frac{u}{\Omega_s} (\bun \cdot \nabla_\bR \times \bun) \left ( u \bun \cdot \nabla_\bR - \mu \bun \cdot \nabla_\bR B \frac{\partial}{\partial u} \right ) g_{s1}^\lw,
\end{eqnarray} 
\begin{eqnarray} \label{eq:Flwncphi1lw}
\fl \Bigg [ \left ( \bv_{Ms} - \frac{c}{B} \nabla_\bR \phi_0 \times \bun \right ) \cdot \nabla_\bR - \frac{u}{\Omega_s} (\bun \times \kappabf) \cdot \left ( \mu \nabla_\bR B + \frac{Z_s e}{m_s} \nabla_\bR \phi_0 \right ) \frac{\partial}{\partial u} \Bigg ] \nonumber \\ \times \left ( \frac{Z_s e \phi_1^\lw}{T_s} f_{Ms} \right ) - \frac{Z_s e}{m_s} \bun \cdot \nabla_\bR \phi_1^\lw \frac{\partial}{\partial u} \Bigg \{ \frac{I u}{\Omega_s} \Bigg [ \frac{1}{p_s} \frac{\partial p_s}{\partial \psi} + \frac{Z_s e}{T_s} \frac{\partial \phi_0}{\partial \psi} \nonumber\\ + \left ( \frac{m_s (u^2 + 2\mu B)}{2T_s} - \frac{5}{2} \right ) \frac{1}{T_s} \frac{\partial T_s}{\partial \psi} \Bigg ] f_{Ms} \Bigg \} + \frac{c}{B} (\nabla_\bR \phi_1^\lw \times \bun )\cdot \nabla_\bR \psi \nonumber\\ \times \Bigg [ \frac{\partial}{\partial \psi} \ln p_s + \left ( \frac{m_s ( u^2 + 2\mu B)}{2T_s} - \frac{5}{2} \right ) \frac{\partial}{\partial \psi} \ln T_i \Bigg ] f_{Ms} \nonumber\\ - \frac{Z_s e}{T_s} \bv_{Ms} \cdot \nabla_\bR \phi_1^\lw f_{Ms} = \left ( u \bun \cdot \nabla_\bR - \mu \bun \cdot \nabla_\bR B \frac{\partial}{\partial u} \right ) \Bigg \{ \frac{Z_s e \phi_1^\lw}{T_s}  \nonumber\\ \times \frac{I u}{\Omega_s} \Bigg [ \frac{1}{p_s} \frac{\partial p_s}{\partial \psi} + \frac{Z_s e}{T_s} \frac{\partial \phi_0}{\partial \psi} + \Bigg ( \frac{m_s (u^2 + 2\mu B)}{2T_s} - \frac{7}{2} \Bigg ) \frac{1}{T_s} \frac{\partial T_s}{\partial \psi} \Bigg ] f_{Ms} \Bigg \},
\end{eqnarray}  
\begin{eqnarray} \label{eq:FlwncIuOmega}
\fl \left [ \left ( \bv_{Ms} - \frac{c}{B} \nabla_\bR \phi_0 \times \bun \right ) \cdot \nabla_\bR - \frac{u}{\Omega_s} (\bun \times \kappabf) \cdot \left ( \mu \nabla_\bR B + \frac{Z_s e}{m_s} \nabla_\bR \phi_0 \right ) \frac{\partial}{\partial u} \right ] \nonumber\\ \times \left \{ \frac{I u}{\Omega_s} \left [ \frac{1}{p_s} \frac{\partial p_s}{\partial \psi} + \frac{Z_s e}{T_s} \frac{\partial \phi_0}{\partial \psi} + \left ( \frac{m_s (u^2 + 2\mu B)}{2T_s} - \frac{5}{2} \right ) \frac{1}{T_s} \frac{\partial T_s}{\partial \psi} \right ] f_{Ms}\right \} \nonumber\\ = \left ( u \bun \cdot \nabla_\bR - \mu \bun \cdot \nabla_\bR B \frac{\partial}{\partial u} \right ) \Bigg \{ \frac{I^2  u^2}{2 \Omega_s^2} \Bigg [ \frac{\partial^2}{\partial \psi^2} \ln p_s + \frac{Z_s e}{T_s} \frac{\partial^2 \phi_0}{\partial \psi^2} \nonumber\\ + \left ( \frac{m_s (u^2 + 2\mu B)}{2T_s} - \frac{5}{2} \right ) \frac{\partial^2}{\partial \psi^2} \ln T_s  - \frac{2 Z_s e}{T_s^2} \frac{\partial \phi_0}{\partial \psi} \frac{\partial T_s}{\partial \psi} \nonumber\\ - \frac{m_s (u^2 + 2\mu B)}{2T_s^3} \left ( \frac{\partial T_s}{\partial \psi} \right )^2 + \Bigg ( \frac{1}{p_s} \frac{\partial p_s}{\partial \psi} + \frac{Z_s e}{T_s} \frac{\partial \phi_0}{\partial \psi} \nonumber\\+ \left ( \frac{m_s (u^2 + 2\mu B)}{2T_s} - \frac{5}{2} \right ) \frac{1}{T_s} \frac{\partial T_s}{\partial \psi}\Bigg )^2 \Bigg ] f_{Ms} \Bigg \} \nonumber\\ - \frac{u}{\Omega_s} (\bun \cdot \nabla_\bR \times \bun) \left ( u \bun \cdot \nabla_\bR - \mu \bun \cdot \nabla_\bR B \frac{\partial}{\partial u} \right ) \nonumber\\ \times \left \{ \frac{I u}{\Omega_s} \left [ \frac{1}{p_s} \frac{\partial p_s}{\partial \psi} + \frac{Z_s e}{T_s} \frac{\partial \phi_0}{\partial \psi} + \left ( \frac{m_s (u^2 + 2\mu B)}{2T_s} - \frac{5}{2} \right ) \frac{1}{T_s} \frac{\partial T_s}{\partial \psi} \right ] f_{Ms} \right \},
\end{eqnarray}
\begin{eqnarray} \label{eq:Flwncphi1lwg1lw}
\fl \frac{Z_s e}{m_s} \bun \cdot \nabla_\bR \phi_1^\lw \frac{\partial g_{s1}^\lw}{\partial u} = \left ( u \bun \cdot \nabla_\bR - \mu \bun \cdot \nabla_\bR B \frac{\partial}{\partial u} \right ) \left ( \frac{Z_s e \phi_1^\lw}{m_s u} \frac{\partial g_{s1}^\lw}{\partial u} \right ) \nonumber\\ - \frac{\partial}{\partial u} \left [ \frac{Z_s e \phi_1^\lw}{m_s u} \left ( u \bun \cdot \nabla_\bR - \mu \bun \cdot \nabla_\bR B \frac{\partial}{\partial u} \right ) g_{s1}^\lw \right ]
\end{eqnarray}
and
\begin{eqnarray} \label{eq:Flwncphi1lw2}
\fl - \frac{Z_s e}{m_s} \bun \cdot \nabla_\bR \phi_1^\lw \frac{\partial}{\partial u} \left ( \frac{Z_s e \phi_1^\lw}{T_s} f_{Ms} \right ) = \left ( u \bun \cdot \nabla_\bR - \mu \bun \cdot \nabla_\bR B \frac{\partial}{\partial u} \right ) \nonumber\\ \times \left [ \frac{1}{2} \left ( \frac{Z_s e \phi_1^\lw}{T_s} \right )^2 f_{Ms} \right ].
\end{eqnarray}
In equations \eq{eq:Flwncg1lw} and \eq{eq:FlwncIuOmega}, the terms proportional to $\bun \cdot \nabla_\bR \times \bun$ are negligible because for $B_p/B \ll 1$, $\bun \cdot \nabla_\bR \times \bun \sim B_p/Ba \ll a^{-1}$. 

Using equations \eq{eq:Flwncg1lw}-\eq{eq:Flwncphi1lw2}, neglecting the terms proportional to $\bun \cdot \nabla_\bR \times \bun$ and employing \eq{eq:lwfirstordergkv2} to rewrite $u\bun \cdot \nabla_\bR g_{s1}^\lw - \mu \bun \cdot \nabla_\bR B (\partial g_{s1}^\lw/\partial u)$ in terms of the collision operator, we finally obtain \eq{eq:Flwnctotal}, with
\begin{eqnarray} \label{eq:Deltahs2nclwdef}
\fl \Delta h_{s2, \nc}^\lw = \frac{I}{\Omega_s} \left ( \frac{Z_s e}{m_s} \frac{\partial \phi_0}{\partial \psi} + \mu \frac{\partial B}{\partial \psi} \right ) \frac{\partial g_{s1}^\lw}{\partial u} - \frac{Iu}{\Omega_s} \frac{\partial g_{s1}^\lw}{\partial \psi} \nonumber\\ + \frac{Z_s e \phi_1^\lw}{T_s} \frac{I u}{\Omega_s} \Bigg [ \frac{1}{p_s} \frac{\partial p_s}{\partial \psi} + \frac{Z_s e}{T_s} \frac{\partial \phi_0}{\partial \psi}  \nonumber\\ + \Bigg ( \frac{m_s (u^2 + 2\mu B)}{2T_s} - \frac{7}{2} \Bigg ) \frac{1}{T_s} \frac{\partial T_s}{\partial \psi} \Bigg ] f_{Ms} \nonumber\\ + \frac{I^2  u^2}{2 \Omega_s^2} \Bigg [ \frac{\partial^2}{\partial \psi^2} \ln p_s + \frac{Z_s e}{T_s} \frac{\partial^2 \phi_0}{\partial \psi^2} \nonumber\\ + \left ( \frac{m_s (u^2 + 2\mu B)}{2T_s} - \frac{5}{2} \right ) \frac{\partial^2}{\partial \psi^2} \ln T_s  - \frac{2 Z_s e}{T_s^2} \frac{\partial \phi_0}{\partial \psi} \frac{\partial T_s}{\partial \psi} \nonumber\\ - \frac{m_s (u^2 + 2\mu B)}{2T_s^3} \left ( \frac{\partial T_s}{\partial \psi} \right )^2 + \Bigg ( \frac{1}{p_s} \frac{\partial p_s}{\partial \psi} + \frac{Z_s e}{T_s} \frac{\partial \phi_0}{\partial \psi} \nonumber\\+ \left ( \frac{m_s (u^2 + 2\mu B)}{2T_s} - \frac{5}{2} \right ) \frac{1}{T_s} \frac{\partial T_s}{\partial \psi}\Bigg )^2 \Bigg ] f_{Ms} \nonumber\\ + \frac{Z_s e \phi_1^\lw}{m_s u} \frac{\partial g_{s1}^\lw}{\partial u} + \left ( \frac{Z_s e \phi_1^\lw}{T_s} \right )^2 f_{Ms}.
\end{eqnarray} 

\section{Equations in the frame rotating with $\Omega_{\zeta, E}$} \label{app:equationsrotating}
Several gyrokinetic codes solve the gyrokinetic equation in the frame rotating with velocity $\Omega_{\zeta, E} = - c (\partial \phi_0/\partial \psi)$ \cite{dorland00, candy03, dannert05, peeters09c}. We are going to rewrite the equations in section \ref{sec:secondordertotal} so that they are valid in the frame rotating with $\Omega_{\zeta, E}$. 

The new parallel velocity in the frame rotating with $\Omega_{\zeta, E}$ is 
\begin{equation}
u - \Omega_{\zeta, E} R \zun \cdot \bun = u + \frac{cI}{B} \frac{\partial \phi_0}{\partial \psi}.
\end{equation}
Using this relation, the long wavelength piece of the distribution function in the laboratory frame, $f_s^\lw$, can be written in terms of the long wavelength piece of the distribution function in the rotating frame, $f_s^{R, \lw}$, 
\begin{equation}
\fl f_s^\lw (\psi (\bR), \theta(\bR), u, \mu, \varphi, t) = f_s^{R,\lw} (\psi (\bR), \theta(\bR), u + (cI/B)(\partial \phi_0/\partial \psi), \mu, \varphi, t).
\end{equation}
Since in our ordering, $(cI/B) (\partial \phi_0/\partial \psi) \sim (B/B_p) \rho_\ast v_{ti} \ll v_{ti}$, we can Taylor expand to find
\begin{equation}
f_s^\lw \simeq f_s^{R, \lw} + \frac{cI}{B} \frac{\partial \phi_0}{\partial \psi} \frac{\partial f_s^{R,\lw}}{\partial u} + \frac{1}{2} \frac{c^2 I^2}{B^2} \left ( \frac{\partial \phi_0}{\partial \psi} \right )^2 \frac{\partial^2 f_s^{R,\lw}}{\partial u^2}.
\end{equation}
Using the expansion in \eq{eq:flwexpansion}, we obtain that $f_s^\lw$ is to lowest order $f_{Ms}$, and to higher order,
\begin{equation} \label{eq:fs1lwprime}
f_{s1}^\lw = f_{s1}^{R, \lw} - \frac{cI}{B} \frac{\partial \phi_0}{\partial \psi} \frac{m_s u}{T_s} f_{Ms}
\end{equation}
and
\begin{equation} \label{eq:fs2lwprime}
f_{s2}^\lw = f_{s2}^{R, \lw} + \frac{cI}{B} \frac{\partial \phi_0}{\partial \psi} \frac{\partial f_{s1}^{R, \lw}}{\partial u} + \frac{m_s}{2T_s} \frac{c^2 I^2}{B^2} \left ( \frac{\partial \phi_0}{\partial \psi} \right )^2 \left ( \frac{m_s u^2}{T_s} - 1 \right ) f_{Ms}.
\end{equation}
The long wavelength potential in the rotating frame is
\begin{equation}
\phi^{R, \lw} = \phi_1^{R, \lw} + \phi_2^{R, \lw},
\end{equation}
that is, there is no lowest order potential. The lowest order potential transforms to zero because $- \nabla \phi_0 + c^{-1} \Omega_{\zeta, E} R \zun \times \bB = 0$. The higher order pieces of the potential are related to the potential in the laboratory frame by
\begin{equation} \label{eq:phi1lwprime}
\phi_1^\lw = \phi_1^{R,\lw}
\end{equation}
and
\begin{equation} \label{eq:phi2lwprime}
\phi_2^\lw = \phi_2^{R, \lw}.
\end{equation}

The turbulent pieces of the distribution function and the potential in the rotating frame, $\phi^{R, \tb}$ and $f_s^{R, \tb}$, depend on $\zeta - \Omega_{\zeta, E} t$. As a result, we can write them as
\begin{equation}
\phi^{R, \tb} = \sum_{k_\psi, k_\alpha} \underline{\phi}^{R, \tb} (k_\psi, k_\alpha, \psi, \theta, t) \exp ( i k_\psi \psi + i k_\alpha (\alpha - \Omega_{\zeta, E} t) )
\end{equation}
and
\begin{equation}
f_{s}^{R, \tb} = \sum_{k_\psi, k_\alpha} \underline{f}_{s}^{R, \tb} (k_\psi, k_\alpha, \psi, \theta, u, \mu, \varphi, t) \exp ( i k_\psi \psi + i k_\alpha (\alpha - \Omega_{\zeta, E} t) ).
\end{equation}
Comparing this result with \eq{eq:phitbscales} and \eq{eq:ftbscales}, we find that the turbulent pieces of the distribution function and the potential in the rotating frame are related to the same pieces in the laboratory frame by
\begin{equation}
\underline{\phi}^\tb (k_\psi, k_\alpha, \psi, \theta, t) = \underline{\phi}^{R, \tb} (k_\psi, k_\alpha, \psi, \theta, t) \exp ( - i k_\alpha \Omega_{\zeta, E} t)
\end{equation}
and
\begin{eqnarray}
\fl \underline{f}_s^\tb (k_\psi, k_\alpha, \psi, \theta, u, \mu, \varphi, t) = \underline{f}_s^{R, \tb} (k_\psi, k_\alpha, \psi, \theta, u + (cI/B)(\partial \phi_0/\partial \psi), \mu, \varphi, t) \nonumber\\ \times \exp ( - i k_\alpha \Omega_{\zeta, E} t).
\end{eqnarray}
Using the expansions \eq{eq:phitbexpansion} and \eq{eq:ftbexpansion}, we find
\begin{equation} \label{eq:phi1tbprime}
\underline{\phi}_1^\tb = \underline{\phi}_1^{R, \tb} \exp \left ( i k_\alpha c \frac{\partial \phi_0}{\partial \psi} t \right )
\end{equation}
and
\begin{equation} \label{eq:phi2tbprime}
\underline{\phi}_2^\tb = \underline{\phi}_2^{R, \tb} \exp\left ( i k_\alpha c \frac{\partial \phi_0}{\partial \psi} t \right )
\end{equation}
for the potential, and
\begin{equation} \label{eq:fs1tbprime}
\underline{f}_{s1}^\tb = \underline{f}_{s1}^{R, \tb} \exp \left ( i k_\alpha c \frac{\partial \phi_0}{\partial \psi} t \right ).
\end{equation}
and
\begin{equation} \label{eq:fs2tbprime}
\underline{f}_{s2}^\tb = \left ( \underline{f}_{s2}^{R, \tb} + \frac{cI}{B} \frac{\partial \phi_0}{\partial \psi} \frac{\partial \underline{f}_{s1}^{R, \tb}}{\partial u} \right ) \exp \left ( i k_\alpha c \frac{\partial \phi_0}{\partial \psi} t \right )
\end{equation}
for the distribution function.

We use all these expressions to find the equations for the distribution function and the potential in the rotating frame. We give the equations for the long wavelength, first order pieces in \ref{sub:lwfirstorderrotating}, the equations for the short wavelength, first order pieces in \ref{sub:tbfirstorderrotating}, the equations for the long wavelength, second order pieces in \ref{sub:lwsecondorderrotating} and the equations for the short wavelength, second order pieces in \ref{sub:tbsecondorderrotating}. We find the momentum flux in \ref{sub:momentumfluxrotating}.

\subsection{Long wavelength, first order equations} \label{sub:lwfirstorderrotating}
Substituting \eq{eq:fs1lwprime} and \eq{eq:phi1lwprime} into \eq{eq:lwfirstordergk} and \eq{eq:lwfirstorderquasineutrality}, and using equations \eq{eq:vMstrick} and
\begin{equation} \label{eq:CsameV}
C_{ss^\prime}^{(\ell)} \left [ \frac{m_s u}{T_s} f_{Ms}; \frac{m_{s^\prime} u}{T_{s^\prime}} f_{Ms^\prime} \right ] = 0, 
\end{equation}
we find
\begin{eqnarray} \label{eq:lwfirstordergkrotating}
\fl \left ( u \bun \cdot \nabla_\bR \theta \frac{\partial}{\partial \theta} - \mu \bun \cdot \nabla_\bR B \frac{\partial}{\partial u} \right ) f_{s1}^{R, \lw} - \sum_{s^\prime} C_{ss^\prime}^{(\ell)} [ f_{s1}^{R, \lw}; f_{s^\prime 1}^{R, \lw} ] = - \bv_{Ms}\cdot \nabla \psi \Bigg [ \frac{\partial}{\partial \psi} \ln p_s \nonumber\\ + \left ( \frac{m_s (u^2 + 2\mu B)}{2T_s} - \frac{5}{2} \right ) \frac{\partial}{\partial \psi} \ln T_s \Bigg ] f_{Ms} - \frac{Z_s e f_{Ms}}{T_s} u \bun \cdot \nabla_\bR \theta \frac{\partial \phi_1^{R, \lw}}{\partial \theta}
\end{eqnarray}
and
\begin{equation} \label{eq:lwfirstorderquasineutralityrotating}
2 \pi \sum_{s \neq e} Z_s \int \dd v_{||} \, \dd \mu_0\, B f_{s1}^{R, \lw} = \frac{e\phi_1^{R, \lw}}{T_e} n_e.
\end{equation}

As in subsection \ref{sub:lwfirstorder}, we define a new function $g_{s1}^{R,\lw}$ that is convenient for some calculations,
\begin{eqnarray} \label{eq:gs1lwdefrotating}
\fl g_{s1}^{R, \lw} = f_{s1}^{R, \lw} + \frac{I u}{\Omega_s} \left [ \frac{1}{p_s} \frac{\partial p_s}{\partial \psi} + \left ( \frac{m_s (u^2 + 2\mu B)}{2T_s} - \frac{5}{2} \right ) \frac{1}{T_s} \frac{\partial T_s}{\partial \psi} \right ] f_{Ms} + \frac{Z_s e \phi_1^{R, \lw}}{T_s} f_{Ms}.
\end{eqnarray}
Note that $g_{s1}^{R,\lw} = g_{s1}^\lw$ and as a result, the equation for $g_{s1}^{R,\lw}$ is \eq{eq:lwfirstordergkv2}.

\subsection{Short wavelength, first order equations} \label{sub:tbfirstorderrotating}
Substituting \eq{eq:phi1tbprime} and \eq{eq:fs1tbprime} into \eq{eq:tbfirstordergk} and \eq{eq:tbfirstorderquasineutrality}, we find
\begin{eqnarray} \label{eq:tbfirstordergkrotating}
\fl \frac{\partial \underline{f}_{s1}^{R, \tb}}{\partial t} + \left ( u \bun \cdot \nabla_\bR \theta \frac{\partial}{\partial \theta} - \mu \bun \cdot \nabla_\bR B \frac{\partial}{\partial u} \right ) \underline{f}_{s1}^{R, \tb} + i \bk_\bot \cdot \bv_{Ms} \underline{f}_{s1}^{R, \tb} \nonumber\\ - \sum_{s^\prime} C_{ss^\prime}^{GK} \left [ \underline{f}_{s1}^{R, \tb}; \underline{f}_{s^\prime 1}^{R, \tb} \right ] + \{ \underline{\phi}_1^{R, \tb} J_0 (\Lambda_s), \underline{f}_{s1}^{R, \tb} \} \nonumber\\ = - f_{Ms} \Bigg [ \frac{Z_s e}{T_s} \left ( u \bun \cdot \nabla_\bR \theta \frac{\partial}{\partial \theta} + i \bk_\bot \cdot \bv_{Ms} \right ) \nonumber\\ + i k_\alpha c \Bigg ( \frac{\partial}{\partial \psi} \ln n_s  + \left ( \frac{m_s (u^2 + 2\mu B)}{2T_s} - \frac{3}{2} \right ) \frac{\partial}{\partial \psi} \ln T_s \Bigg ) \Bigg ] \underline{\phi}_1^{R, \tb} J_0 (\Lambda_s) \nonumber\\ + \sum_{s^\prime} C_{ss^\prime}^{GK} \left [ \frac{Z_s e \underline{\phi}_1^{R, \tb}}{T_s} J_0 (\lambda_s) f_{Ms}; \frac{Z_{s^\prime} e \underline{\phi}_1^{R, \tb}}{T_{s^\prime}} J_0 (\lambda_{s^\prime}) f_{Ms^\prime} \right ]
\end{eqnarray}
and
\begin{eqnarray} \label{eq:tbfirstorderquasineutralityrotating}
\fl 2 \pi \sum_s Z_s \int \dd v_{||}\, \dd \mu_0\, B J_0 (\lambda_s) \underline{f}_{s1}^{R, \tb} - \sum_{s} \frac{Z_s^2 e \underline{\phi}_1^{R, \tb} }{T_s} n_s ( 1 - \Gamma_0(b_s)) = 0.
\end{eqnarray}

\subsection{Long wavelength, second order equations} \label{sub:lwsecondorderrotating}
As in subsection \ref{sub:lwsecondordertotal}, we define the function 
\begin{equation}
h_{s2}^{R, \lw} = \langle f_{s2}^{R,\lw} \rangle + \frac{Z_s e \phi_2^{R,\lw}}{T_i} f_{Ms}
\end{equation}
for ionic species ($s \neq e$). Since in subsection \ref{sub:lwsecondordertotal}, $h_{s2}^\lw$ is split into pieces of different physical origin (see \eq{eq:splith2total}), we do the same for $h_{s2}^{R, \lw}$,
\begin{equation} \label{eq:splith2rotating}
h_{s2}^{R,\lw} = h_{s2, \nc}^{R,\lw} + h_{s2, \tb}^{R,\lw} + h_{s2, \Delta T}^{R,\lw} + h_{s2, Q}^{R,\lw}.
\end{equation}
We define the different pieces of $h_{s2}^{R, \lw}$ in terms of the corresponding pieces in the laboratory frame using \eq{eq:fs2lwprime} and \eq{eq:phi2lwprime},
\begin{equation}
\fl h_{s2,\nc}^\lw = h_{s2,\nc}^{R,\lw} + \frac{cI}{B} \frac{\partial \phi_0}{\partial \psi} \frac{\partial f_{s1}^{R, \lw}}{\partial u} + \frac{m_s}{2T_s} \frac{c^2 I^2}{B^2} \left ( \frac{\partial \phi_0}{\partial \psi} \right )^2 \left ( \frac{m_s u^2}{T_s} - 1 \right ) f_{Ms}
\end{equation}
and
\begin{equation}
h_{s2,\beta}^\lw = h_{s2,\beta}^{R,\lw}
\end{equation}
for $\beta = \tb, \Delta T, Q$. Substituting these expressions for $h_{s2,\beta}^\lw$ into \eq{eq:eqh2betatotal}, and using \eq{eq:fs1lwprime}, \eq{eq:phi1lwprime}, \eq{eq:phi1tbprime}, \eq{eq:fs1tbprime}, \eq{eq:CsameV}, 
\begin{eqnarray}
\fl C_{ss^\prime}^{(\ell)} \left [ \frac{m_s}{2T_s} \left ( \frac{m_s u^2}{T_s} - 1 \right ) f_{Ms}; \frac{m_{s^\prime}} {2T_{s^\prime}} \left ( \frac{m_{s^\prime} u^2}{T_{s^\prime}} - 1 \right ) f_{Ms^\prime} \right ] \nonumber\\ + C_{ss^\prime} \left [ \frac{m_s u}{T_s} f_{Ms},  \frac{m_{s^\prime} u}{T_{s^\prime}} f_{Ms^\prime} \right ] = 0
\end{eqnarray}
and the expression
\begin{equation} \label{eq:dducollision}
\fl C_{ss^\prime}^{(\ell)} \left [ \frac{\partial g_s}{\partial u}; \frac{\partial g_{s^\prime}}{\partial u} \right ] - C_{ss^\prime} \left [ g_s,  \frac{m_{s^\prime} u}{T_{s^\prime}} f_{Ms^\prime} \right ] - C_{ss^\prime} \left [ \frac{m_s u}{T_s} f_{Ms}, g_{s^\prime} \right ] = \frac{\partial}{\partial u} \left ( C_{ss^\prime}^{(\ell)} [ g_s;  g_{s^\prime} ] \right ),
\end{equation}
valid for any set of functions $g_s$, we find that the equation for $h_{s2, \beta}^{R,\lw}$ is \eq{eq:eqh2betatotal} with $h_{s2,\beta}^\lw$ and $F_{s2, \beta}^\lw$ replaced by $h_{s2, \beta}^{R,\lw}$ and $F_{s2, \beta}^{R,\lw}$. The functions $F_{s2, \Delta T}^{R,\lw}$ and $F_{s2, Q}^{R,\lw}$ are equal to $F_{s2, \Delta T}^\lw$ and $F_{s2, Q}^\lw$, given in \eq{eq:FlwDeltaTtotal} and \eq{eq:FlwQtotal}, $F_{s2,\tb}^{R,\lw}$ is defined by \eq{eq:Flwtbtotal} with $\underline{f}_{s1}^\tb$ and $\underline{\phi}_1^\tb$ replaced by $\underline{f}_{s1}^{R,\tb}$ and $\underline{\phi}_1^{R,\tb}$, and
\begin{eqnarray} \label{eq:Flwncrotating}
\fl F_{s2, \nc}^{R,\lw} = \left ( u \bun \cdot \nabla_\bR - \mu \bun \cdot \nabla_\bR B \frac{\partial}{\partial u} \right ) \Delta h_{s2, \nc}^{R,\lw} \nonumber\\ + \bun \cdot \nabla_\bR \theta \frac{\partial}{\partial \psi} \left [ \frac{1}{\bun \cdot \nabla_\bR \theta}\frac{I u}{\Omega_s} \sum_{s^\prime \neq e} C_{ss^\prime}^{(\ell)} [ f_{s1}^{R, \lw}; f_{s^\prime 1}^{R, \lw} ] \right ] \nonumber\\ - \frac{\partial}{\partial u} \left [ \left ( \frac{ I \mu}{\Omega_s} \frac{\partial B}{\partial \psi} + \frac{Z_s e \phi_1^\lw}{m_s u} \right ) \sum_{s^\prime \neq e} C_{ss^\prime}^{(\ell)} [ f_{s1}^{R, \lw}; f_{s^\prime 1}^{R, \lw} ] \right ] \nonumber\\ + \sum_{s^\prime \neq e} C_{ss^\prime} \left [ f_{s1}^{R, \lw}, f_{s^\prime 1}^{R, \lw} \right ].
\end{eqnarray}
In this last equation, 
\begin{eqnarray} \label{eq:Deltahs2nclwdefrotating}
\fl \Delta h_{s2, \nc}^{R, \lw} = \Delta h_{s2, \nc}^\lw - \frac{cI}{B} \frac{\partial \phi_0}{\partial \psi} \frac{\partial f_{s1}^{R, \lw}}{\partial u}  - \frac{m_s}{2T_s} \frac{c^2 I^2}{B^2} \left ( \frac{\partial \phi_0}{\partial \psi} \right )^2 \left ( \frac{m_s u^2}{T_s} - 1 \right ) f_{Ms} = \nonumber\\  \frac{I \mu}{\Omega_s} \frac{\partial B}{\partial \psi} \frac{\partial g_{s1}^{R,\lw}}{\partial u} - \frac{Iu}{\Omega_s} \frac{\partial g_{s1}^{R,\lw}}{\partial \psi} \nonumber\\ +\frac{cI^2}{B\Omega_s} \frac{\partial \phi_0}{\partial \psi} \left [ \frac{1}{p_s} \frac{\partial p_s}{\partial \psi} + \Bigg ( \frac{m_s (u^2 + 2\mu B)}{2T_s} - \frac{5}{2} \Bigg ) \frac{1}{T_s} \frac{\partial T_s}{\partial \psi} \right ] f_{Ms} \nonumber\\ + \frac{Z_s e \phi_1^{R,\lw}}{T_s} \frac{I u}{\Omega_s} \Bigg [ \frac{1}{p_s} \frac{\partial p_s}{\partial \psi} + \Bigg ( \frac{m_s (u^2 + 2\mu B)}{2T_s} - \frac{7}{2} \Bigg ) \frac{1}{T_s} \frac{\partial T_s}{\partial \psi} \Bigg ] f_{Ms} \nonumber\\ + \frac{I^2  u^2}{2 \Omega_s^2} \Bigg [ \frac{\partial^2}{\partial \psi^2} \ln p_s + \frac{Z_s e}{T_s} \frac{\partial^2 \phi_0}{\partial \psi^2} \nonumber\\ + \left ( \frac{m_s (u^2 + 2\mu B)}{2T_s} - \frac{5}{2} \right ) \frac{\partial^2}{\partial \psi^2} \ln T_s \nonumber\\ - \frac{m_s (u^2 + 2\mu B)}{2T_s^3} \left ( \frac{\partial T_s}{\partial \psi} \right )^2 + \Bigg ( \frac{1}{p_s} \frac{\partial p_s}{\partial \psi} \nonumber\\+ \left ( \frac{m_s (u^2 + 2\mu B)}{2T_s} - \frac{5}{2} \right ) \frac{1}{T_s} \frac{\partial T_s}{\partial \psi}\Bigg )^2 \Bigg ] f_{Ms} \nonumber\\ + \frac{Z_s e \phi_1^{R,\lw}}{m_s u} \frac{\partial g_{s1}^{R,\lw}}{\partial u} + \left ( \frac{Z_s e \phi_1^{R,\lw}}{T_s} \right )^2 f_{Ms} + \frac{m_s}{2T_s} \frac{c^2 I^2}{B^2} \left ( \frac{\partial \phi_0}{\partial \psi} \right )^2 f_{Ms}.
\end{eqnarray}

\subsection{Short wavelength, second order equations} \label{sub:tbsecondorderrotating}
In subsection \ref{sub:tbsecondordertotal} we split $\underline{\phi}_2^\tb$ and $\langle \underline{f}_{s2}^\tb \rangle$ into pieces of different physical origin (see \eq{eq:splitf2tbtotal} and \eq{eq:splitphi2tbtotal}). We do the same for $\underline{\phi}_2^{R,\tb}$ and $\langle \underline{f}_{s2}^{R, \lw} \rangle$,
\begin{equation}
\underline{\phi}_2^{R, \tb} = \underline{\phi}_{2, \nc}^{R, \tb} + \underline{\phi}_{2,\grad}^{R, \tb} + \underline{\phi}_{2,\acc}^{R, \tb}
\end{equation}
and
\begin{equation}
\langle \underline{f}_{s2}^{R, \tb} \rangle = \underline{f}_{s2, \nc}^{R, \tb} + \underline{f}_{s2,\grad}^{R, \tb} + \underline{f}_{s2,\acc}^{R, \tb}.
\end{equation}
We write the different pieces in terms of the corresponding functions in the laboratory frame using \eq{eq:phi2tbprime} and \eq{eq:fs2tbprime},
\begin{equation}
\underline{\phi}_{2,\beta}^\tb = \underline{\phi}_{2,\beta}^{R,\tb} \exp \left ( i k_\alpha c \frac{\partial \phi_0}{\partial \psi} t \right )
\end{equation}
for $\beta = \nc, \grad, \acc$,
\begin{equation} \label{eq:f2nctbprime}
\underline{f}_{s2,\nc}^\tb = \left ( \underline{f}_{s2,\nc}^{R,\tb} + \frac{cI}{B} \frac{\partial \phi_0}{\partial \psi} \frac{\partial \underline{f}_{s1}^{R, \tb}}{\partial u} \right ) \exp \left ( i k_\alpha c \frac{\partial \phi_0}{\partial \psi} t \right )
\end{equation}
and
\begin{equation}
\underline{f}_{s2,\beta}^\tb = \underline{f}_{s2,\beta}^{R,\tb} \exp \left ( i k_\alpha c \frac{\partial \phi_0}{\partial \psi} t \right )
\end{equation}
for $\beta = \grad, \acc$. Substituting these expressions into \eq{eq:eqf2betatbtotal} and \eq{eq:eqphi2betatbtotal}, we find the equations for $\underline{f}_{s2, \beta}^{R,\tb}$ and $\underline{\phi}_{2, \beta}^{R, \tb}$. When substituting \eq{eq:f2nctbprime} into equation \eq{eq:eqf2betatbtotal}, we find a kinetic equation with several unintuitive terms that contain $\partial \underline{f}_{s1}^\tb/\partial u$. To eliminate these terms, we use
\begin{equation}
(\mathrm{Equation\,\, \eq{eq:eqf2betatbtotal}\,\,for\,\,}\beta = \nc) - \frac{cI}{B} \frac{\partial \phi_0}{\partial \psi} \frac{\partial}{\partial u} (\mathrm{Equation\,\, \eq{eq:tbfirstordergk}}) = 0,
\end{equation}
where equation \eq{eq:tbfirstordergk} is taken in the limit $B_p/B \ll 1$ with $k_\bot \rho_i \sim B_p/B$. We also use \eq{eq:fs1lwprime}, \eq{eq:phi1lwprime}, \eq{eq:phi1tbprime}, \eq{eq:fs1tbprime}, \eq{eq:dducollision}, and the approximation
\begin{equation}
\nabla \left ( \frac{I}{B} \right ) \simeq \nabla R \simeq - R \kappabf
\end{equation}
valid for $B_p/B \ll 1$.

The equations for $\underline{f}_{s2, \beta}^{R,\tb}$ are equations \eq{eq:eqf2betatbtotal} and \eq{eq:eqphi2betatbtotal} with $\underline{f}_{s2, \beta}^\tb$, $\underline{\phi}_{2, \beta}^\tb$, $\underline{F}_{s2, \beta}^\tb$ and $\Delta \underline{n}_{s2, \beta}^\tb$ replaced by $\underline{f}_{s2, \beta}^{R, \tb}$, $\underline{\phi}_{2, \beta}^{R, \tb}$, $\underline{F}_{s2, \beta}^{R, \tb}$ and $\Delta \underline{n}_{s2, \beta}^{R, \tb}$. The function $\underline{F}_{s2, \acc}^\tb$ is \eq{eq:F2acctbtotal} with $\underline{f}_{s1}^\tb$ and $\underline{\phi}_1^\tb$ replaced by $\underline{f}_{s1}^{R, \tb}$ and $\underline{\phi}_1^{R, \tb}$. The functions $\underline{F}_{s2, \nc}^{R,\tb}$ and $\underline{F}_{s2, \grad}^{R,\tb}$ are
\begin{eqnarray} \label{eq:F2nctbrotating}
\fl \underline{F}_{s2, \nc}^{R,\tb} = \Bigg [ \frac{c}{B} (i \bk_\bot \times \bun ) \cdot \nabla_\bR f_{s1}^{R, \lw} + \frac{Z_s e}{m_s} \frac{\partial f_{s1}^{R, \lw}}{\partial u} \Bigg ( \bun \cdot \nabla_\bR \theta \frac{\partial}{\partial \theta} + \frac{u}{\Omega_s} i \bk_\bot \cdot (\bun \times \kappabf) \Bigg ) \nonumber\\ - f_{Ms} \Bigg ( \frac{Z_s e}{T_s} i \bk_\bot \cdot \bv_{Co,s} + i k_\alpha c \frac{\partial \Omega_{\zeta, E}}{\partial \psi} \frac{I}{B} \frac{m_s u}{T_s} \Bigg ) \Bigg ] \underline{\phi}_1^{R, \tb} J_0 (\Lambda_s) \nonumber\\ - i \bk_\bot \cdot \left ( \bv_{Co,s} - \frac{c}{B} \nabla_\bR \phi_1^{R, \lw} \times \bun \right ) \underline{f}_{s1}^{R, \tb} \nonumber\\ + \left [ \frac{Z_s e}{m_s} \bun \cdot \nabla_\bR \phi_1^{R, \lw} + \frac{u \mu}{\Omega_s} (\bun \times \kappabf) \cdot \nabla_\bR B \right ] \frac{\partial \underline{f}_{s1}^{R,\tb}}{\partial u} \nonumber\\ + \sum_{s^\prime} \Bigg ( C_{ss^\prime} \left [ \underline{f}_{s1}^{R,\tb} , f_{s^\prime 1}^{R,\lw} \right ] + C_{ss^\prime} \left [ f_{s1}^{R,\lw}, \underline{f}_{s^\prime 1}^{R,\tb} \right ] \nonumber\\ +\sum_{k_\psi^\prime, k_\alpha^\prime} C_{ss^\prime} \left [ (\underline{f}_{s1}^{R, \tb})^\prime, (\underline{f}_{s^\prime 1}^{R, \tb})^{\prime\prime} \right ] \Bigg ),
\end{eqnarray}
where
\begin{equation}
\bv_{Co,s} = \frac{2u\Omega_{\zeta,E}}{\Omega_s} (\nabla R \times \zun)_\bot \simeq \frac{2uR\Omega_{\zeta,E}}{\Omega_s} \bun \times \kappabf
\end{equation}
is the Coriolis drift, and
\begin{eqnarray} \label{eq:F2gradtbrotating}
\fl \underline{F}_{s2, \grad}^{R,\tb} = - f_{Ms} \Bigg [ \frac{Z_s e}{T_s} \bv_{Ms} \cdot \Bigg ( \nabla_\bR \psi \left (- i k_\alpha \frac{\partial \Omega_{\zeta, E}}{\partial \psi} t + \frac{\partial}{\partial \psi} \right )+ \nabla_\bR \theta \frac{\partial}{\partial \theta} \Bigg ) + \Bigg ( \frac{\partial}{\partial \psi} \ln n_s  \nonumber\\ + \left ( \frac{m_s (u^2 + 2\mu B)}{2T_s} - \frac{3}{2} \right ) \frac{\partial}{\partial \psi} \ln T_s \Bigg ) \frac{c}{B} (\nabla_\bR \psi \times \bun) \cdot \nabla_\bR \theta \frac{\partial}{\partial \theta} \Bigg ] \underline{\phi}_1^{R, \tb} \nonumber\\ - \bv_{Ms} \cdot \left ( \nabla_\bR \psi \left (- i k_\alpha \frac{\partial \Omega_{\zeta,E}}{\partial \psi} t + \frac{\partial}{\partial \psi} \right ) + \nabla_\bR \theta \frac{\partial}{\partial \theta} \right ) \underline{f}_{s1}^{R, \tb} \nonumber\\ + \sum_{k_\psi^\prime, k_\alpha^\prime} \Bigg [ \frac{c (\underline{\phi}_1^{R,\tb})^\prime}{B} i (\bk^\prime_\bot \times \bun) \cdot \Bigg ( \nabla_\bR \psi \left (- i k_\alpha^{\prime\prime} \frac{\partial \Omega_{\zeta,E}}{\partial \psi} t + \frac{\partial}{\partial \psi} \right ) \nonumber\\ + \nabla_\bR \theta \frac{\partial}{\partial \theta} \Bigg ) (\underline{f}_{s1}^{R, \tb})^{\prime\prime} - \frac{c (\underline{f}_{s1}^{R,\tb})^\prime}{B} i (\bk_\bot^\prime \times \bun) \cdot \Bigg ( \nabla_\bR \psi \Bigg (- i k_\alpha^{\prime\prime} \frac{\partial \Omega_{\zeta,E}}{\partial \psi} t \nonumber\\ + \frac{\partial}{\partial \psi} \Bigg )  + \nabla_\bR \theta \frac{\partial}{\partial \theta} \Bigg ) (\underline{\phi}_1^{R,\tb})^{\prime\prime} \Bigg ].
\end{eqnarray}
Here a prime on a Fourier coefficient indicates that it depends on $k_\psi^\prime$ and $k_\alpha^\prime$, and two primes that it depends on $k_\psi^{\prime\prime} = k_\psi - k_\psi^\prime$ and $k_\alpha^{\prime\prime} = k_\alpha - k_\alpha^\prime$. The functions $\Delta \underline{n}_{s2, \grad}^\tb$ and $\Delta \underline{n}_{s2, \acc}^\tb$ are zero, and $\Delta \underline{n}_{s2, \nc}^\tb$ is given by \eq{eq:Deltan2nctbtotal} with $\underline{\phi}_1^\tb$ and $f_{s1}^\lw$ replaced by $\underline{\phi}_1^{R, \tb}$ and $f_{s1}^{R, \lw}$. To evaluate $\underline{F}_{s2, \grad}^{R,\tb}$, we need $\partial \underline{f}_{s1}^{R,\tb}/\partial \psi$ and $\partial \underline{\phi}_1^{R,\tb}/\partial \psi$. These derivatives can be calculated substituting
\begin{equation}
\frac{\partial \underline{f}_{s1}^\tb}{\partial \psi} = \left ( \frac{\partial \underline{f}_{s1}^{R,\tb}}{\partial \psi} + i k_\alpha c \frac{\partial^2 \phi_0}{\partial \psi^2} t\, \underline{f}_{s1}^{R,\tb} \right ) \exp \left ( i k_\alpha c \frac{\partial \phi_0}{\partial \psi} t \right )
\end{equation}
and
\begin{equation}
\frac{\partial \underline{\phi}_1^\tb}{\partial \psi} = \left ( \frac{\partial \underline{\phi}_1^{R,\tb}}{\partial \psi} + i k_\alpha c \frac{\partial^2 \phi_0}{\partial \psi^2} t\, \underline{\phi}_1^{R,\tb} \right ) \exp \left ( i k_\alpha c \frac{\partial \phi_0}{\partial \psi} t \right )
\end{equation}
into \eq{eq:radialderivativegktotal} and \eq{eq:radialderivativequasineutralitytotal}. We also need to use \eq{eq:tbfirstordergkrotating} and \eq{eq:tbfirstorderquasineutralityrotating} in the limit $B_p/B \ll 1$ with $k_\bot \rho_i \sim B_p/B$ to cancel several terms. The final equations are
\begin{eqnarray} \label{eq:radialderivativegkrotating}
\fl \frac{\partial}{\partial t} \left ( \frac{\partial \underline{f}_{s1}^{R,\tb}}{\partial \psi} \right )+ \left ( u \bun \cdot \nabla_\bR \theta \frac{\partial}{\partial \theta} - \mu \bun \cdot \nabla_\bR B \frac{\partial}{\partial u} \right ) \frac{\partial \underline{f}_{s1}^{R, \tb}}{\partial \psi} + i \bk_\bot \cdot \bv_{Ms} \frac{\partial \underline{f}_{s1}^{R,\tb}}{\partial \psi} \nonumber\\ - \sum_{s^\prime} C_{ss^\prime}^{(\ell)} \left [  \frac{\partial \underline{f}_{s1}^{R,\tb}}{\partial \psi}; \frac{\partial \underline{f}_{s1}^{R,\tb}}{\partial \psi} \right ] + \left \{ \frac{\partial \underline{\phi}_1^{R,\tb}}{\partial \psi}, \underline{f}_{s1}^{R,\tb} \right \} + \left \{ \underline{\phi}_1^{R,\tb} , \frac{\partial \underline{f}_{s1}^{R,\tb}}{\partial \psi} \right \}\nonumber\\ = - f_{Ms} \Bigg [ \frac{Z_s e}{T_s} \left ( u \bun \cdot \nabla_\bR \theta \frac{\partial}{\partial \theta} + i \bk_\bot \cdot \bv_{Ms} \right ) \nonumber\\ + i k_\alpha c \Bigg ( \frac{\partial}{\partial \psi} \ln n_s  + \left ( \frac{m_s (u^2 + 2\mu B)}{2T_s} - \frac{3}{2} \right ) \frac{\partial}{\partial \psi} \ln T_s \Bigg ) \Bigg ] \frac{\partial \underline{\phi}_1^{R,\tb}}{\partial \psi} \nonumber\\ - u \frac{\partial}{\partial \psi} ( \bun \cdot \nabla_\bR \theta ) \frac{\partial \underline{f}_{s1}^{R,\tb}}{\partial \theta} + \mu \frac{\partial}{\partial \psi}( \bun \cdot \nabla_\bR B ) \frac{\partial \underline{f}_{s1}^{R,\tb}}{\partial u} \nonumber\\ - \underline{f}_{s1}^{R,\tb} \frac{\partial}{\partial \psi} \left ( i \bk_\bot \cdot \bv_{Ms} \right ) + \sum_{s^\prime} \frac{\partial C_{ss^\prime}^{(\ell)}}{\partial \psi} \left [  \underline{f}_{s1}^{R,\tb}; \underline{f}_{s1}^{R,\tb} \right ] \nonumber\\ - \frac{\partial }{\partial \psi} \left ( \frac{Z_s e f_{Ms}}{T_s} u \bun \cdot \nabla_\bR \theta \right ) \frac{\partial \underline{\phi}_1^{R,\tb}}{\partial \theta} \nonumber\\- \underline{\phi}_1^{R,\tb} \frac{\partial }{\partial \psi} \Bigg [ \frac{Z_s e f_{Ms}}{T_s} i \bk_\bot \cdot \bv_{Ms} + i k_\alpha c f_{Ms} \Bigg ( \frac{\partial}{\partial \psi} \ln n_s \nonumber\\ + \left ( \frac{m_s (u^2 + 2\mu B)}{2T_s} - \frac{3}{2} \right ) \frac{\partial}{\partial \psi} \ln T_s \Bigg )  \Bigg ]
\end{eqnarray}
and quasineutrality equation \eq{eq:radialderivativequasineutralitytotal} with $\partial \underline{f}_{s1}^\tb/\partial \psi$ and $\underline{f}_{s1}^\tb$ replaced by $\partial \underline{f}_{s1}^{R,\tb}/\partial \psi$ and $\underline{f}_{s1}^{R,\tb}$.

Note that $\underline{F}_{s2, \grad}^{R,\tb}$ contains terms linear in time, such as $i k_\alpha (\partial \Omega_\zeta/\partial \psi) t \, \bv_{Ms} \cdot \nabla_\bR \psi \underline{f}_{s1}^{R,\tb}$. In gyrokinetic equations derived in the frame rotating with $\Omega_{\zeta, E}$, these terms correspond to the use of the perpendicular wavevector 
\begin{equation} \label{eq:kbotrotating}
\bk_\bot = \left ( k_\psi - k_\alpha \frac{\partial \Omega_{\zeta, E}}{\partial \psi} t \right ) \nabla \psi + k_\alpha \nabla \alpha
\end{equation}
instead of the wavevector $\bk_\bot$ defined \eq{eq:kbotdef}. Equation \eq{eq:kbotrotating} is a convenient representation of the shearing of turbulent structures due to the background $\bE \times \bB$ velocity. In our formulation, these terms are second order in $\rho_\ast$ because we have ordered $R \Omega_{\zeta, E} \sim (B/B_p) \rho_\ast v_{ti}$. We do not need to consider the time dependent term in our finite gyroradius terms because due to our ordering, the background $\bE \times \bB$ shear is negligible unless the turbulent eddies are of the order of the ion poloidal gyroradius, in which case finite gyroradius effects are unimportant.

\subsection{Momentum flux} \label{sub:momentumfluxrotating}
Using \eq{eq:fs1lwprime}, \eq{eq:fs2lwprime}, \eq{eq:phi1lwprime}, \eq{eq:phi2lwprime} and \eq{eq:phi1tbprime} - \eq{eq:fs2tbprime} in the expressions for the momentum flux given in subsection \ref{sub:momentumfluxtotal}, we find
\begin{eqnarray} \label{eq:Pirotating}
\fl \Pi = \Pi_{-1, \ud}^{R,\tb} + \Pi_{-1,\ud}^{R,\nc} + \Pi^{R,\tb}_{0, \nc} + \Pi^{R,\tb}_{0, \grad} + \Pi^{R,\tb}_{0,\acc} + \Pi_0^{R,\nc} \nonumber\\ +\Pi_{0, \tb}^{R,\FOW} + \Pi_{0, \Delta T}^{R,\FOW} + \Pi_{0, Q}^{R,\FOW}  + \Omega_\zeta (\Gamma_\zeta^\tb + \Gamma_\zeta^\nc).
\end{eqnarray}
The pieces $\Gamma_\zeta^\tb$ and $\Gamma_\zeta^\nc$ give the flux of angular momentum due to particle flux of turbulent and neoclassical origin, and they are defined in \eq{eq:Gammamtb} and \eq{eq:Gammamnc}. To calculate $\Gamma_\zeta^\tb$ and $\Gamma_\zeta^\nc$ in the rotating frame, we just need to replace $f_{s1}^\lw$, $\phi_1^\lw$, $\underline{f}_{s1}^\tb$ and $\underline{\phi}_1^\tb$ by $f_{s1}^{R,\lw}$, $\phi_1^{R,\lw}$, $\underline{f}_{s1}^{R,\tb}$ and $\underline{\phi}_1^{R,\tb}$ in \eq{eq:Gammamtb} and \eq{eq:Gammamnc}. Similarly, the other pieces of the momentum flux in \eq{eq:Pirotating} can be obtained from the corresponding expressions \eq{eq:Piminusudtbtotal}, \eq{eq:Piminusudnctotal}, \eq{eq:Pi0nctotal}, \eq{eq:Pi0betatbtotal} and \eq{eq:Pi0tbFOWtotal} - \eq{eq:Pi0QFOWtotal} by replacing $f_{s1}^\lw$, $\phi_1^\lw$, $\underline{f}_{s1}^\tb$, $\underline{\phi}_1^\tb$, $h_{s2, \beta}^\lw$, $F_{s2, \beta}^\lw$, $\underline{f}_{s2}^\tb$ and $\underline{\phi}_2^\tb$ by $f_{s1}^{R,\lw}$, $\phi_1^{R,\lw}$, $\underline{f}_{s1}^{R, \tb}$, $\underline{\phi}_1^{R,\tb}$, $h_{s2, \beta}^{R,\lw}$, $F_{s2, \beta}^{R,\lw}$, $\underline{f}_{s2}^{R,\tb}$ and $\underline{\phi}_2^{R,\tb}$, respectively.

\section*{References}

\end{document}